\RequirePackage[mathlines,displaymath]{lineno}
\documentclass[aps,prd,twocolumn,showpacs,amsmath,superscriptaddress,groupedaddress]{revtex4}  
\usepackage{graphicx}  
\usepackage{dcolumn}   
\usepackage{bm}        
\usepackage{amssymb}   

\usepackage{mathptmx}
\usepackage{lineno}
\usepackage{hyperref}
\usepackage{xspace}
\usepackage{multirow}
\usepackage{epsfig}
\usepackage{psfrag}
\usepackage{lscape}
\usepackage{comment}
\usepackage[normalem]{ulem}
\usepackage{color}
\usepackage{overpic}
\usepackage{textcomp}
\usepackage{afterpage} 
\usepackage{longtable}

\newcommand{\dzero}{D0\xspace}

\newcommand{\ppbar}{\ensuremath{p\bar{p}}\xspace}
\newcommand{\ttbar}{\ensuremath{t\bar{t}}\xspace}
\newcommand{\etmiss}{\ensuremath{E \kern-0.6em\slash_{\rm T}}\xspace}
\newcommand{\etmissx}{\ensuremath{E \kern-0.6em\slash_{\rm x}}\xspace}
\newcommand{\etmissy}{\ensuremath{E \kern-0.6em\slash_{\rm y}}\xspace}

\newcommand{\alpgen}{{\sc alpgen}\xspace}
\newcommand{\apythia}{{\sc alpgen+pythia}\xspace}
\newcommand{\aherwig}{{\sc alpgen+herwig}\xspace}
\newcommand{\mcnlo}{{\sc mc@nlo}\xspace}
\newcommand{\powheg}{{\sc powheg}\xspace}
\newcommand{\pythia}{{\sc pythia}\xspace}
\newcommand{\mcfm}{{\sc mcfm}\xspace}
\newcommand{\herwig}{{\sc herwig}\xspace}
\newcommand{\vecbos}{{\sc vecbos}\xspace}

\newcommand{\ejets}{\ensuremath{e\!+\!{\rm jets}}\xspace}
\newcommand{\mujets}{\ensuremath{\mu\!+\!{\rm jets}}\xspace}
\newcommand{\ljets}{\ensuremath{\ell\!+\!{\rm jets}}\xspace}
\newcommand{\wjets}{\ensuremath{W\!+\!{\rm jets}}\xspace}
\newcommand{\zjets}{\ensuremath{Z\!+\!{\rm jets}}\xspace}

\newcommand{\pevt}{\ensuremath{P_{\rm evt}}\xspace}
\newcommand{\psig}{\ensuremath{P_{\rm sig}}\xspace}
\newcommand{\pbkg}{\ensuremath{P_{\rm bkg}}\xspace}

\newcommand{\etj}{\ensuremath{E_T^j}\xspace}

\newcommand{\x}{\ensuremath{\vec x}\xspace}
\newcommand{\X}{\ensuremath{\mathbb{X}}\xspace}
\newcommand{\y}{\ensuremath{\vec y}\xspace}

\newcommand{\eps}{\varepsilon}

\newcommand{\GeV}{\ensuremath{\textnormal{GeV}}\xspace}

\newcommand{\dif}{\ensuremath{{\rm d}}}

\newcommand{\met}{\ensuremath{p\!\!\!\!/_T}\xspace}

\newcommand{\pb}{\ensuremath{{\rm pb}^{-1}}\xspace}
\newcommand{\fb}{\ensuremath{{\rm fb}^{-1}}\xspace}
\newcommand{\mw}{\ensuremath{M_W}\xspace}
\newcommand{\mt}{\ensuremath{m_t}\xspace}
\newcommand{\mtfit}{\ensuremath{m_t^{\rm fit}}\xspace}
\newcommand{\dmtfit}{\ensuremath{\sigma_{m_t}^{\rm fit}}\xspace}
\newcommand{\dmt}{\ensuremath{\sigma_{m_t}}\xspace}
\newcommand{\mtgen}{\ensuremath{m_t^{\rm gen}}\xspace}
\newcommand{\kjes}{\ensuremath{k_{\rm JES}}\xspace}
\newcommand{\kjesfit}{\ensuremath{k_{\rm JES}^{\rm fit}}\xspace}
\newcommand{\dkjesfit}{\ensuremath{\sigma_{k_{\rm JES}}^{\rm fit}}\xspace}
\newcommand{\dkjes}{\ensuremath{\sigma_{k_{\rm JES}}}\xspace}
\newcommand{\kjesgen}{\ensuremath{k_{\rm JES}^{\rm gen}}\xspace}
\newcommand{\fmax}{\ensuremath{\hat f}\xspace}
\newcommand{\ffit}{\ensuremath{f^{\rm fit}}\xspace}
\newcommand{\fgen}{\ensuremath{f^{\rm gen}}\xspace}
\newcommand{\wjet}{\ensuremath{W_{\rm jet}}\xspace}

\newcommand{\lumi}{\ensuremath{\mathcal{L}}\xspace}

\newcommand{\pt}{\ensuremath{p_T}\xspace}

\newcommand{\like}{\ensuremath{{\cal L}}\xspace}

\newcommand{\stt}{\ensuremath{\sigma_{t\bar t}}\xspace}

\begin{document}

\hspace{5.2in} \mbox{FERMILAB-PUB-15-021-E}

\title{
Precision measurement of the top-quark mass in lepton$+$jets final states
}
\affiliation{LAFEX, Centro Brasileiro de Pesquisas F\'{i}sicas, Rio de Janeiro, Brazil}
\affiliation{Universidade do Estado do Rio de Janeiro, Rio de Janeiro, Brazil}
\affiliation{Universidade Federal do ABC, Santo Andr\'e, Brazil}
\affiliation{University of Science and Technology of China, Hefei, People's Republic of China}
\affiliation{Universidad de los Andes, Bogot\'a, Colombia}
\affiliation{Charles University, Faculty of Mathematics and Physics, Center for Particle Physics, Prague, Czech Republic}
\affiliation{Czech Technical University in Prague, Prague, Czech Republic}
\affiliation{Institute of Physics, Academy of Sciences of the Czech Republic, Prague, Czech Republic}
\affiliation{Universidad San Francisco de Quito, Quito, Ecuador}
\affiliation{LPC, Universit\'e Blaise Pascal, CNRS/IN2P3, Clermont, France}
\affiliation{LPSC, Universit\'e Joseph Fourier Grenoble 1, CNRS/IN2P3, Institut National Polytechnique de Grenoble, Grenoble, France}
\affiliation{CPPM, Aix-Marseille Universit\'e, CNRS/IN2P3, Marseille, France}
\affiliation{LAL, Universit\'e Paris-Sud, CNRS/IN2P3, Orsay, France}
\affiliation{LPNHE, Universit\'es Paris VI and VII, CNRS/IN2P3, Paris, France}
\affiliation{CEA, Irfu, SPP, Saclay, France}
\affiliation{IPHC, Universit\'e de Strasbourg, CNRS/IN2P3, Strasbourg, France}
\affiliation{IPNL, Universit\'e Lyon 1, CNRS/IN2P3, Villeurbanne, France and Universit\'e de Lyon, Lyon, France}
\affiliation{III. Physikalisches Institut A, RWTH Aachen University, Aachen, Germany}
\affiliation{Physikalisches Institut, Universit\"at Freiburg, Freiburg, Germany}
\affiliation{II. Physikalisches Institut, Georg-August-Universit\"at G\"ottingen, G\"ottingen, Germany}
\affiliation{Institut f\"ur Physik, Universit\"at Mainz, Mainz, Germany}
\affiliation{Ludwig-Maximilians-Universit\"at M\"unchen, M\"unchen, Germany}
\affiliation{Panjab University, Chandigarh, India}
\affiliation{Delhi University, Delhi, India}
\affiliation{Tata Institute of Fundamental Research, Mumbai, India}
\affiliation{University College Dublin, Dublin, Ireland}
\affiliation{Korea Detector Laboratory, Korea University, Seoul, Korea}
\affiliation{CINVESTAV, Mexico City, Mexico}
\affiliation{Nikhef, Science Park, Amsterdam, the Netherlands}
\affiliation{Radboud University Nijmegen, Nijmegen, the Netherlands}
\affiliation{Joint Institute for Nuclear Research, Dubna, Russia}
\affiliation{Institute for Theoretical and Experimental Physics, Moscow, Russia}
\affiliation{Moscow State University, Moscow, Russia}
\affiliation{Institute for High Energy Physics, Protvino, Russia}
\affiliation{Petersburg Nuclear Physics Institute, St. Petersburg, Russia}
\affiliation{Instituci\'{o} Catalana de Recerca i Estudis Avan\c{c}ats (ICREA) and Institut de F\'{i}sica d'Altes Energies (IFAE), Barcelona, Spain}
\affiliation{Uppsala University, Uppsala, Sweden}
\affiliation{Taras Shevchenko National University of Kyiv, Kiev, Ukraine}
\affiliation{Lancaster University, Lancaster LA1 4YB, United Kingdom}
\affiliation{Imperial College London, London SW7 2AZ, United Kingdom}
\affiliation{The University of Manchester, Manchester M13 9PL, United Kingdom}
\affiliation{University of Arizona, Tucson, Arizona 85721, USA}
\affiliation{University of California Riverside, Riverside, California 92521, USA}
\affiliation{Florida State University, Tallahassee, Florida 32306, USA}
\affiliation{Fermi National Accelerator Laboratory, Batavia, Illinois 60510, USA}
\affiliation{University of Illinois at Chicago, Chicago, Illinois 60607, USA}
\affiliation{Northern Illinois University, DeKalb, Illinois 60115, USA}
\affiliation{Northwestern University, Evanston, Illinois 60208, USA}
\affiliation{Indiana University, Bloomington, Indiana 47405, USA}
\affiliation{Purdue University Calumet, Hammond, Indiana 46323, USA}
\affiliation{University of Notre Dame, Notre Dame, Indiana 46556, USA}
\affiliation{Iowa State University, Ames, Iowa 50011, USA}
\affiliation{University of Kansas, Lawrence, Kansas 66045, USA}
\affiliation{Louisiana Tech University, Ruston, Louisiana 71272, USA}
\affiliation{Northeastern University, Boston, Massachusetts 02115, USA}
\affiliation{University of Michigan, Ann Arbor, Michigan 48109, USA}
\affiliation{Michigan State University, East Lansing, Michigan 48824, USA}
\affiliation{University of Mississippi, University, Mississippi 38677, USA}
\affiliation{University of Nebraska, Lincoln, Nebraska 68588, USA}
\affiliation{Rutgers University, Piscataway, New Jersey 08855, USA}
\affiliation{Princeton University, Princeton, New Jersey 08544, USA}
\affiliation{State University of New York, Buffalo, New York 14260, USA}
\affiliation{University of Rochester, Rochester, New York 14627, USA}
\affiliation{State University of New York, Stony Brook, New York 11794, USA}
\affiliation{Brookhaven National Laboratory, Upton, New York 11973, USA}
\affiliation{Langston University, Langston, Oklahoma 73050, USA}
\affiliation{University of Oklahoma, Norman, Oklahoma 73019, USA}
\affiliation{Oklahoma State University, Stillwater, Oklahoma 74078, USA}
\affiliation{Brown University, Providence, Rhode Island 02912, USA}
\affiliation{University of Texas, Arlington, Texas 76019, USA}
\affiliation{Southern Methodist University, Dallas, Texas 75275, USA}
\affiliation{Rice University, Houston, Texas 77005, USA}
\affiliation{University of Virginia, Charlottesville, Virginia 22904, USA}
\affiliation{University of Washington, Seattle, Washington 98195, USA}
\author{V.M.~Abazov} \affiliation{Joint Institute for Nuclear Research, Dubna, Russia}
\author{B.~Abbott} \affiliation{University of Oklahoma, Norman, Oklahoma 73019, USA}
\author{B.S.~Acharya} \affiliation{Tata Institute of Fundamental Research, Mumbai, India}
\author{M.~Adams} \affiliation{University of Illinois at Chicago, Chicago, Illinois 60607, USA}
\author{T.~Adams} \affiliation{Florida State University, Tallahassee, Florida 32306, USA}
\author{J.P.~Agnew} \affiliation{The University of Manchester, Manchester M13 9PL, United Kingdom}
\author{G.D.~Alexeev} \affiliation{Joint Institute for Nuclear Research, Dubna, Russia}
\author{G.~Alkhazov} \affiliation{Petersburg Nuclear Physics Institute, St. Petersburg, Russia}
\author{A.~Alton$^{a}$} \affiliation{University of Michigan, Ann Arbor, Michigan 48109, USA}
\author{A.~Askew} \affiliation{Florida State University, Tallahassee, Florida 32306, USA}
% ...
%
%
\collaboration{The D0 Collaboration\footnote{with visitors from
$^{a}$Augustana College, Sioux Falls, SD, USA,
$^{b}$The University of Liverpool, Liverpool, UK,
$^{c}$DESY, Hamburg, Germany,
$^{d}$Universidad Michoacana de San Nicolas de Hidalgo, Morelia, Mexico
$^{e}$SLAC, Menlo Park, CA, USA,
$^{f}$University College London, London, UK,
$^{g}$Centro de Investigacion en Computacion - IPN, Mexico City, Mexico,
$^{h}$Universidade Estadual Paulista, S\~ao Paulo, Brazil,
$^{i}$Karlsruher Institut f\"ur Technologie (KIT) - Steinbuch Centre for Computing (SCC),
D-76128 Karlsruhe, Germany,
$^{j}$Office of Science, U.S. Department of Energy, Washington, D.C. 20585, USA,
$^{k}$American Association for the Advancement of Science, Washington, D.C. 20005, USA,
$^{l}$Kiev Institute for Nuclear Research, Kiev, Ukraine,
$^{m}$University of Maryland, College Park, Maryland 20742, USA
and
$^{n}$European Orgnaization for Nuclear Research (CERN), Geneva, Switzerland
}} \noaffiliation
\vskip 0.25cm
\date{January 26, 2015}

\begin{abstract}
We measure the mass of the top quark in lepton$+$jets final states using the full sample of $p\bar{p}$ collision data collected by the D0 experiment in Run~II of the Fermilab Tevatron Collider at \mbox{$\sqrt s=1.96~$TeV}, corresponding to $9.7~\fb$ of integrated luminosity. We use a matrix element technique that calculates the probabilities for each event to result from $\ttbar$ production or background. The overall jet energy scale is constrained {\it in situ} by the mass of the $W$ boson. We measure 
$m_t=174.98\pm0.76~\GeV$. 
This constitutes the most precise single measurement of the top-quark mass.
\end{abstract}
\pacs{14.65.Ha}
\maketitle


\section{Introduction}

Since the discovery of the top quark~\cite{bib:discoverydzero,bib:discoverycdf}, the determination of its properties has been one of the main goals of the Fermilab Tevatron Collider, and more recently also of the CERN Large Hadron Collider (LHC). The measurement of the top-quark mass \mt, a fundamental parameter of the standard model (SM), has received particular attention, since \mt, the mass of the $W$ boson, \mw, and the mass of the Higgs boson are related through radiative corrections that provide an internal consistency check of the SM~\cite{bib:lepewwg}. Furthermore, \mt dominantly affects the stability of the SM Higgs potential, which has cosmological implications~\cite{bib:vstab1,bib:vstab2,bib:vstab3}.

The world-average $\mt=173.34\pm0.76~\GeV$ has reached a precision of about 0.5\%~\cite{bib:combitevprd,bib:combitev,bib:combiworld}, which has been limited for some time by systematic uncertainties. The main systematic uncertainties arise from the calibration of the jet energy scale (JES) and the modeling of \ttbar events in Monte Carlo (MC) simulations.

In this manuscript, we present a measurement of \mt based on the full set of $p\bar p$ data at $\sqrt s=1.96~$TeV recorded by the D0 detector in Run~II of the Fermilab Tevatron Collider, corresponding to an integrated luminosity of $9.7~\fb$. The analysis  focuses on \ttbar events identified in \ljets final states with $\ell$ representing either an electron or a muon. This includes contributions from leptonic $\tau\to\ell\nu_\ell\nu_\tau$ decays. The top and antitop quark are assumed to always decay into a $W$ boson and a $b$~quark~\cite{bib:topdecay}.
Thus, at leading order (LO) in perturbative quantum chromodynamics (QCD), one of the $W$ bosons in the $W^{+}W^{-}b\bar{b}$ final state decays into a charged lepton and a neutrino, while the other decays into a pair of quarks, and all four quarks ($q\bar{q}^{\prime}b\bar{b}$) evolve into jets. Such \ljets events are characterized by an isolated electron or muon with large transverse momentum $\pt$,  a large imbalance in transverse momentum caused by the undetected neutrino, and four high-$\pt$ jets. We exploit this distinct signature in the event selection to discriminate \ttbar events from background.

To extract the value of \mt, we utilize a matrix element (ME) technique that determines the probability of observing an event under both the $\ttbar$ signal and background hypotheses described by the relevant ME~\cite{bib:mtopnature}. The overall JES is calibrated {\it in situ} through a constant factor \kjes by taking into account all the kinematic information in a given event, while constraining the reconstructed invariant mass of the $W\to q\bar q'$ decay products to $\mw=80.4$~GeV~\cite{bib:wmass}.

We provide a detailed description of the measurement that was presented previously in a Letter~\cite{bib:prl}, and is an update and supersedes a previous D0 measurement of $\mt=174.94\pm1.14\thinspace({\rm stat+JES})\pm0.96\thinspace({\rm syst})~\GeV$~\cite{bib:mt36} using 3.6~\fb of integrated luminosity. In the present measurement, we use not only a larger data sample to improve the statistical precision, but also refine the estimation of systematic uncertainties, resulting in a reduction of the uncertainty by 50\%. This improvement is achieved through an updated detector calibration, in particular through improvements to $b$~quark JES corrections~\cite{bib:jes}, and through the use of recent improvements in modeling the \ttbar signal. In the following, we point out the most important improvements of this analysis with respect to the previous measurement. The analysis is performed ``blinded'', i.e., with no reference to the resulting \mt, until the final approval of the analysis methodology.

As in all direct measurements of \mt, this analysis relies on simulated \ttbar MC events, which are used for the absolute calibration of the response of the ME technique. Therefore, the extracted \mt corresponds to the generated $\mt^{\rm gen}$ parameter implemented in the LO MC event generator used to simulate \ttbar events, which in our case is \alpgen~\cite{bib:alpgen}. Although $\mt^{\rm gen}$ is not well defined at LO, it is expected to correspond within $\approx1~\GeV$ to the \mt defined in the pole-mass scheme~\cite{bib:pdg}. There is currently no universal MC event generator available for simulating \ttbar events at next-to-leading order (NLO) that includes the effects of parton showering and hadronization. Programs such as \mcnlo~\cite{bib:mcnlo} or \powheg~\cite{bib:powheg}, commonly referred to as NLO generators, simulate only the $q\bar q\to\ttbar$, $gg\to\ttbar$, and $qg\to\ttbar$ processes at NLO, whereas the decays of the top quark are simulated at LO through the narrow-width approximation.

This manuscript is arranged as follows. A brief description of the D0 detector is
given in Section~\ref{sec:detector}, which is followed in Section~\ref{sec:mc} by a summary of simulations used to model signal and background events. The selection of \ttbar candidate events is reviewed and the sample composition presented in Section~\ref{sec:selection}. Section~\ref{sec:method} discusses the ME technique used to extract \mt, with particular emphasis placed on the improvements since the previous publication~\cite{bib:mt36}. This is followed in Section~\ref{sec:f} by the description of the measurement of the signal fraction $f$ in data using the ME technique, including its calibration. The measurement of \mt and the calibration of the response of the ME technique in \mt is discussed in Section~\ref{sec:mt}. After describing the evaluation of systematic uncertainties in Section~\ref{sec:syst}, we compare our measurement with previous results in Section~\ref{sec:comparison}, and conclude and give the final result with statistical and systematic uncertainties in Section~\ref{sec:summary}. Appendix~\ref{app:isr} discusses the issue of initial and final state radiation, and Appendix~\ref{app:tf} summarizes the parametrization of the detector response through transfer functions.

\section{
The D0 detector
}
\label{sec:detector}
The D0 detector~\cite{bib:run2det,bib:run1det} contains a magnetic central tracking system,
calorimetry, and a muon system.  The central tracking system comprises a
silicon microstrip tracker (SMT) and a central fiber  tracker (CFT), both
located within a 2~T superconducting solenoidal  magnet. The
SMT~\cite{bib:run2smt} has $\approx 800\thinspace000$ individual strips,  with typical pitch
of $50-80$ $\mu$m, and a design optimized for track and vertex finding within
$|\eta|<2.5$~\cite{bib:coor}.
The system has a six-barrel longitudinal structure, each with a set  of four layers arranged axially around
the beam pipe, and interspersed  with radial disks.  In 2006, a fifth layer,
referred to as Layer~0, was installed close to the beam
pipe~\cite{bib:run2lyr0}.  The CFT has eight thin coaxial barrels, each 
supporting two doublets of overlapping scintillating fibers of 0.835~mm 
diameter, one doublet being parallel to the collision axis, and the  other
alternating by $\pm 3^{\circ}$ relative to the axis. Light signals  are
transferred via clear fibers to solid-state visible-light photon counters (VLPCs)  that have
$\approx 80$\% quantum efficiency.

Central and forward preshower detectors, located just outside of the
superconducting coil (in front of the calorimetry), are constructed of several
layers of extruded triangular scintillator strips that are read  out using
wavelength-shifting fibers and VLPCs.  These detectors provide initial
sampling of electromagnetic showers, and thereby help distinguish between incident
electrons and jets.  The next layer of  detection involves three
liquid-argon/uranium calorimeters: a central  section (CC) covering up
to  $|\eta|\approx 1.1$, and two end calorimeters (EC) that extend coverage to
$|\eta|\approx 4.2$, housed in separate cryostats.  The electromagnetic
(EM) section of the calorimeter is segmented into four layers, with transverse
segmentation of the cells in pseuodorapidity and azimuth of
$\Delta\eta\times\Delta\phi=0.1\times0.1$, except for the third layer, where the
segmentation is $0.05\times0.05$.  The hadronic portion of the calorimeter is
located after the EM sections and consists of fine hadron-sampling layers,
followed by more coarse hadronic layers.  In addition, scintillators between the
CC and EC cryostats provide  sampling of developing showers for
$1.1<|\eta|<1.4$.

A muon system~\cite{bib:run2muon} is located beyond the calorimetry, and consists of
a  layer of tracking detectors and scintillation trigger counters  before 1.8~T
iron toroid magnets, followed by two similar layers after the toroids. Tracking for
$|\eta|<1$ relies on 10~cm wide drift tubes, while 1~cm mini-drift tubes are
used for $1<|\eta|<2$.

Luminosity is measured using plastic scintillator arrays located in front  of
the EC cryostats, covering $2.7 < |\eta| < 4.4$.  The trigger and data
acquisition systems are designed to accommodate  the high instantaneous
luminosities of the Tevatron~\citep{bib:run2det,bib:run2l1cal}. Based on coarse
information from  tracking, calorimetry, and muon systems, the output of the
first level  of the trigger is used to limit the rate for accepted events to 
$\approx$ 2~kHz.  At the next trigger stage, with more refined  information, the
rate is reduced further to $\approx$ 1~kHz. These first two levels of triggering
rely mainly on hardware and firmware. The third and final level of the trigger,
with access to all of the event  information, uses software algorithms and a
computing farm, and reduces  the output rate to $\approx$ 100~Hz, which is
recorded.

\section{
The data sample and simulations
}
\label{sec:mc}

The sample of $p\bar p$ collision data considered in this analysis was collected between April 2002 and September 2011 in Run~II of the Fermilab Tevatron Collider. The sample is split into four data-taking epochs: ``Run~IIa'', ``Run~IIb1'', ``Run~IIb2'', and ``Run~IIb3'' corresponding to 1.1~\fb, 1.2~\fb, 3.0~\fb, and 4.4~\fb of integrated luminosity, respectively, after implementation of data-quality requirements. All MC simulations are split accordingly into epochs to model variations in the response of the detector, such as the insertion of Layer~0 into the SMT detector between Run~IIa and Run~IIb1~\cite{bib:run2lyr0}, and the dependence of the track reconstruction efficiency on instantaneous luminosity~\cite{bib:muid}.

Simulated MC events for  \ttbar signal and the dominant \wjets background, which are used to verify the agreement between data and simulations and to calibrate the measurement (described in Sections~\ref{sec:calibf} and~\ref{sec:calibmt}), are generated using the MC generator \alpgen~\cite{bib:alpgen} to simulate the hard-scattering process 
with up to five additional partons,
and \pythia~\cite{bib:pythia1,bib:pythia2} to simulate hadronization and shower evolution. The MLM matching scheme~\cite{bib:mlm} is employed to avoid overlaps between components of the event belonging to the hard process, implemented through a ME, and parton evolution (showering) into jets. The \wjets background samples are divided into two categories: (i) $W\!+\!\rm hf$ which represents $W\!+\!c\bar c$ and  $W\!+\!b\bar b$ production, possibly in association with $u$, $d$, $s$ quarks, or gluons, and (ii) $W\!+\!\rm lf$ which represents all other flavor configurations. The cross section for the \wjets contribution provided by \alpgen is multiplied by a factor of 1.3 to match the NLO calculations of the \mcfm MC generator~\cite{bib:mcfm}. In addition, the heavy-flavor contributions of category (i) are multiplied by a factor of 1.47, using the same calculations. 

We also consider other small background contributions on the order of few percent to improve the agreement between data and simulations. The \zjets background is simulated in the same way as \wjets with \alpgen, taking into account contributions from $Z\!+\!\rm hf$ production, with rates normalized to NLO calculations~\cite{bib:mcfm}. \pythia is used to simulate diboson ($WW$, $WZ$, or $ZZ$) production, while {\sc comphep+pythia}~\cite{bib:comphep} is used for single top quark processes that are simulated for $\mt=172.5~\GeV$.  The hard-scattering process for all signal and background MC samples is simulated using the CTEQ6L1 set of parton distribution functions (PDF), except for {\sc comphep}, which uses the NLO CTEQ6M set~\cite{bib:cteq}. More details on the MC simulations and their normalization can be found in Ref.~\cite{bib:diffxsec}.

The simulation of parton showers with \pythia uses tune~A in combination with the CTEQ6L1 PDF set and with $\Lambda_{\rm QCD}=0.26~\GeV$, also referred to as ``D0 tune A''. The detector response is fully simulated through {\sc geant3}~\cite{bib:geant}. To simulate the effects from additional \ppbar interactions (pileup), events taken with minimal trigger requirements, selected from random \ppbar crossings at the same instantaneous luminosity as the data, are overlaid on the fully simulated MC events. These events are then reconstructed using the same algorithms as used on data.

Events from multijet (MJ) production can pass our selection criteria, which happens typically when a jet mimics an electron or when a muon arising from semileptonic decay of a $b$ or $c$ quark appears isolated. The kinematic distributions from the MJ background are modeled using data events for which lepton isolation requirements are inverted. The absolute contribution of this background to each of the \ejets and \mujets final states is estimated using the ``matrix method,'' described in Ref.~\cite{bib:diffxsec}. This method uses the number of events with leptons originating from the hard interaction, $N^{\ttbar,W}_{\rm loose}$, and from MJ production, $N^{\rm MJ}_{\rm loose}$, both with less restrictive lepton-identification criteria, and relates them to the contributions to the sample of events that fulfil standard lepton identification criteria. This is calculated using the relation $N=\eps^{\ttbar,W}N^{\ttbar,W}_{\rm loose}+\eps^{\rm MJ}N^{\rm MJ}_{\rm loose}$, where $\eps^{\ttbar,W}$ and $\eps^{\rm MJ}$ represent the efficiency of events that pass the loosened lepton identification criteria to also pass the standard identification criteria. The parameters  $\eps^{\ttbar,W}$ and $\eps^{\rm MJ}$ are measured, respectively, in control regions dominated by leptons originating from the hard interaction and by MJ events.

\section{
Event selection and sample composition
}
\label{sec:selection}

The trigger selects \ljets events by requiring either at least one lepton (electron or muon), or at least one lepton and at least one jet, resulting in an efficiency of 95\% for \ttbar candidate events containing an electron and 80\% for events containing a muon, before any offline selections are applied.

Selected \ttbar candidate events are required to have at least one primary $p\bar p$ collision vertex (PV) with three or more tracks reconstructed within 60~cm of the center of the detector in the coordinate along the beam axis. 
In addition, there must be exactly four jets with $\pt > 20~\GeV$ within $|\eta|<2.5$, with the leading jet further satisfying $\pt > 40~\GeV$.
Jets are reconstructed using an iterative cone algorithm~\cite{bib:cone} with a cone parameter of $R_{\rm cone}=0.5$. Jet energies are corrected to the particle level using calibrations derived from exclusive $\gamma+$jet, $Z+$jet, and dijet events~\cite{bib:jes}. These calibrations account for differences in detector response to jets originating from a gluon, a $b$~quark, or $u,d,s,$ and $c$~quarks. 
Furthermore, we require the presence of one isolated electron~\cite{bib:emid} or muon~\cite{bib:muid} with transverse momentum $\pt>20~\GeV$ and $|\eta|<1.1$ or $|\eta|<2$, respectively. The electrons and muons are required to originate from within 1~cm of the PV in the coordinate along the beam axis and to be separated in $\Delta R\equiv\sqrt{(\Delta\eta)^2+(\Delta\phi)^2}$ from the closest jet by $\Delta R>0.5$, where $\Delta\eta$ is the difference in pseudorapidity and $\Delta\phi$ in azimuth (in rad) between the lepton and the jet. 
At least one neutrino is expected in the \ljets final state, and an imbalance in transverse momentum of $\met>20~\GeV$ is required. The \met variable is defined as the opposite of the vector sum of the transverse energies in each calorimeter cell, corrected for the energy carried by identified muons and after applying the jet energy scale calibration. The kinematic selections are summarized in Table~\ref{tab:selection}. 

Events containing an additional isolated charged lepton, i.e., an electron with $\pt>10$~GeV and $|\eta|<2.5$ or a muon with $\pt>10$~GeV and $|\eta|<2$, are rejected. In addition, any $\mu+$jets events with an invariant mass $M_{\mu\mu}$ of the isolated muon and a second {\em loosely} isolated muon with $\pt>15$~GeV within $70<M_{\mu\mu}<110$~GeV are rejected to suppress \zjets events.

\begin{table}
\caption{\label{tab:selection}
A summary of kinematic event selections.
}
\begin{ruledtabular}
\begin{tabular}{lll}
\multirow{2}{*}{One charged lepton}
                          & \multirow{2}{*}{$\pt > 20~\GeV$} & $|\eta| < 1.1$ ($e$)\\
                          &  & $|\eta| < 2.0$ ($\mu$)\\
Four jets            & $\pt > 20~\GeV$ & $|\eta| < 2.5$\\
Jet with highest \pt        & $\pt > 40~\GeV$ & $|\eta| < 2.5$\\
{Imbalance in transverse momentum}
                          & $\met > 20\,~\GeV$ \\
\end{tabular}
\end{ruledtabular}
\end{table}

To reject \mujets events where the momentum of the muon has been mismeasured, we require $\met<250~\GeV$ and $M_T^W<250~\GeV$, where $M_T^W$ is the transverse mass of the $W$ boson~\cite{bib:mtw}. To further reduce contributions from such events, which is found at the \% level, we use additional requirements on the significance of the curvature  of the muon track, $S_\kappa$. It is defined as the curvature $\kappa\equiv 1/\pt$ divided by its uncertainty, where $Q$ is the electric charge of the muon. We reject events with muons within one or both of the two regions where they are likely to be mismeasured, $S_\kappa < \Delta\phi(\mu,\met)\times25.5-70$ or $S_\kappa < \Delta\phi(\mu,\met)\times4.4+8.8$, where $\phi(\mu,\met)$ is the difference in azimuth between the muon momentum and the direction of \met.

To reduce the contribution from MJ production caused by the misidentification of a quark or gluon jet as a lepton, which changes the calculated \met, we require $\Delta\phi(\ell,\met)>0.5$. Additional reduction of the MJ background is achieved by requiring $\Delta\phi(e,\met) > 2.2-\met\times0.045~\GeV^{-1}$ in the \ejets final state and $\Delta\phi(\mu,\met) > 2.1 - \met\times0.035~\GeV^{-1}$ in the \mujets final state. 

Since two $b$-quark jets are expected in the final state, at least one jet per event is required to be tagged as originating from a $b$~quark ($b$-tagged) through the use of a multivariate algorithm~\cite{bib:bid}. The tagging efficiency in this analysis is $\approx65\%$ for $b$-quark jets, while the mistag rate for gluons and for $u,d,s$ quark jets is $\approx5\%$.

We select 1502 events in the \ejets final state in data, while 1286 are selected in the \mujets final state.
To verify that the simulations agree with the data following all selections, we use the NNLO calculation, including next-to-next-to-leading-logarithmic (NNLL) corrections~\cite{bib:xsec}, to normalize the \ttbar signal. The contribution from $\wjets$ events is adjusted to normalize the absolute yield from our signal and background models to data before applying $b$~jet identification criteria. The relative yields of contributions from $W\!+\!(b\bar b,n\rm lp)$ and $W\!+\!(c\bar c,n\rm lp)$ production and contributions from $W\!+\!n\rm lp$ production are further adjusted using data, as described in Ref.~\cite{bib:diffxsec}. The number of observed events is compared to expectations in Table~\ref{tab:prefit}, considering statistical uncertainties only.

\begin{table}
\caption{
\label{tab:prefit}
Comparison of the number of observed data events after all selections with the expectation from simulations, assuming $\sigma_{\ttbar}=7.24~\pb$~\cite{bib:xsec} and using $\mtgen=172.5~\GeV$. The category ``other backgrounds'' encompasses $\zjets$, $WW$, $WZ$, $ZZ$, and single top quark production, as well as dileptonic \ttbar decays.  The production of a $W$ boson in association with at least one $b$ or $c$ quark is denoted as ``$W+{\rm hf}$'', while ``$W+{\rm lf}$'' stands for all other flavor configurations. The uncertainties are purely statistical.
}
\begin{center} 
\begin{ruledtabular}
\begin{tabular}{l|rcr|rcr} 
Contribution&         \multicolumn{3}{c|}{\ejets} & \multicolumn{3}{c}{\mujets} \\ 
\hline                                                       
\ttbar&              $918.1$ & $\pm$ & $ 3.6$    & $824.9$ & $\pm$ & $ 3.5$     \\
Other backgrounds&   $97.8$ &  $\pm$ & $ 0.5$    & $79.2$ & $\pm$ & $ 0.9$      \\
$W+{\rm hf}$&        $126.0$ & $\pm$ & $ 2.1$    & $162.2$ & $\pm$ & $ 2.8$     \\
$W+{\rm lf}$&        $77.9$ & $\pm$ & $ 2.1$     & $101.0$ & $\pm$ & $ 2.9$     \\
Multijet&            $144.4$ & $\pm$ & $ 24.2$   & $48.2$ & $\pm$ & $ 16.1$     \\
\hline
Expected&            $1364.1$ & $\pm$ & $ 24.7$  & $1215.5$ & $\pm$ & $ 17.0$   \\ 
Observed&            \multicolumn{3}{c|}{$\mathbf{1502}$} & \multicolumn{3}{c}{$\mathbf{1286}$}  \\ 
\end{tabular} 
\end{ruledtabular}
 \end{center} 
\end{table}

The distribution in the invariant mass $m_{\ttbar}$ of the \ttbar system is shown in Fig.~\ref{fig:mtt}~(a) and~(b) in Section~\ref{sec:postfit}, together with predictions from our signal and background models. The distribution is generated with a mass of the top quark of $\mtgen=172.5~\GeV$.
The mass $m_{\ttbar}$ is constructed using \met as an initial estimate for $\pt^\nu$ of the neutrino, assuming $M_W=80.4~\GeV$, and solving the resulting quadratic equation for $p_{\rm z}^\nu$. When no solution exists, $\pt^\nu$ is scaled to obtain a solution. The resulting neutrino momentum is then combined with the momenta of the four jets and the charged lepton to obtain $m_{\ttbar}$. 
The invariant mass, $\mw$, of the two jets that match one of the $W$ bosons is presented in Fig.~\ref{fig:mwh}~(a) and~(b) in Section~\ref{sec:postfit}. The reconstruction of $M_W$ in Fig.~\ref{fig:mwh} is unique for events with two $b$-tagged jets, as the two other jets are assigned to the $W$ boson. If only one jet is $b$-tagged, all possible assignments of jets to partons in the LO approximation of \ttbar decays are tried, and the one yielding the smallest value of $|m_t-m_{\bar t}|$ is retained. If the $b$-tagged jet is matched to the top quark decaying into three jets, the two untagged jets matched to the same top quark form the $W$ boson. If all three jets that are matched to this top quark are not $b$-tagged, the two lowest $\pt$ jets are matched to the $W$ boson. A similar matching algorithm is used if there are more than two $b$-tagged jets.
The uncertainties in the predictions include the dominant sources of systematic uncertainty that change the shape of the kinematic distributions: JES, jet energy resolution, jet identification, flavor dependence of the detector response, $b$~tagging, lepton identification, lepton momentum scale, trigger, as well as hadronization and higher-order effects for simulated \ttbar events. Good agreement is observed between data and simulations. We verify that the agreement improves using the values for the signal fraction $f$, \mt, and \kjes  measured in this analysis, as shown in Section~\ref{sec:postfit}.


\section{
Analysis method
}
\label{sec:method}

The extraction of \mt is based on the full kinematic and topological information in the event, and is performed through a likelihood technique using probability densities (PDs) for each event, described by the ME of the processes contributing to the observed events.

\subsection{
Matrix element technique
}
\label{sec:me}

The ME technique was first suggested by Kondo~\cite{bib:kondo}, and was experimentally applied for the first time at the Tevatron in the measurement of \mt~\cite{bib:mtopnature}, and later in the determination of the helicity of the $W$ boson~\cite{bib:whel} and in the first evidence obtained for production of single top quarks~\cite{bib:singletopcdf,bib:singletopd0}. 

Assuming two non-interfering contributing processes, \ttbar and $W+{\rm jets}$ production, the per-event PD is:
\begin{linenomath}
\begin{eqnarray}
\pevt &=& A(\x)[ f\psig(\x; \mt,\kjes)\nonumber\\
      &+& (1-f)\pbkg(\x;\kjes) ]\,,\label{eq:pevt}
\end{eqnarray}
\end{linenomath}
where the observed parameters $f$, \mt, and 
$\kjes$, are to be determined from data, $\x$~\cite{bib:coor2} represents a vector of the measured four-momenta of the four jets and the charged lepton, but not \met due to its limited experimental resolution, and $A(\x)$ accounts for acceptance and efficiencies. The function \psig describes the PD for \ttbar production, while \pbkg describes the PD for $W+{\rm jets}$ production. $W+{\rm jets}$ events, i.e., the sum of $W+{\rm hf}$ and $W+{\rm lf}$ categories in Table~\ref{tab:prefit}, represent the dominant background contribution of 14\% in the \ejets and 20\% in the \mujets final states, derived with the normalization procedure described  in Section~\ref{sec:selection}. 
The next-to-dominant background contribution is from MJ events. It is accounted for in \pevt through \pbkg, since $W+{\rm jets}$ and MJ production have similar PDs in the selected kinematic region. The similarity of PDs for \wjets and MJ production can be understood in that the main difference between the two processes arises from the presence of the $W$ boson in the $W+{\rm jets}$ production, while an energetic jet which is misidentified as an isolated lepton is present in MJ production.  At the same time, both processes are characterized by the presence of four additional jets, which, together with the $W$ boson for $W+{\rm jets}$ production or the jet misidentified as an isolated lepton for MJ production, set a similar momentum-transfer scale of the hard interaction process. The effect of including MJ production through \pbkg is accounted for by the calibration procedure outlined in Sections~\ref{sec:f} and~\ref{sec:mt}. The combined contribution from all other backgrounds amounts to about 6\% in both final states (cf. Table~\ref{tab:prefit}), and is therefore not included in the calculation of \pevt.

\subsection{
The signal probability
}

\label{sec:psig}
The probability density for \ttbar production to yield a given set of partonic final-state four-momenta $\y$~\cite{bib:coor2} in the hard scattering of two massless quarks with four-momenta $q_1$ and $q_2$ is proportional to the differential cross section $\dif\sigma$ of the corresponding process, given by
\begin{linenomath}
\begin{equation}
\label{eq:dsigma}
\dif\sigma_{q\bar q\to\ttbar}(\y;\mt) = \frac{(2\pi)^{4}\left|{\cal M}_{q\bar q\to \ttbar}\right|^{2}}
                                             {2g_{\mu\nu}q^\mu_{1} q^\nu_{2}}
                                        \dif\Phi_{6}\,,
\end{equation}
\end{linenomath}
where ${\cal M}_{q\bar q\to \ttbar}$ denotes the ME for the $q\bar{q}\to t\bar{t}\to b(\ell\nu)\bar b(q\bar q')$ process, $g_{\mu\nu}$ is the tensor of the Minkowski metric, and $\dif\Phi_{6}$ is an infinitesimal element of six-body phase space.

To obtain the differential cross section in $p\bar p$ collisions, the expression in Eq.~(\ref{eq:dsigma}) is convoluted with the corresponding PD for all possible flavor combinations of the colliding quark and antiquark, yielding
\begin{linenomath}
\begin{eqnarray}
\label{eq:dsigmapp}
\dif\sigma_{p\bar p\to\ttbar}(\y;\mt) &=& \sum_{{}^{\rm quark}_{\rm flavors}}~\int\limits_{\vec q_{1},\,\vec q_{2}}\dif \vec q_{1}\dif \vec q_{2} f(\vec q_{1})f(\vec q_{2})\nonumber\\
&&\hspace{1.1cm}\times~~\dif\sigma_{q\bar q\to\ttbar}(\y;\,\mt)\,.
\end{eqnarray}
\end{linenomath}
The longitudinal momentum PDFs, $f(q_{i,{\rm z}})$, are taken from CTEQ6L1~\cite{bib:cteq}, while the dependence of $f(q_{i,{\rm x}})$ and $f(q_{i,{\rm y}})$ on transverse momentum is taken from the PD in the \pythia simulation. Only quark-antiquark annihilation at LO is taken into account in ${\cal M}_{q\bar q\to \ttbar}$, and the differential cross section $\dif\sigma_{p\bar p\to\ttbar}$ therefore does not represent the full differential cross section for $\ttbar$ production in $p\bar p$ collisions. Effects from gluon-gluon and quark-gluon induced \ttbar production, accounting for about 15\% of the cross section, as well as from higher-order corrections from real emissions, are accounted for in the calibration procedure described in Sections~\ref{sec:calibf} and~\ref{sec:calibmt}.

In general, the set $\x$ of measured four-momenta is not identical to the set of corresponding partonic variables $\y$ because of finite detector resolution and the evolution of quarks into jets. As will be discussed in Section~\ref{sec:tf}, this is taken into account through the  transfer function (TF) $W(\x,\y,\kjes)$, representing the probability density for the measured set \x to have arisen from the partonic set \y. Thus, the probability to observe a given reconstructed \ttbar event characterized by \x is obtained through a convolution with the TF in the calculation of the differential cross section
\begin{linenomath}
\begin{eqnarray}
\label{eq:dsigmax}
\dif\sigma_{p\bar p\to\ttbar}(\x;\,\mt,\kjes)
           &=& \int_{\vec y}\dif\vec y~\dif\sigma_{p\bar p\to\ttbar}(\y;\,\mt)\nonumber\\
           &\times& W(\x,\y;\,\kjes)\,.
\end{eqnarray}
\end{linenomath}

The PD to observe a \ttbar event with a set of kinematic quantities \x observed in the detector is given by
\begin{linenomath}
\begin{eqnarray}
\label{eq:psig}
\psig(\x;\mt,\kjes) &=& \frac{\dif\sigma_{p\bar p\to\ttbar}(\x;\,\mt,\kjes)}
                            {\sigma_{p\bar p\to\ttbar,\rm obs}(\mt,\kjes)}\,,
\end{eqnarray}
\end{linenomath}
where the total cross section for \ttbar production observed in the detector, defined as
\begin{linenomath}
\begin{equation}
\label{eq:sigmaobs}
\sigma_{p\bar p\to\ttbar,\rm obs}(\mt,\kjes)
            = \int_{\x}\dif\x~A(\x)\dif\sigma_{p\bar p\to\ttbar}(\x;\,\mt,\kjes)\,,
\end{equation}
\end{linenomath}
ensures a proper normalization of $A(\x)\psig$. Using Eqs.~(\ref{eq:dsigma}), (\ref{eq:dsigmapp}), and~(\ref{eq:dsigmax}), the PD to observe a \ttbar event with four-momenta \x in the detector can be written explicitly as
\begin{linenomath}
\begin{eqnarray}
\psig(\x;\mt,\kjes) &=& \frac{1}{\sigma_{p\bar p\to\ttbar,\rm obs}(\mt,\kjes)}\nonumber\\
                   &\times& \sum_{{}^{\rm quark}_{\rm flavors}}~\int\limits_{\vec q_{1},\,\vec q_{2}}
                            \dif \vec q_{1}\dif \vec q_{2} f(\vec q_{1})f(\vec q_{2})\nonumber\\
                   &\times& \frac{(2\pi)^{4}\left|{\cal M}_{q\bar q\to \ttbar}\right|^{2}}
                                 {2g_{\mu\nu}q_{1}^\mu q_{2}^\nu}\dif\Phi_{6}\nonumber\\
                   &\times& W(\x,\y;\,\kjes)\,.
\label{eq:psigfull}
\end{eqnarray}
\end{linenomath}
The parametrization of $|{\cal M}_{q\bar q\to \ttbar}|^{2}$ and $\sigma_{p\bar p\to\ttbar,\rm obs}$ are discussed below, and the calculation of \psig is given in Section~\ref{sec:psigcalc}.

\subsubsection{
Calculation of $|{\cal M}_{q\bar q\to \ttbar}|^2$
}

The LO matrix element for the $q\bar q\to\ttbar$ process~\cite{bib:melo1,bib:melo2} is given by
\begin{linenomath}
\begin{equation}
\left|{\cal M}_{q\bar q\to \ttbar}\right|^{2}=\frac{16\pi^2\alpha_{s}^{2}}{9}F\bar{F}\cdot\left(2-\beta^{2}\sin^2(\theta_{qt})\right)\,,
\label{eq:me}
\end{equation}
\end{linenomath}
where $\alpha_s$ is the strong coupling, $\beta=|\vec p_{t}|/E_{t}$ represents the velocity of the $t$ (or $\bar{t}$) quark in the $q\bar q$ rest frame, and $\theta_{qt}$ denotes the angle between the incoming parton and the outgoing top quark in the $q\bar q$ rest frame. The form factors $F\bar F$ are identical to those given in Eqs.~(24) and~(25) of Ref.~\cite{bib:mt2006}.

\subsubsection{
Calculation of $\sigma_{p\bar p\to\ttbar,\rm obs}$
}

\begin{figure}
\centering
\begin{overpic}[width=0.99\columnwidth]{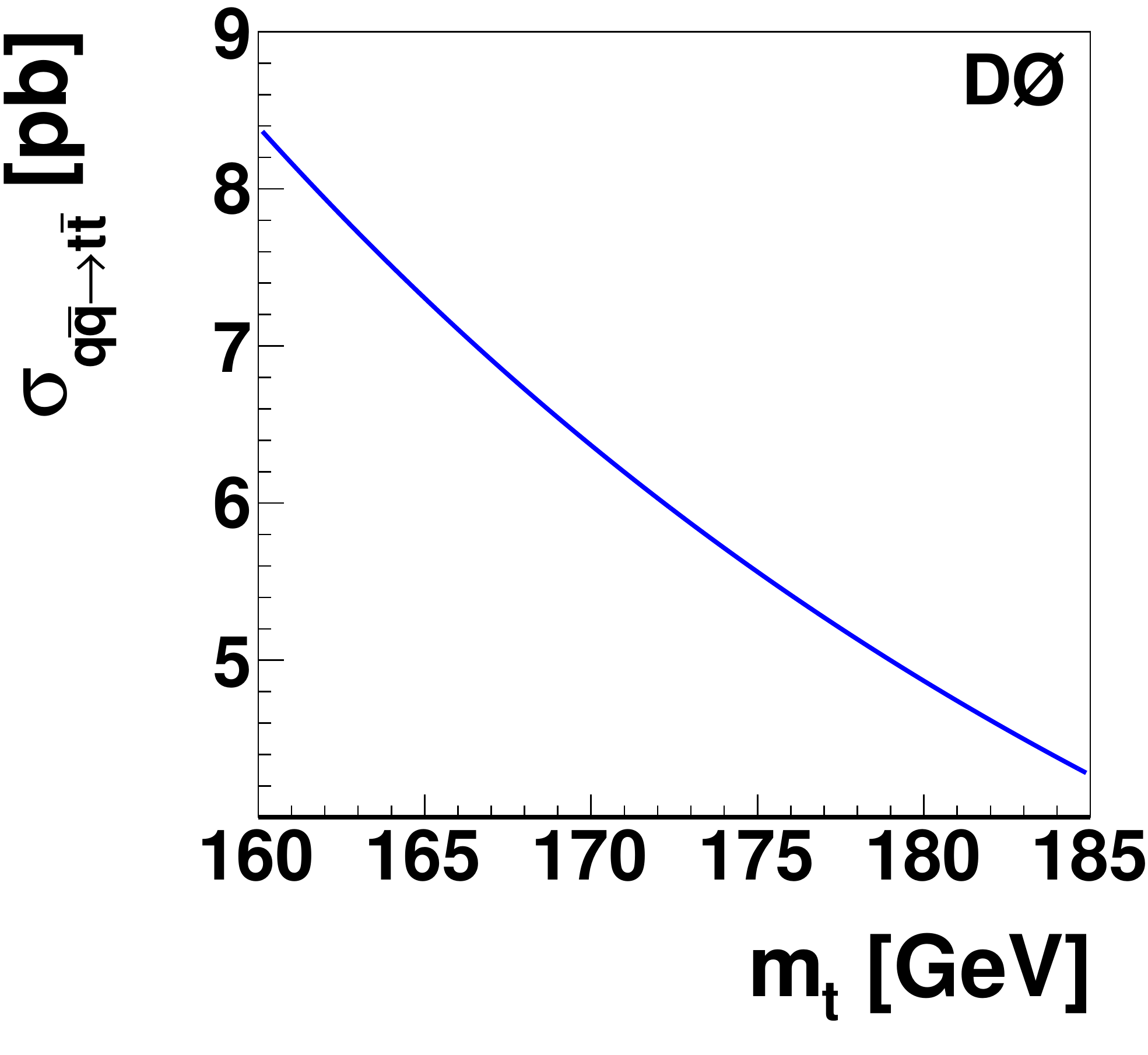}
\end{overpic}
\caption{\label{fig:xsec}
The total partonic cross section for \ttbar production in $p\bar p$ collisions through quark-antiquark annihilation at LO, as defined in Eq.~(\ref{eq:sigmatot}).
}
\end{figure}

The calculation of the total cross section for \ttbar production observed in the detector defined in Eq.~(\ref{eq:sigmaobs}) is challenging, because a double integral has to be performed over all possible partonic and measured configurations for each set of $(\mt,\kjes)$. We therefore apply a MC integration technique, and factorize the partonic and measured parts of the calculation to obtain:
\begin{linenomath}
\begin{eqnarray}
\sigma_{p\bar p\to\ttbar,\rm obs}(\mt)
  &=& \int_{\y}\dif\y~\dif\sigma_{p\bar p\to\ttbar}(\y;\,\mt,\kjes)A(\x)W(\x,\y;\,\kjes)\nonumber \\
  &\approx& \sigma_{p\bar p\to\ttbar,\rm tot}(\mt)\times\langle A(\mt,\kjes)\rangle_{\x}\,,
\end{eqnarray}
\end{linenomath}
where 
\begin{linenomath}
\begin{equation}
\label{eq:sigmatot}
\sigma_{p\bar p\to\ttbar,\rm tot}(\mt)=\int_{\y}\dif\y~\dif\sigma_{p\bar p\to\ttbar}(\y;\,\mt,\kjes)
\end{equation}
\end{linenomath}
is the total partonic cross section for \ttbar production in $p\bar p$ collisions as shown in Fig.~\ref{fig:xsec}, and $\langle A(\mt,\kjes)\rangle_{\x}$ is the mean acceptance averaged over all possible configurations of \x for a given set of $(\mt,\kjes)$ parameters.


\subsubsection{
Calculation of the mean acceptance
\label{sec:acc}
}

\begin{figure}
\centering
\begin{overpic}[width=0.49\columnwidth]{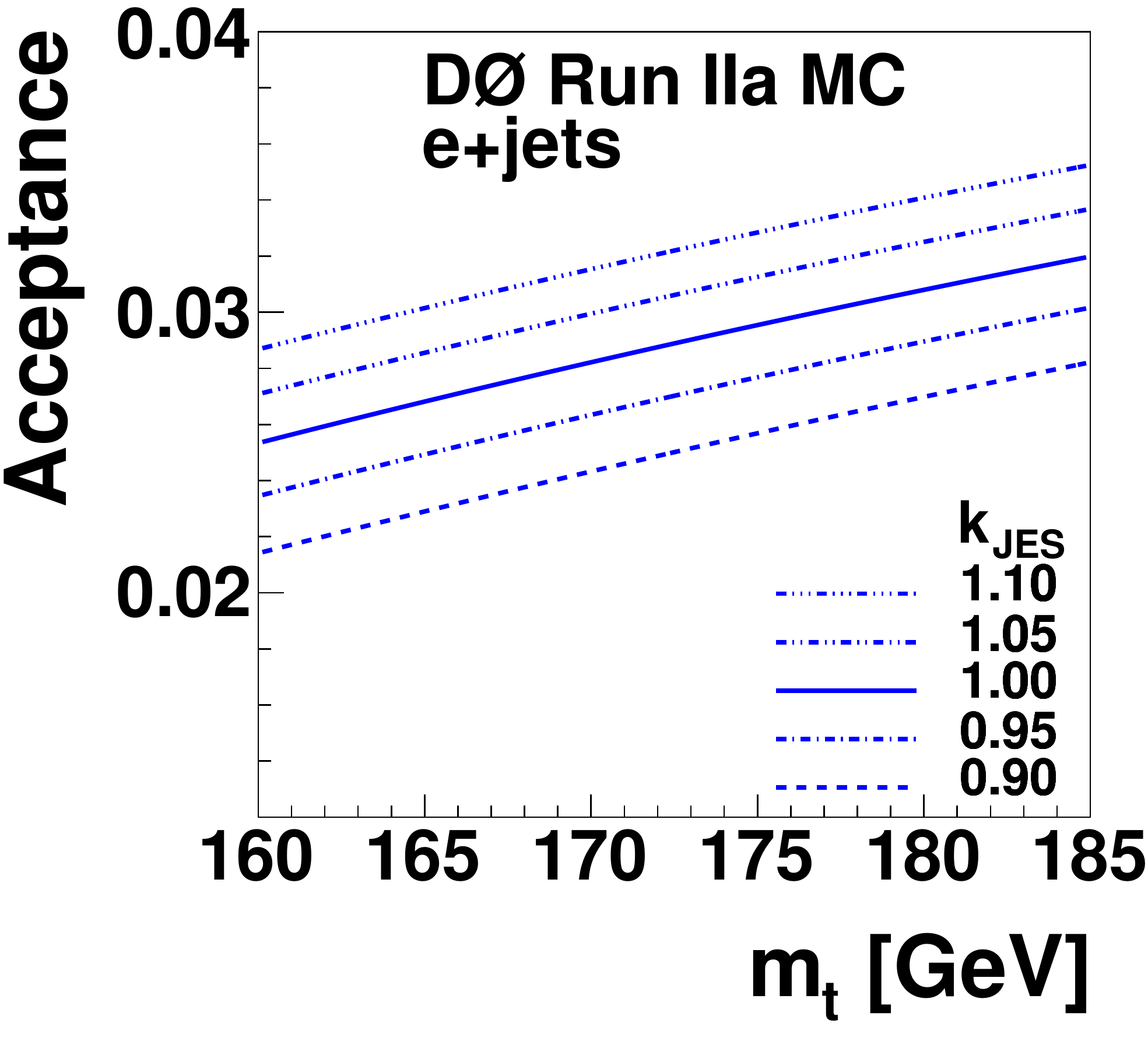}
\put(26,82){\footnotesize\textsf{\textbf{(a)}}}
\end{overpic}
\begin{overpic}[width=0.49\columnwidth]{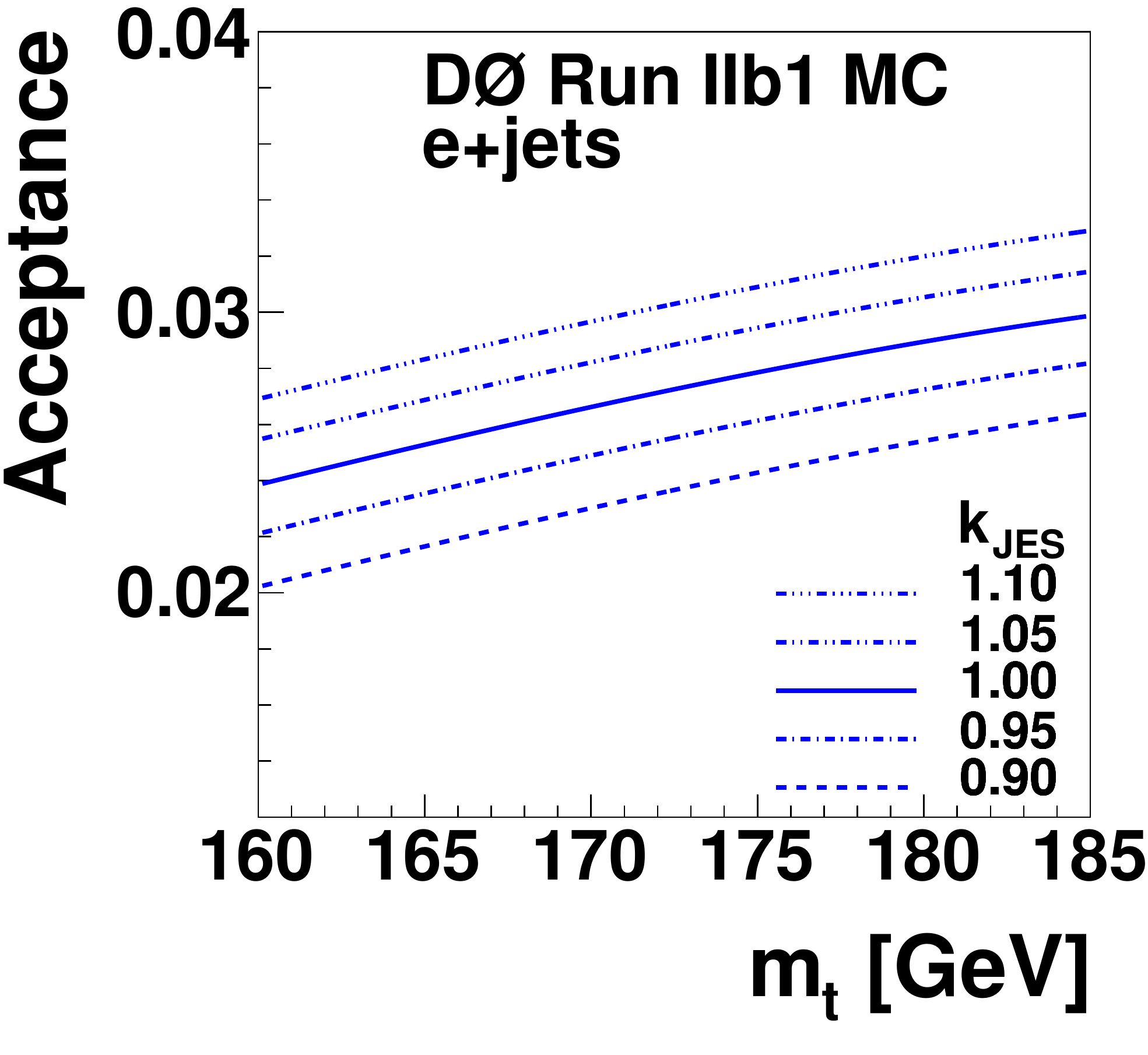}
\put(26,82){\footnotesize\textsf{\textbf{(b)}}}
\end{overpic}
\begin{overpic}[width=0.49\columnwidth]{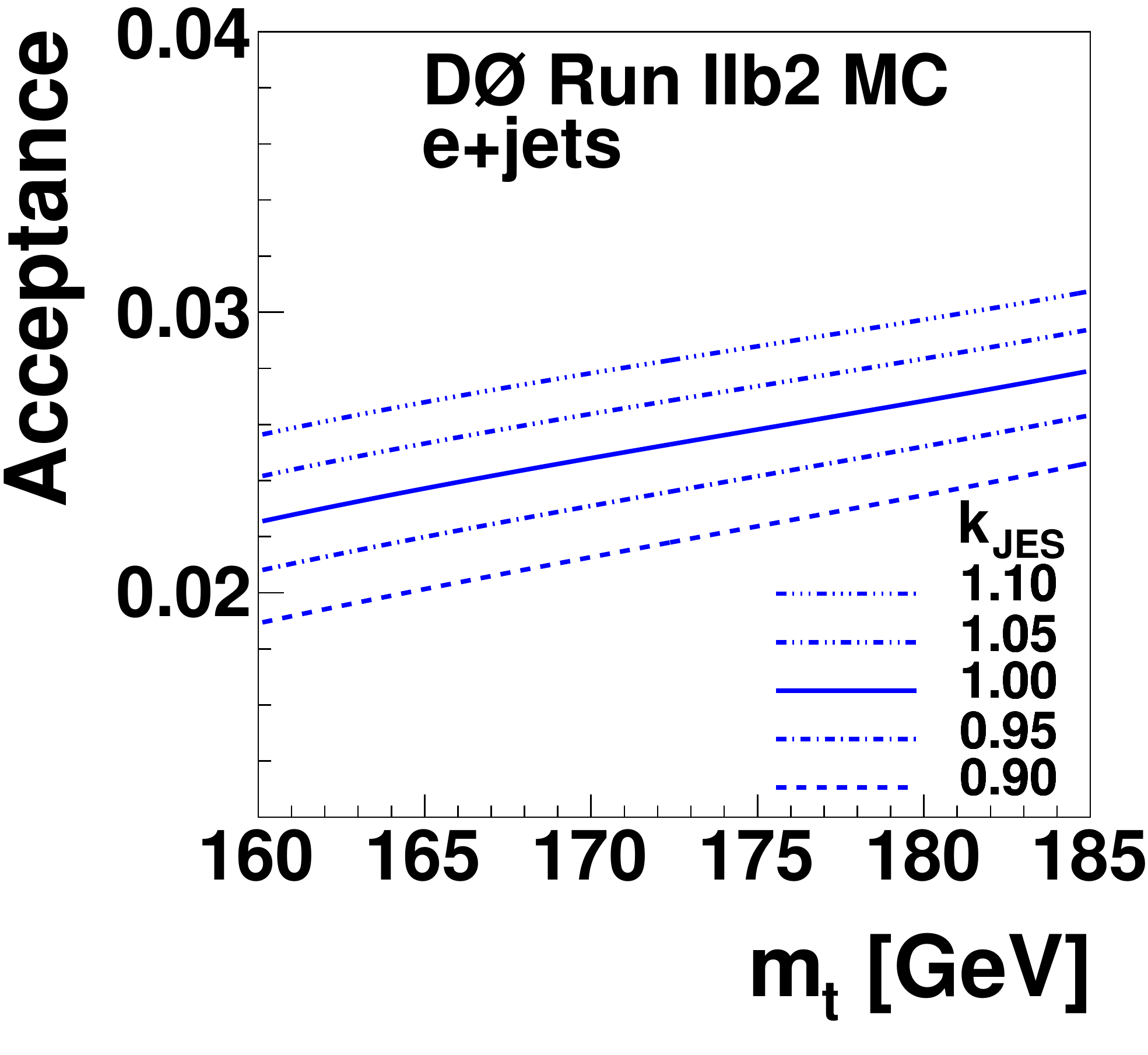}
\put(26,82){\footnotesize\textsf{\textbf{(c)}}}
\end{overpic}
\begin{overpic}[width=0.49\columnwidth]{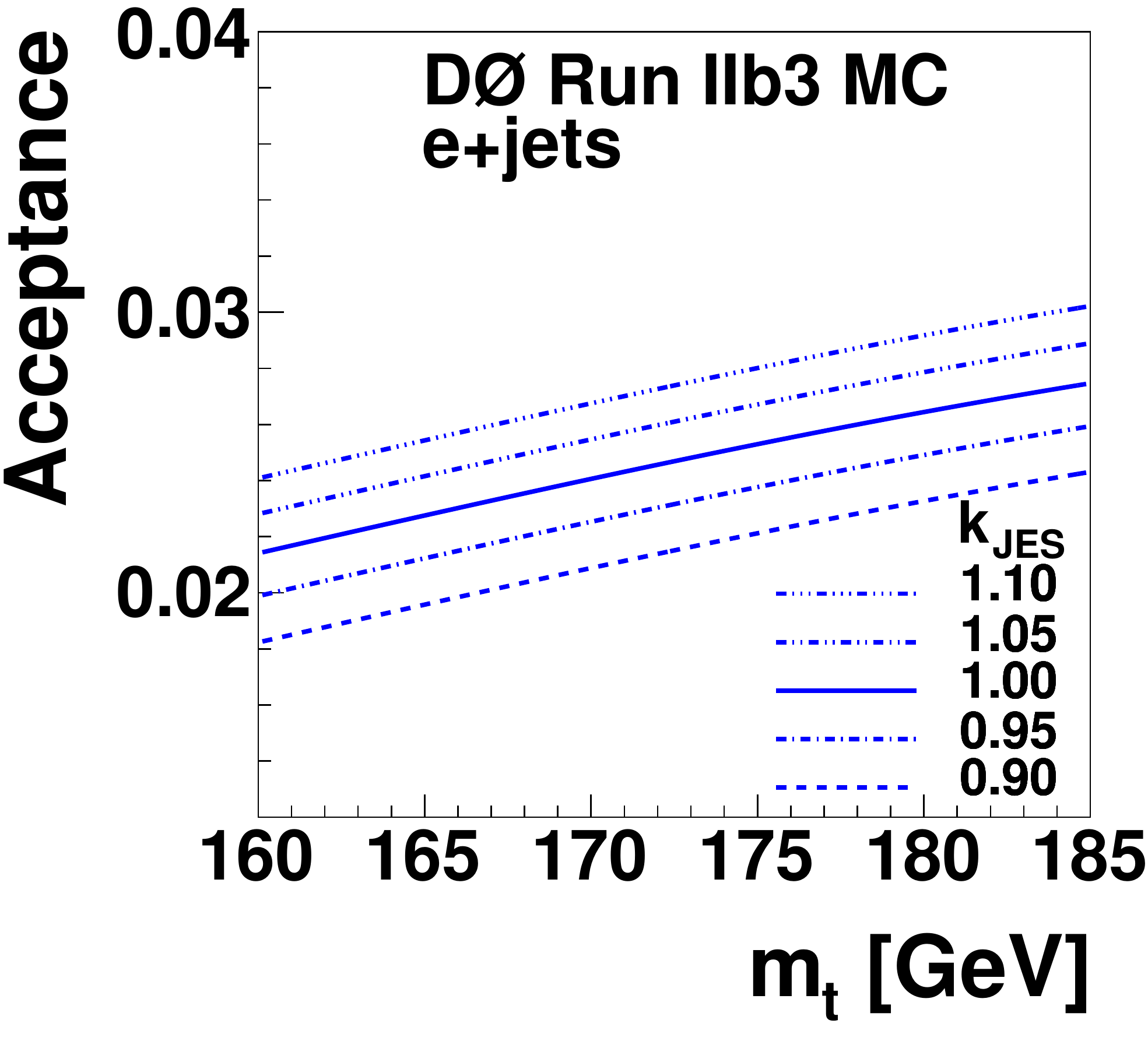}
\put(26,82){\footnotesize\textsf{\textbf{(d)}}}
\end{overpic}
\caption{\label{fig:accem}
Mean acceptance $\langle A(\mt,\kjes\equiv1)\rangle_{\x}$ with respect to the total partonic cross section for $p\bar p\to\ttbar$ production, as a function of \mt for different \kjes values in the \ejets final state, for (a)~Run IIa, (b)~Run~IIb1, (c)~Run~IIb2, and (d)~Run~IIb3. The acceptance tends to decrease for later epochs due to a decreased reconstruction efficiency, which is mainly caused by the higher instantaneous luminosity in the second half of Run~II and detector aging.
}
\end{figure}

\begin{figure}
\centering
\begin{overpic}[width=0.49\columnwidth]{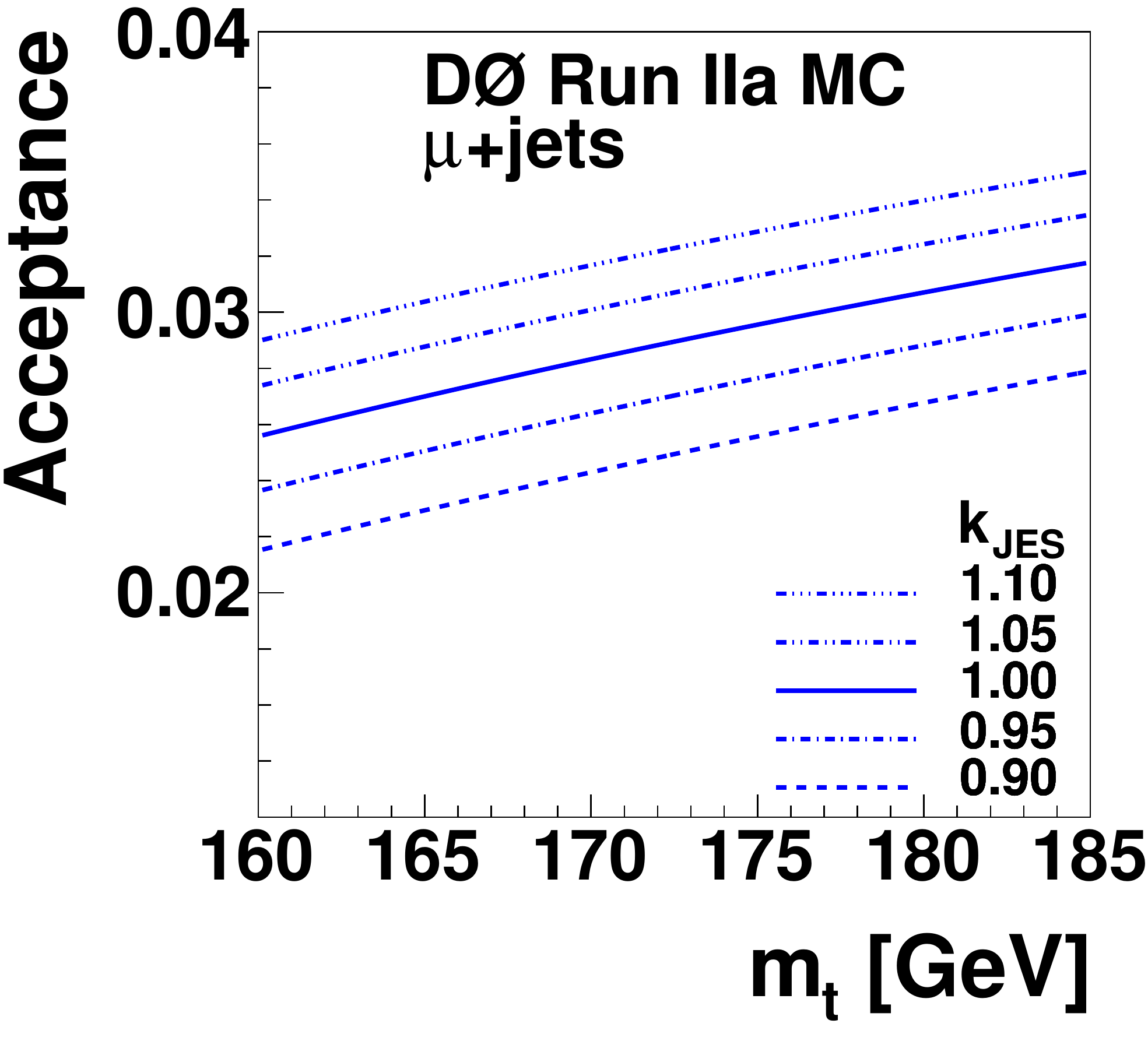}
\put(26,82){\footnotesize\textsf{\textbf{(a)}}}
\end{overpic}
\begin{overpic}[width=0.49\columnwidth]{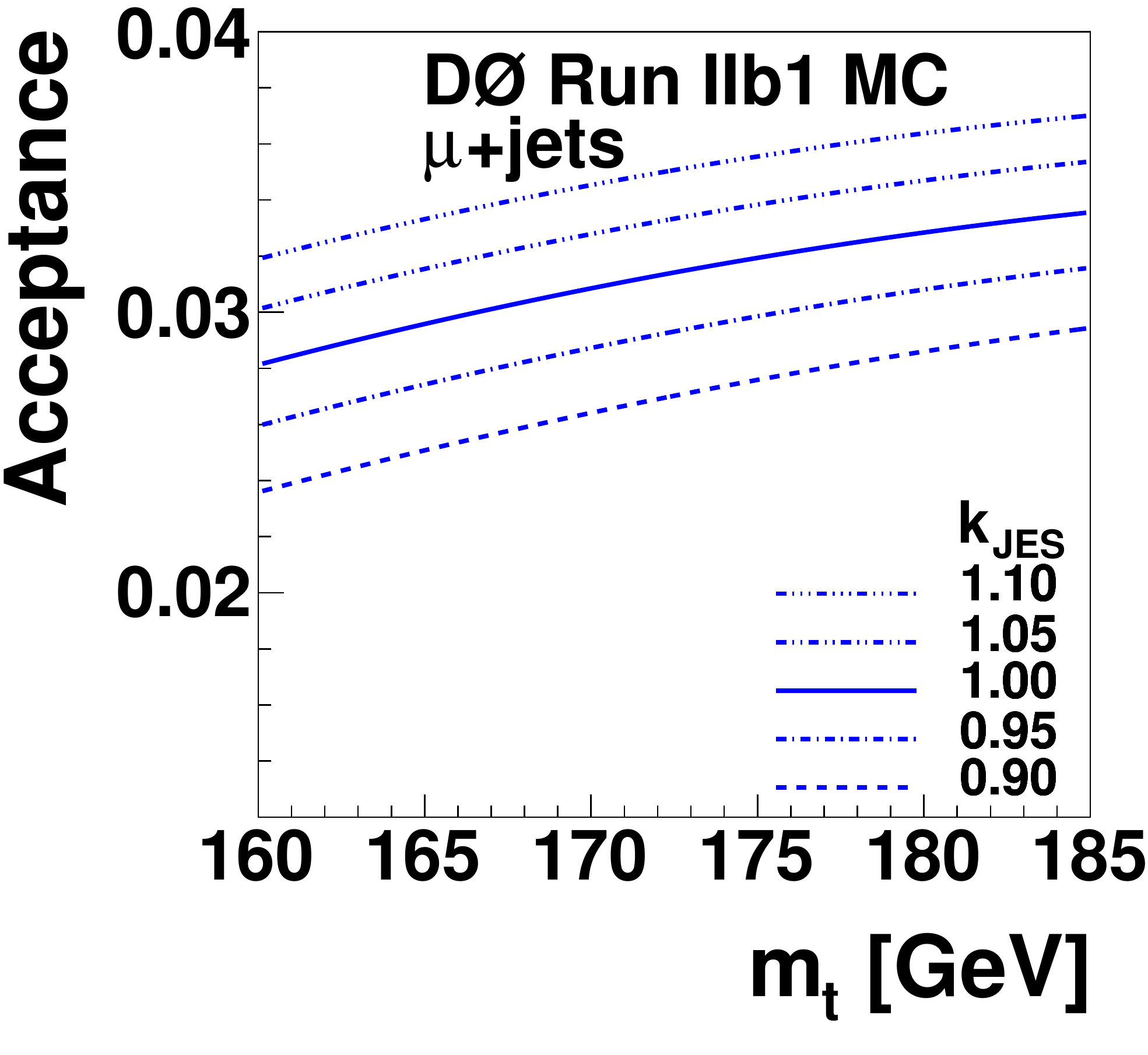}
\put(26,82){\footnotesize\textsf{\textbf{(b)}}}
\end{overpic}
\begin{overpic}[width=0.49\columnwidth]{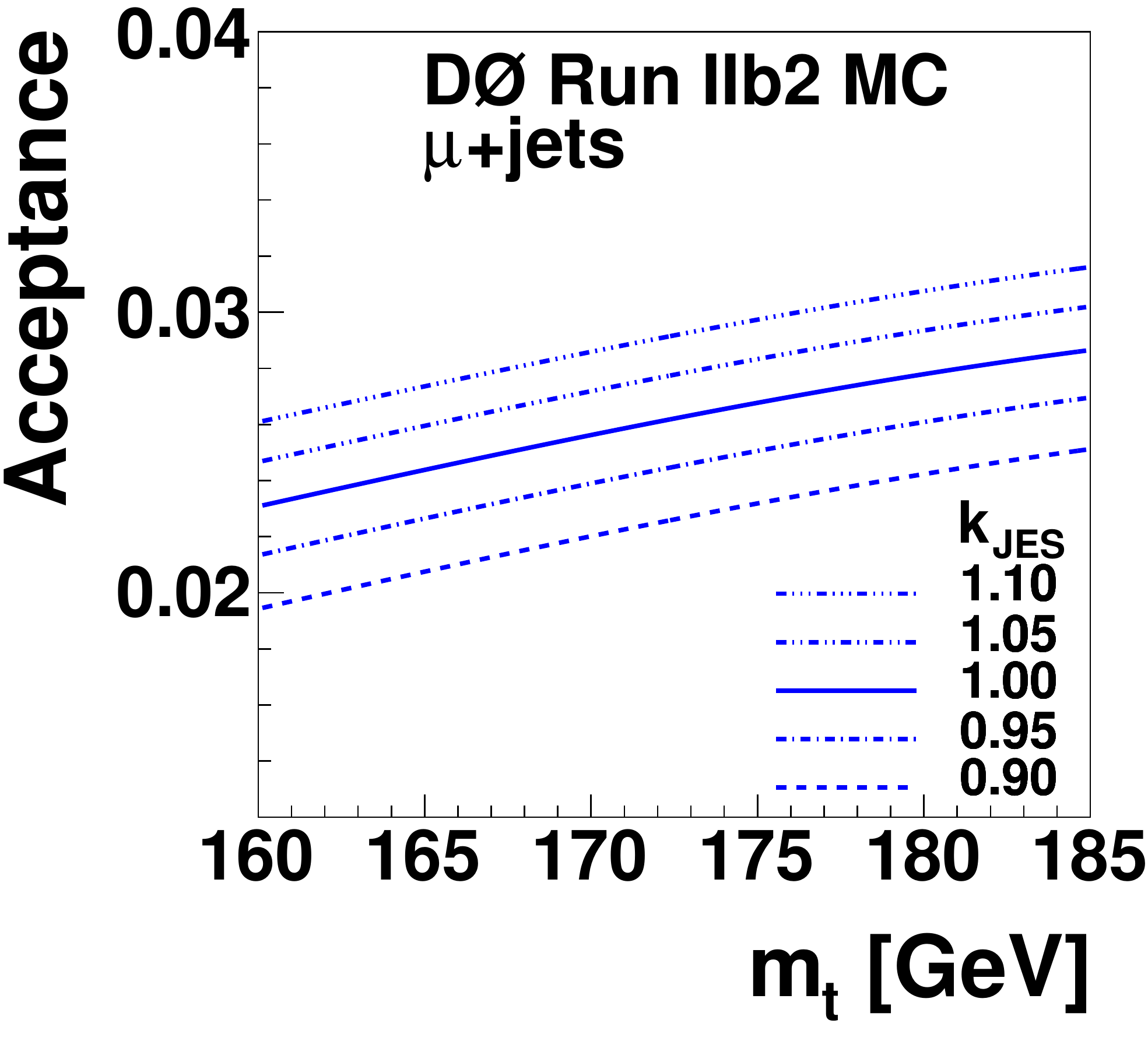}
\put(26,82){\footnotesize\textsf{\textbf{(c)}}}
\end{overpic}
\begin{overpic}[width=0.49\columnwidth]{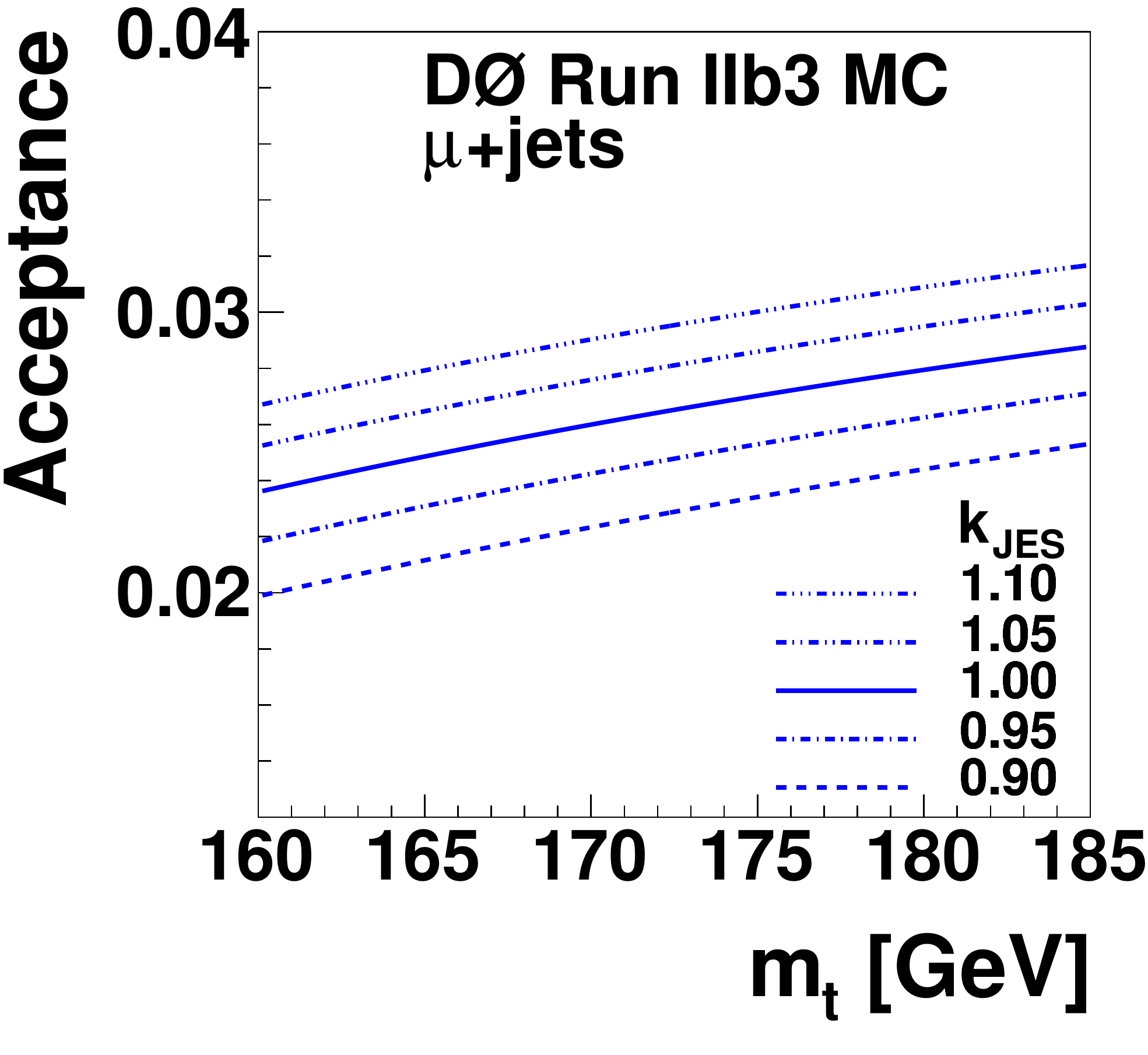}
\put(26,82){\footnotesize\textsf{\textbf{(d)}}}
\end{overpic}
\caption{\label{fig:accmu}
Same as Fig.~\ref{fig:accem}, but for the \mujets final state for (a)~Run IIa, (b)~Run~IIb1, (c)~Run~IIb2, and (d)~Run~IIb3.
}
\end{figure}

The calculation of the mean acceptance $\langle A(\mt,\kjes)\rangle_{\x}$ proceeds in two steps:
\begin{enumerate}
\item
First, $\langle A(\mt,\kjes\equiv1)\rangle_{\x}$ is determined using MC events processed with the full simulation of the detector, according to
\begin{linenomath}
\begin{equation}
\label{eq:meanacc}
\langle A(\mt,\kjes\equiv1)\rangle_{\x} = \frac{\sum_{i=1}^{N_{\rm acc}}w_{\rm MC}w_{\rm exp}}{\sum_{i=1}^{N_{\rm gen}}w_{\rm MC}}\,.
\end{equation}
\end{linenomath}
In this expression, $N_{\rm gen}$ denotes the total number of generated events, and $N_{\rm acc}$ represents the total number of events from within detector acceptance, while $w_{\rm MC}$ is the MC weight from {\alpgen}, and $w_{\rm exp}$ the weight that accounts for experimental factors such as the trigger and identification efficiencies.
\item
In the second step, the dependence of the mean acceptance on \kjes is derived using simulated parton level events, where the detector response is taken into account through the TF  $W(\x,\y;\,\kjes)$ for the corresponding \kjes value.
\end{enumerate}
The mean acceptance is shown for each data-taking epoch in Fig.~\ref{fig:accem} for \ejets and in Fig.~\ref{fig:accmu} for \mujets final states.

\subsection{
The background probability
}
\label{sec:pbkg}

The differential partonic cross section for \pbkg is calculated according to
\begin{linenomath}
\begin{eqnarray}
\pbkg(\x;\kjes) &=& \frac{1}{\sigma_{p\bar p\to\wjets,\rm obs}(\kjes)}\nonumber\\
                   &\times& \sum_{{}^{\rm quark}_{\rm flavors}}
                            ~\int\limits_{q_{1,\rm z},q_{2,\rm z}}
                            \dif q_{1,\rm z}\dif q_{2,\rm z} f(q_{1,\rm z})f(q_{2,\rm z})\nonumber\\
                   &\times& \frac{(2\pi)^{4}\left|{\cal M}_{q\bar q\to \wjets}\right|^{2}}
                                 {2g_{\mu\nu}q_{1}^\mu q_{2}^\nu}\dif\Phi_{6}\nonumber\\
                   &\times& W(\x,\y;\,\kjes)\,,
\label{eq:pbkgfull}
\end{eqnarray}
\end{linenomath}
which is similar to \psig in Eq.~(\ref{eq:psigfull}), with the exception that the initial-state partons are all assumed to have $\pt=0$. The ME ${\cal M}_{q\bar q\to \wjets}$ is calculated at LO using the ME for  $W\!+\!4{\rm~jets}$ production as implemented in the \vecbos~\cite{bib:vecbos} program. Since \vecbos treats the final state quarks as massless, the same ME is used to model the production of $W$ bosons in association with $u,d,s,c,$ and $b$-quark jets. By definition, \pbkg is independent of \mt. The dependence of \pbkg on \kjes is taken into account, as before, through the TF.

\subsection{
Modeling of detector response
}
\label{sec:tf}

Here, we describe the modeling of the detector response through the TF $W(\x,\y;\,\kjes)$. We use a parameterized TF, as a full simulation of the detector would not be computationally feasible. 

In constructing the TF, we assume that the functions for individual final-state particles are not correlated. We therefore factorize the TF into contributions from the four jets and the charged lepton used to calculate \psig. The resolution of the \met is worse than that of the \pt of jets and leptons, and is therefore not used in defining event probabilities. We assume that the directions of the $e$, $\mu$, and jets in $(\eta,\phi)$ space are well-measured. The effect of finite detector resolutions in $\eta$ and $\phi$ is studied for jets and found to be negligible. The TF for jet and charged lepton directions are  therefore represented by $\delta$ functions, $\delta^2(\eta,\phi)\equiv\delta(\eta_y-\eta_x)\delta(\phi_y-\phi_x)$, thereby reducing the number of integration variables by  $5\times2=10$.  The finite resolutions on the energies of jets and charged leptons are explicitly taken into account, as described in Sections~\ref{sec:tfjet},~\ref{sec:tfem}, and~\ref{sec:tfmu}. 

There is an inherent ambiguity in assigning measured jets to partons from $\ttbar$ decay. Consequently, all 24 jet-parton assignments that are possible in LO are considered. The inclusion of $b$~tagging information provides improved identification of the correct jet-parton assignment through weights. The weight $w_i$ for the $i$-th jet-parton assignment is defined by a product of individual weights $w_i^j$ for each jet $j$. For $b$~tag\-ged jets, $w_i^j$ is equal to the per-jet  tagging efficiency $\epsilon_{\rm tag}(\alpha_k;\ \etj,\,\eta^j)$, where $\alpha_k$ labels the configurations where the jet is assigned to either (i)~a~$b$~quark, (ii)~a~$c$~quark, or (iii)~a~$u,d,s$ quark or gluon. For untagged jets, the $w_i^j$ factors are equal to  $1-\epsilon_{\rm tag}(\alpha_k;\ \etj,\,\eta^j)$. Hence, the weights $w_i$ are close to unity for those configurations that assign $b$-tagged jets to $b$~quarks and untagged jets to $u,d,s$ quarks or gluons, and $w_i\ll1$ otherwise. 
Since the contributions to \wjets processes are parameterized by $\mathcal{M}_{q\bar q\to\wjets}$ without regard to heavy-flavor content, the weights $w_i$ for each of the 24 permutations in the background probability are all equal.
The sum of weights is always normalized to unity: $\sum_{i=1}^{24}\!w_i = 1$.

We define the TF as
\begin{eqnarray}
    \lefteqn{W(\x,\y;\,\kjes) = W_{\ell}(E_x,E_y)\delta_\ell^2(\eta,\phi)} \nonumber\\
             & & \times\sum_{i=1}^{24}\!w_i \left\{ \prod_{j=1}^{4}\,\delta_{ij}^2(\eta,\phi) \wjet(E^i_x,E^j_y;\kjes) \!\!\right\}\!,\label{eq:tf}
\end{eqnarray}
where $E$ is the energy and \wjet is the TF appropriate for the energy resolution of jets, as discussed in Section~\ref{sec:tfjet}, and $W_{\ell}$ is either $W_e$, the energy resolution for electrons, as discussed in Section~\ref{sec:tfem}, or $W_\mu$, the resolution in the transverse momentum of muons, as discussed in Section~\ref{sec:tfmu}.

\subsubsection{
Modeling of jet momentum
}
\label{sec:tfjet}

The TF for jets represents the probability that the  jet energy $E_{x}$ measured in the detector corresponds to a parent quark of energy $E_{y}$. It is parameterized in terms of a double Gaussian function whose means and widths depend on $E_y$.  For $\kjes\equiv1$, it is given by
\begin{linenomath}
\begin{eqnarray}
\lefteqn{\wjet\left(E_{x},E_{y};\,\kjes=1\right) = \frac{1}{\sqrt{2\pi}(p_{2}+p_{3}p_{5})}}\nonumber\\
&&\times \left\{\exp\left({-\frac{[(E_{x}-E_{y})-p_{1}]^{2}}{2p_{2}^{2}}}\right)\right.\nonumber\\
&&+ \left.p_{3}\exp\left({-\frac{[(E_{x}-E_{y})-p_{4}]^{2}}{2p_{5}^{2}}}\right)\right\}\,,
\end{eqnarray}
\end{linenomath}
where the $p_{i}$ are linear functions of the energy of the  parent quark:
\begin{linenomath}
\begin{equation}
\label{eq:aibi}
p_{i}=a_{i}+b_{i}\,E_{y}\,.
\end{equation}
\end{linenomath}
For $\kjes\neq1$, the jet TF changes to
\begin{linenomath}
\begin{equation}
\label{eq:kjesdef}
\wjet\left(E_{x},E_{y};\,\kjes\right)=\frac{\wjet\left(\frac{E_{x}}{\kjes},E_{y};\,1\right)}{\kjes}\,,
\end{equation}
\end{linenomath}
where the \kjes factor in the denominator preserves the normalization $\int\dif E_x~\wjet(E_x,E_y;\kjes) = 1$.

\begin{figure}
\centering
\begin{overpic}[width=0.49\columnwidth]{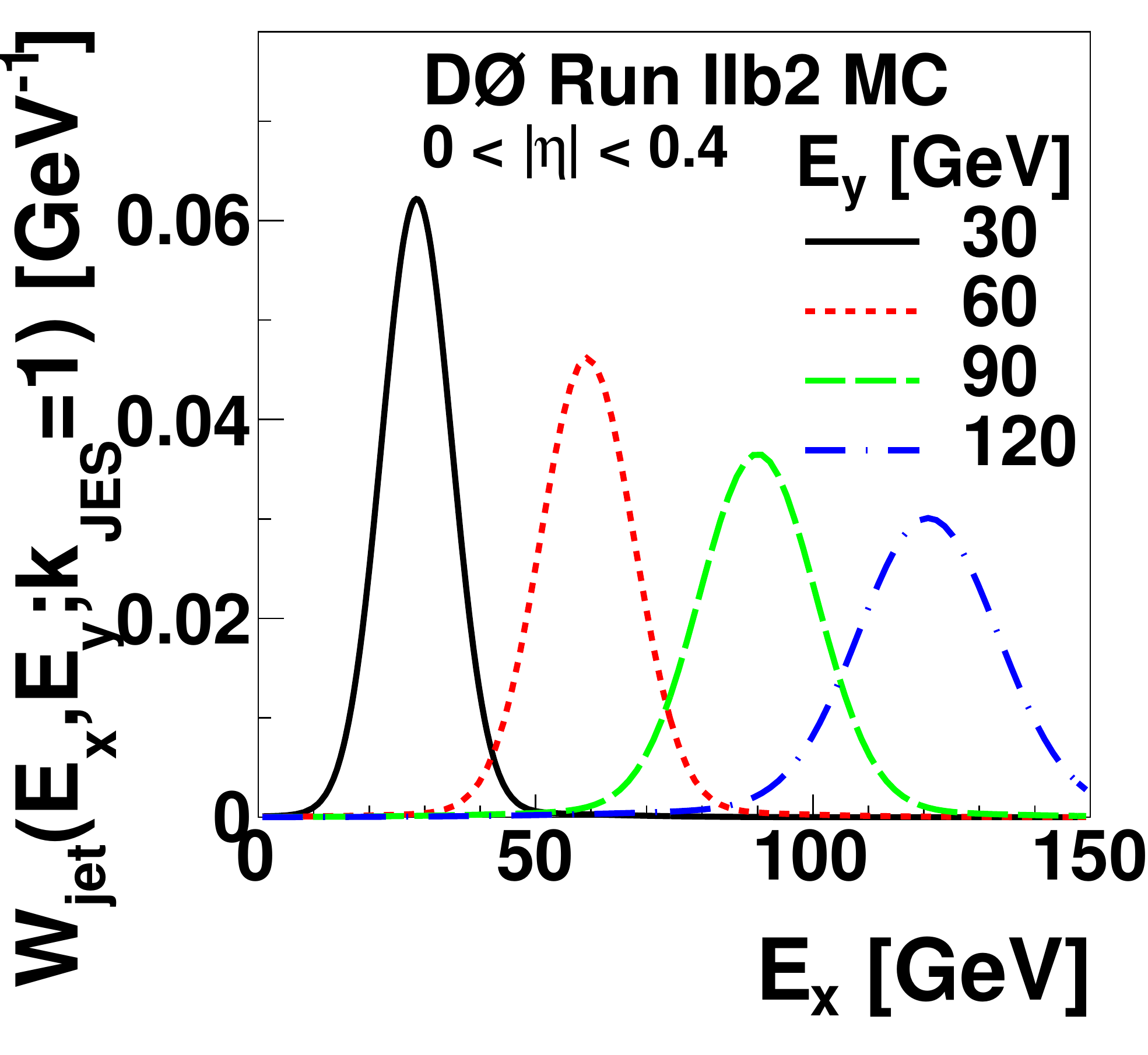}
\put(26,82){\footnotesize\textsf{\textbf{(a)}}}
\end{overpic}
\begin{overpic}[width=0.49\columnwidth]{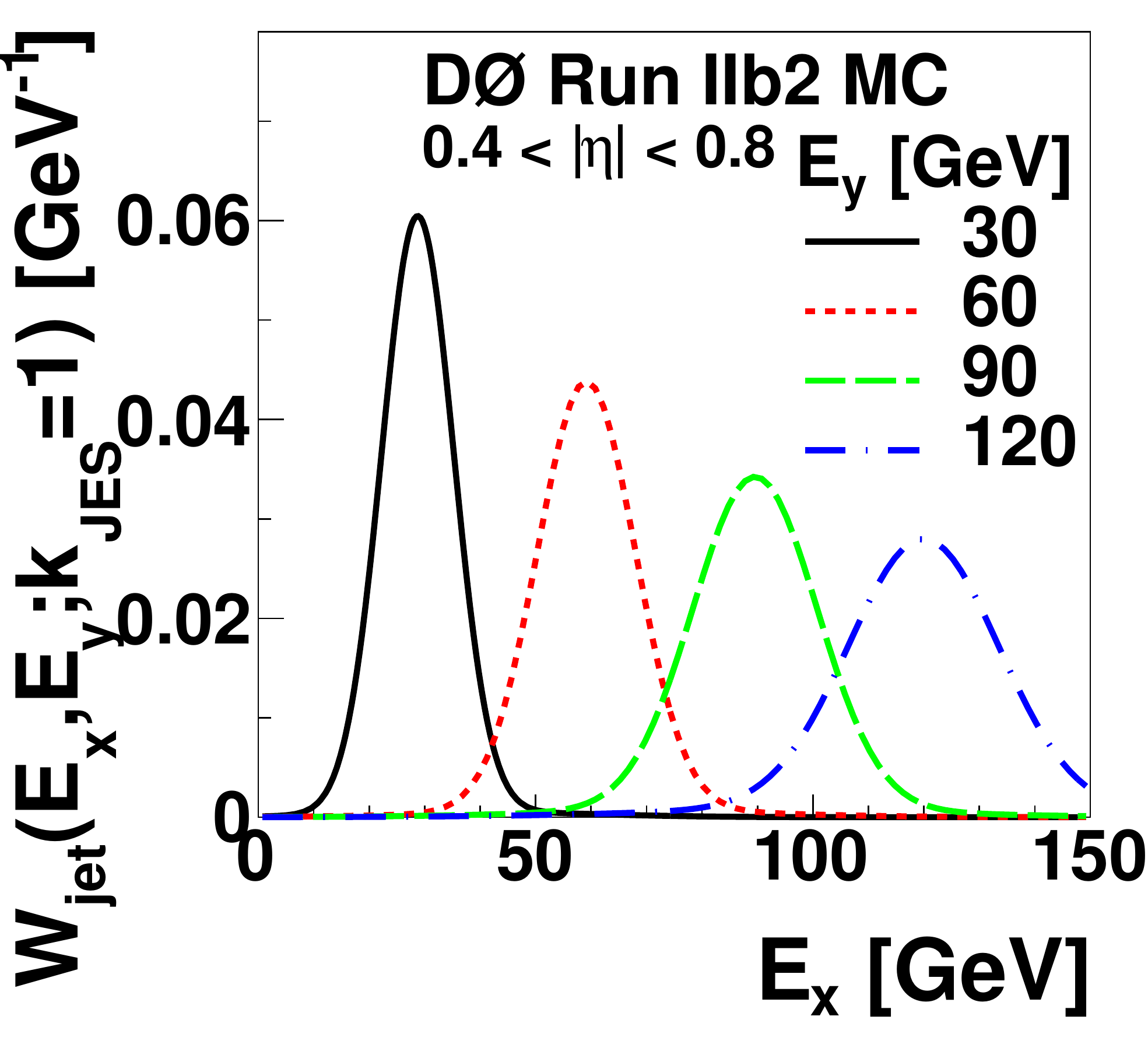}
\put(26,82){\footnotesize\textsf{\textbf{(b)}}}
\end{overpic}
\begin{overpic}[width=0.49\columnwidth]{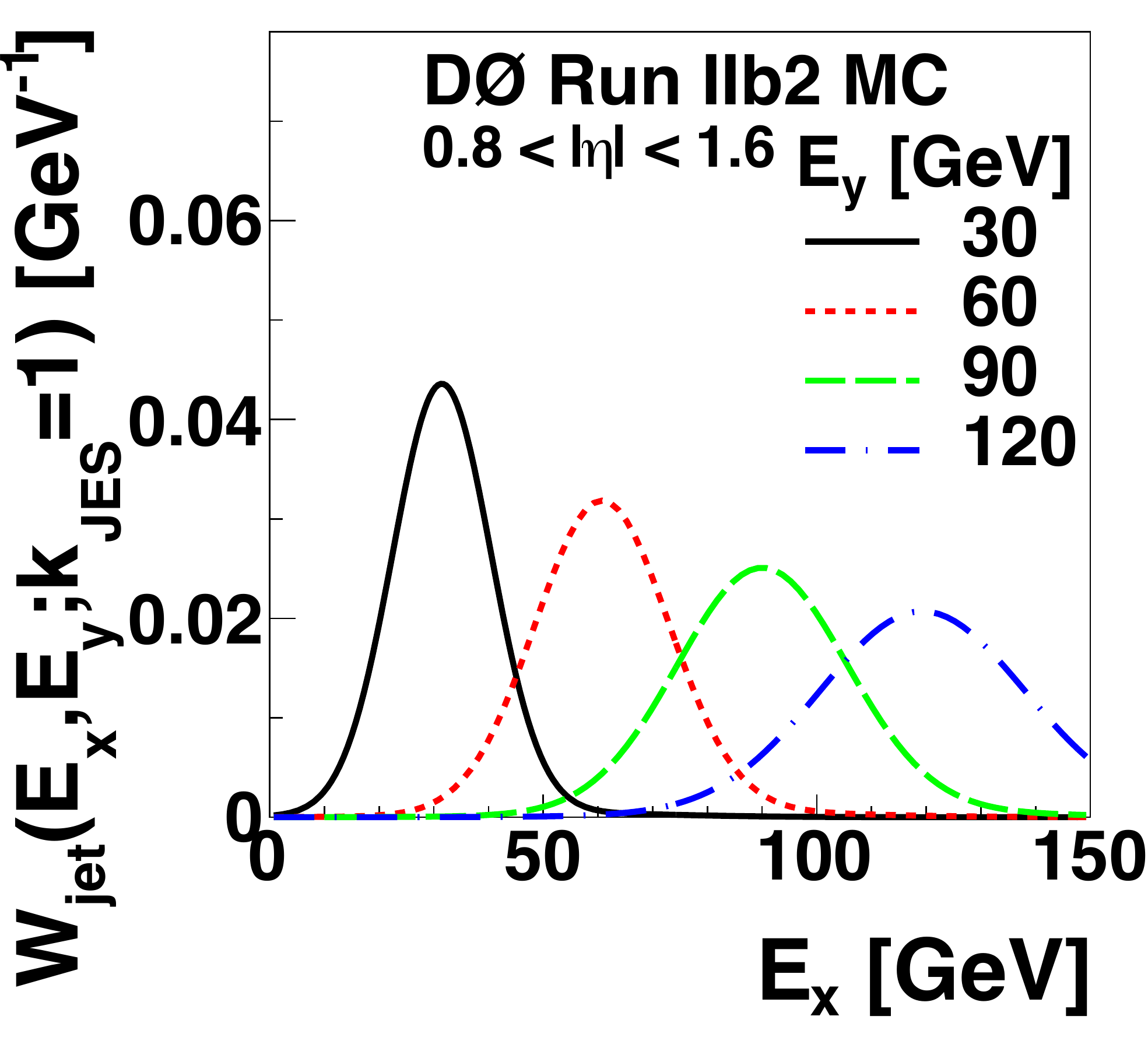}
\put(26,82){\footnotesize\textsf{\textbf{(c)}}}
\end{overpic}
\begin{overpic}[width=0.49\columnwidth]{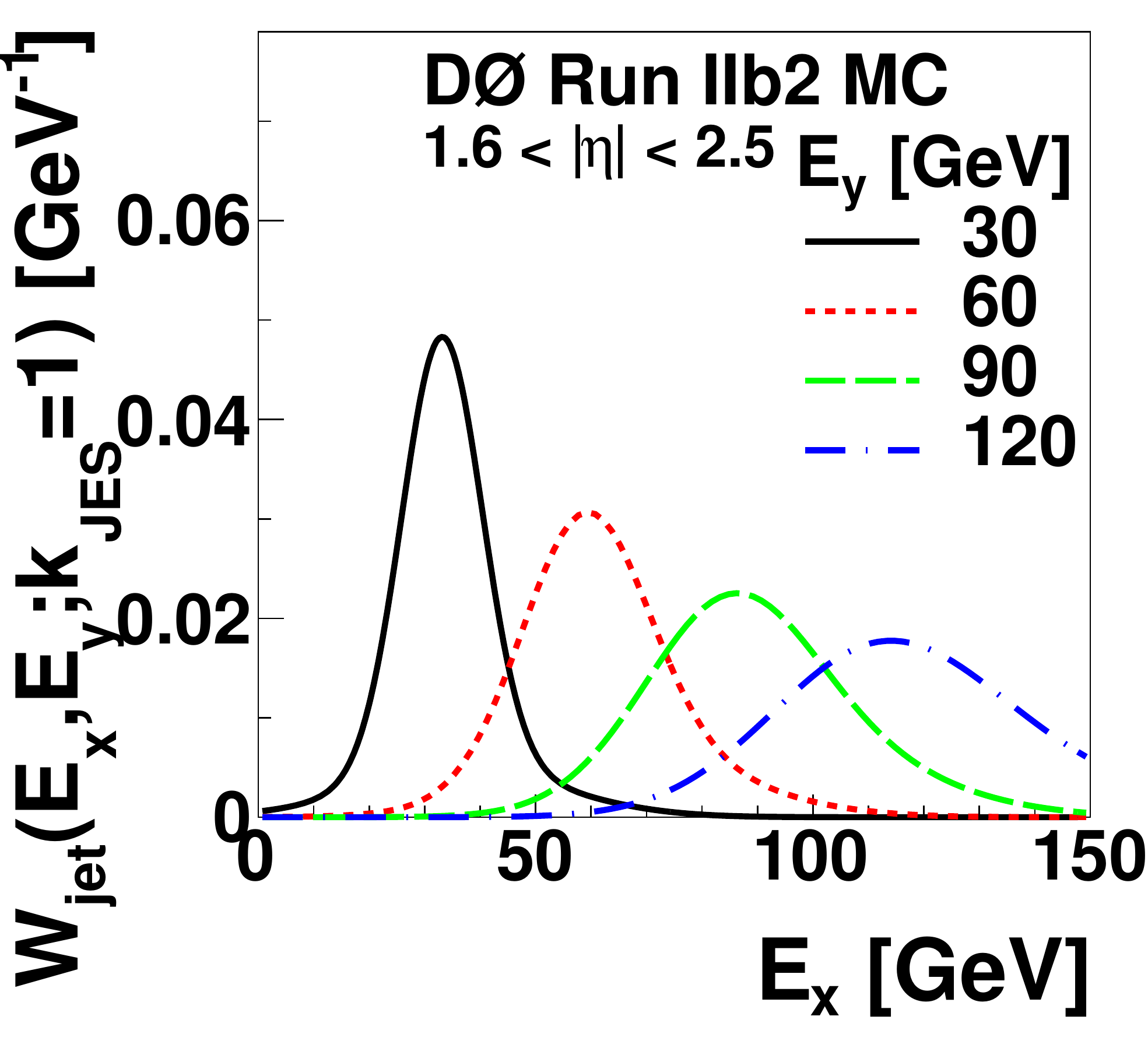}
\put(26,82){\footnotesize\textsf{\textbf{(d)}}}
\end{overpic}
\caption{\label{fig:tfl}
The TF $\wjet(E_x,E_y;\kjes)$ for light-quark jets at $\kjes\equiv1$ in Run~IIb2, as a function of the measured jet energy $E_x$, for different parton energies $E_y$, as indicated in the legend. The TF are shown in four pseudorapidity regions: (a) $|\eta|\leq0.4$, (b) $0.4<|\eta|\leq0.8$, (c) $0.8<|\eta|\leq1.6$, and (d) $1.6<|\eta|\leq2.5$. Energies $E_x>150~\GeV$ are also considered, but not shown.
}
\end{figure}

\begin{figure}
\centering
\begin{overpic}[width=0.49\columnwidth]{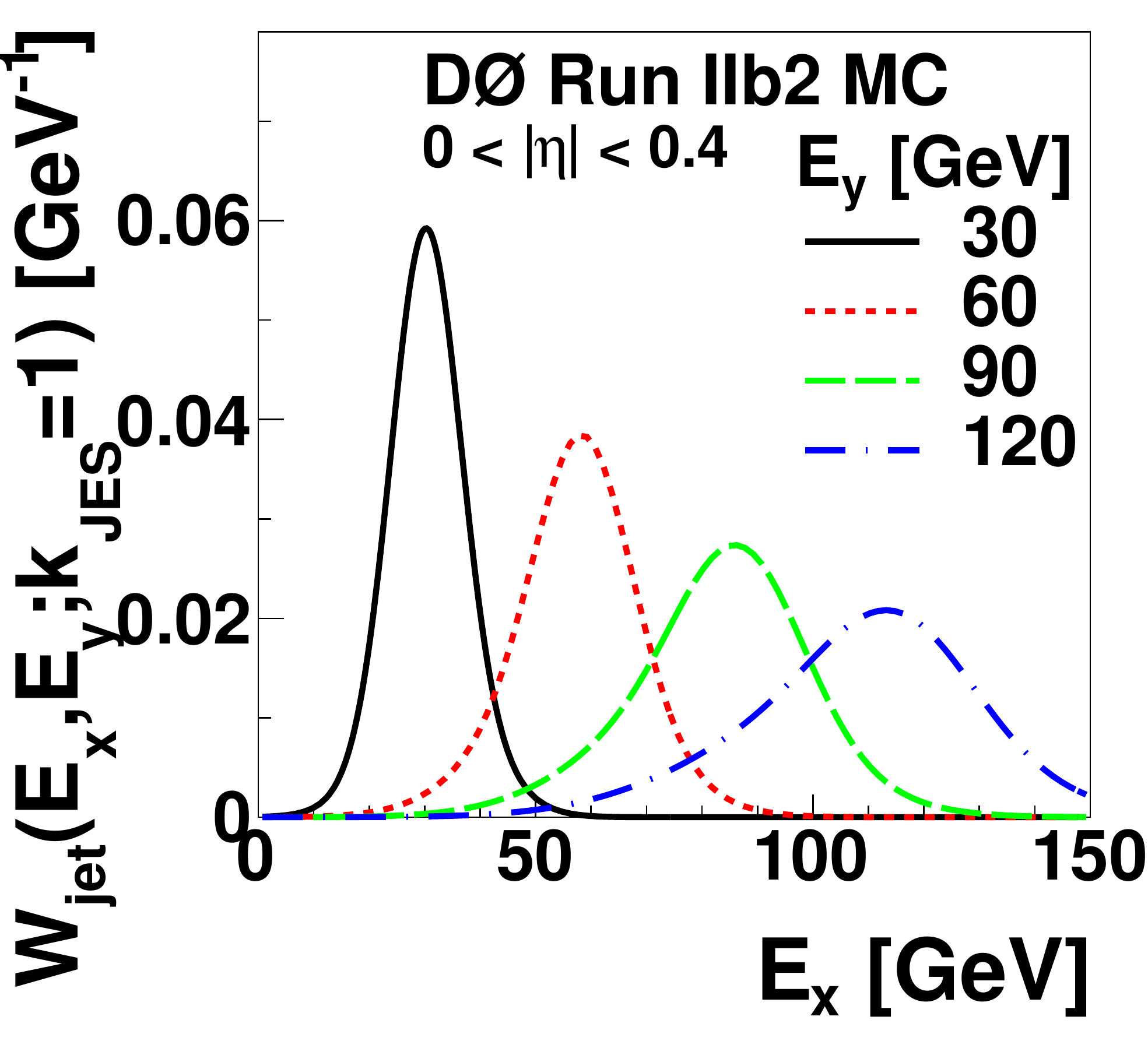}
\put(26,82){\footnotesize\textsf{\textbf{(a)}}}
\end{overpic}
\begin{overpic}[width=0.49\columnwidth]{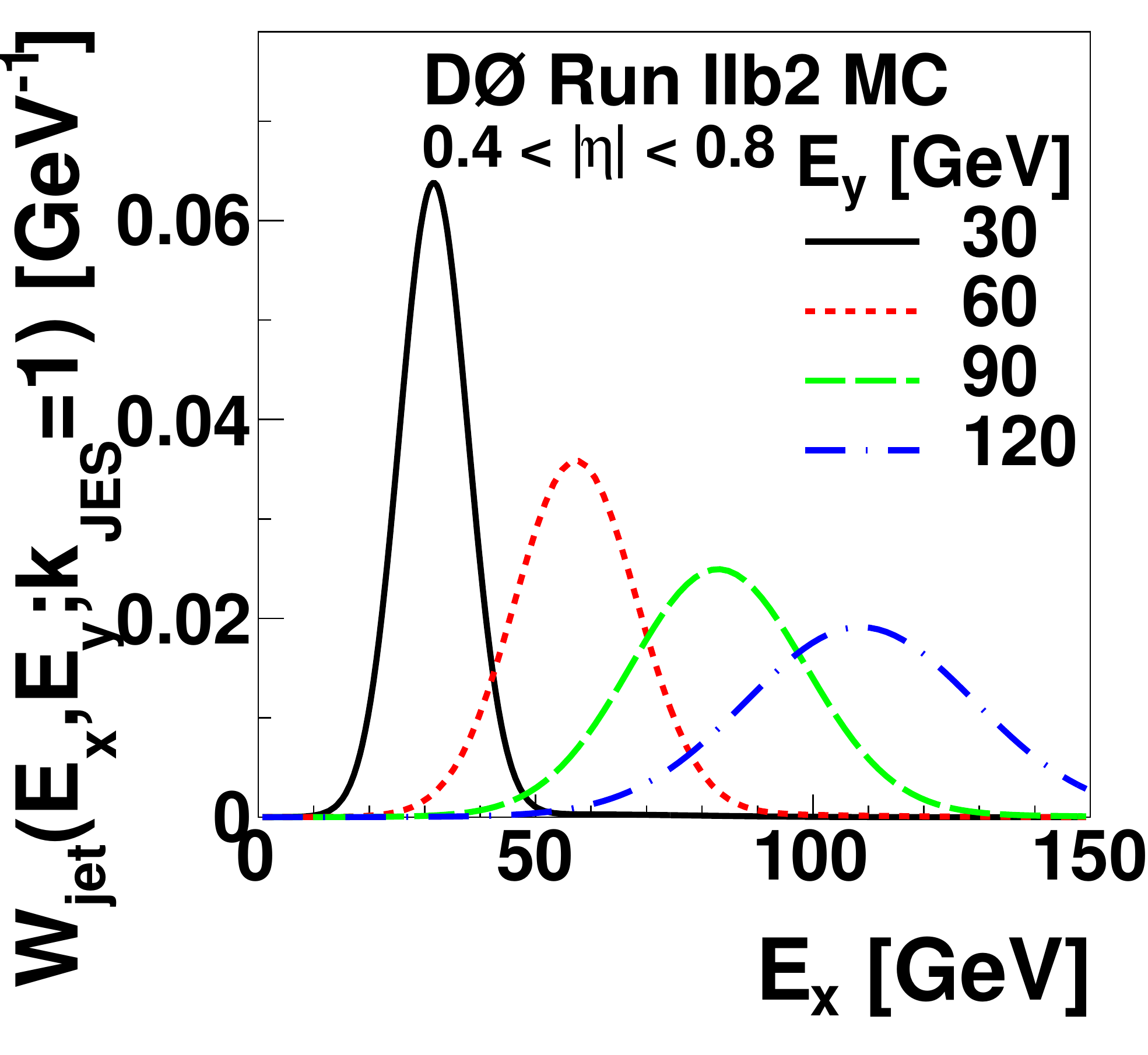}
\put(26,82){\footnotesize\textsf{\textbf{(b)}}}
\end{overpic}
\begin{overpic}[width=0.49\columnwidth]{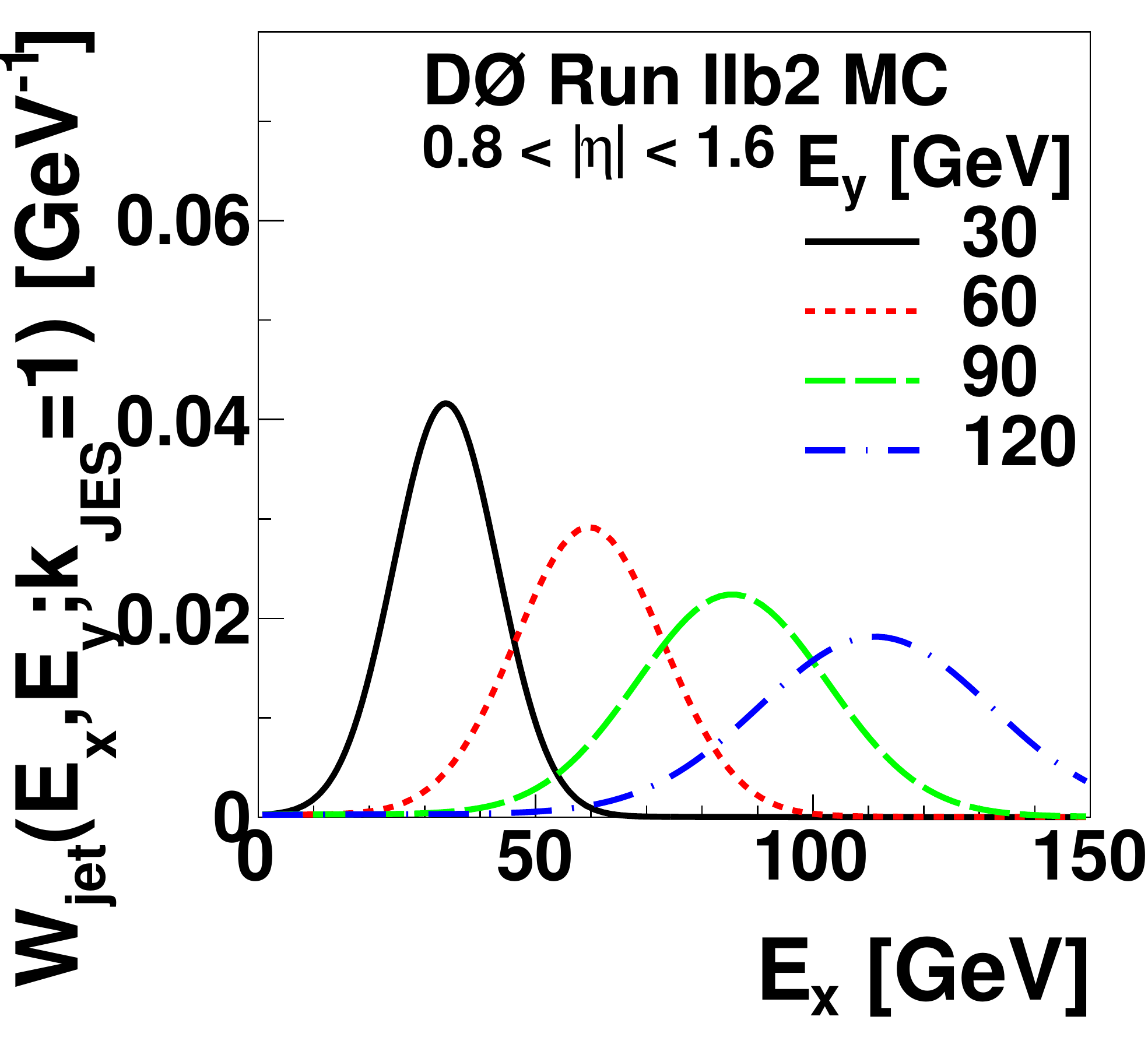}
\put(26,82){\footnotesize\textsf{\textbf{(c)}}}
\end{overpic}
\begin{overpic}[width=0.49\columnwidth]{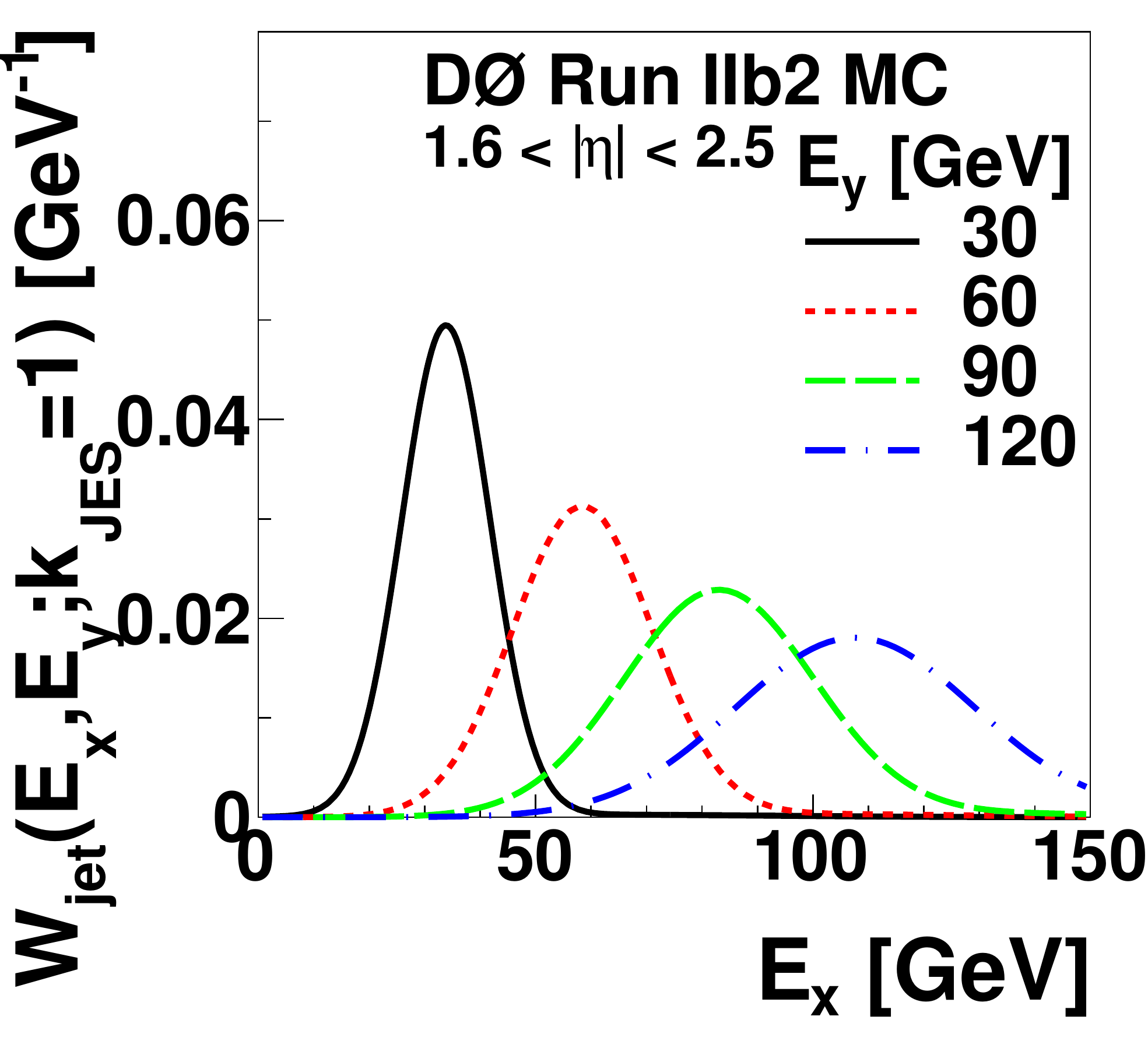}
\put(26,82){\footnotesize\textsf{\textbf{(d)}}}
\end{overpic}
\caption{\label{fig:tfs}
Same as Fig.~\ref{fig:tfl}, but for $b$-quark jets with at least one muon inside the jet cone. The TF are shown in four pseudorapidity regions: (a) $|\eta|\leq0.4$, (b) $0.4<|\eta|\leq0.8$, (c) $0.8<|\eta|\leq1.6$, and (d) $1.6<|\eta|\leq2.5$.
}
\end{figure}

\begin{figure}
\centering
\begin{overpic}[width=0.49\columnwidth]{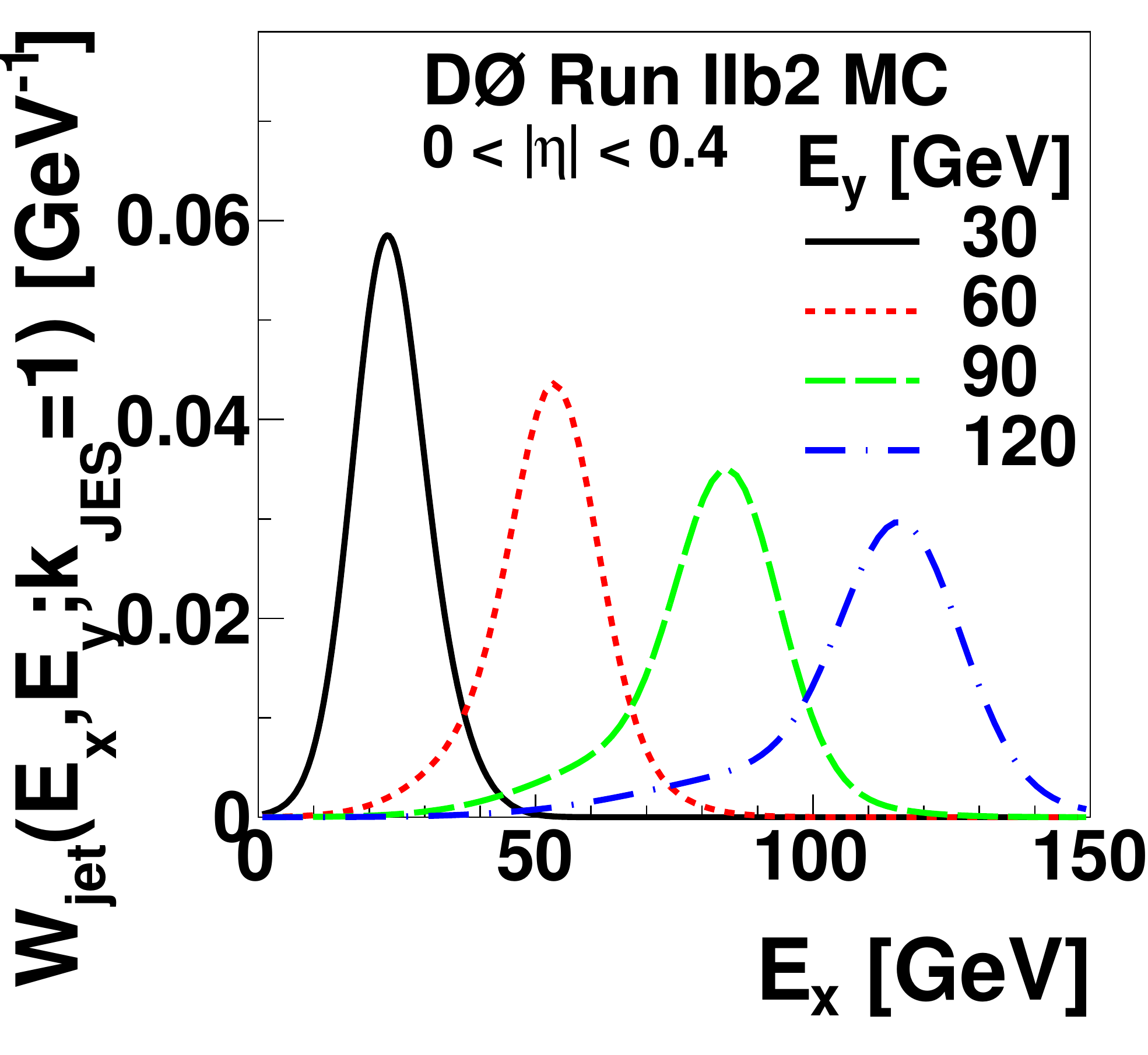}
\put(26,82){\footnotesize\textsf{\textbf{(a)}}}
\end{overpic}
\begin{overpic}[width=0.49\columnwidth]{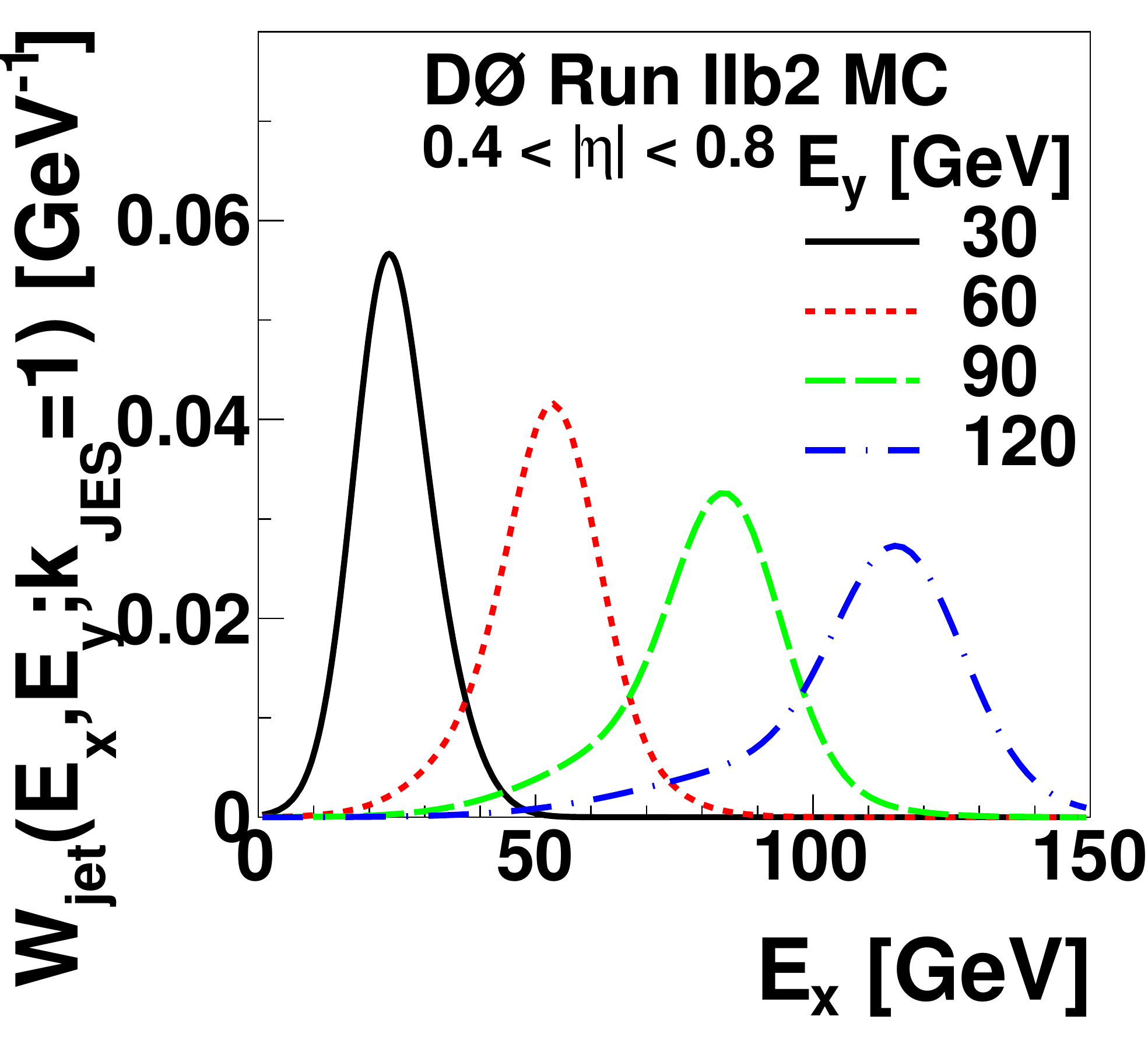}
\put(26,82){\footnotesize\textsf{\textbf{(b)}}}
\end{overpic}
\begin{overpic}[width=0.49\columnwidth]{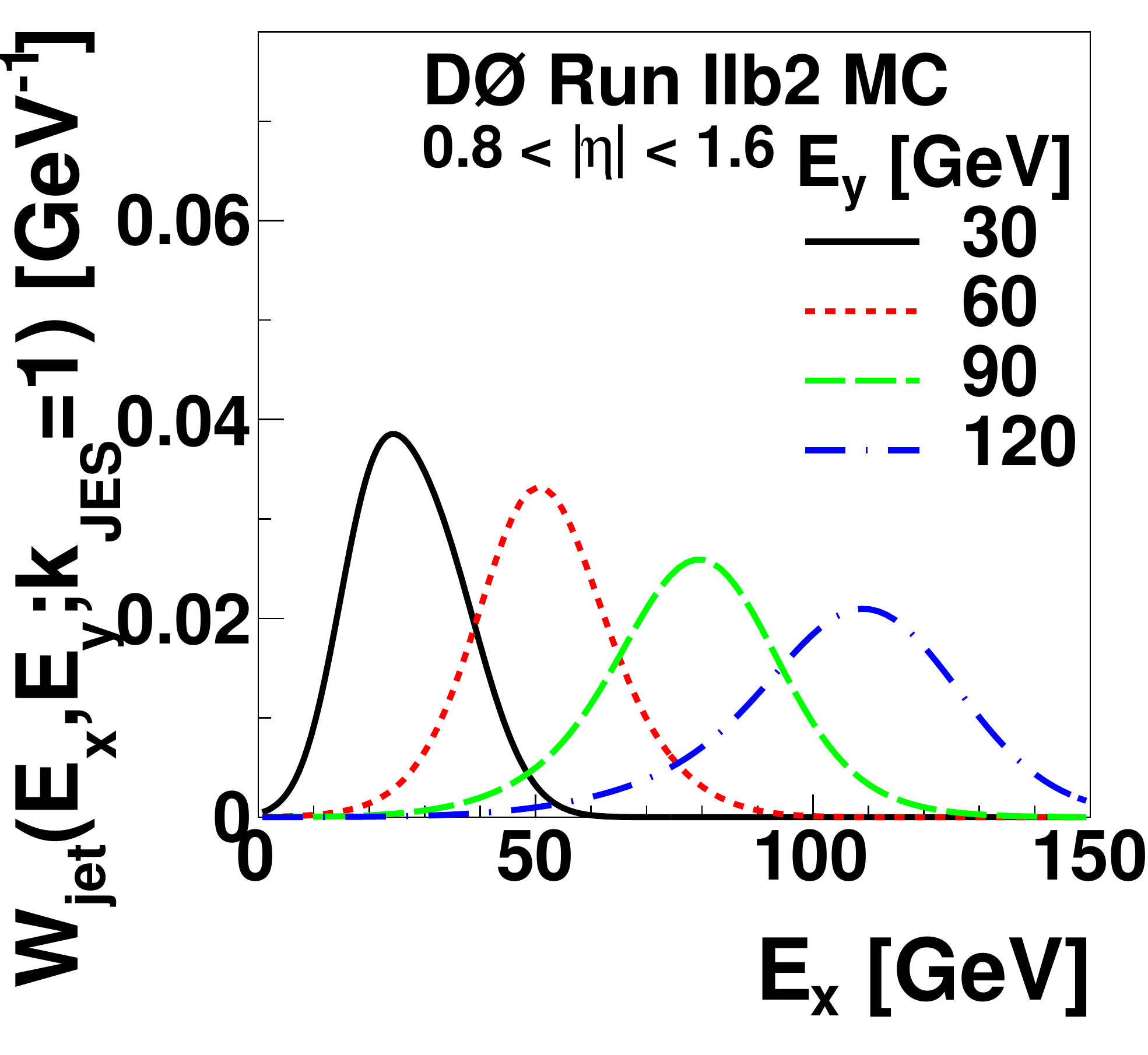}
\put(26,82){\footnotesize\textsf{\textbf{(c)}}}
\end{overpic}
\begin{overpic}[width=0.49\columnwidth]{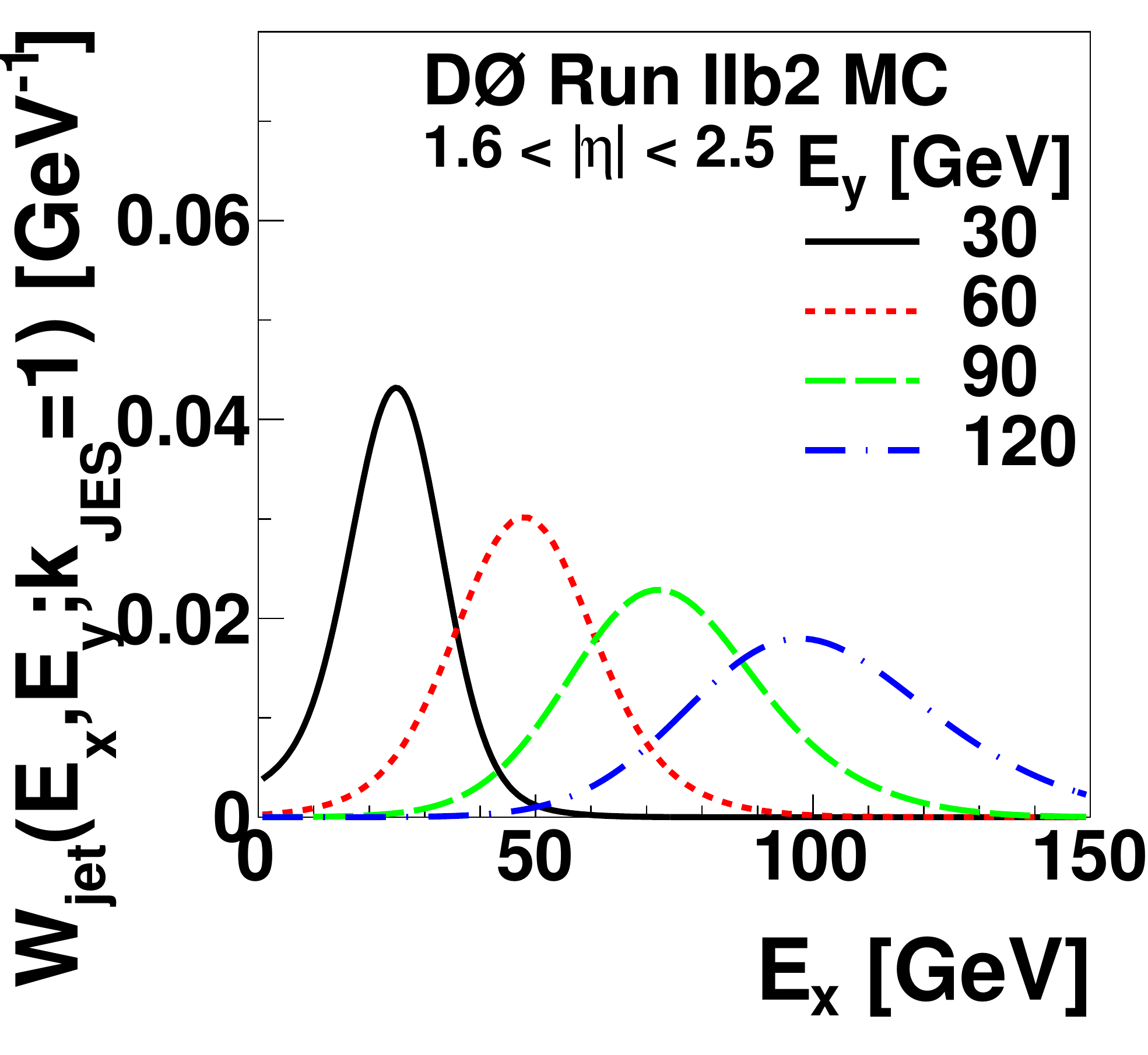}
\put(26,82){\footnotesize\textsf{\textbf{(d)}}}
\end{overpic}
\caption{\label{fig:tfb}
Same as Fig.~\ref{fig:tfl}, but for $b$-quark jets without muons inside the jet cone. The TF are shown in four pseudorapidity regions: (a) $|\eta|\leq0.4$, (b) $0.4<|\eta|\leq0.8$, (c) $0.8<|\eta|\leq1.6$, and (d) $1.6<|\eta|\leq2.5$.
}
\end{figure}

The parameters $a_{i}$ and $b_{i}$ in Eq.~(\ref{eq:aibi}) are determined from fully simulated \ttbar events for each data-taking epoch, after applying the jet energy corrections and resolution smearing, which are required to match the detector response in data. To avoid potential bias from the kinematic selections, all the requirements described in Section~\ref{sec:selection} are applied at detector level, except for having exactly four jets, at least one of which is $b$-tagged, and except the $\pt$ threshold for the jets, which is lowered to 15~GeV. The parameters $a_i$ and $b_i$ are extracted from MC generated \ttbar events with $\mt$ values ranging between $150$ and $190~\GeV$, for each data-taking epoch. We use an unbinned likelihood fit that maximizes the product of the $\wjet$ terms for all jets in all events as a function of  $a_{i}$ and $b_{i}$. Separate sets of $a_i$ and $b_i$ parameters are derived for three categories of jets: (i)~from light quarks ($u$, $d$, $s$) and charm, (ii)~from $b$~quarks with a muon of a relatively low \pt within the jet cone (soft muon tag), and (iii)~for all other $b$~quarks. This is done for four regions of $\eta$: $|\eta|\leq0.4$, $0.4<|\eta|\leq0.8$, $0.8<|\eta|\leq1.6$, and $1.6<|\eta|\leq2.5$ to take into account the $\eta$~dependence of the detector response. These same four regions are used to derive the correction for the flavor-dependence of the detector response to jets~\cite{bib:jes}. When deriving the TF, the matching of jets to partons is performed using a simple cone-matching algorithm, where a jet is considered uniquely matched to a parton if $\Delta R({\rm jet},{\rm closest~parton})<R_{\rm cone}/2=0.25$. The cone-matching parameter is chosen to maximize the matching efficiency, while minimizing the mismatching rate of quarks to jets from pile up. The probability of a detector-level jet not to be matched to a particle-level jet is 2\% per jet.  Figure~\ref{fig:tfl} illustrates the TF for jets from category (i) as a function of $E_{x}$ for different values of $E_{y}$ for Run~IIb2, which is representative of all four data-taking epochs. The TF for jets from category~(ii) is shown in Fig.~\ref{fig:tfs}, and for category~(iii) in Fig.~\ref{fig:tfb}. The values of the $a_{i}$ and $b_{i}$ parameters for each of the four data-taking epochs are summarized in Appendix~\ref{app:tf}.

\begin{figure}
\centering
\begin{overpic}[width=0.49\columnwidth]{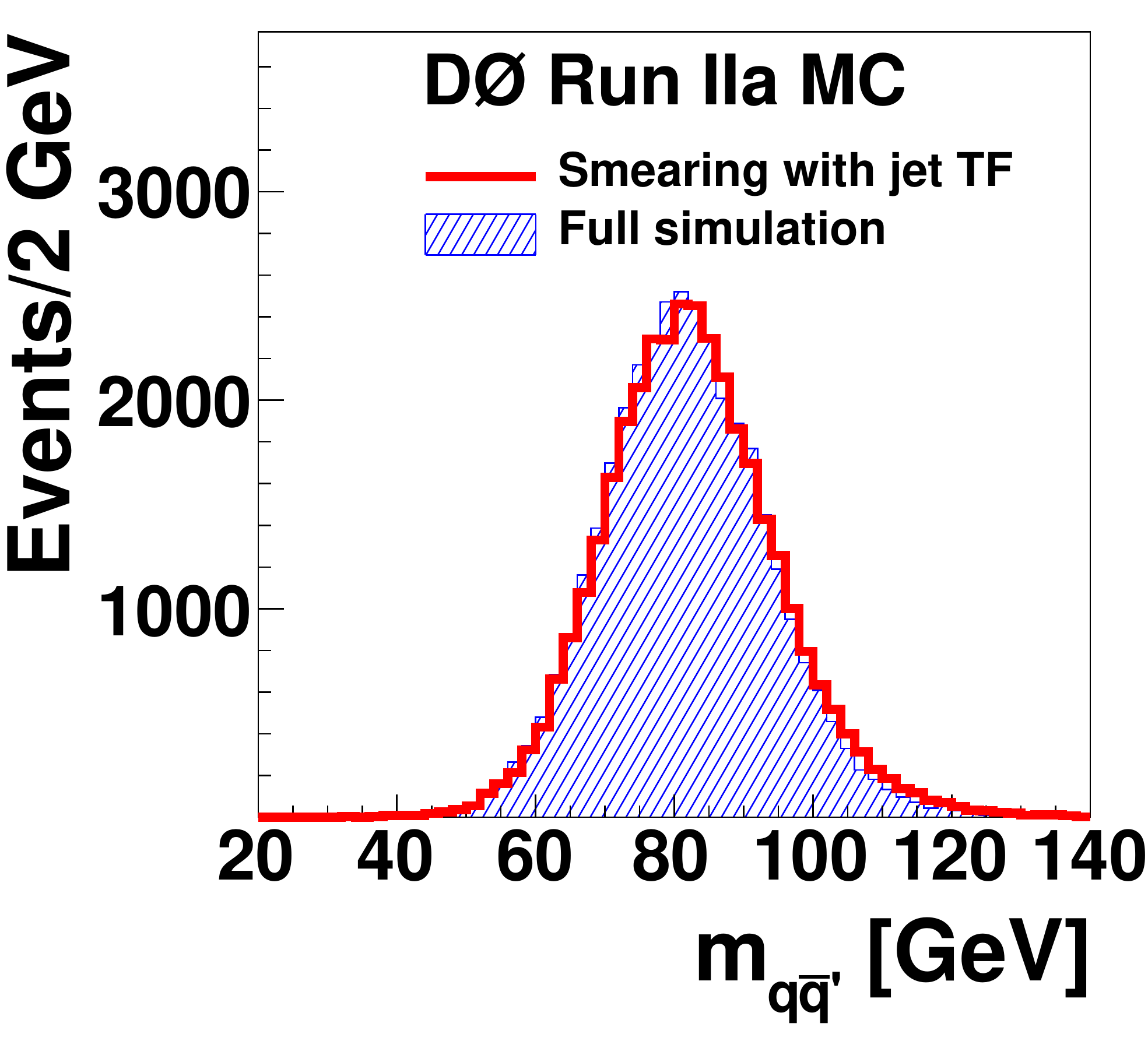}
\put(26,82){\footnotesize\textsf{\textbf{(a)}}}
\end{overpic}
\begin{overpic}[width=0.49\columnwidth]{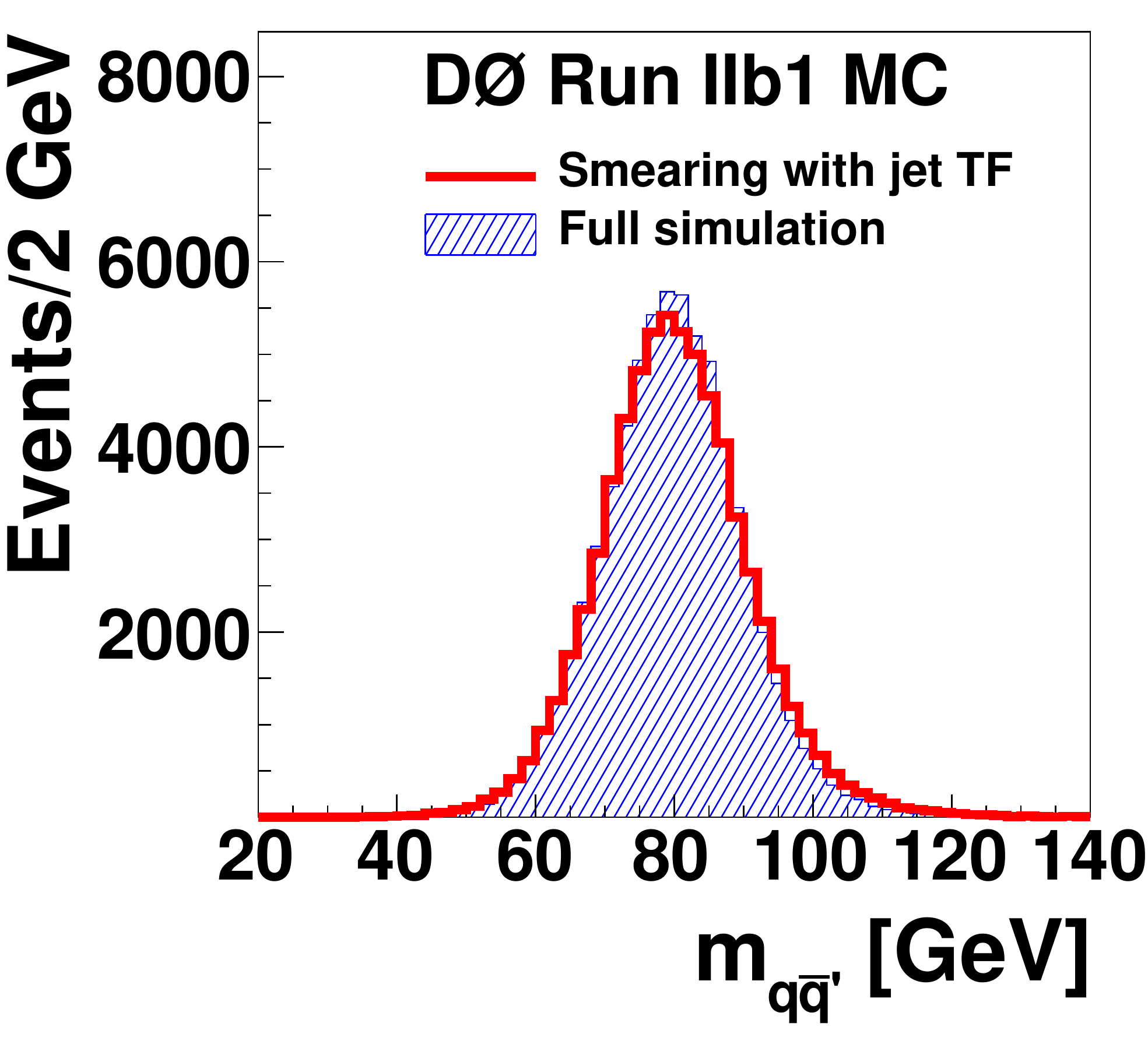}
\put(26,82){\footnotesize\textsf{\textbf{(b)}}}
\end{overpic}
\begin{overpic}[width=0.49\columnwidth]{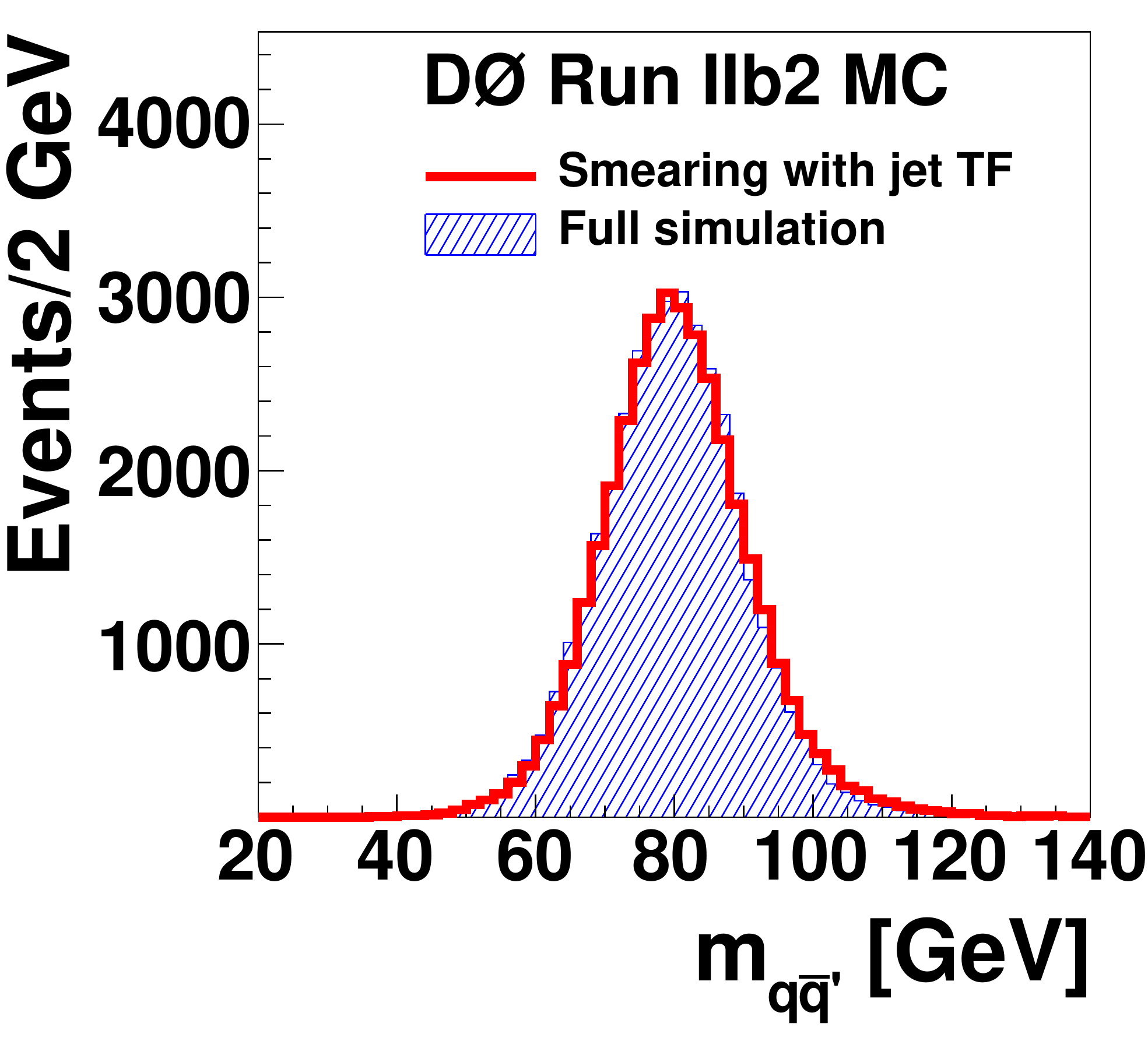}
\put(26,82){\footnotesize\textsf{\textbf{(c)}}}
\end{overpic}
\begin{overpic}[width=0.49\columnwidth]{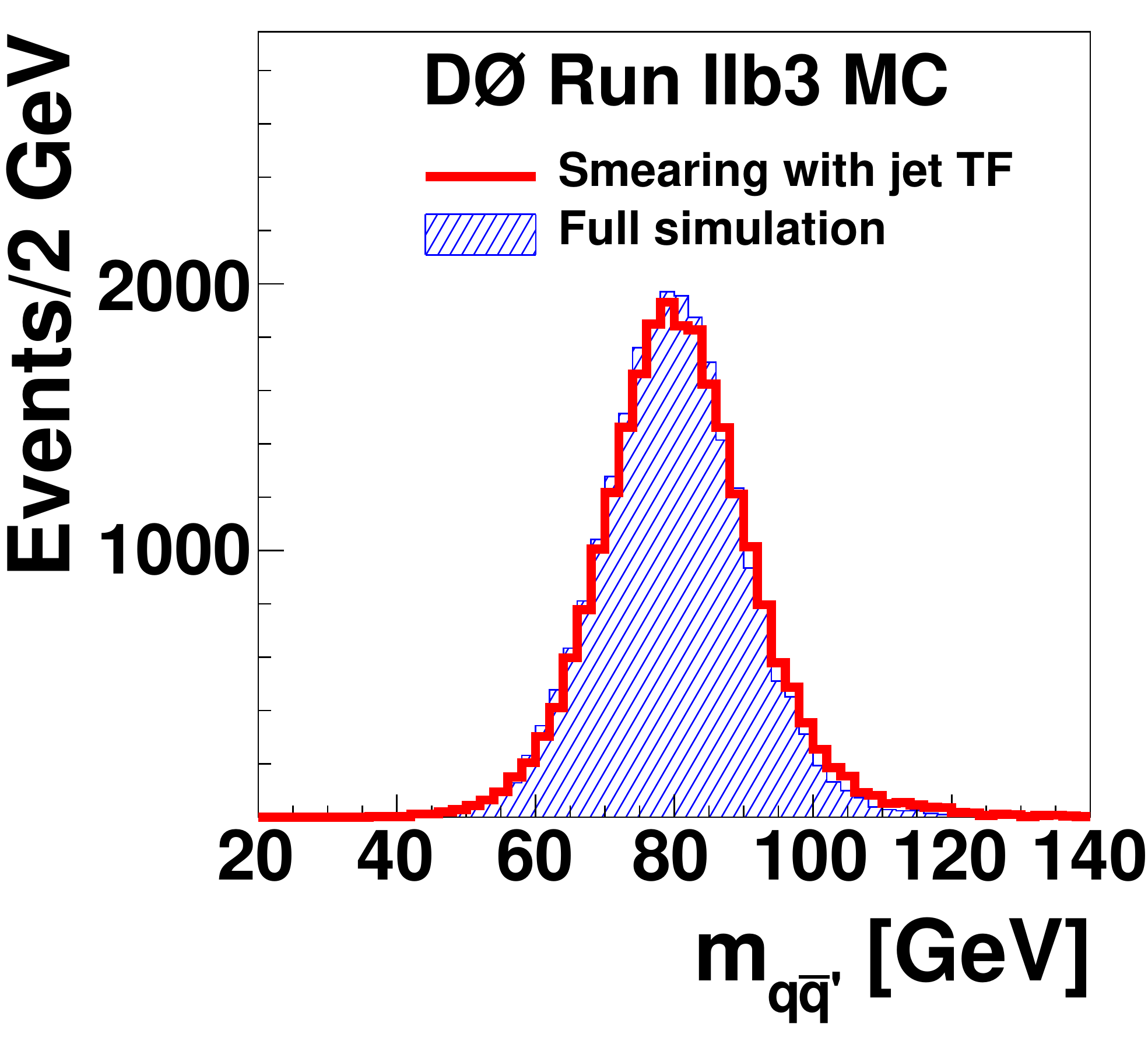}
\put(26,82){\footnotesize\textsf{\textbf{(d)}}}
\end{overpic}
\caption{\label{fig:tf2j}
Invariant mass of the dijet system matched to the $W\to q\bar q'$ decay for (a)~Run~IIa, (b)~Run~IIb1, (c)~Run~IIb2, and (d)~Run~IIb3.
Fully simulated jet-parton matched MC events passing all selection requirements are compared to partons with energies smeared using the jet TF, and then applying the same standard event selection criteria.}
\end{figure}

\begin{figure}
\centering
\begin{overpic}[width=0.49\columnwidth]{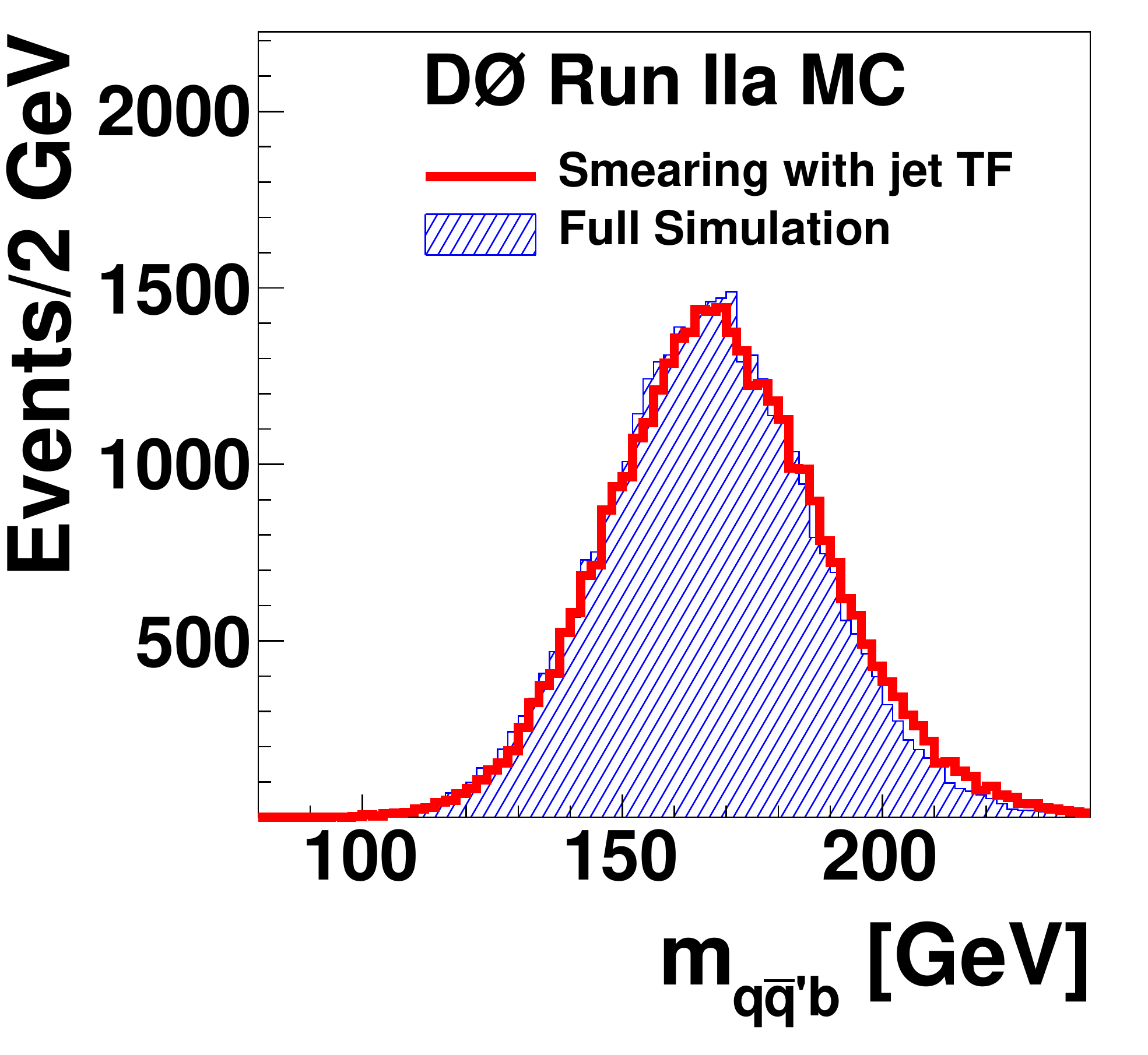}
\put(26,82){\footnotesize\textsf{\textbf{(a)}}}
\end{overpic}
\begin{overpic}[width=0.49\columnwidth]{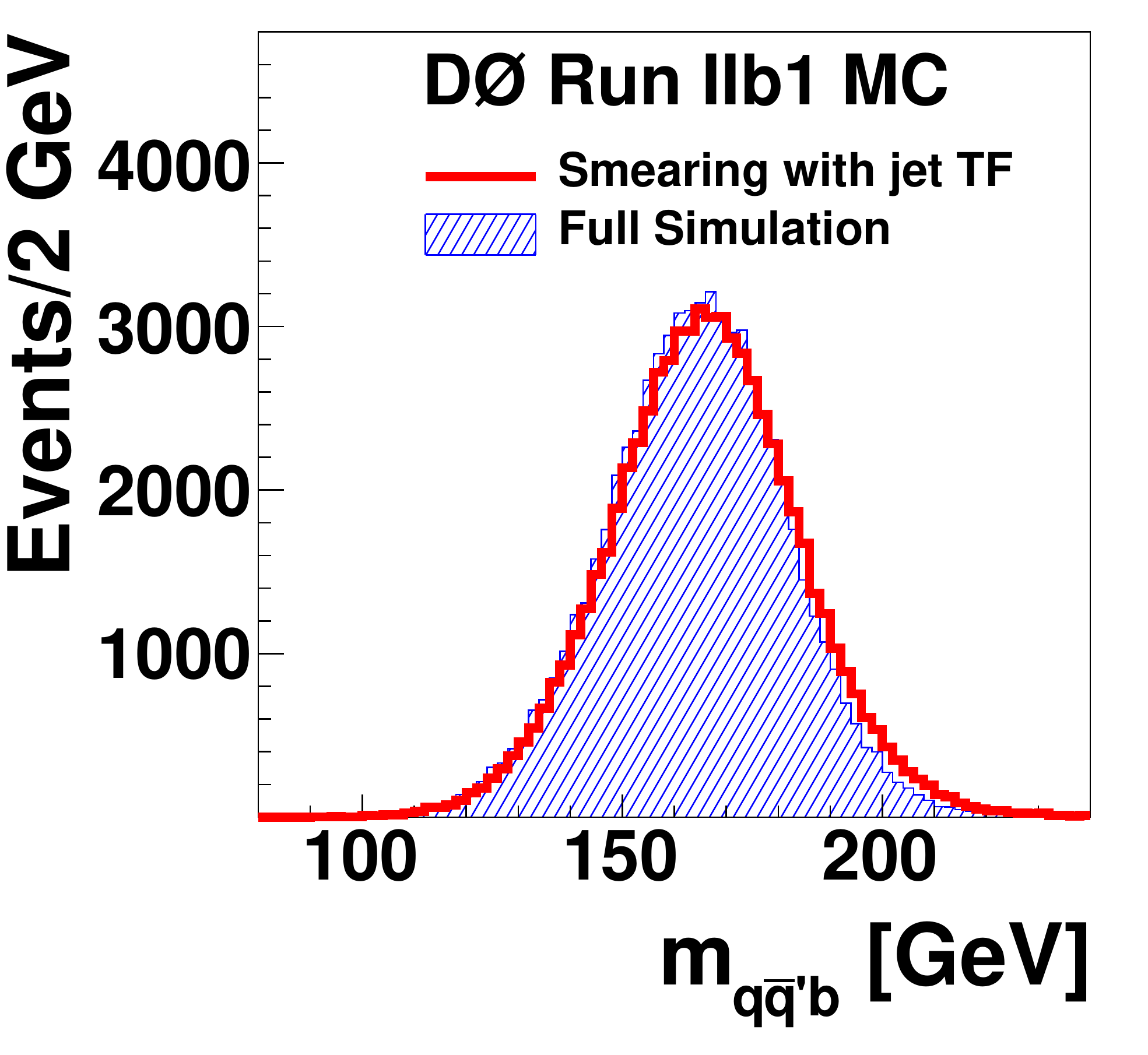}
\put(26,82){\footnotesize\textsf{\textbf{(b)}}}
\end{overpic}
\begin{overpic}[width=0.49\columnwidth]{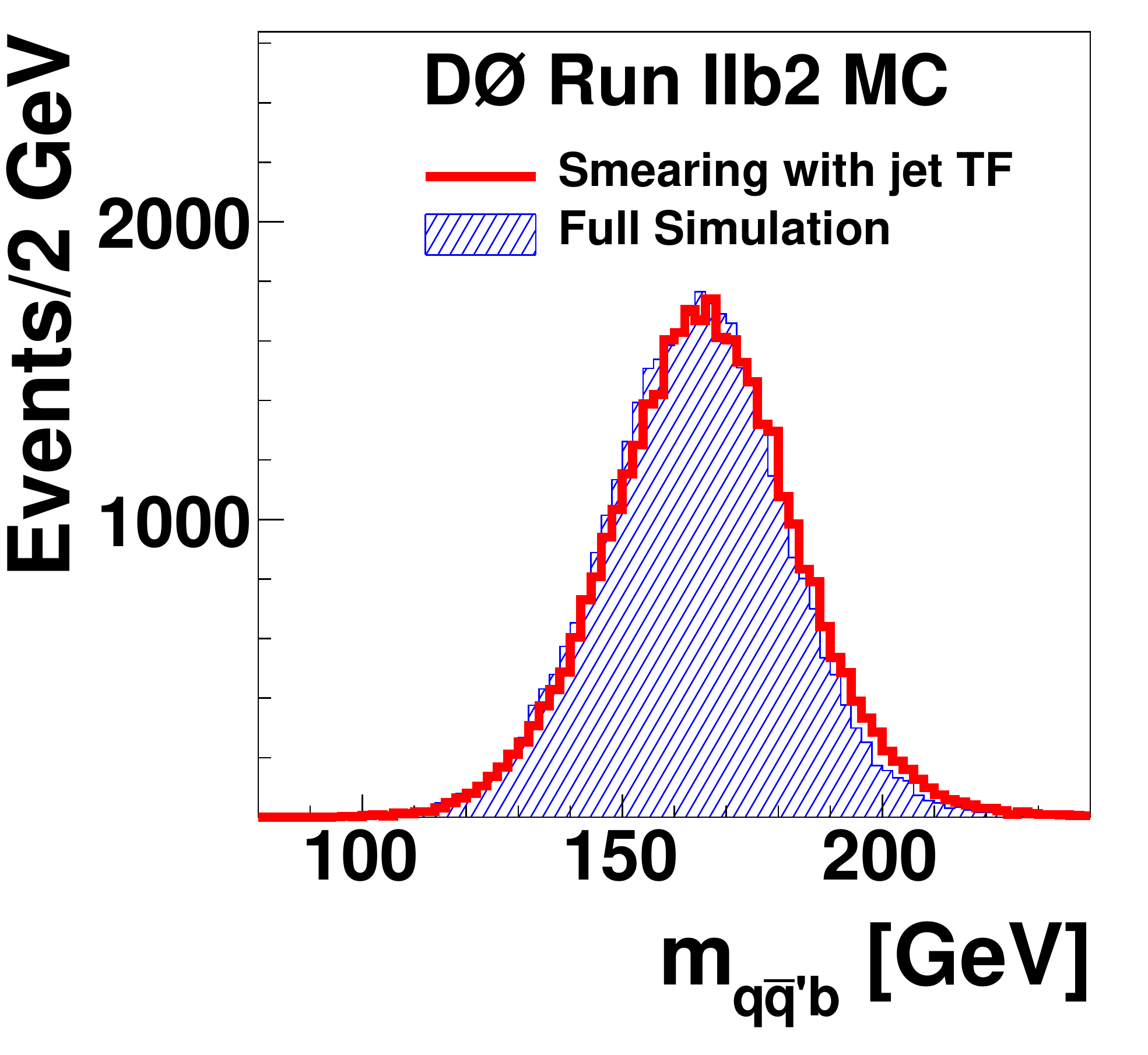}
\put(26,82){\footnotesize\textsf{\textbf{(c)}}}
\end{overpic}
\begin{overpic}[width=0.49\columnwidth]{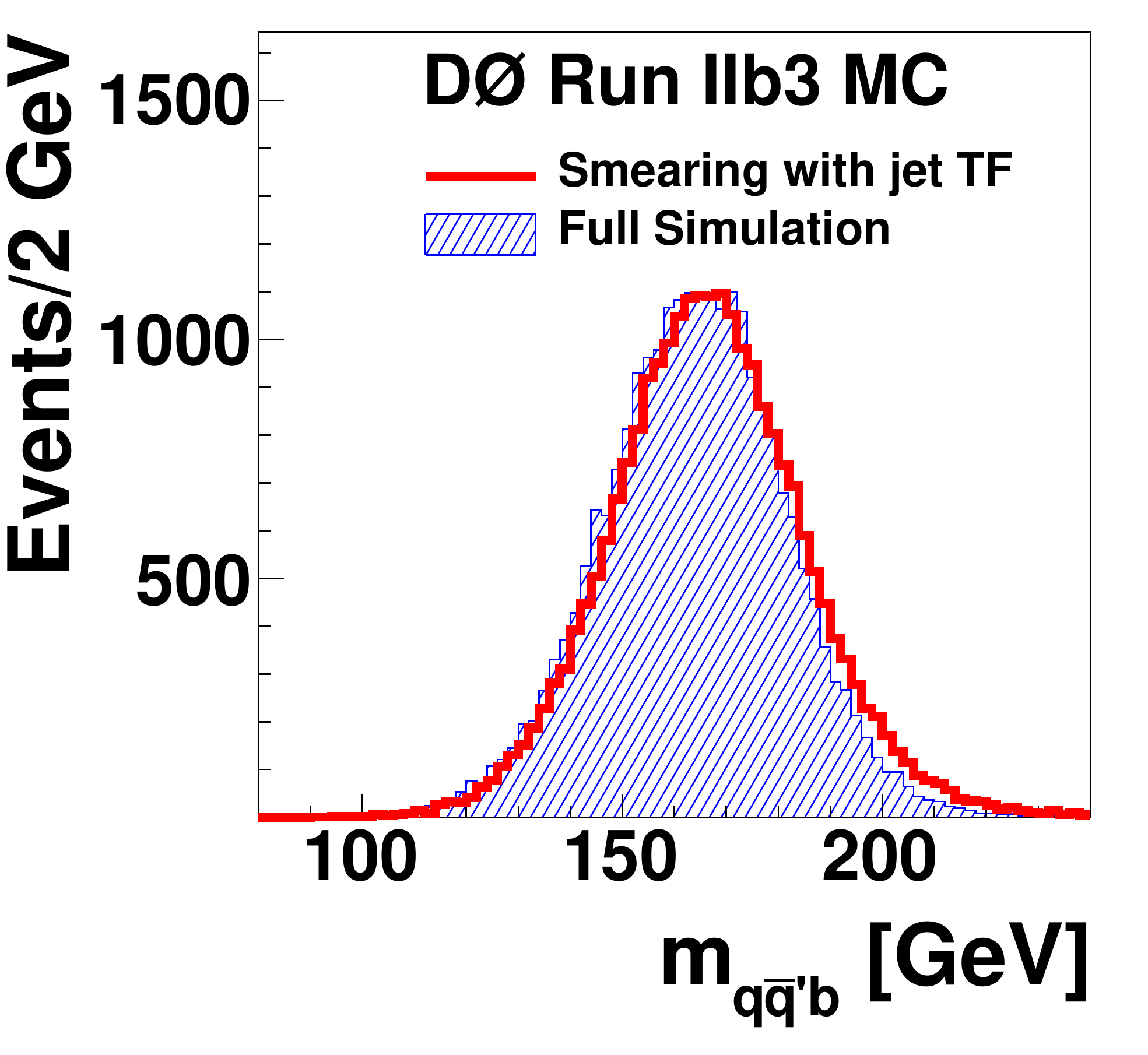}
\put(26,82){\footnotesize\textsf{\textbf{(d)}}}
\end{overpic}
\caption{\label{fig:tf3j}
Same as Fig.~\ref{fig:tf2j}, but for the invariant mass of the trijet system matched to the hadronic decay of the top quark for (a)~Run~IIa, (b)~Run~IIb1, (c)~Run~IIb2, and (d)~Run~IIb3.
}
\end{figure}

To verify that the TF describes the detector response to jets, we use fully simulated \ttbar MC events with all jet energy corrections and smearing needed to match resolutions in data. They must pass all selections summarized in Section~\ref{sec:selection}, and each of the four jets needs to be uniquely matched to a quark. In Fig.~\ref{fig:tf2j} we compare the distributions in the invariant mass of the dijet system that matches the two quarks from the $W\to q\bar q'$ decay, and the invariant mass of those two quarks, smeared with the jet TF. We compare the distributions in the invariant mass of the trijet system matched to one of the top quarks and the corresponding quarks smeared with the jet TF in Fig.~\ref{fig:tf3j}. Both tests indicate that the jet TF well describe the jet resolutions. Any residual disagreements are taken into account through the calibration procedures, described  in Sections~\ref{sec:calibf} and~\ref{sec:calibmt}.

\subsubsection{
Modeling of electron momentum
}
\label{sec:tfem}

The electron energy resolution is parameterized by the TF
\begin{linenomath}
\begin{equation}
W_{e}\left(E_{x},E_{y}\right)=\frac{1}{\sqrt{2\pi}\sigma}\exp\left[-\frac{1}{2}\left(\frac{E_{x}-E_{y}^\prime}{\sigma}\right)^{2}\right],
\end{equation}
\end{linenomath}
with
\begin{linenomath}
\begin{eqnarray}
E_{y}^\prime & = & E_y + 0.324~\GeV\,,\\
\sigma & = & \sqrt{(0.028\cdot E_{y}^\prime)^{2}+(S\cdot E_{y}^\prime)^{2}+(0.4~{\rm GeV})^{2}}\,,\\
S & = & \frac{0.164~{\rm GeV}^{\frac{1}{2}}}{\sqrt{E_y^\prime}}
    + \frac{0.122~{\rm GeV}}{E_y^\prime}
\exp\left(\frac{C}{\sin\theta_e}\right)-C\,,\hspace{10pt}\\
C & = & 1.3519-\frac{2.0956\,{\rm GeV}}{E_{y}^\prime}-\frac{6.9858\,{\rm GeV}^2}{E_{y}^{\prime2}}\,,
\end{eqnarray}
\end{linenomath}
where $E_x$ is the reconstructed electron energy, $E_y$ is the electron energy at parton level, and $\theta_e$ is the polar angle of the electron relative to the proton beam direction.  These parameters are obtained from the modeling of electron energy response and resolution in Ref.~\cite{bib:wmassprl} for Run~IIa, and are used for all four data-taking epochs because they changed only marginally between the epochs.

\subsubsection{
Modeling of muon momentum
}
\label{sec:tfmu}

The resolution of the central tracker, which provides the measurement of the muon \pt, is described through the uncertainty on the curvature $\kappa\equiv 1/\pt$ of the muon track. The muon TF is parameterized as
\begin{linenomath}
\begin{equation}
W_{\mu}\left(\kappa_x,\kappa_y\right)=\frac{1}{\sqrt{2\pi}\sigma}
\exp\left[-\frac{1}{2}\left(\frac{\kappa_{x}-\kappa_{y}}{\sigma}\right)^{2}\right]\,,
\end{equation}
\end{linenomath}
where $\kappa_x$ is the reconstructed muon curvature, and $\kappa_y$ is the corresponding quantity at parton level.

The resolutions
\begin{linenomath}
\begin{equation}
\sigma = 
\left\{
\begin{array}{cc}
s & {\rm for}\ |\eta|\le1.4\\
\sqrt{s^{2}+\left\{c\cdot\left(|\eta|-1.4\right)\right\}^{2}} & {\rm for}\
|\eta|>1.4
\end{array}\right\}
\end{equation}
\end{linenomath}
are obtained from muon tracks in simulated events, where the $s$ and $c$
parameters are linear functions of $\kappa_y$:
\begin{linenomath}
\begin{eqnarray}
s & = & s_0+s_1\,\kappa_y\,,\\
c & = & c_0+c_1\,\kappa_y\,.
\end{eqnarray}
\end{linenomath}
The values of these coefficients, which are the same for all four data-taking epochs are given in Table \ref{tab:mu_tfpar} for muon tracks with at least three associated hits in the SMT, and for all other tracks.
This simplified parameterization of the momentum resolution is valid at high transverse momenta ($p_{T}>20$~GeV) where the limitations in coordinate resolution dominate over the effects of multiple scattering. 

\begin{table}
\caption{
\label{tab:mu_tfpar} 
Parameters of the muon TF for muons with at least three associated hits in the SMT and for all other tracks. These values are the same for all the four data-taking epochs.
}
\centering
\begin{ruledtabular}
\begin{tabular}{lcc}
\multirow{2}{*}{Parameter~}
          & $\geq 3$ hits & $< 3$ hits \\
          & in the SMT & in the SMT \\
\hline\vspace{-3mm}\\
$s_0 ~({\rm GeV}^{-1})$ &~$2.082\times10^{-3}$~&~$3.620\times10^{-3}$ \\
$s_1$             & $1.125\times10^{-2}$ & $1.388\times10^{-2}$ \\
$c_0~({\rm GeV}^{-1})$       & $7.668\times10^{-3}$ & $2.070\times10^{-2}$ \\
$c_1$                   & $7.851\times10^{-2}$ & $7.042\times10^{-2}$ \\
\end{tabular}
\end{ruledtabular}
\end{table}

\subsection{
Extraction of the parameters of interest
}
\label{sec:extract}

In this Section, we discuss how the parameters $f$, $\mt$, and $\kjes$ are extracted from a sample of $N$ selected events that are characterized by their measured quantities $\X_N=\{\x_1,\x_2,...,\x_N\}$.  A likelihood function, $\like(\X_N;\mt,\kjes,f)$, is constructed at each grid point from the product of the individual \pevt values according to 
\begin{linenomath}
\begin{equation}
\like(\X_N;\,\mt,\kjes,f)\equiv\prod_{i=1}^{N}\pevt(\x_i;\,\mt,\kjes,f)\,.\label{eq:likejoint}
\end{equation}
\end{linenomath}

To construct this likelihood from \pevt, \psig is calculated on a grid in $(\mt,\kjes)$ with spacings of $(1~\GeV,0.01)$. For simulated $\ttbar$ events, it covers 25 points in \mt ranging between $\mt^{\rm gen}-12~\GeV$ and $\mt^{\rm gen}+12~\GeV$, and 21 points in \kjes between $\kjes^{\rm gen}-0.1$ and $\kjes^{\rm gen}+0.1$.  The ranges are chosen to ensure that they contain the numerically relevant part of the likelihood for all pseudo-experiments (cf. Sections~\ref{sec:calibf} and~\ref{sec:calibmt}). For data, the grid is extended to $[153~\GeV,192~\GeV]$  in \mt and to $[0.85,1.15]$ in \kjes. This choice is made because the values of \mt and \kjes in data are not known {\em a priori}, as the analysis is performed blinded, following the procedure to be described in Section~\ref{sec:resultmt}. For simulated background events, the same grid is chosen as for data.  Since \pbkg does not depend on \mt, it is calculated on a grid in \kjes only, with the same spacing and ranges as \psig.

\subsubsection{
Extraction of the signal fraction
}

The likelihood is determined by the value of the signal fraction $\fmax(\mt,\kjes)$ that maximizes \like for every assumed pair of $(\mt,\kjes)$ values from the grid on which \like is calculated. As will be detailed in Sections~\ref{sec:f} and~\ref{sec:mt}, the measurement of the signal fraction is performed prior to the measurement of \mt and \kjes. To extract $\ffit$ independently, the parameters \mt and \kjes are integrated over a uniform prior
\begin{linenomath}
\begin{equation}
\ffit = \frac{\int\dif\mt\dif\kjes~\fmax\like(\X_N;\,\mt,\kjes,\fmax)}
             {\int\dif\mt\dif\kjes~     \like(\X_N;\,\mt,\kjes,\fmax)}\,,
\end{equation}
\end{linenomath}
by summing over all points $(\mt,\kjes)$ on the grid that reflect the numerically relevant part of the likelihood, cf.\ Section~\ref{sec:extract}.

\subsubsection{
Extraction of the top-quark mass and $\kjes$
}

To obtain the best unbiased estimate of \mtfit, the two-dimensional likelihood $\like(\X_N;\,\mt,\kjes)=\like(\X_N;\,\mt,\kjes,\fmax(\mt,\kjes))$ is projected onto the \mt axis
\begin{linenomath}
\begin{equation}
\label{eq:likemt}
\like(\X_N;\,\mt)=\int\dif\kjes~\like(\X_N;\,\mt,\kjes)
\end{equation}
\end{linenomath}
using Simpson's rule~\cite{bib:simpson}. The best unbiased estimate for \mtfit is then given by
\begin{linenomath}
\begin{equation}
\label{eq:mtfit}
\mtfit = \frac{\int\dif\mt~\mt\like(\X_N;\,\mt)}
              {\int\dif\mt~\like(\X_N;\,\mt)}\,,
\end{equation}
\end{linenomath}
and its uncertainty by
\begin{linenomath}
\begin{equation}
\label{eq:dmtfit}
\dmtfit = \left\{\frac{\int\dif\mt~(\mt-\mtfit)^2\like(\X_N;\,\mt)}
               {\int\dif\mt~\like(\X_N;\,\mt)}\right\}^{\frac12}\,.
\end{equation}
\end{linenomath}
Thus, by construction, the statistical uncertainty on \mt includes the impact of the statistical uncertainty on \kjes. To  obtain the best unbiased estimates for \kjesfit and \dkjesfit, the same procedure is applied as to estimate \mtfit and \dmtfit, by replacing  $\kjes\leftrightarrow\mt$ in Eqs.~(\ref{eq:likemt}-\ref{eq:dmtfit}).

\subsection{
Calculation of the signal probability
\label{sec:psigcalc}
}

In the calculation of \psig in  Eq.~(\ref{eq:psigfull}), a total of 24 variables are associated with the integration over the momenta of the two colliding quarks  $\dif\vec q_{1}\dif\vec q_{2}$ and over the phase space of the six-body final state, $\dif\Phi_6$. After taking into account the $\delta^2(\eta,\phi)$ functions for the directions of the four jets and the charged lepton according to the TF of Eq.~(\ref{eq:tf}), the dimensionality of the integral is reduced by $2\times5=10$. Imposing energy-momentum conservation provides 4 additional constraints, resulting in a 10-dimensional integral that must be evaluated. The transverse momentum of the neutrinos is inferred from momentum conservation, resulting in up to eight solutions for the neutrino kinematics, which are averaged. The measurement of \met is not used because of its limited experimental resolution. 

The integration in Eq.~(\ref{eq:psigfull}) is performed numerically using MC integration~\cite{bib:mcmethod}. To reduce the computational demands of the integration, importance sampling~\cite{bib:numrec} is used. To fully exploit importance sampling, we perform a Jacobian transformation of the nominal 10 integration variables to optimized variables where prior information is either known or can be obtained easily.  This prior information is then used in the importance sampling. Thus, the integration is performed over the masses of the top and antitop quarks, which are assumed to be equal, the masses of the $W^+$ and $W^-$ bosons, the energy (curvature) of the electron (muon),  $E_q/(E_q+E_{\bar q'})$ for the quarks from the $W\to q\bar q'$ decay in the LO approximation, and the transverse momenta $q_{1,\rm x}, q_{1,\rm y}, q_{2,\rm x}, q_{2,\rm y}$ of the colliding quarks. The constraint $M_W = 80.4$~GeV for the {\it in-situ} JES calibration is imposed by integrating over $M_W$ according to a prior given by a Breit-Wigner distribution. More details about the motivation for the choice of integration variables can be found in Ref.~\cite{bib:nim}.

The computation of \psig in the previous version of this analysis~\cite{bib:mt36}, which is based on about 1/4 of the full Run~II integrated luminosity, and uses only one set of MC simulations, required two million CPU-hours~\cite{bib:cpuhours}.
To make our ME technique applicable to the full Run~II data sample of 9.7~\fb, and to perform a calibration of the method with dedicated MC simulations for each of the four data-taking epochs, we implement two new approaches to reduce computational demand. These approaches are described in detail in Ref.~\cite{bib:nim}, and are briefly summarized below.
\begin{enumerate}
\item
We utilize low-discrepancy sequences for the numerical MC integration of \psig according to Eq.~(\ref{eq:psigfull}), instead of conventional pseudo-random numbers in order to benefit from the key property of low-discrepancy sequences of sampling the integration phase space in a maximally uniform way. This leads to a convergence rate of the estimated integral inversely proportional to ${N_{\rm sample}}$~\cite{bib:bossert}, which is superior to ${\sqrt{N_{\rm sample}}}$ expected for pseudo-random numbers~\cite{bib:numrec}, where $N_{\rm sample}$ is the number of samplings of the integrand. We utilize the implementation of Bratley and Fox~\cite{bib:brat88} of the Sobol sequence~\cite{bib:sobol}, which is among the best performing low-discrepancy sequences.
\\
An accurate, and computationally efficient estimate of the numerical precision of the integral is a key ingredient of numerical MC integration. Since the  standard error estimator based on the variance of the integrand is too pessimistic for low-discrepancy sequences, because of their superior convergence rate, we implement a dedicated estimator of the numerical uncertainty according to Ref.~\cite{bib:warnock}, which represents a novel treatment in the context of the ME technique. 
\\
The utilization of low-discrepancy sequences is of  particular interest, as it is a procedure universally applicable to MC integration and independent of the computing hardware. Their implementation in our ME technique provides a reduction of about one order of magnitude in computation time, compared to pseudo-random numbers, for the same numerical accuracy.
\item
We factorize the \kjes parameter from the ME computation, a novelty in the context the measurements of \mt with the ME technique. As described in Section~\ref{sec:extract}, \psig has to be calculated for $25\times21=525$ grid points in $(\mt,\kjes)$ for simulated \ttbar events.
Instead of recalculating \psig for each point in $(\mt,\kjes)$, we calculate the integral
\begin{linenomath}
\begin{equation}
\sum_{{}^{\rm quark}_{\rm flavors}}~\int\limits_{\vec q_{1},\,\vec q_{2}}
                            \dif \vec q_{1}\dif \vec q_{2} f(\vec q_{1})f(\vec q_{2})
                            \frac{(2\pi)^{4}\left|{\cal M}_{q\bar q\to \ttbar}\right|^{2}}
                                 {2g_{\mu\nu}q_{1}^\mu q_{2}^\nu}\dif\Phi_{6}
\end{equation}
\end{linenomath}
in Eq.~(\ref{eq:psigfull}) only once for a given \mt, and use the result to calculate \psig for all $k_{{\rm JES},i=1,2,...,21}$ through multiplication by $W(\x,\y;k_{{\rm JES},i})$. This yields another order of magnitude reduction in computation time.
\end{enumerate}
In total, we obtain a reduction of computation time from about 2~h per event, averaged over the sample of simulated \ttbar events for $\mt=172.5~\GeV$, to about 80~s per event, i.e., by about two orders of magnitude. Both improvements are verified~\cite{bib:nim} to provide a performance of the ME technique consistent with that in Ref.~\cite{bib:mt36}. They prove essential for a dramatic improvement in the precision of this analysis by allowing to increase the number of simulated MC events used for the calibration of the ME technique (cf.\ Sections~\ref{sec:calibf} and~\ref{sec:calibmt}), and for the evaluation of systematic uncertainties (cf.\ Section~\ref{sec:syst}).

\section{
Measurement of the signal fraction
}
\label{sec:f}

The ME technique calculates \pevt from first principles with a LO ME and a parameterized description of the detector response through $W(\x,\y;\kjes)$. The calculation of the parameters of interest, in this case $f$, has to be therefore calibrated through simulations. The calibration of the response of the ME technique in $f$ and the subsequent extraction of $f$ from data are the subject of this Section.

\subsection{
Calibration of the method
}
\label{sec:calibf}
Fully simulated \ttbar events, as described in Section~\ref{sec:mc}, at $\mt^{\rm gen}=172.5~\GeV$ and $\kjes^{\rm gen}=1$, expected from the standard JES calibration procedure~\cite{bib:jes}, are used for calibrating the response of the ME technique in $f$. Similarly, fully simulated \wjets events at $\kjes^{\rm gen}=1$ are included to model the dominant background contribution from \wjets production. We  refine the calibration procedure relative to that of Ref.~\cite{bib:mt36} by explicitly accounting for the next-to-dominant contribution from MJ production by using a sample of events in data obtained as described in Section~\ref{sec:mc}.  Since the detector response, and therefore its parametrization, changes between data-taking epochs and final states, this calibration is obtained for each epoch and final state.  

At each of the five generated $f^{\rm gen}$ points, ranging between 0.5 and 0.9 in steps of 0.1, 1000 pseudo-experiments (PEs) are constructed, each with the same number of events as observed in data.  Since the matrix method provides a normalization of the background contribution from MJ production, each PE contains the same number of randomly drawn MJ events as determined through the matrix method, as shown in Table~\ref{tab:prefit}. The remaining number of background events is then drawn from the sample of simulated \wjets events,  according to a binomial distribution in $1-f^{\rm gen}$.

\begin{figure}
\centering
\begin{overpic}[width=0.49\columnwidth]{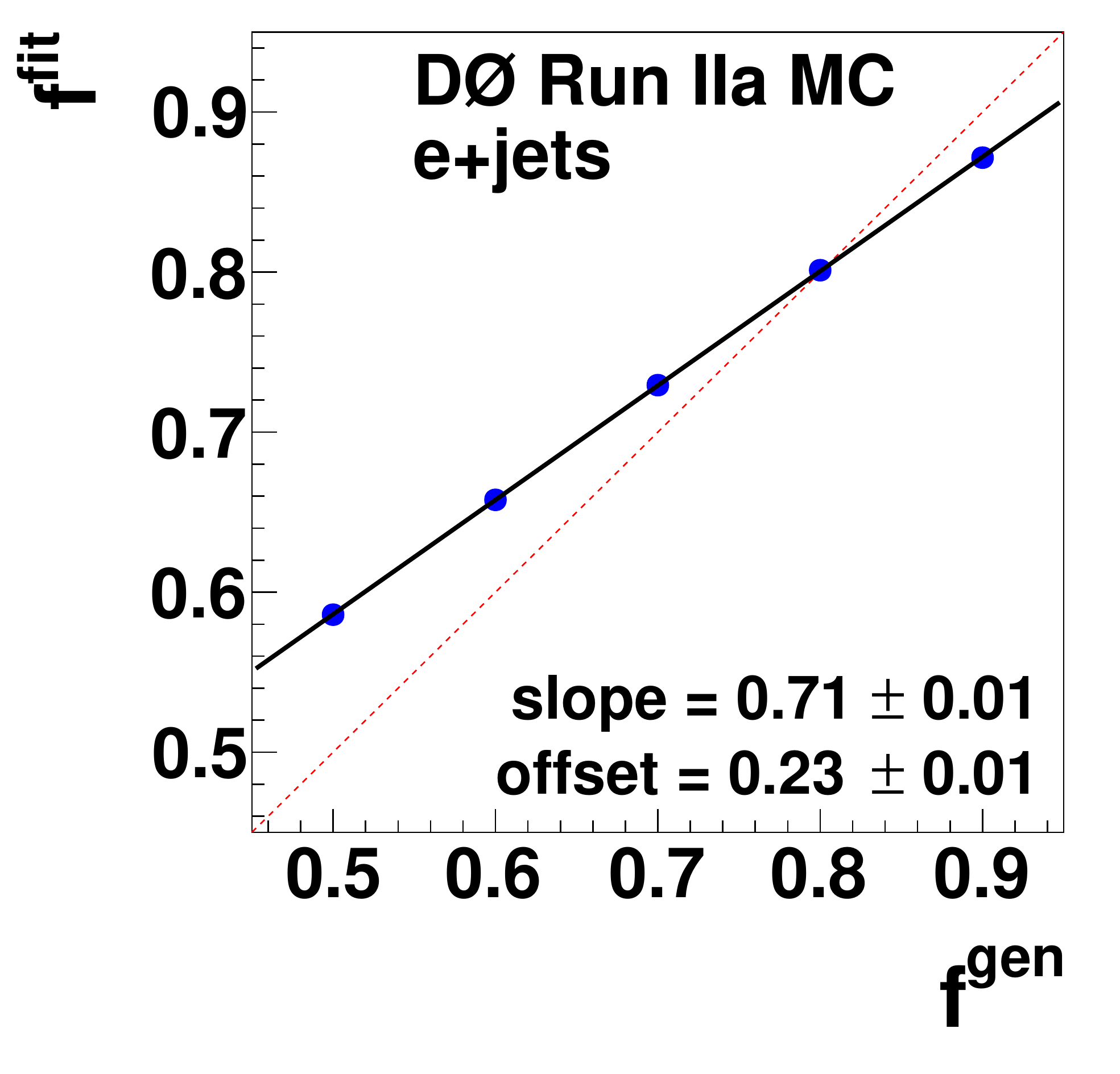}
\put(26,86){\footnotesize\textsf{\textbf{(a)}}}
\end{overpic}
\begin{overpic}[width=0.49\columnwidth]{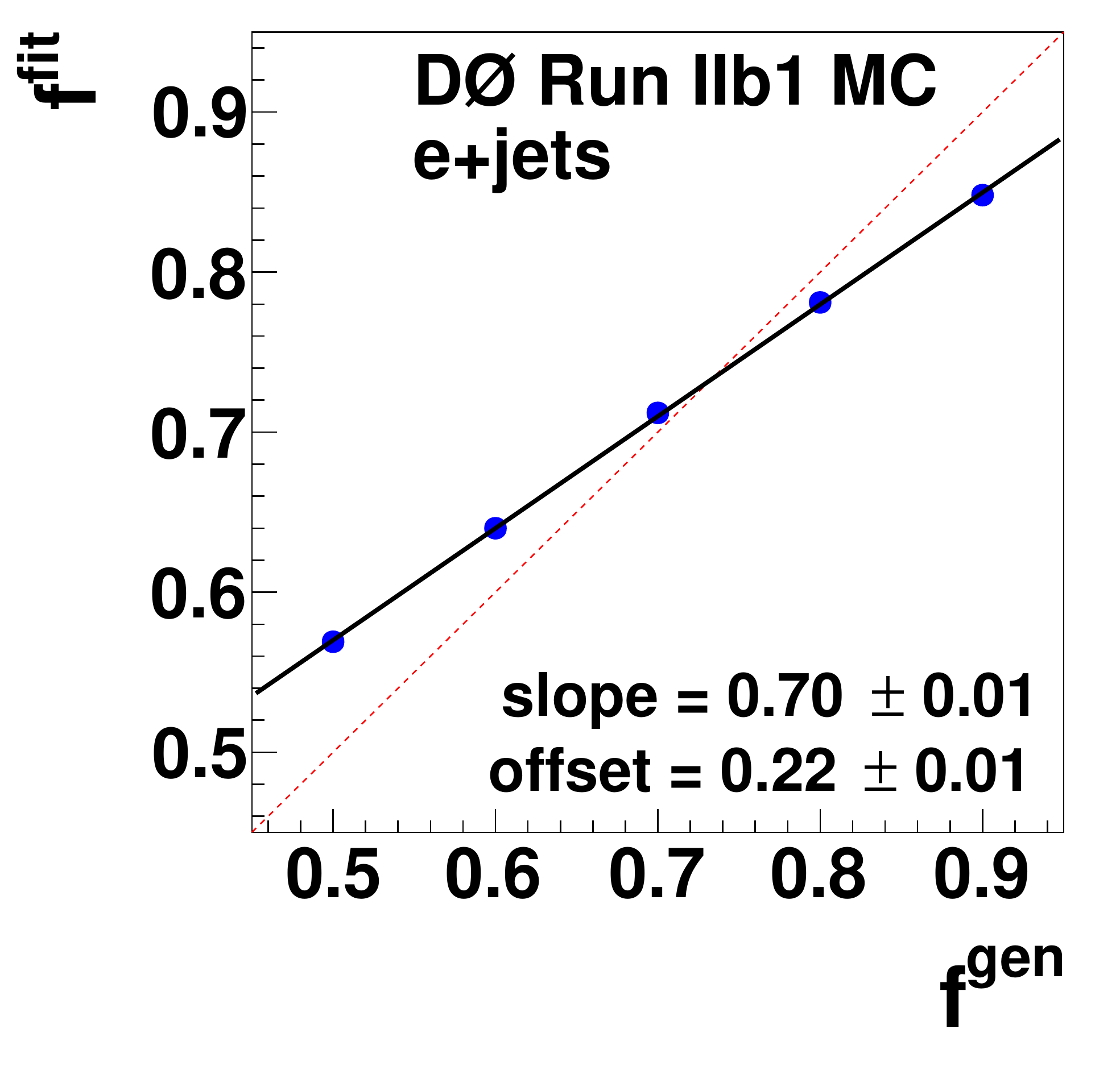}
\put(26,86){\footnotesize\textsf{\textbf{(b)}}}
\end{overpic}
\begin{overpic}[width=0.49\columnwidth]{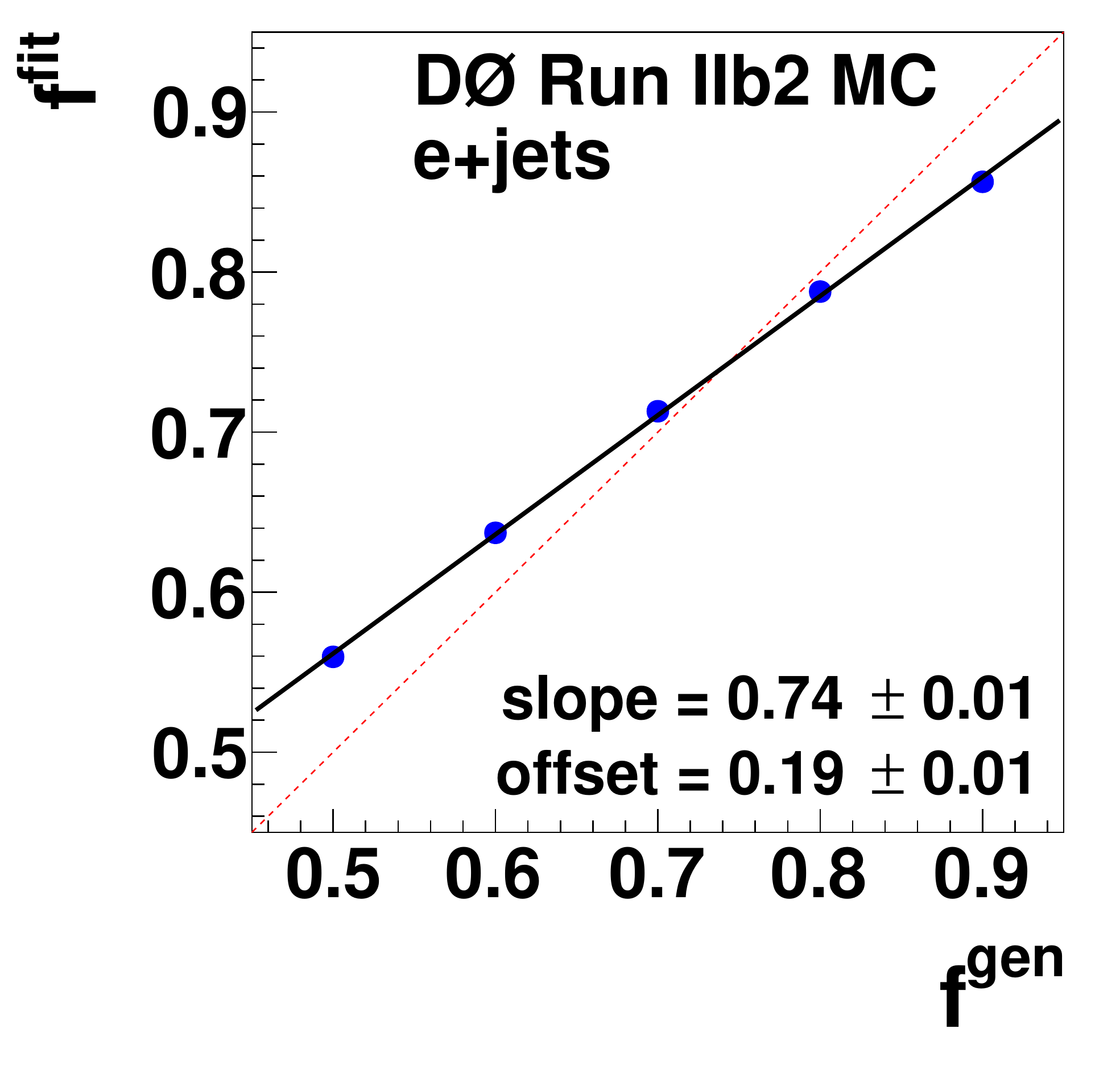}
\put(26,86){\footnotesize\textsf{\textbf{(c)}}}
\end{overpic}
\begin{overpic}[width=0.49\columnwidth]{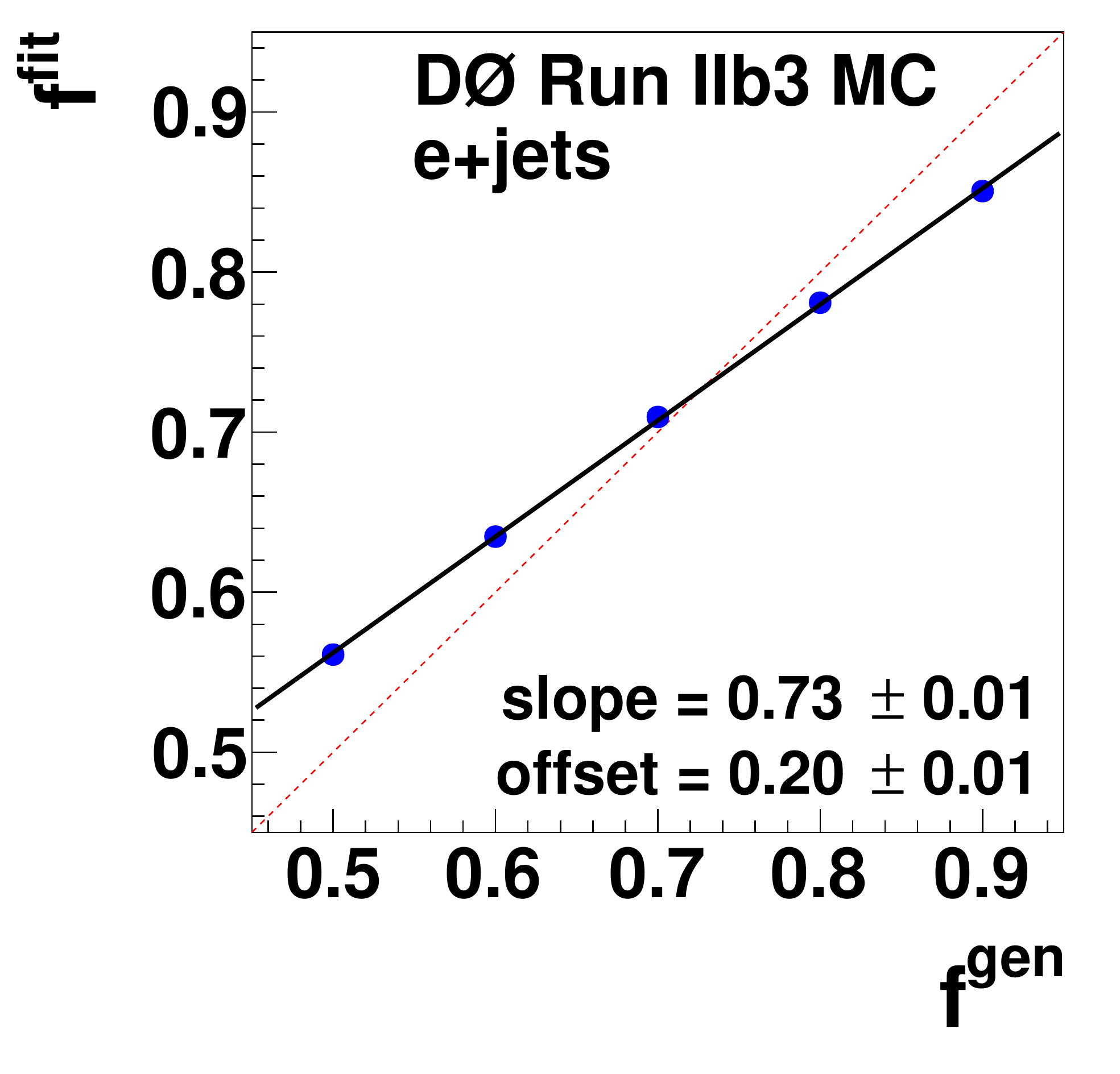}
\put(26,86){\footnotesize\textsf{\textbf{(d)}}}
\end{overpic}
\caption{\label{fig:fem}
The response of the ME technique in $f$, derived from the average of \ffit over 1000 PEs at each \fgen point, for (a)~Run~IIa, (b)~Run~IIb1, (c)~Run~IIb2, and (d)~Run~IIb3 in the \ejets final state. The response is fitted with a linear function shown as a solid black line, whose slope and offset values are given at the bottom of the plots. The broken red line represents an ideal response.
}
\end{figure}

\begin{figure}
\centering
\begin{overpic}[width=0.49\columnwidth]{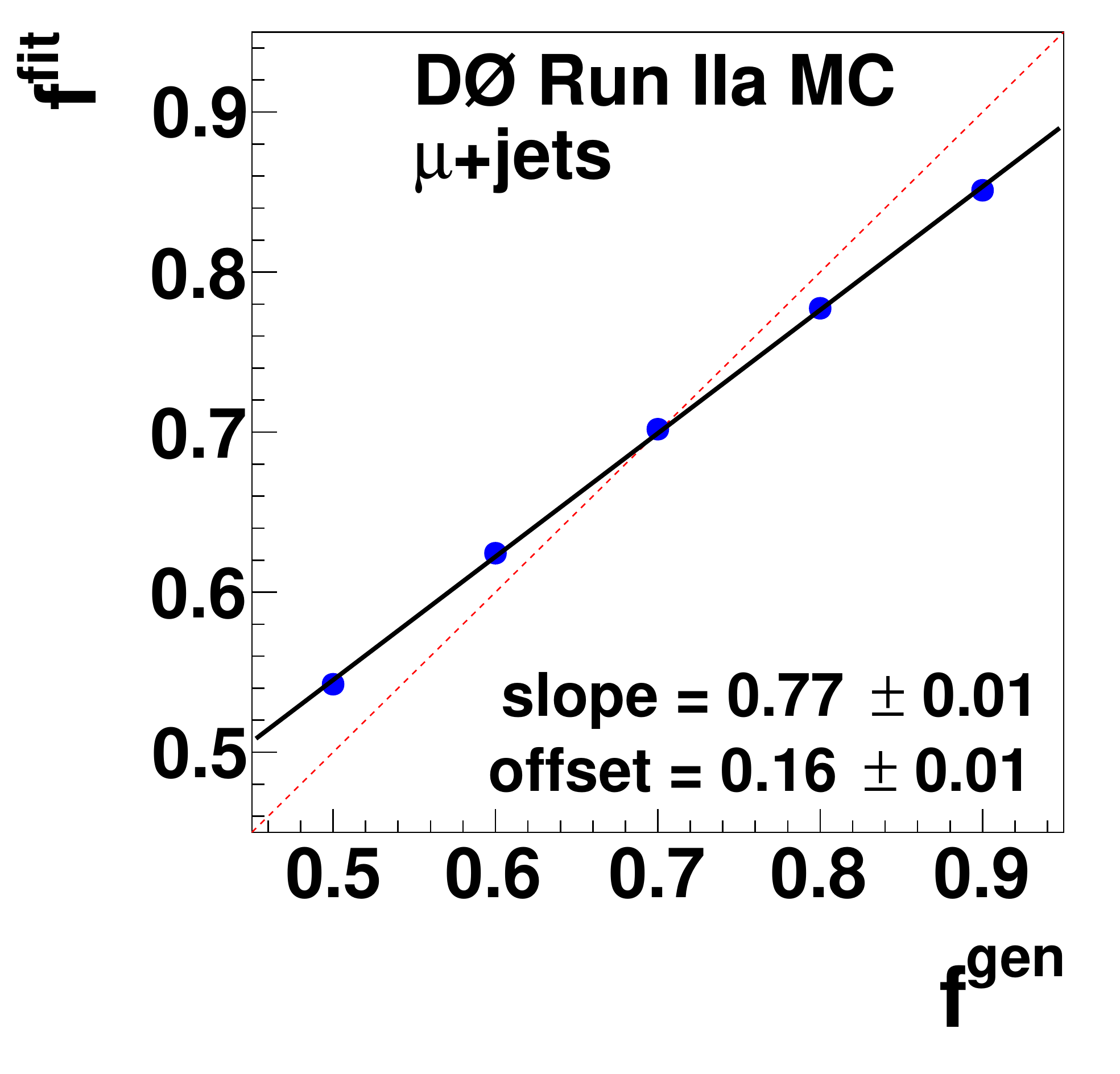}
\put(26,86){\footnotesize\textsf{\textbf{(a)}}}
\end{overpic}
\begin{overpic}[width=0.49\columnwidth]{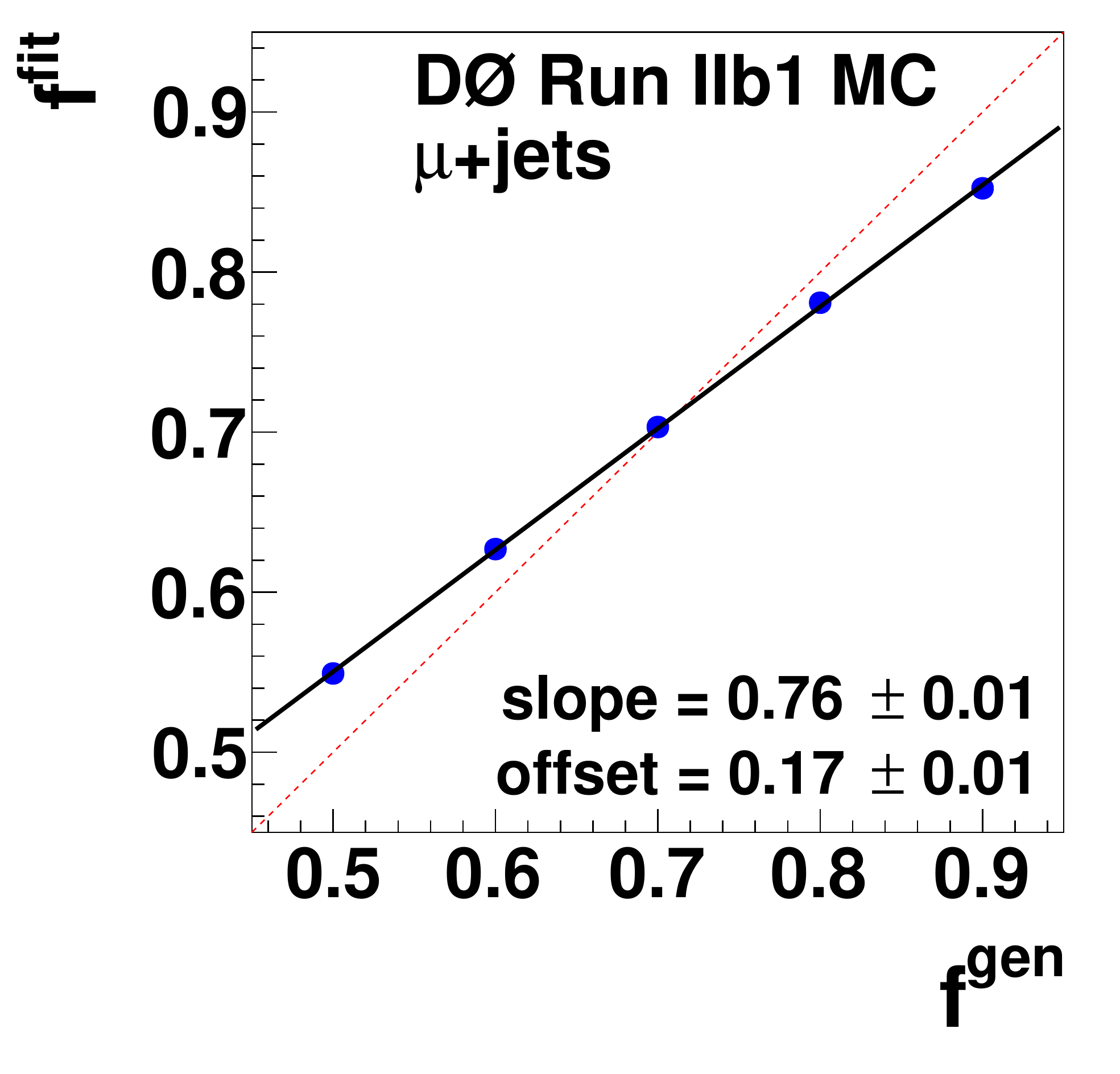}
\put(26,86){\footnotesize\textsf{\textbf{(b)}}}
\end{overpic}
\begin{overpic}[width=0.49\columnwidth]{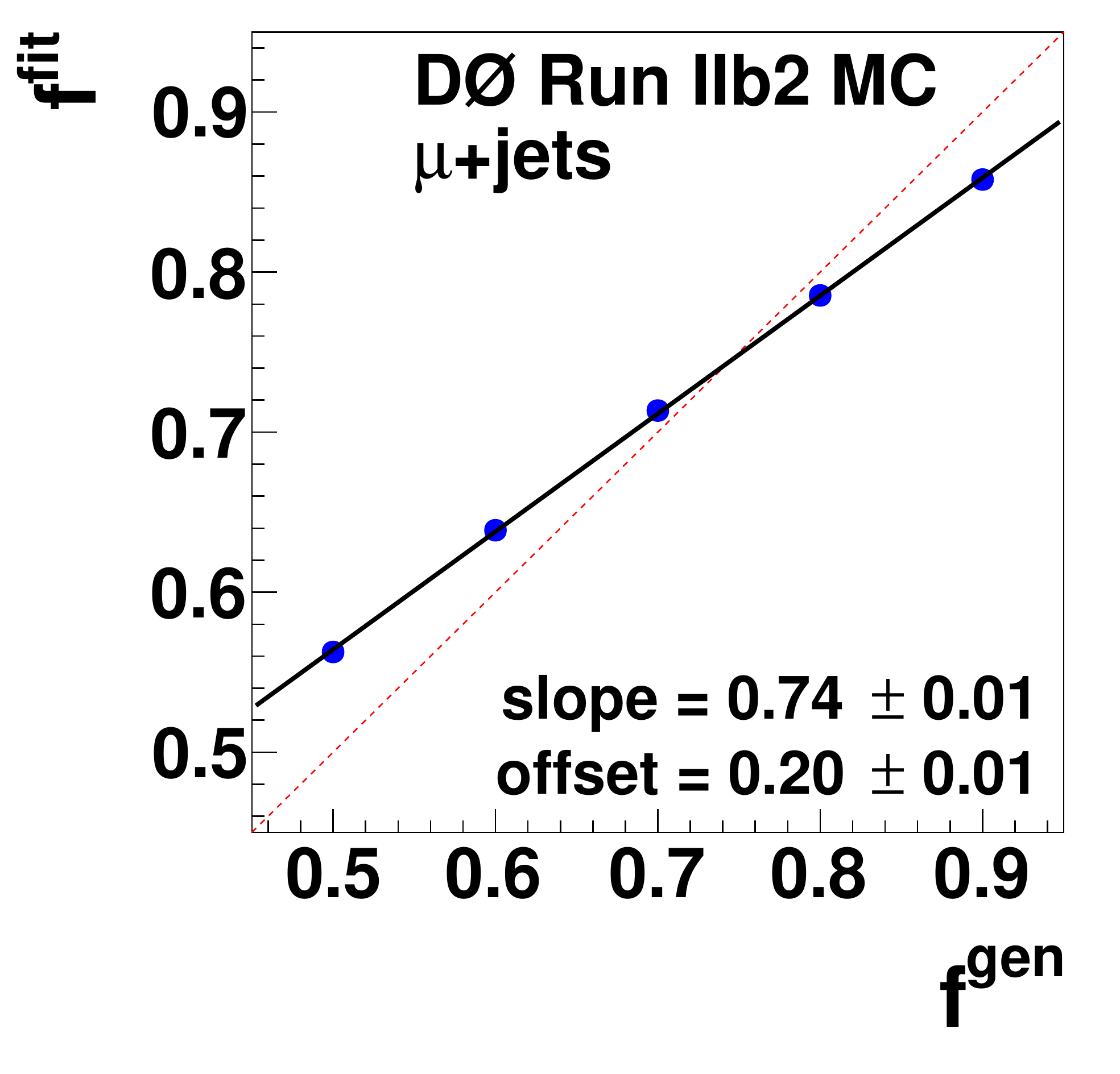}
\put(26,86){\footnotesize\textsf{\textbf{(c)}}}
\end{overpic}
\begin{overpic}[width=0.49\columnwidth]{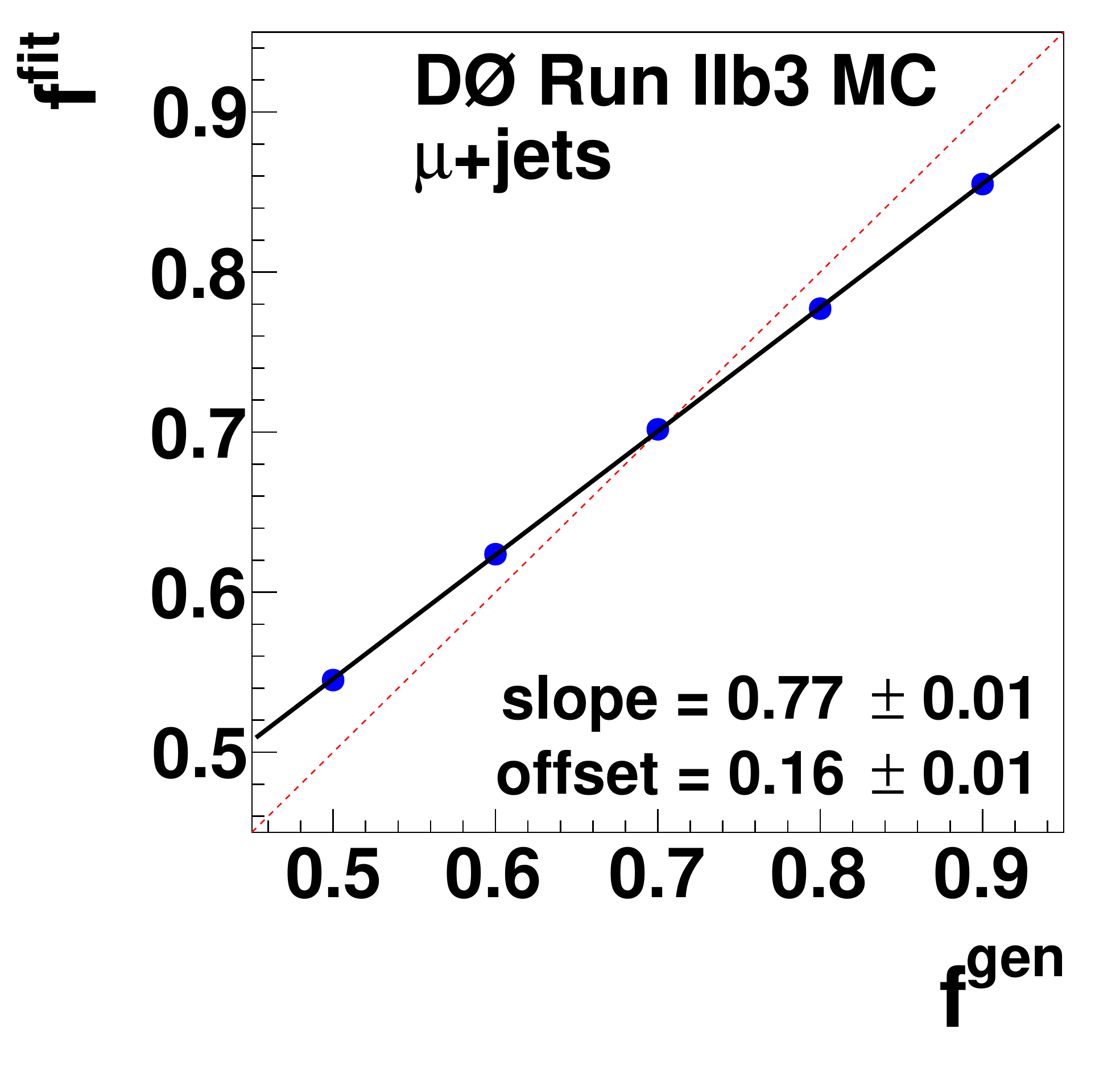}
\put(26,86){\footnotesize\textsf{\textbf{(d)}}}
\end{overpic}
\caption{\label{fig:fmu}
Same as Fig.~\ref{fig:fem}, but in the \mujets final state, for (a)~Run~IIa, (b)~Run~IIb1, (c)~Run~IIb2, and (d)~Run~IIb3.
}
\end{figure}

The PE ensembles at each of the five $\fgen$ points are used  to obtain a linear calibration for the response of the ME technique in $f$. This is achieved by parametrizing the signal fraction $\ffit$, extracted as for data in Section~\ref{sec:extract} and averaged over the PEs for each given \fgen point, versus \fgen. The resulting linear parameterizations of the response of the ME technique in $f$ are shown for each data-taking epoch in Fig.~\ref{fig:fem} for the \ejets, and in Fig.~\ref{fig:fmu} for the \mujets final states. Typical slope parameters are close to 0.75, while typical offsets are few percent at $\fgen\approx0.70$, as expected from Section~\ref{sec:selection}.

\subsection{
Results
}
\label{sec:resultf}

The extracted $\ffit$ in data is  calibrated using the results from the previous Section for each final state and epoch according to
\begin{equation}
f = \frac{\ffit-o_f}{s_f}\,,
\end{equation}
where $o_f$ and $s_f$ are the offset and slope parameters, respectively, as given in Figs.~\ref{fig:fem} and~\ref{fig:fmu}. The uncertainties on $o_f$ and $s_f$ are propagated to $f$ assuming they are Gaussian distributed. The resulting $f$ are given in Table~\ref{tab:postfit}, split by final state and by epoch. Averaging $f$ over the entire Run~II, weighting the contributions from data-taking epochs by their respective integrated luminosities, yields $f=63\%$ in the \ejets and $f=70\%$ in the \mujets final states, with an absolute uncertainty  of 1\% from the finite size of the data sample and the calibration. 

\begin{table}
\caption{
\label{tab:postfit}
Signal fractions $f$ measured in data after calibration, split by final state, and for each data-taking epoch as well as combined over the entire Run~II. Also shown are the resulting measured cross sections for \ttbar production. The combined uncertainty from limited statistical accuracy in data and from the finite number of simulated events used for calibration amounts to 1\% for $f$. 
}
\begin{center} 
\begin{ruledtabular}
\begin{tabular}{llcc}
Epoch & Final state & Signal fraction & $\sigma_{\ttbar}$~(pb)\tabularnewline
\hline 
\multirow{2}{*}{Run IIa}    & \ejets    & 0.72 & 8.9 \\
                            & \mujets  & 0.65 & 7.8 \\
\hline
\multirow{2}{*}{Run IIb1}   & \ejets    & 0.77 & 7.6 \\
                            & \mujets  & 0.66 & 6.8 \\
\hline
\multirow{2}{*}{Run IIb2}   & \ejets    & 0.68 & 7.8 \\
                            & \mujets  & 0.66 & 7.5 \\
\hline
\multirow{2}{*}{Run IIb3}   & \ejets    & 0.56 & 7.6 \\
                            & \mujets  & 0.75 & 8.0 \\
\hline
\hline
\multirow{2}{*}{Run II} & \ejets    & 0.63 & 7.8 \\
                            & \mujets   & 0.70 & 7.6 \\
\end{tabular}
\end{ruledtabular}
\end{center}
\end{table}

As a cross-check we translate the measured $f$ values into cross sections for \ttbar production, which are also reported in  Table~\ref{tab:postfit}. Combining them in a similar fashion to $f$, we find $\stt=7.8\pm0.1\,(\rm stat)$~pb in the \ejets and $\stt=7.6\pm0.1\,(\rm stat)$~pb in the \mujets final states. The cross sections determined with the ME technique are in good agreement with the prediction of $\stt=7.24^{+0.23}_{-0.27}~$pb (scale+PDF) from a NNLO calculation at $\mt=172.5~\GeV$, including NNLL corrections~\cite{bib:xsec}, and using the MSTW2008NNLO set of PDFs~\cite{bib:mstwnlo}. This is also in agreement with $\stt=7.78^{+0.77}_{-0.64}$~pb measured by the D0 Collaboration in 5.3~\fb of integrated luminosity~\cite{bib:xsec54}.

\section{
Measurement of the top-quark mass
}
\label{sec:mt}

As described in Section~\ref{sec:f}, the response of the ME technique  in \mt and \kjes has to be calibrated prior to the measurement of those parameters. In this Section, we describe the calibration procedure, report the measured \mt and \kjes values, demonstrate that simulations with our measured values of $f$, \mt, and \kjes describe the data well,
and draw comparisons to other measurements of \mt.

\subsection{
Calibration of the method
}
\label{sec:calibmt}

The calibration of the response of the ME technique in \mt and \kjes proceeds similarly to the calibration in $f$ described in Section~\ref{sec:f}. It uses seven simulated samples of \ttbar events, five at $\mt^{\rm gen}=165, 170,$ $172.5, 175, 180~\GeV$ for $\kjes^{\rm gen}=1$, and two at $\kjes^{\rm gen}=0.95, 1.05$ for $\mt^{\rm gen}=172.5~\GeV$. These are complemented by three simulated samples of $W+{\rm jets}$ events at $\kjes^{\rm gen}=0.95,1,$ and $1.05$. The calibration procedure is refined relative to Ref.~\cite{bib:mt36} by accounting explicitly for the next-to-dominant contribution from MJ production.  Together, the \ttbar, $W+{\rm jets}$, and MJ samples are used  to obtain a linear calibration for the response of the ME technique in \mt and \kjes. The calibration is applied separately for each data-taking epoch and final state to account for varying detector response and sample composition. 

At each of the generated $(\mt^{\rm gen},\kjes^{\rm gen})$ points, 1000 PEs  are constructed, each containing the same number of events as observed in data, by randomly drawing simulated signal and background events according to the measured $f$  for each data-taking epoch and final state, cf.~Table~\ref{tab:postfit}. For each PE, the fraction of signal events is changed randomly according to a binomial distribution around the $f$ value measured in data. 
Since the background contribution from MJ production is modeled using data, it is not changed for different generated $(\mt^{\rm gen},\kjes^{\rm gen})$ points.

\begin{figure}
\centering
\begin{overpic}[width=0.49\columnwidth]{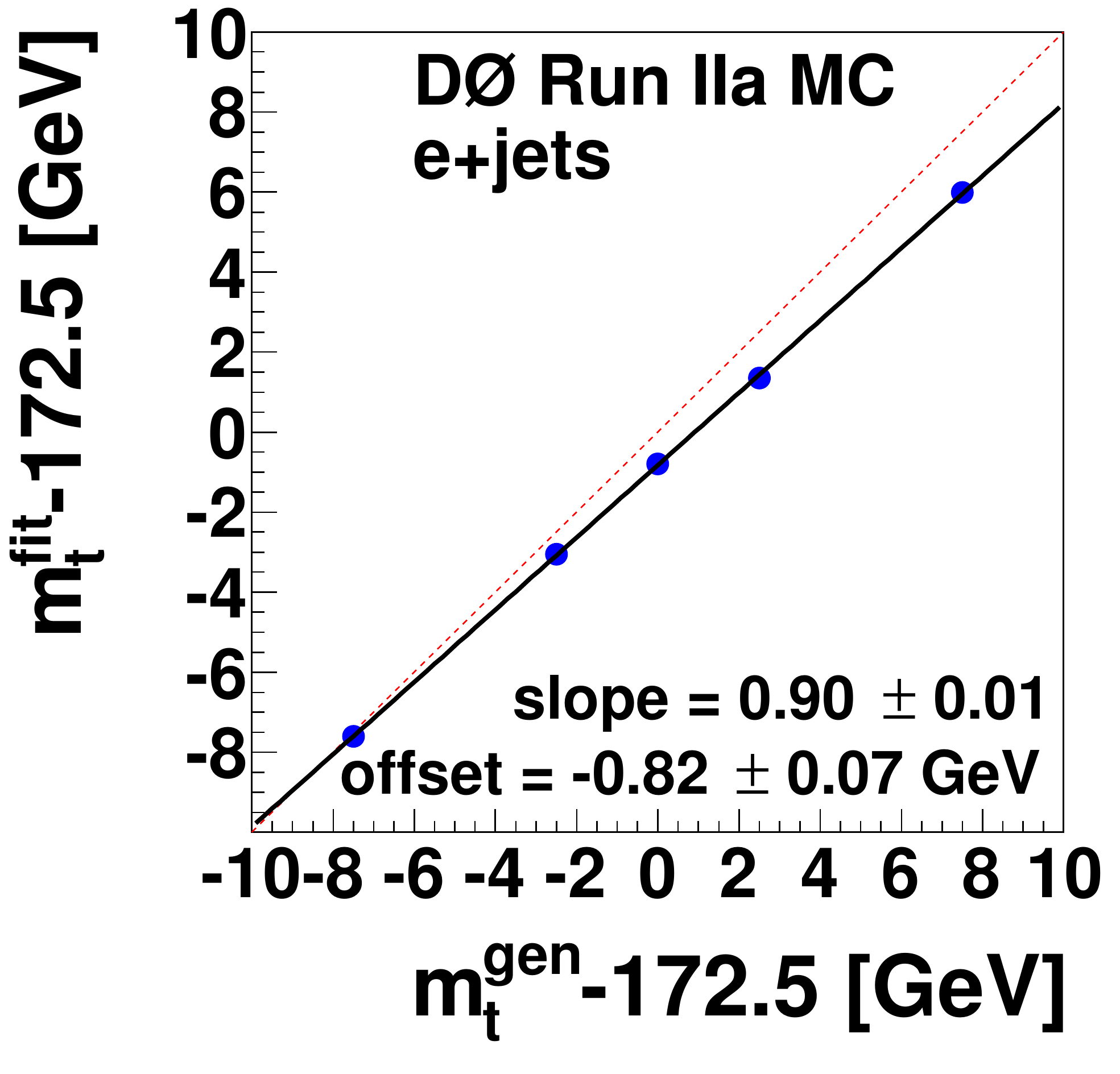}
\put(26,86){\footnotesize\textsf{\textbf{(a)}}}
\end{overpic}
\begin{overpic}[width=0.49\columnwidth]{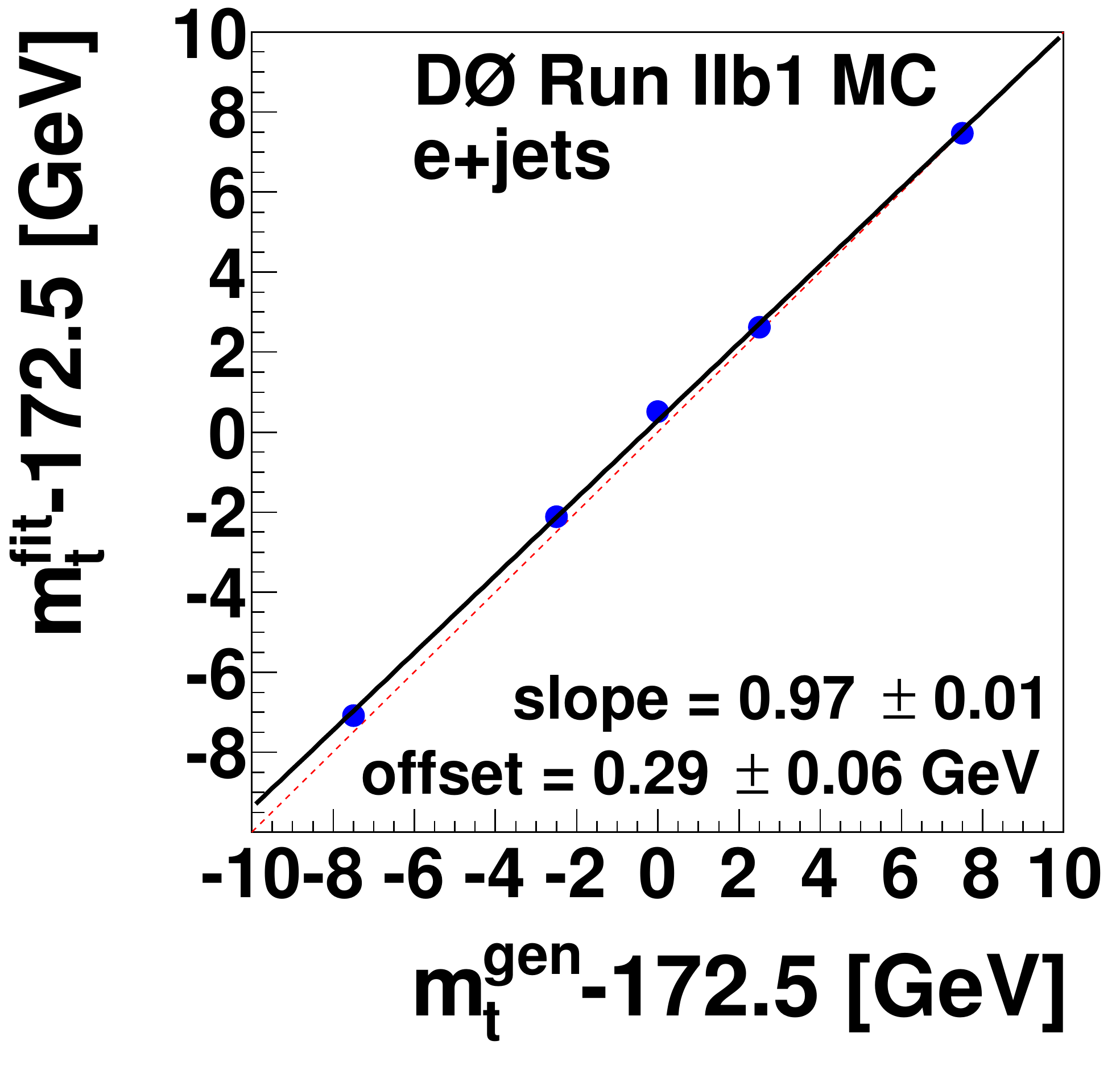}
\put(26,86){\footnotesize\textsf{\textbf{(b)}}}
\end{overpic}
\begin{overpic}[width=0.49\columnwidth]{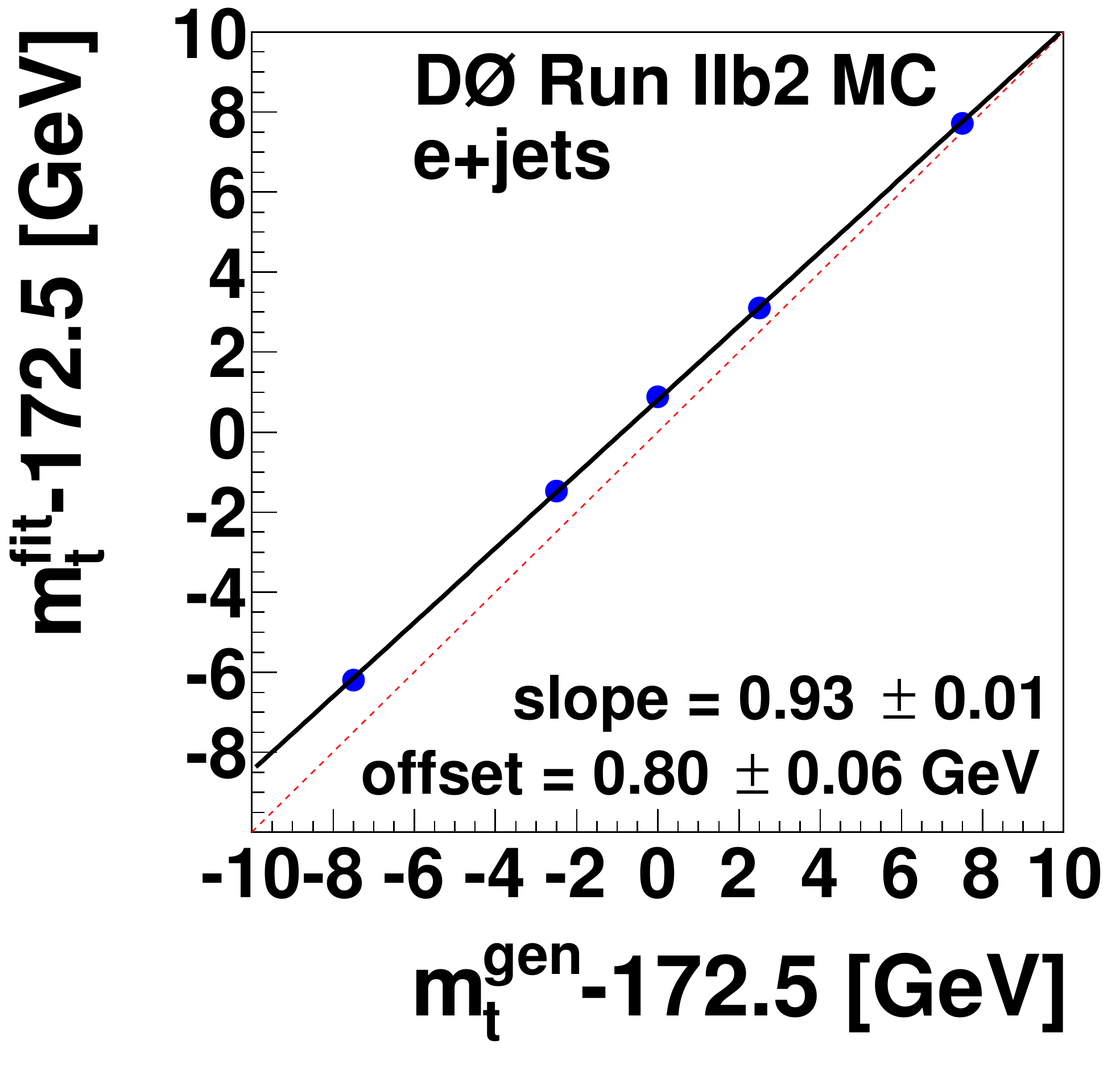}
\put(26,86){\footnotesize\textsf{\textbf{(c)}}}
\end{overpic}
\begin{overpic}[width=0.49\columnwidth]{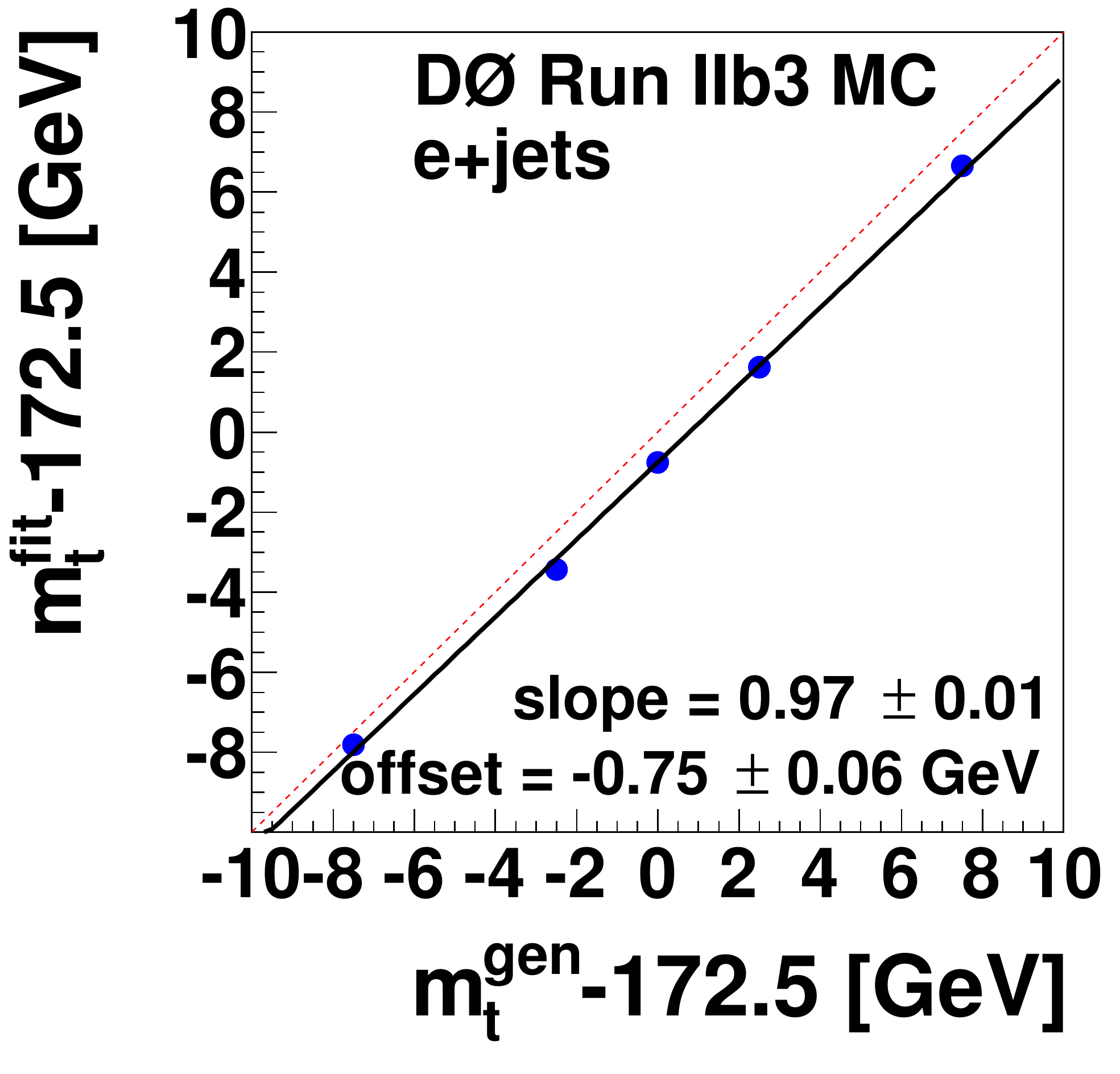}
\put(26,86){\footnotesize\textsf{\textbf{(d)}}}
\end{overpic}
\caption{\label{fig:mtem}
The response of the ME technique in $\mt$, obtained using simulations from the average of \mtfit in 1000 PEs at each \mtgen point, as a function of \mtgen, for (a)~Run~IIa, (b)~Run~IIb1, (c)~Run~IIb2, and (d)~Run~IIb3 in the \ejets final state. The response is fitted with a linear function shown as a solid black line, whose slope and offset values are summarized at the bottom of the plots. The red broken line represents an ideal response.
}
\end{figure}

\begin{figure}
\centering
\begin{overpic}[width=0.49\columnwidth]{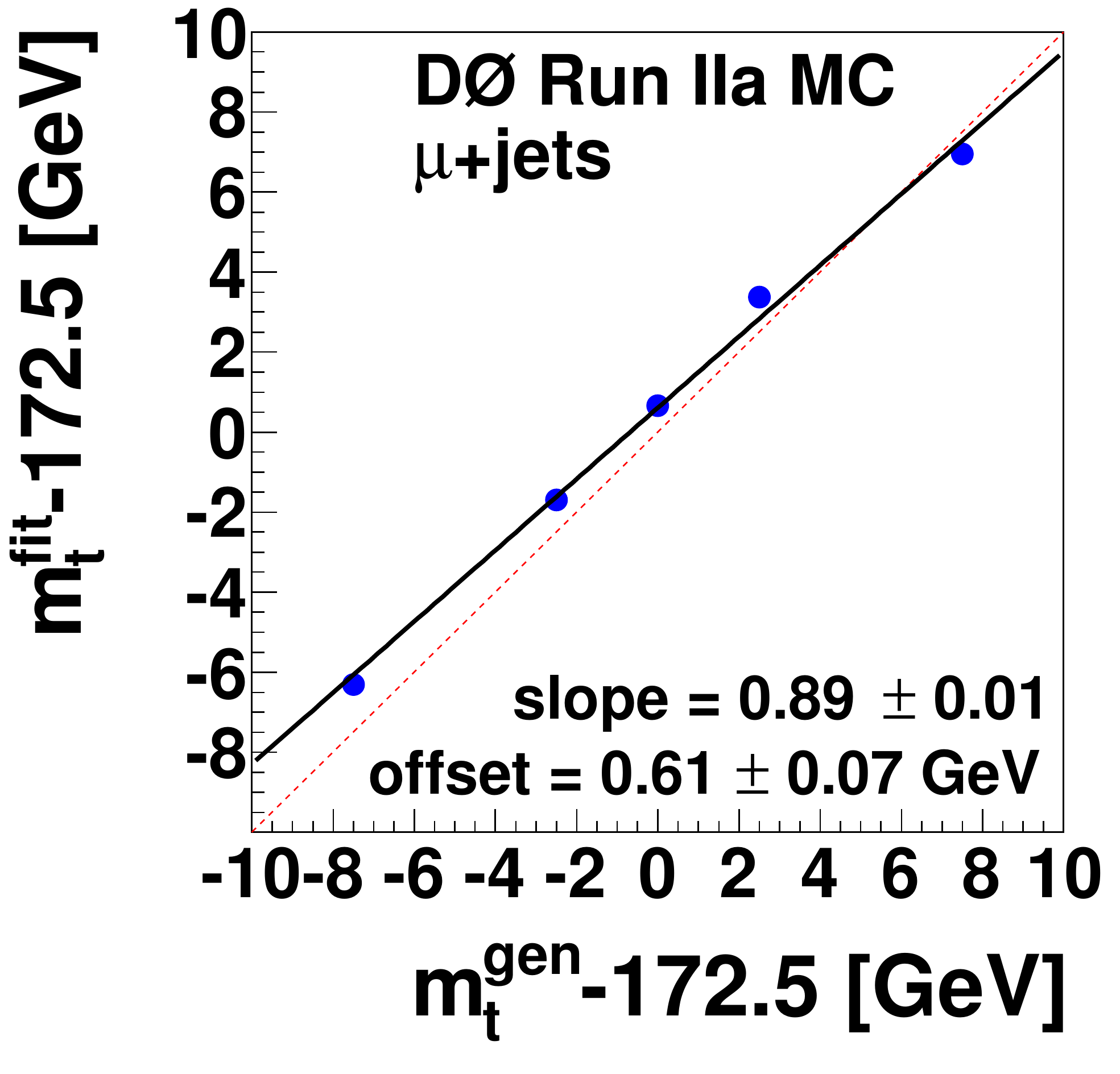}
\put(26,86){\footnotesize\textsf{\textbf{(a)}}}
\end{overpic}
\begin{overpic}[width=0.49\columnwidth]{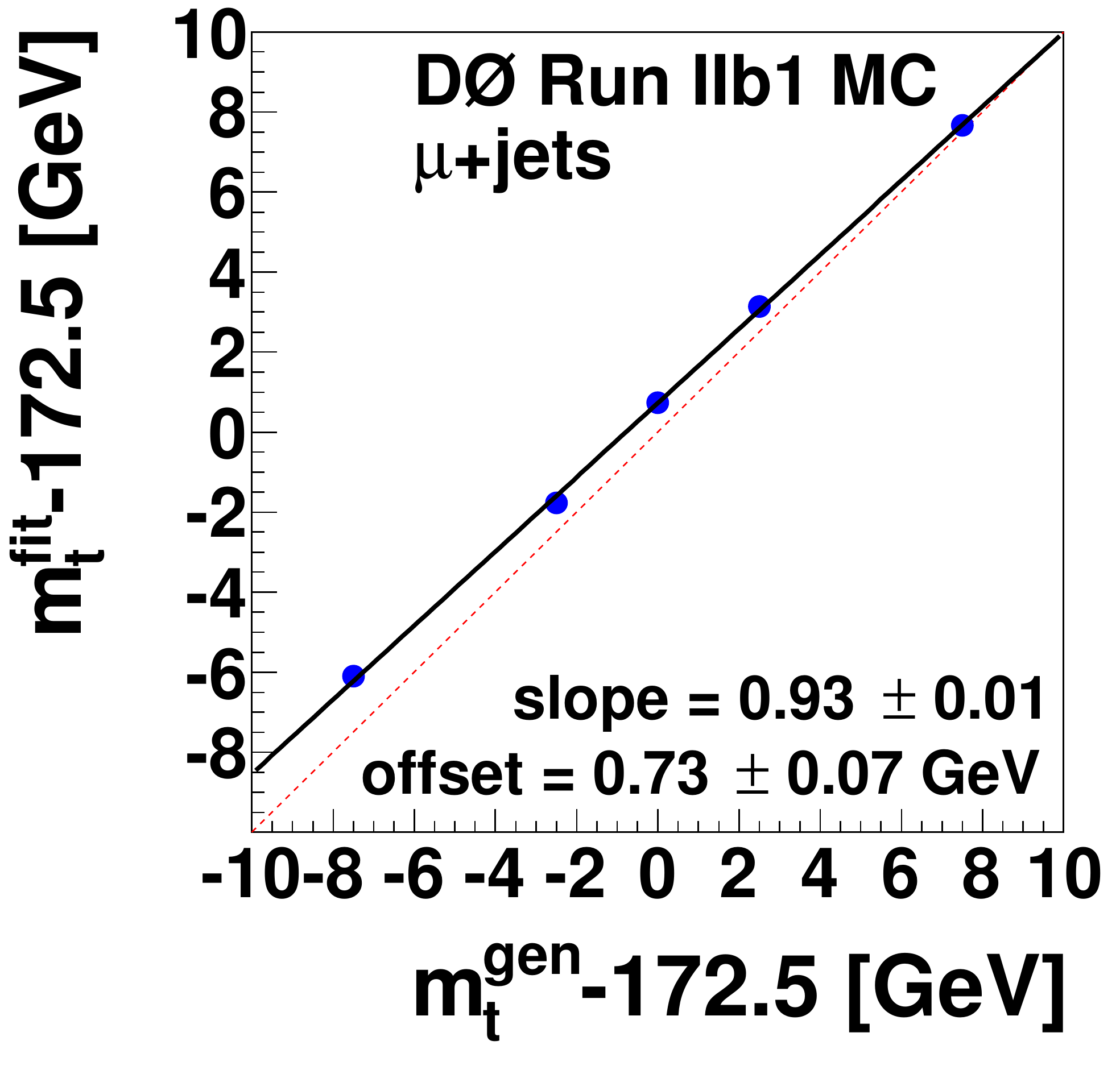}
\put(26,86){\footnotesize\textsf{\textbf{(b)}}}
\end{overpic}
\begin{overpic}[width=0.49\columnwidth]{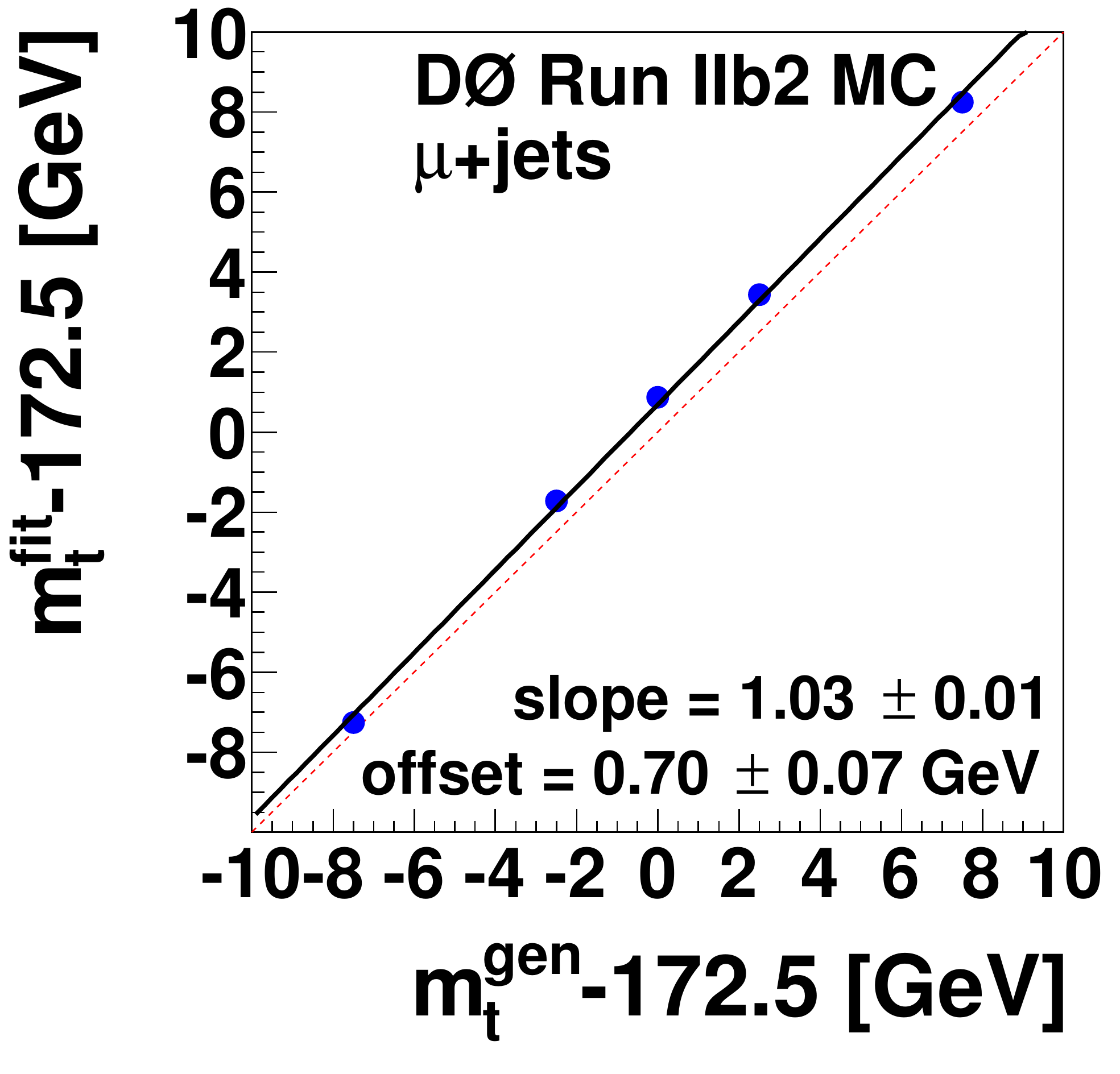}
\put(26,86){\footnotesize\textsf{\textbf{(c)}}}
\end{overpic}
\begin{overpic}[width=0.49\columnwidth]{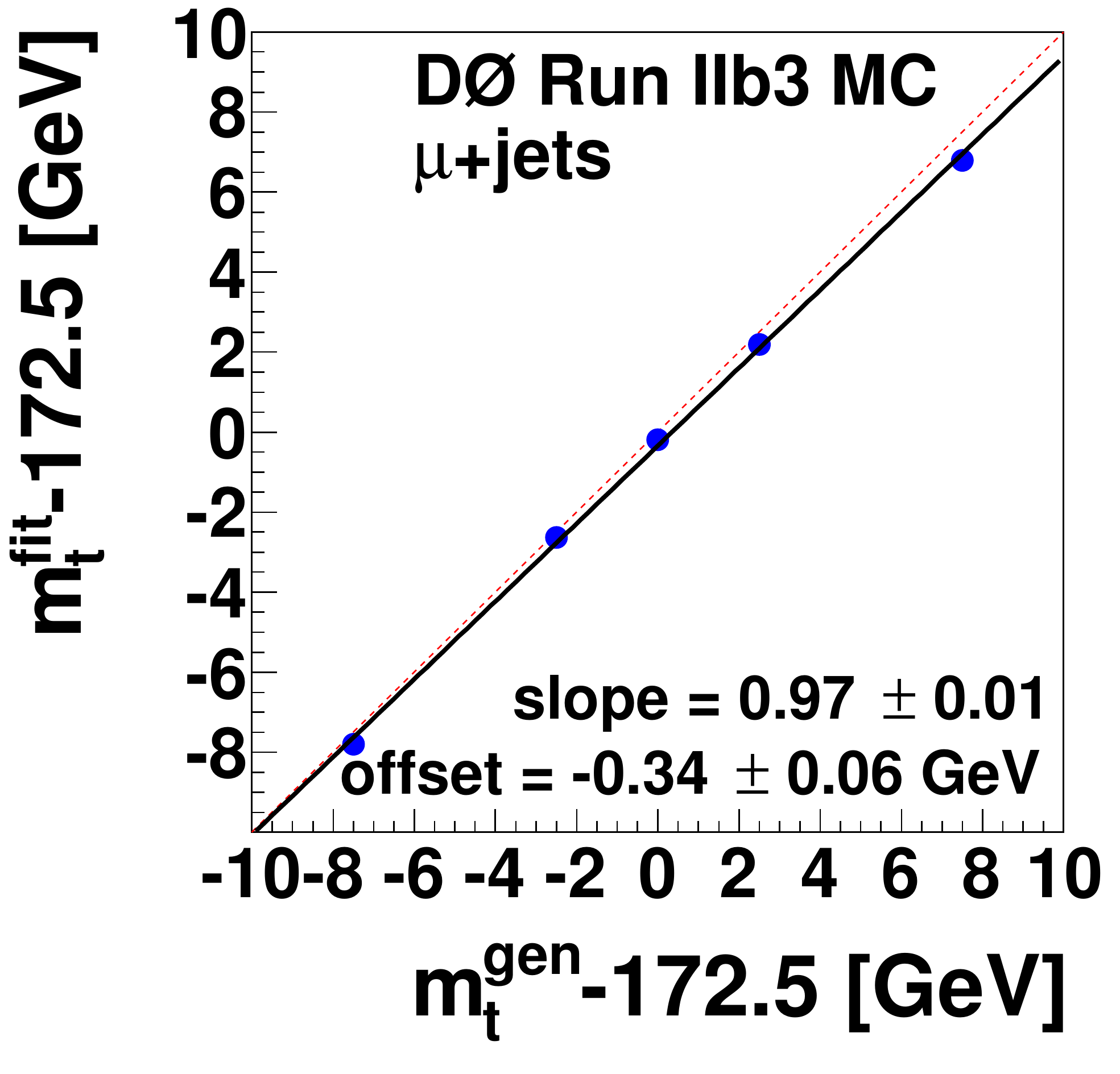}
\put(26,86){\footnotesize\textsf{\textbf{(d)}}}
\end{overpic}
\caption{\label{fig:mtmu}
Same as Fig.~\ref{fig:mtem}, but in the \mujets final state, for (a)~Run~IIa, (b)~Run~IIb1, (c)~Run~IIb2, and (d)~Run~IIb3.
}
\end{figure}

\begin{figure}
\centering
\begin{overpic}[width=0.49\columnwidth]{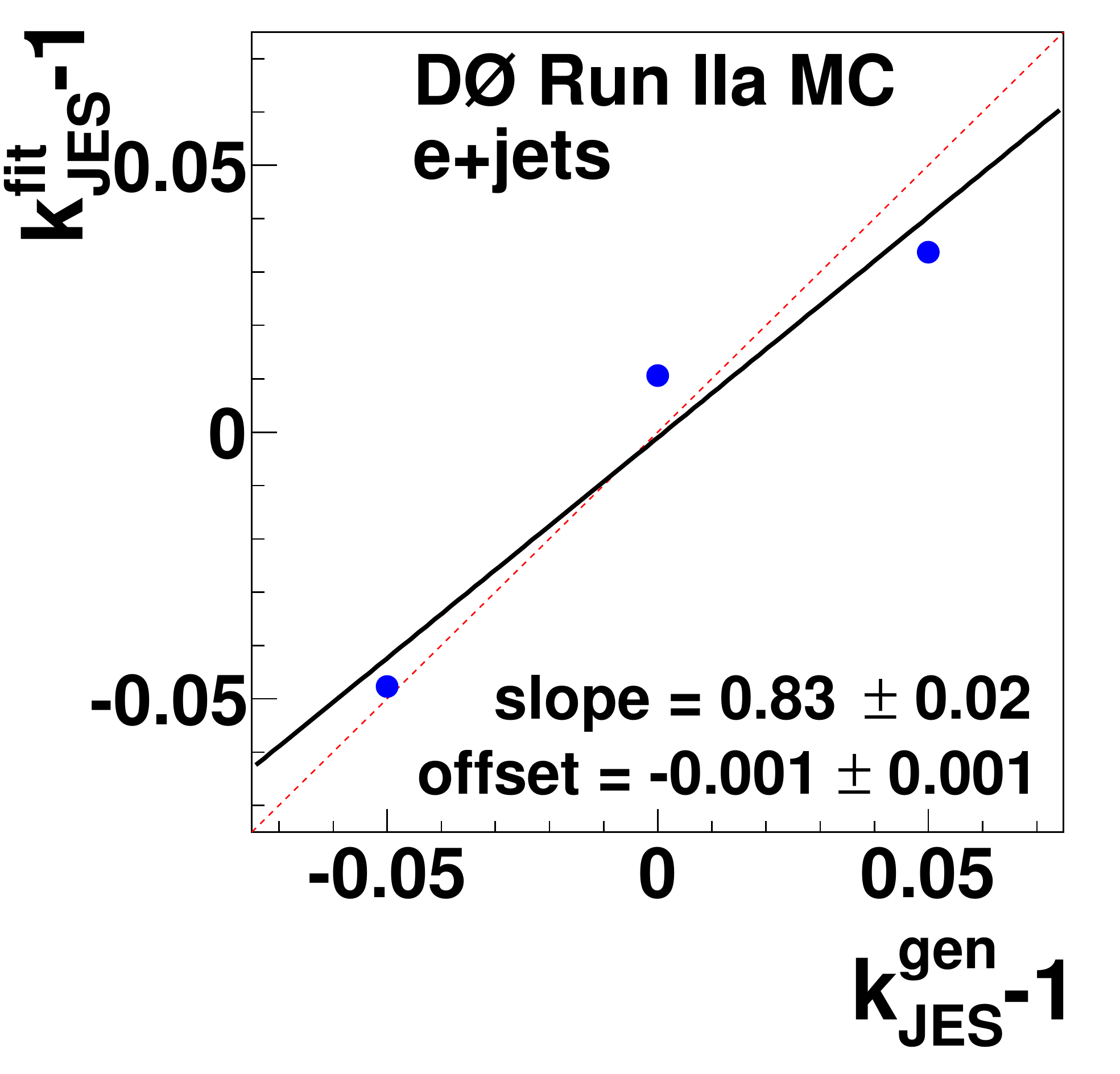}
\put(26,86){\footnotesize\textsf{\textbf{(a)}}}
\end{overpic}
\begin{overpic}[width=0.49\columnwidth]{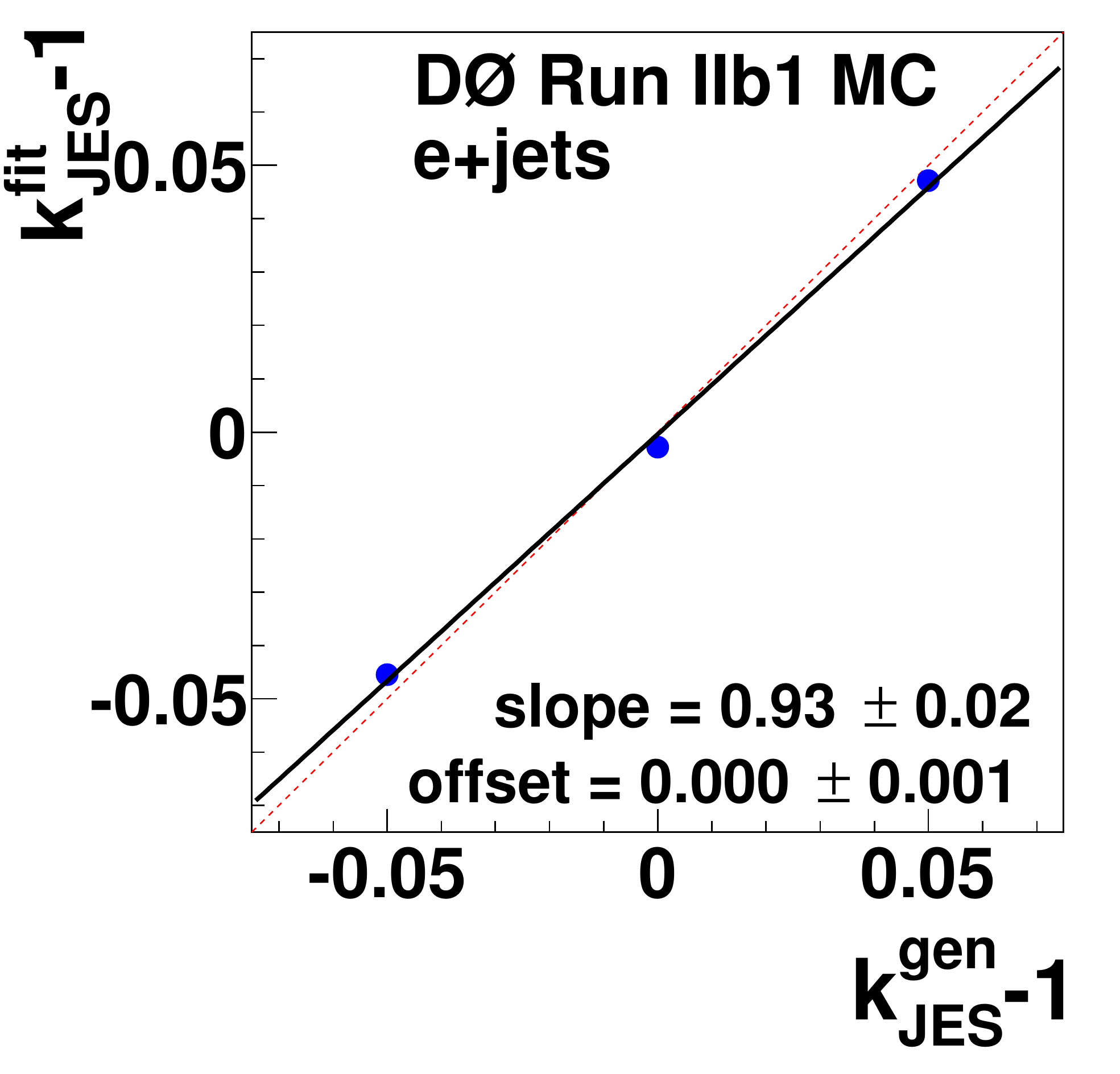}
\put(26,86){\footnotesize\textsf{\textbf{(b)}}}
\end{overpic}
\begin{overpic}[width=0.49\columnwidth]{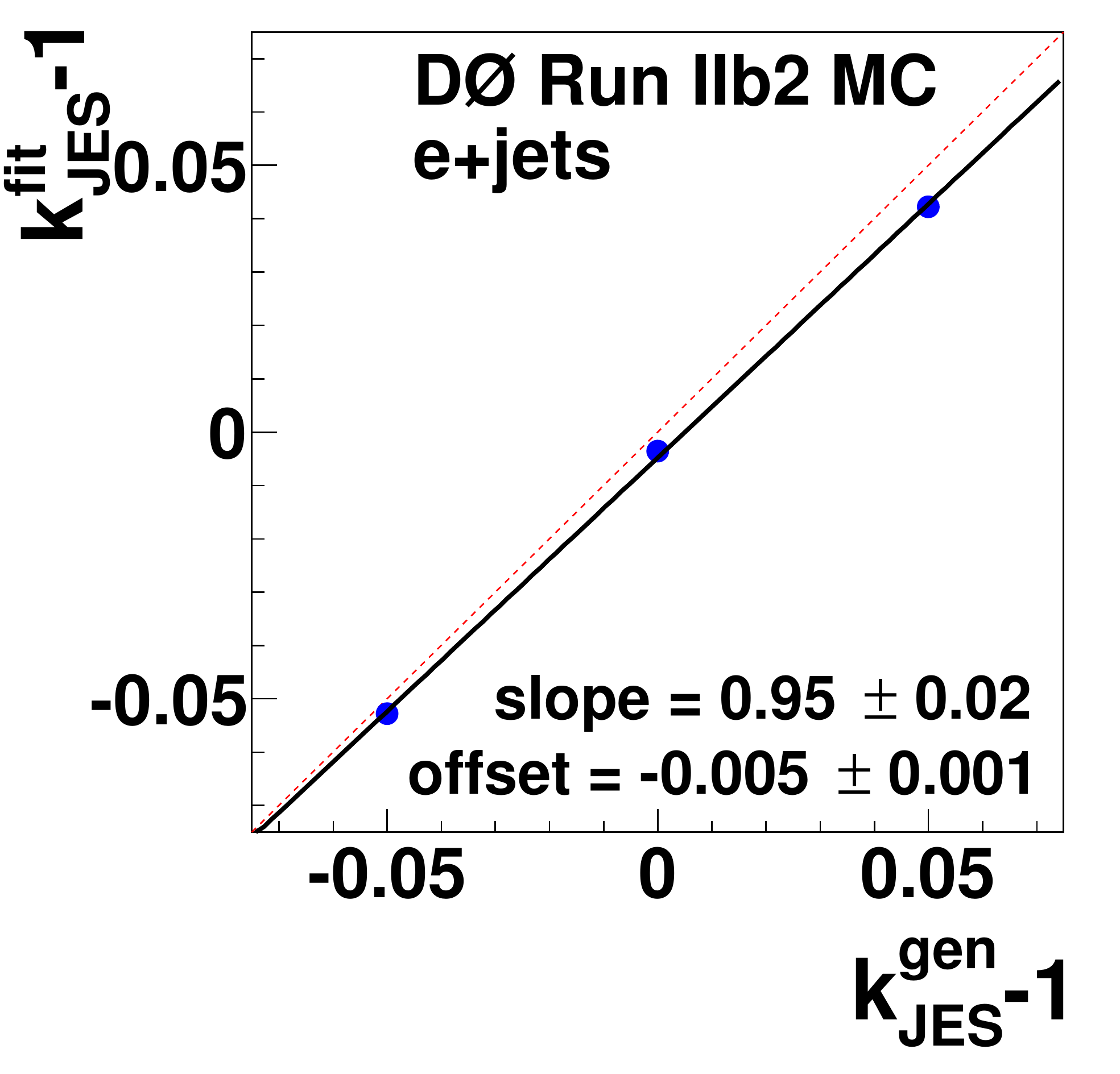}
\put(26,86){\footnotesize\textsf{\textbf{(c)}}}
\end{overpic}
\begin{overpic}[width=0.49\columnwidth]{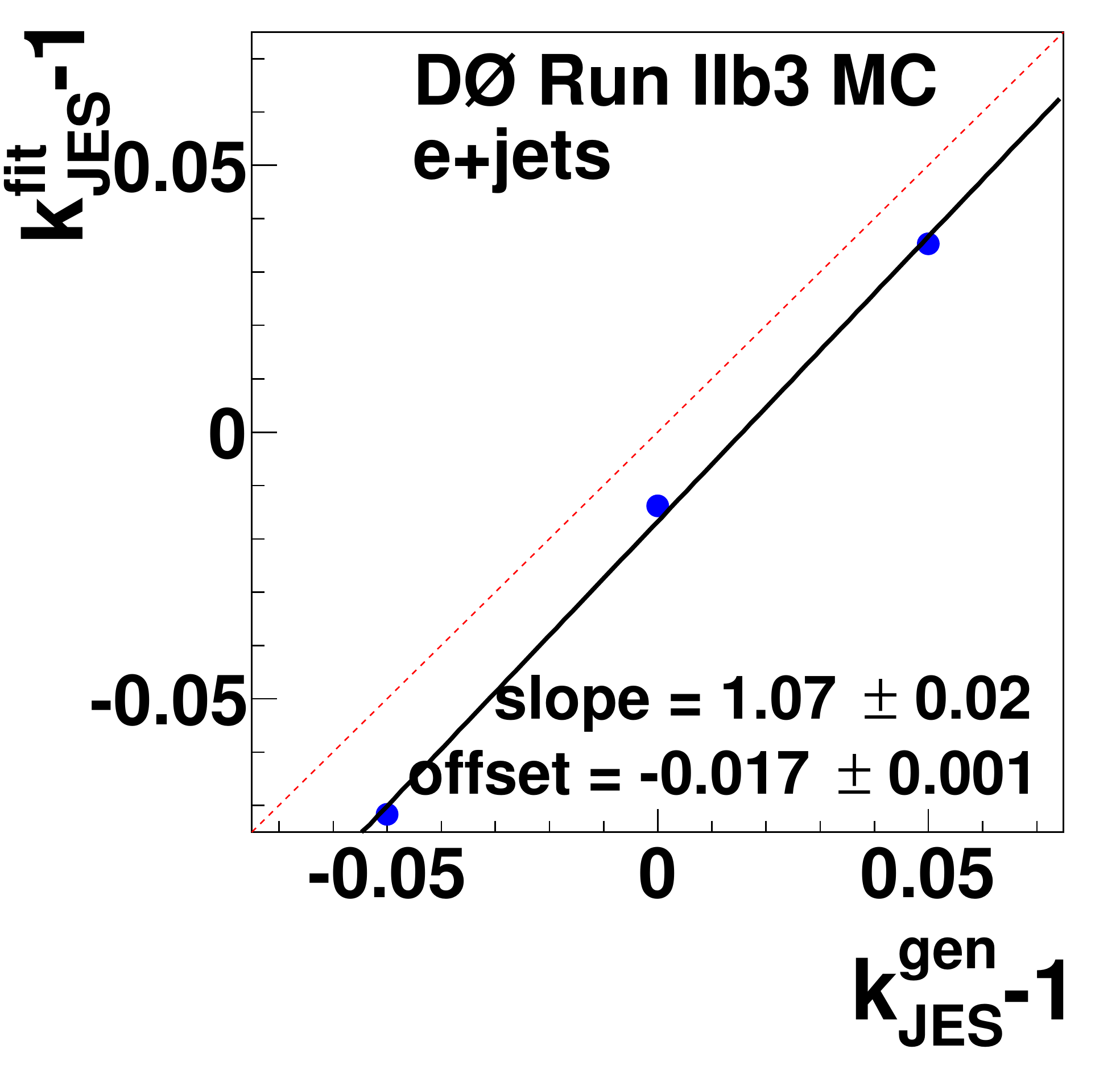}
\put(26,86){\footnotesize\textsf{\textbf{(d)}}}
\end{overpic}
\caption{\label{fig:kjesem}
The response of the ME technique in $\kjes$, obtained using simulations from the average of \kjesfit in 1000 PEs at each \kjesgen point, as a function of \kjesgen, for (a)~Run~IIa, (b)~Run~IIb1, (c)~Run~IIb2, and (d)~Run~IIb3 in the \ejets final state. The response is fitted with a linear function shown as a solid black line, whose slope and offset values are summarized at the bottom of the plots. The red broken line represents an ideal response. The poor compatibility of data with the linear fit displayed in panel~(a) likely to occur if many fits, in our case, 16, are performed.
}
\end{figure}

\begin{figure}
\centering
\begin{overpic}[width=0.49\columnwidth]{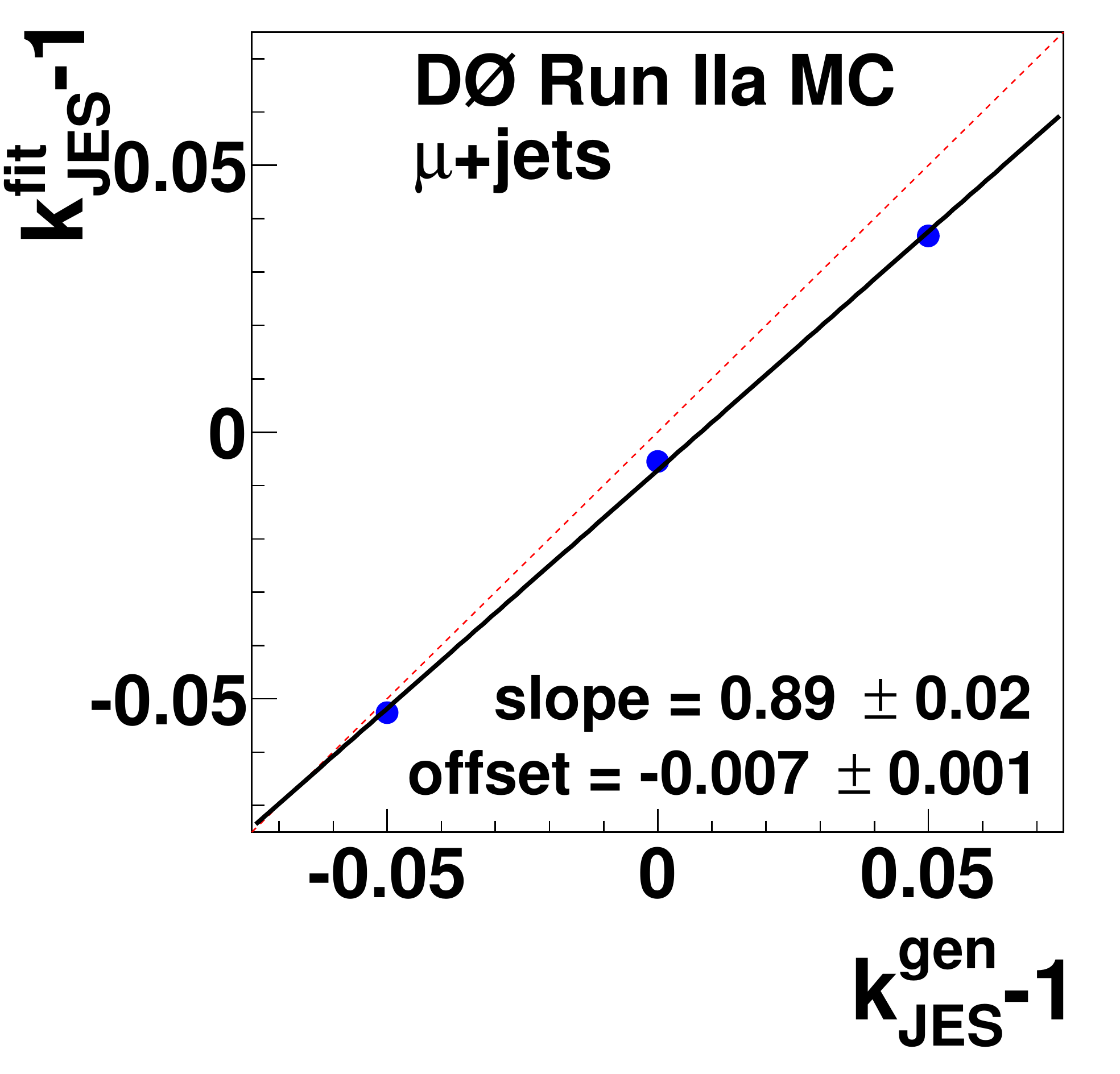}
\put(26,86){\footnotesize\textsf{\textbf{(a)}}}
\end{overpic}
\begin{overpic}[width=0.49\columnwidth]{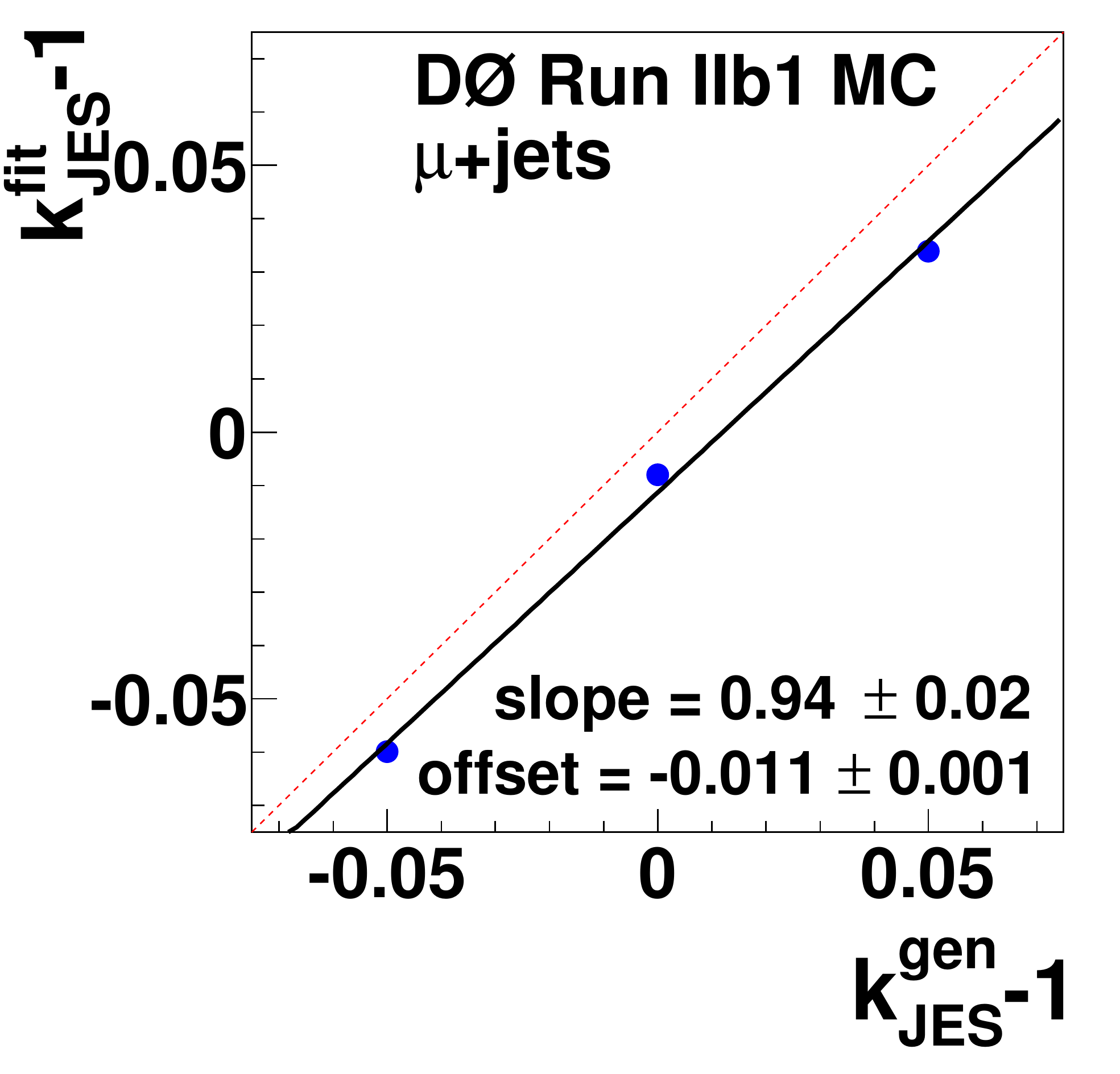}
\put(26,86){\footnotesize\textsf{\textbf{(b)}}}
\end{overpic}
\begin{overpic}[width=0.49\columnwidth]{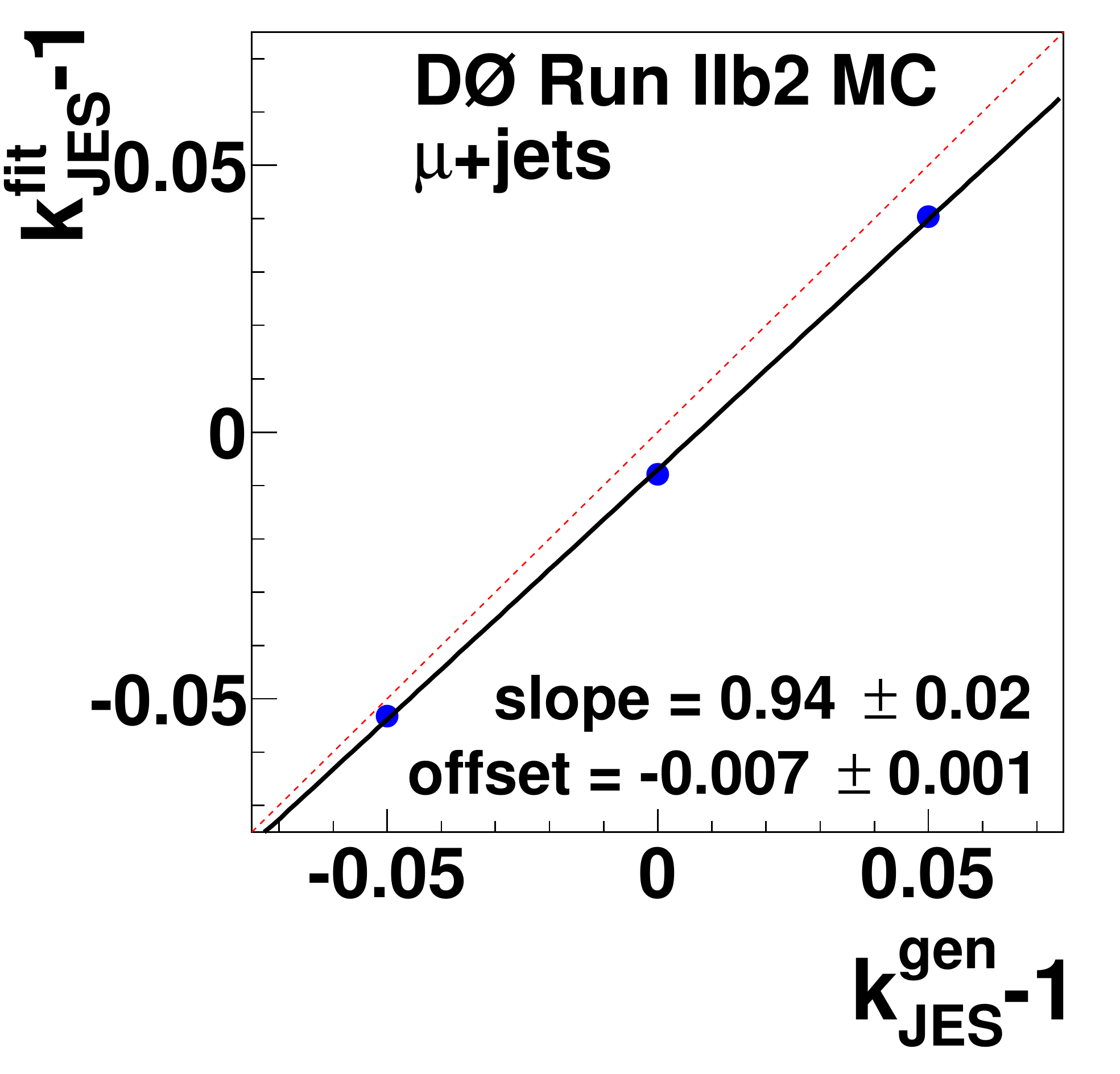}
\put(26,86){\footnotesize\textsf{\textbf{(c)}}}
\end{overpic}
\begin{overpic}[width=0.49\columnwidth]{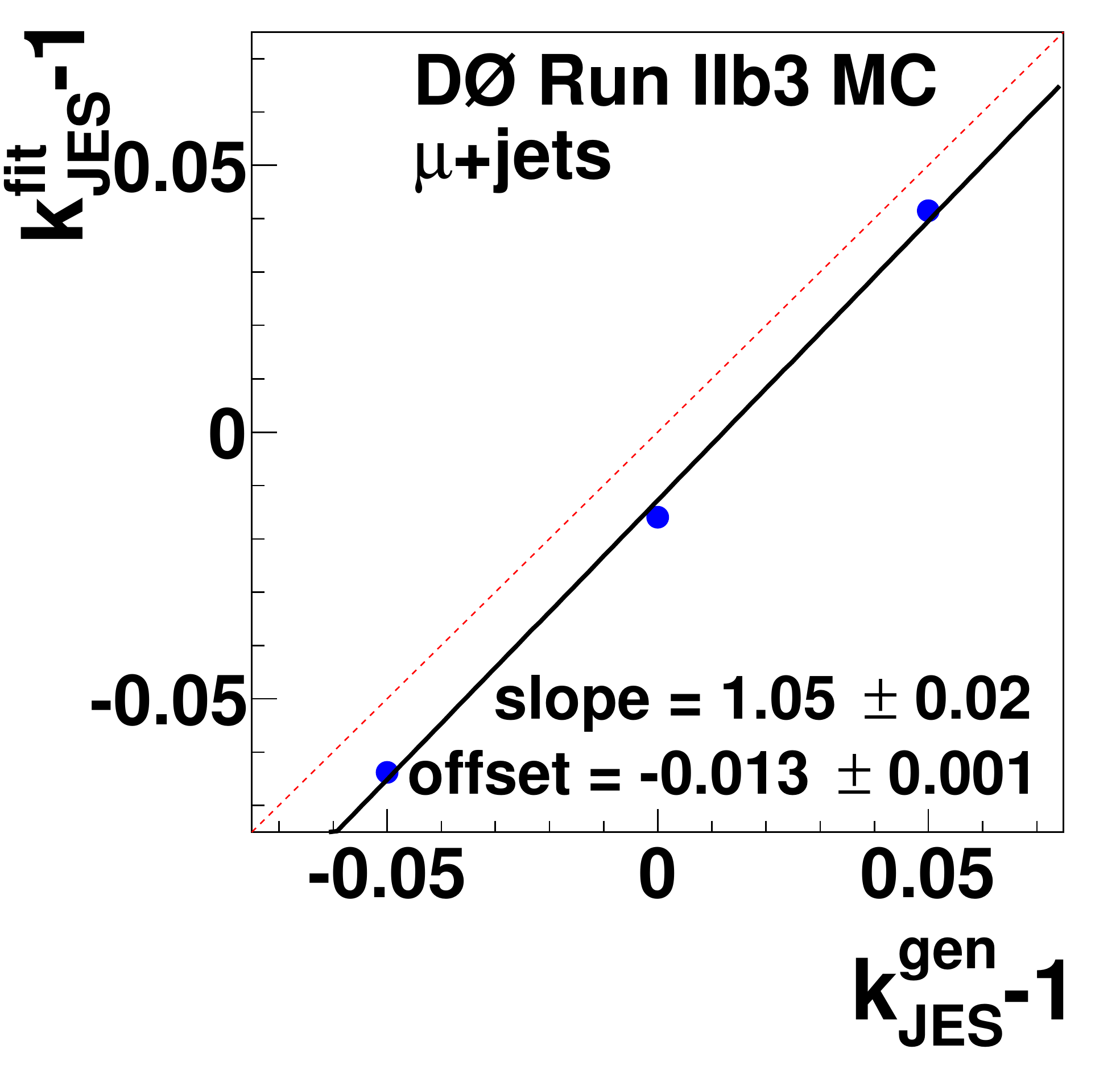}
\put(26,86){\footnotesize\textsf{\textbf{(d)}}}
\end{overpic}
\caption{\label{fig:kjesmu}
Same as Fig.~\ref{fig:kjesem}, but in the \mujets final state, for (a)~Run~IIa, (b)~Run~IIb1, (c)~Run~IIb2, and (d)~Run~IIb3.
}
\end{figure}

The response of the ME technique in \mt and \kjes is obtained by extracting \mtfit and \kjesfit for each pseudo-experiment as in data, i.e., according to the procedure described in Section~\ref{sec:extract}. The averages $\langle\mtfit\rangle$ and $\langle\kjesfit\rangle$ are determined from Gaussian fits to distributions in \mtfit and \kjesfit, respectively. A linear parametrization of the response in \mt is obtained by fitting $\langle\mtfit\rangle$ as a function of $\mtgen$. Those parameterizations are shown for each epoch in Fig.~\ref{fig:mtem} for the \ejets and in Fig.~\ref{fig:mtmu} for the \mujets final states, together with the offset and slope of the linear fits. The response in \mt is close to the ideal case of no offset and unit slope. At $\mtgen=172.5~\GeV$, which is close to the current world average of $\mt=173.34~\GeV$~\cite{bib:combiworld}, the difference between the obtained and ideal response in \mt is $\approx0.5~\GeV$, i.e., at the level of $\approx0.25\%$ -- a very good performance given that a LO ME with parameterized detector response is used in the ME calculation. We verify that the ideal response of the ME technique is retained through ensemble studies using parton-level \ttbar events generated  with \pythia using a LO ME. In a similar fashion, we obtain a linear parametrization of the response of the ME technique in \kjes, as shown in Fig.~\ref{fig:kjesem} for the \ejets and in Fig.~\ref{fig:kjesmu} for the \mujets final states. As for the case of \mt, the response is close to ideal. At $\kjesgen=1$ closest to the expectation from the standard JES correction~\cite{bib:jes}, the difference between the obtained and ideal response is at the level of 1\%.

\begin{figure}
\centering
\begin{overpic}[width=0.49\columnwidth]{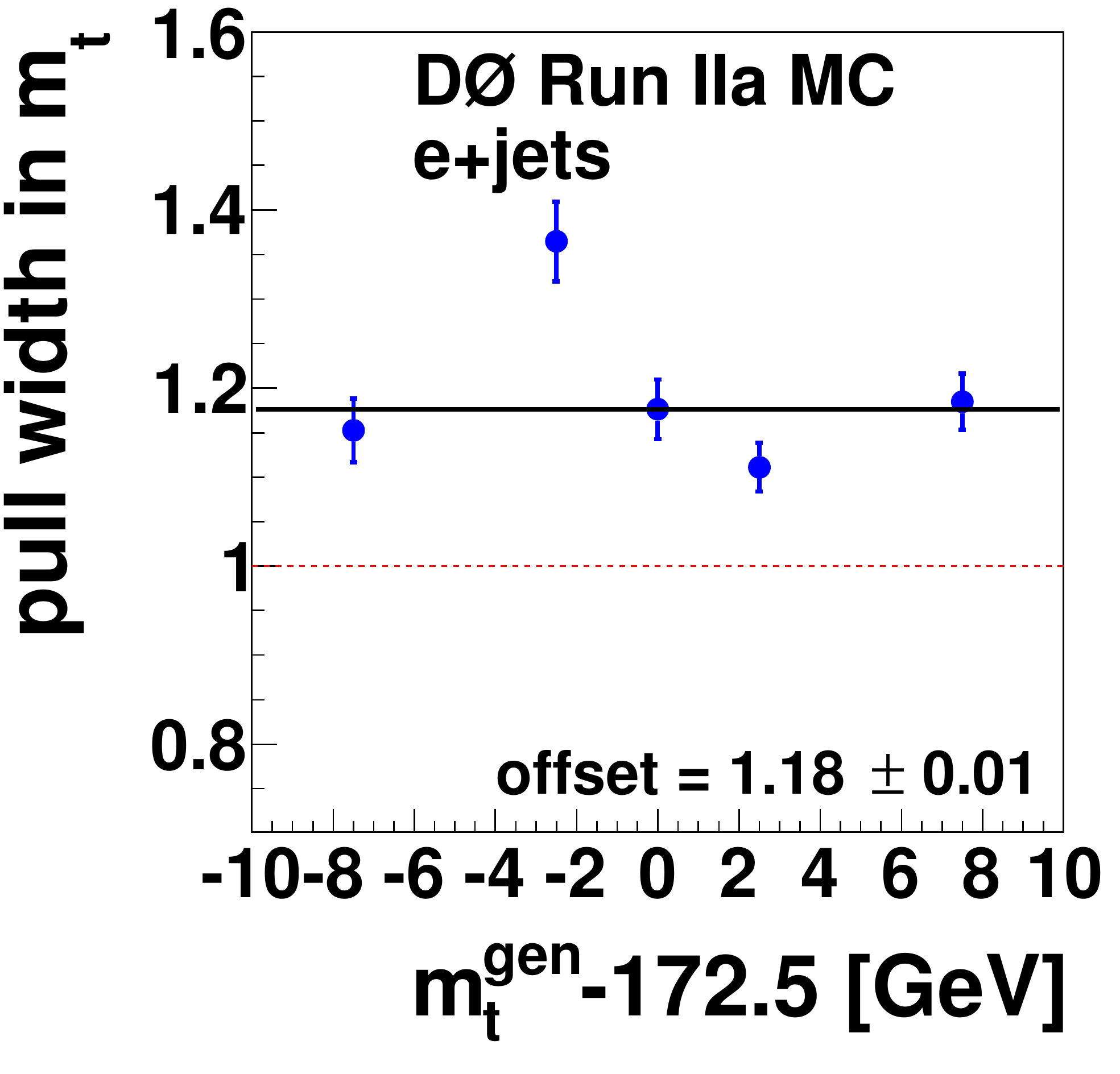}
\put(26,86){\footnotesize\textsf{\textbf{(a)}}}
\end{overpic}
\begin{overpic}[width=0.49\columnwidth]{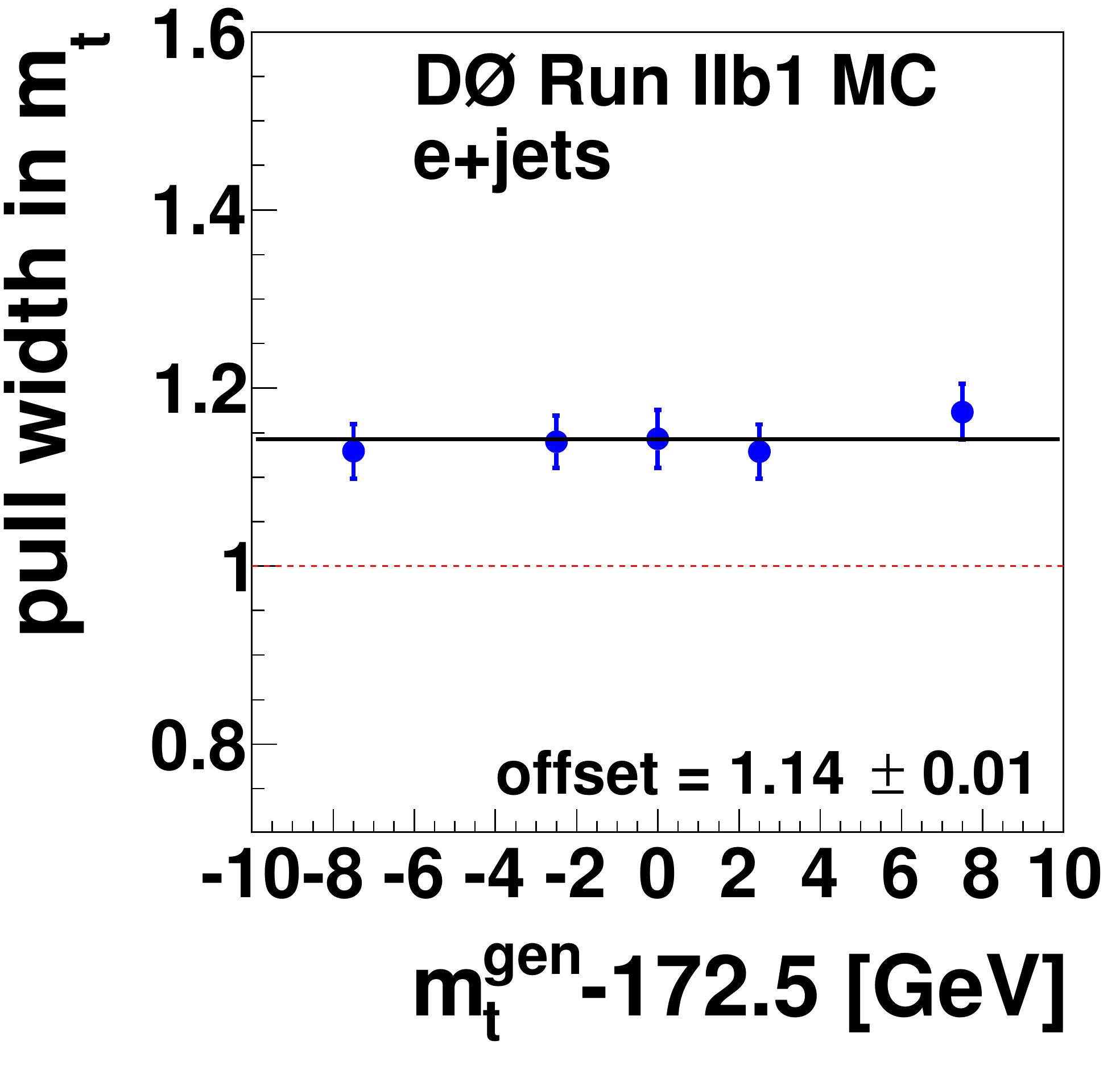}
\put(26,86){\footnotesize\textsf{\textbf{(b)}}}
\end{overpic}
\begin{overpic}[width=0.49\columnwidth]{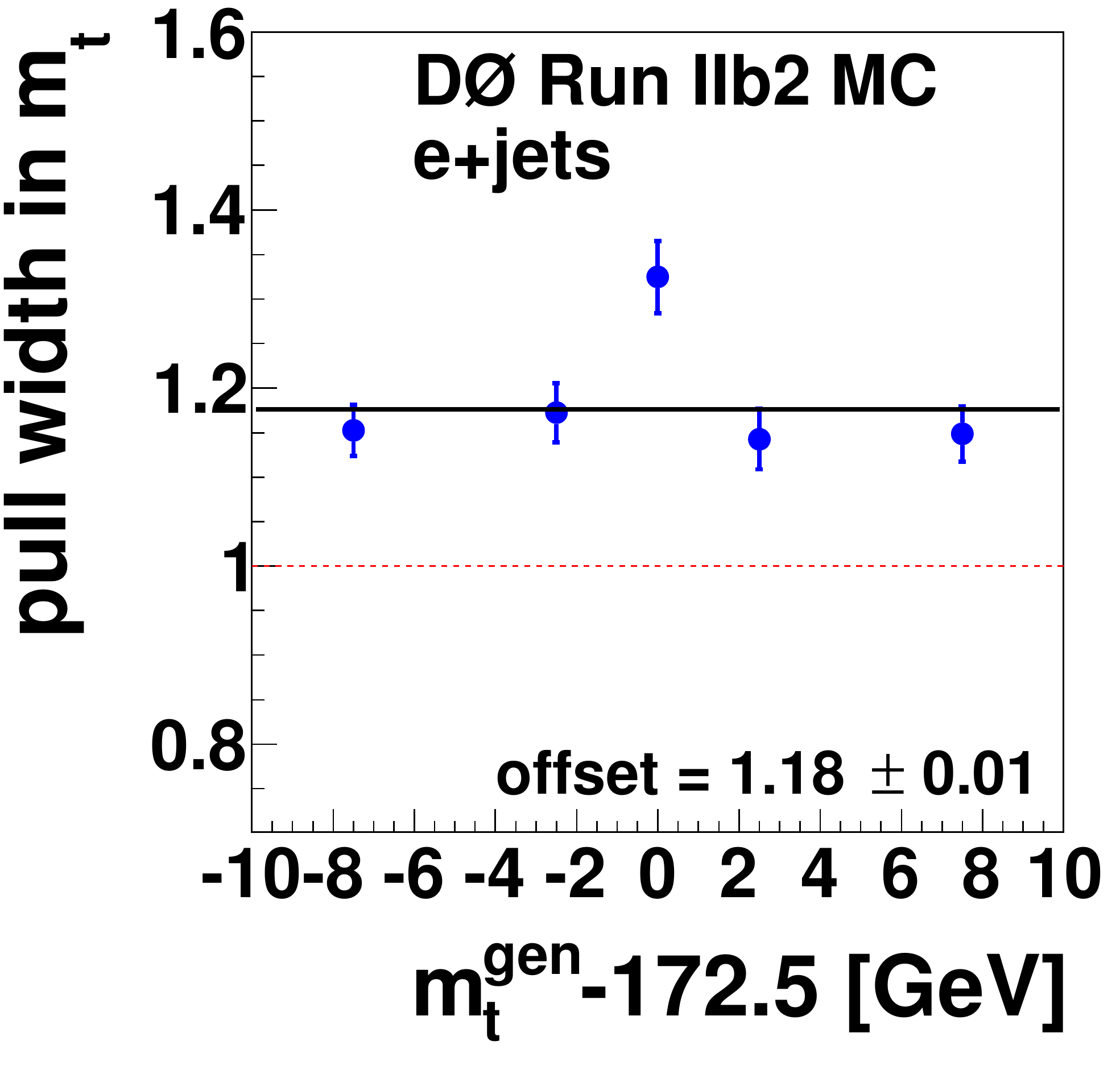}
\put(26,86){\footnotesize\textsf{\textbf{(c)}}}
\end{overpic}
\begin{overpic}[width=0.49\columnwidth]{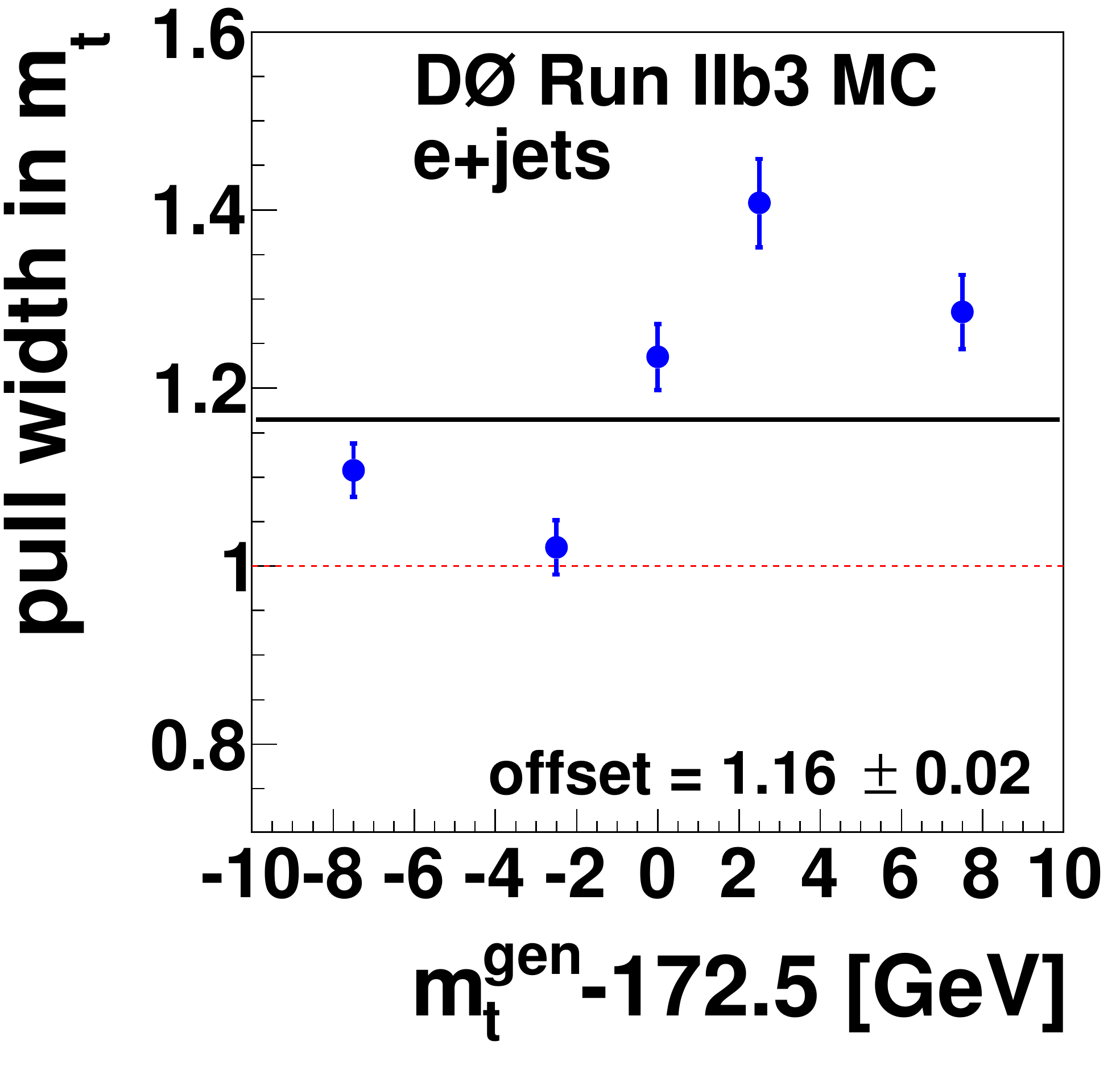}
\put(26,86){\footnotesize\textsf{\textbf{(d)}}}
\end{overpic}
\caption{\label{fig:pullmtem}
The width of the distribution of the pull in $\mt$, obtained using simulations at each \mtgen point, as a function of \mtgen, for (a)~Run~IIa, (b)~Run~IIb1, (c)~Run~IIb2, and (d)~Run~IIb3 in the \ejets final state. The error bars are given by the uncertainty on the width of the distribution in $\pi_{\mt}$, which is not corrected for the effect that the same event can appear in many PEs, and therefore somewhat underestimated. The pull is fitted with a constant shown as a solid black line, whose value is given at the bottom of the plots. The red broken line represents an ideal pull width of unity.
}
\end{figure}

\begin{figure}
\centering
\begin{overpic}[width=0.49\columnwidth]{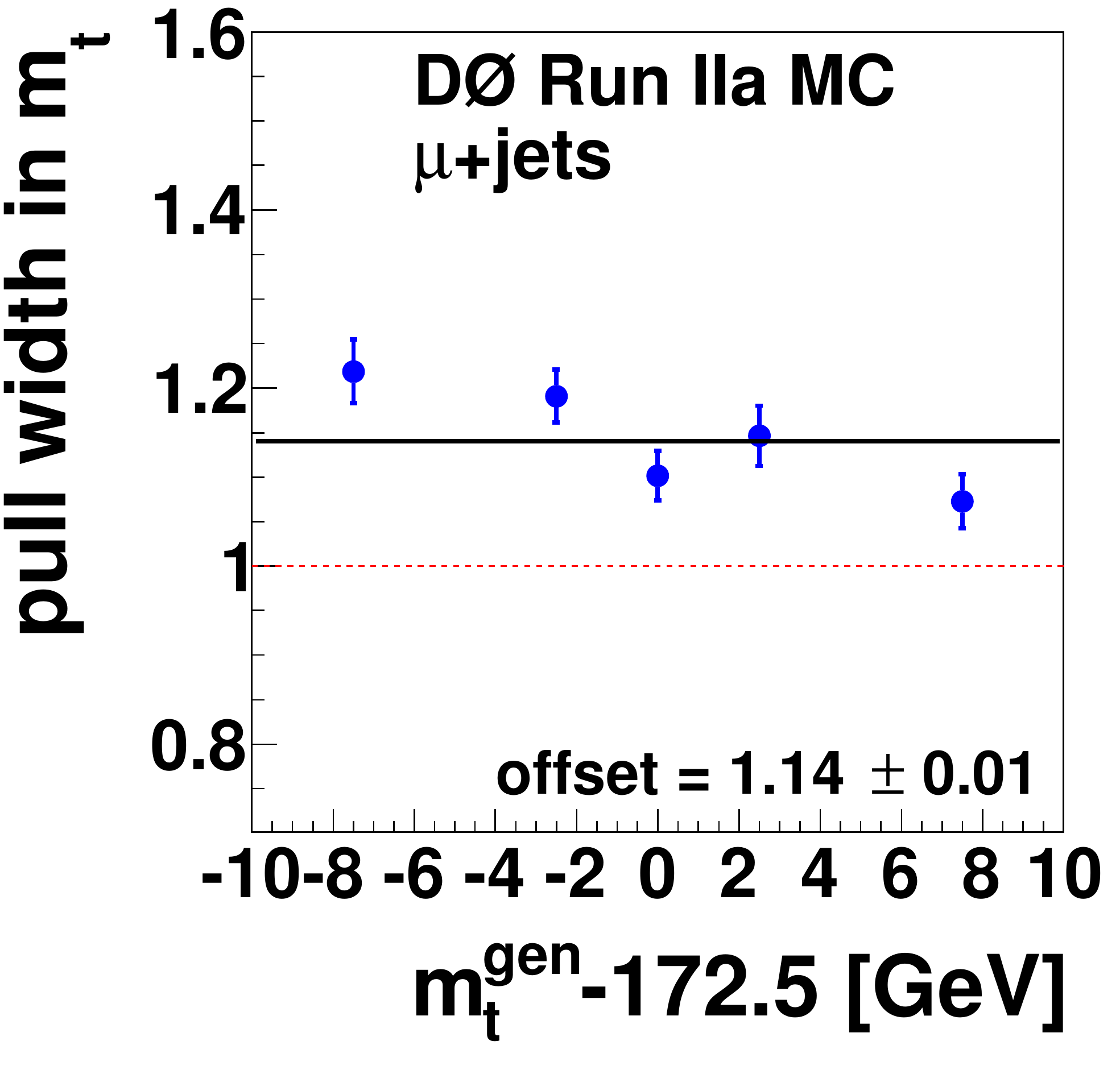}
\put(26,86){\footnotesize\textsf{\textbf{(a)}}}
\end{overpic}
\begin{overpic}[width=0.49\columnwidth]{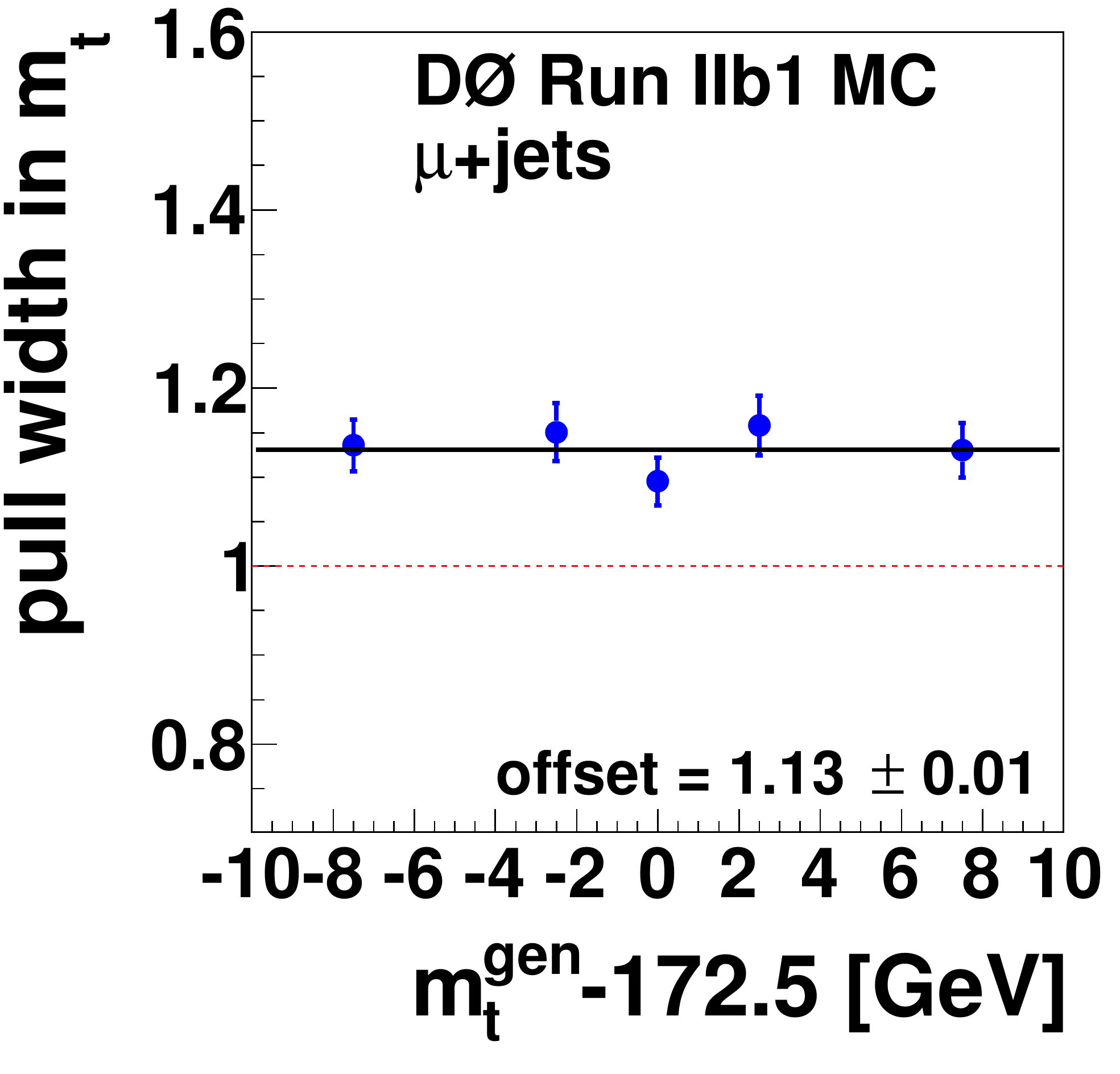}
\put(26,86){\footnotesize\textsf{\textbf{(b)}}}
\end{overpic}
\begin{overpic}[width=0.49\columnwidth]{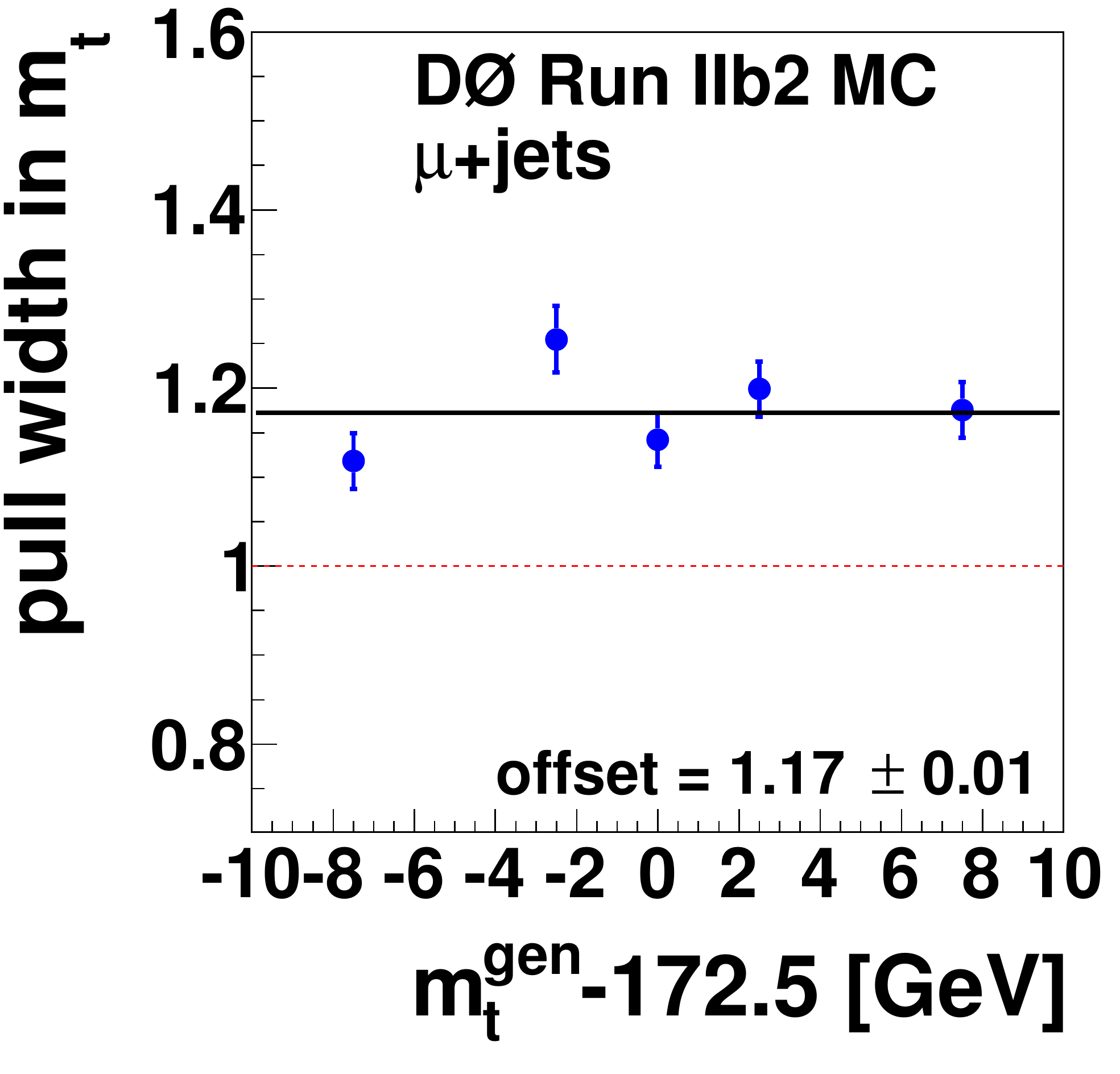}
\put(26,86){\footnotesize\textsf{\textbf{(c)}}}
\end{overpic}
\begin{overpic}[width=0.49\columnwidth]{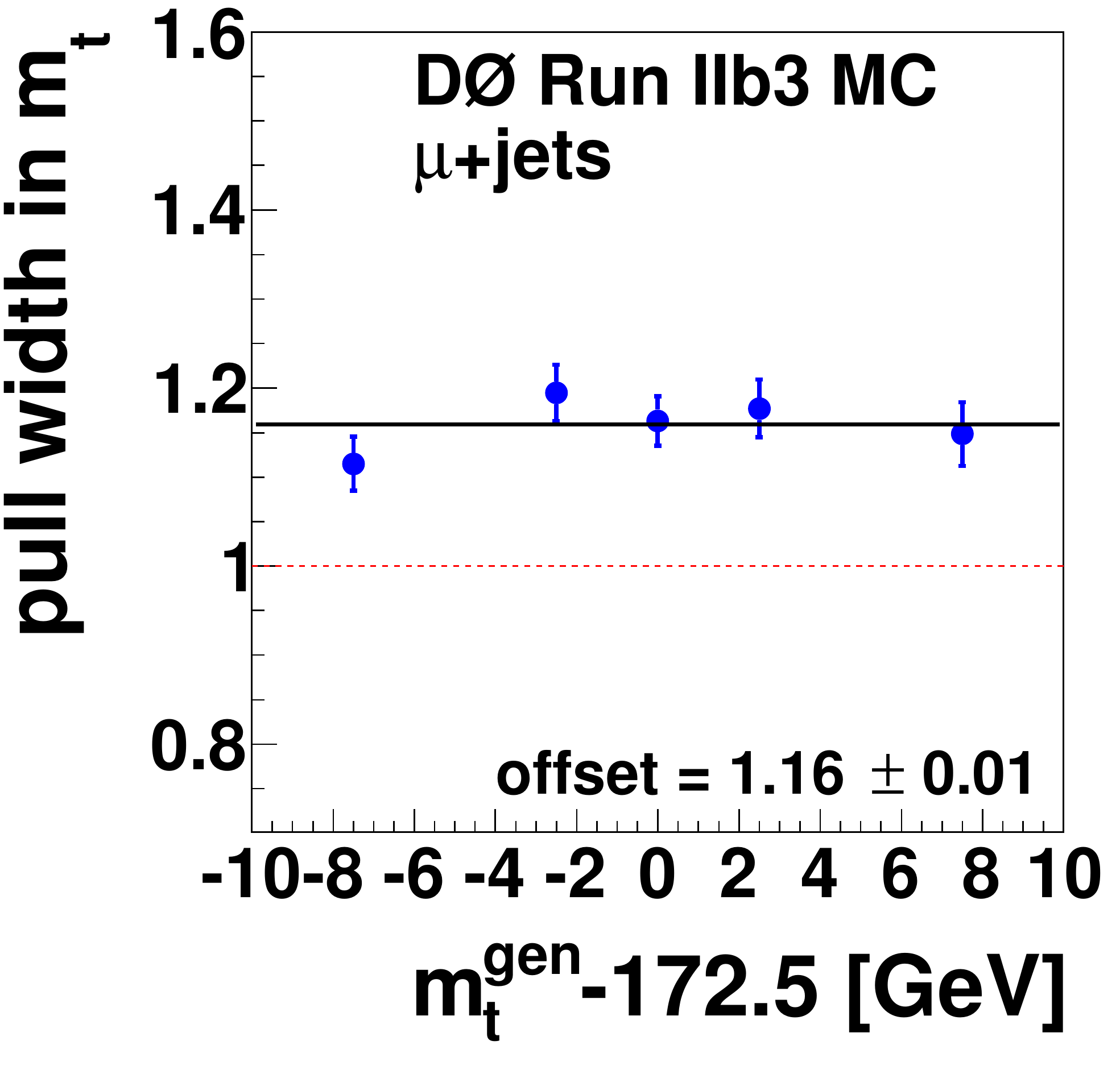}
\put(26,86){\footnotesize\textsf{\textbf{(d)}}}
\end{overpic}
\caption{\label{fig:pullmtmu}
Same as Fig.~\ref{fig:pullmtem}, but in the \mujets final state, for (a)~Run~IIa, (b)~Run~IIb1, (c)~Run~IIb2, and (d)~Run~IIb3.
}
\end{figure}

\begin{figure}
\centering
\begin{overpic}[width=0.49\columnwidth]{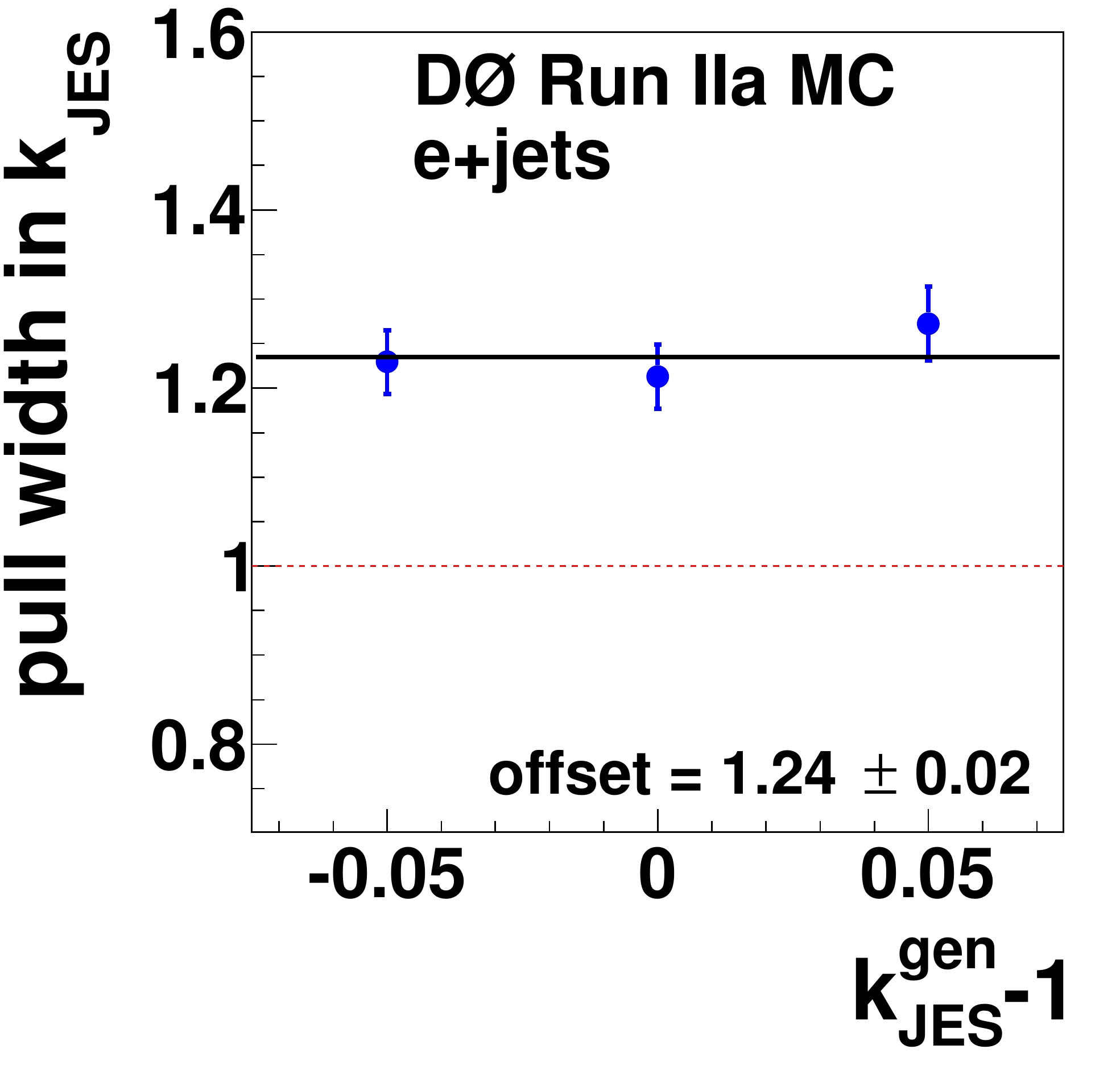}
\put(26,86){\footnotesize\textsf{\textbf{(a)}}}
\end{overpic}
\begin{overpic}[width=0.49\columnwidth]{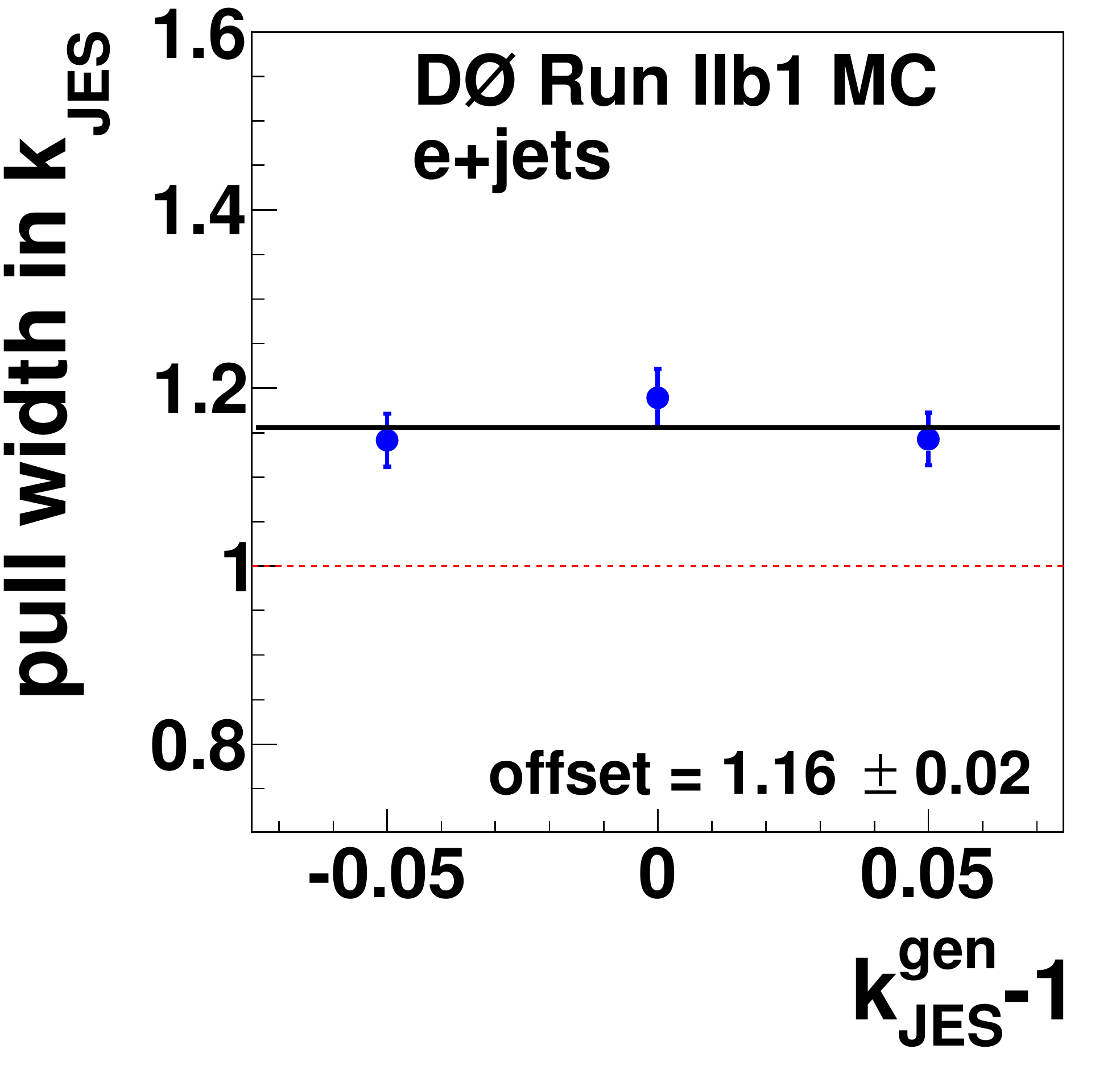}
\put(26,86){\footnotesize\textsf{\textbf{(b)}}}
\end{overpic}
\begin{overpic}[width=0.49\columnwidth]{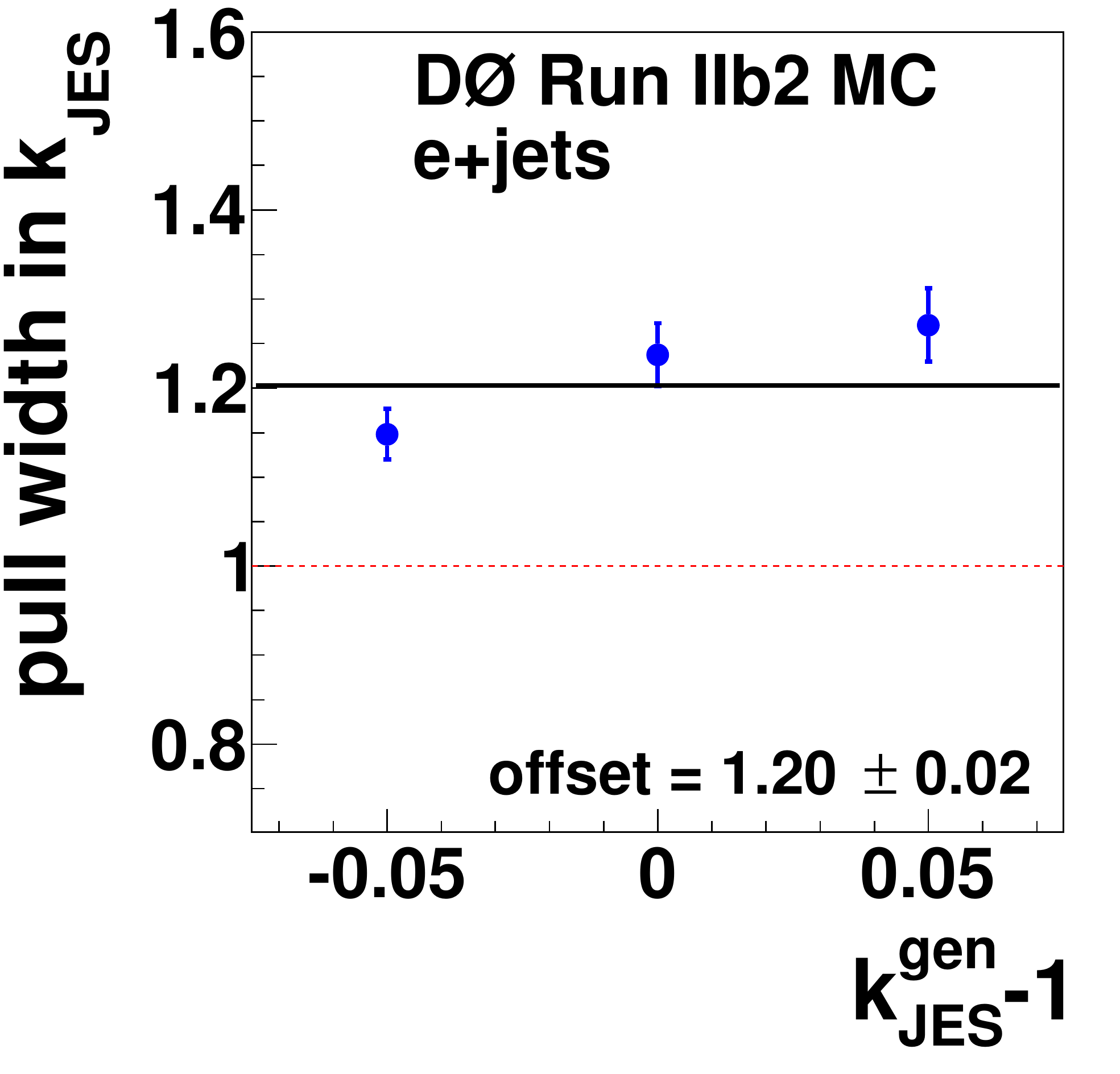}
\put(26,86){\footnotesize\textsf{\textbf{(c)}}}
\end{overpic}
\begin{overpic}[width=0.49\columnwidth]{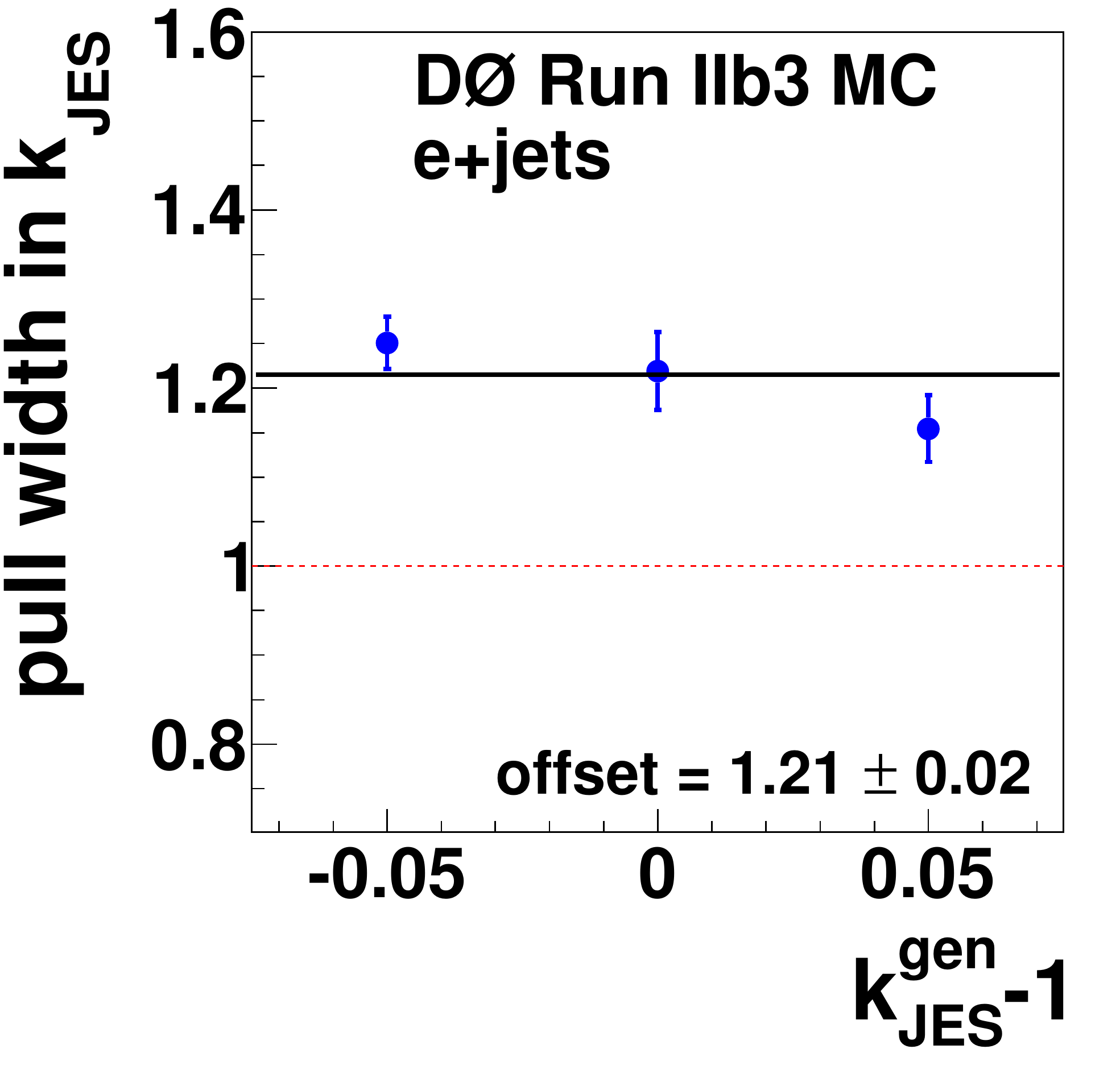}
\put(26,86){\footnotesize\textsf{\textbf{(d)}}}
\end{overpic}
\caption{\label{fig:pullkjesem}
The width of the distribution of the pull in $\kjes$, derived using simulations at each \kjesgen point, as a function of \kjesgen, for (a)~Run~IIa, (b)~Run~IIb1, (c)~Run~IIb2, and (d)~Run~IIb3 in the \ejets final state. The error bars are given by the uncertainty on the width of the distribution in $\pi_{\kjes}$, which is not corrected for the effect that the same event can appear in many PEs, and therefore somewhat underestimated. The pull is fitted with a constant shown as a solid black line, whose value is given at the bottom of the plots. The broken line represents the case of ideal pull width of unity.
}
\end{figure}

\begin{figure}
\centering
\begin{overpic}[width=0.49\columnwidth]{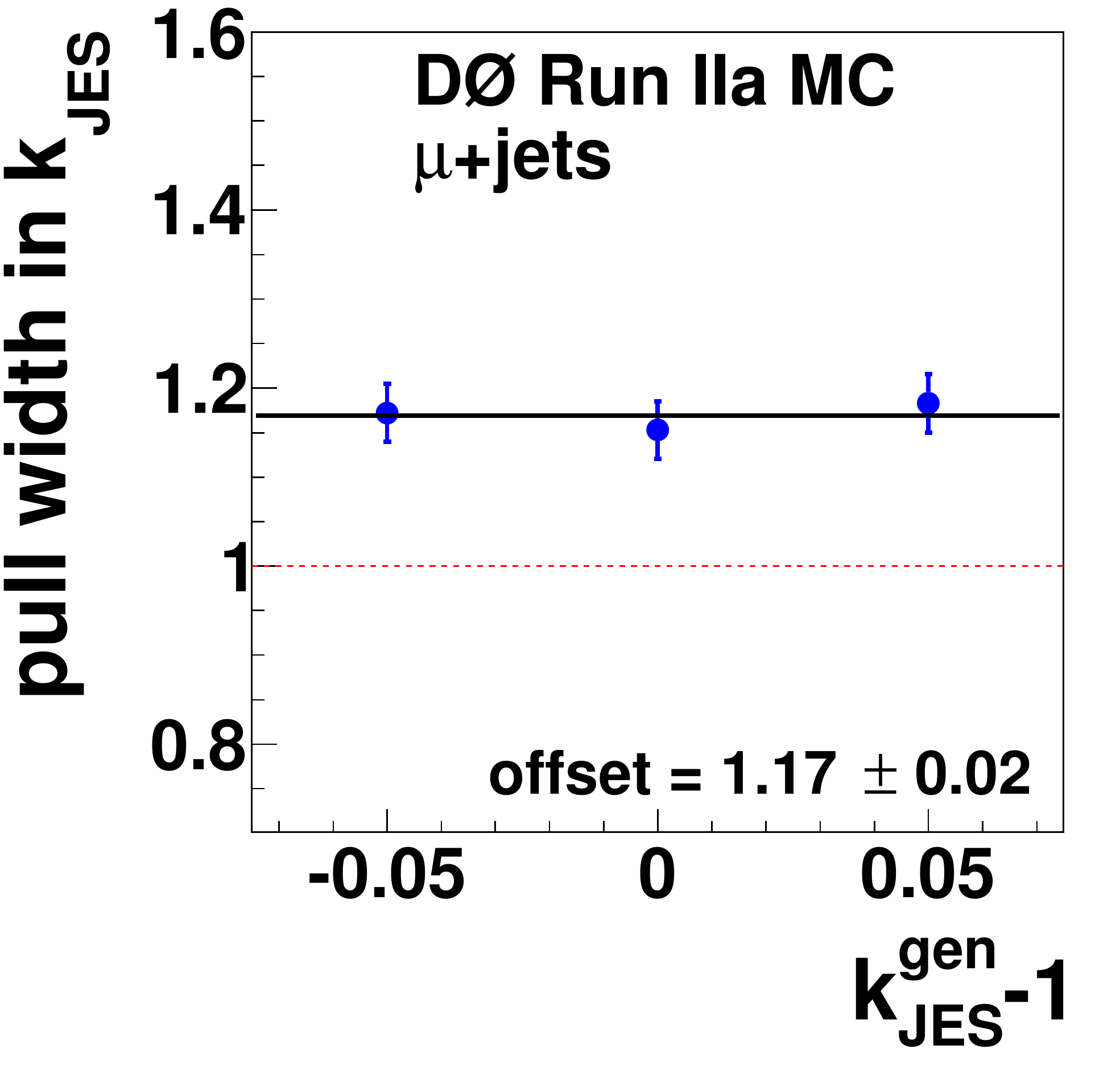}
\put(26,86){\footnotesize\textsf{\textbf{(a)}}}
\end{overpic}
\begin{overpic}[width=0.49\columnwidth]{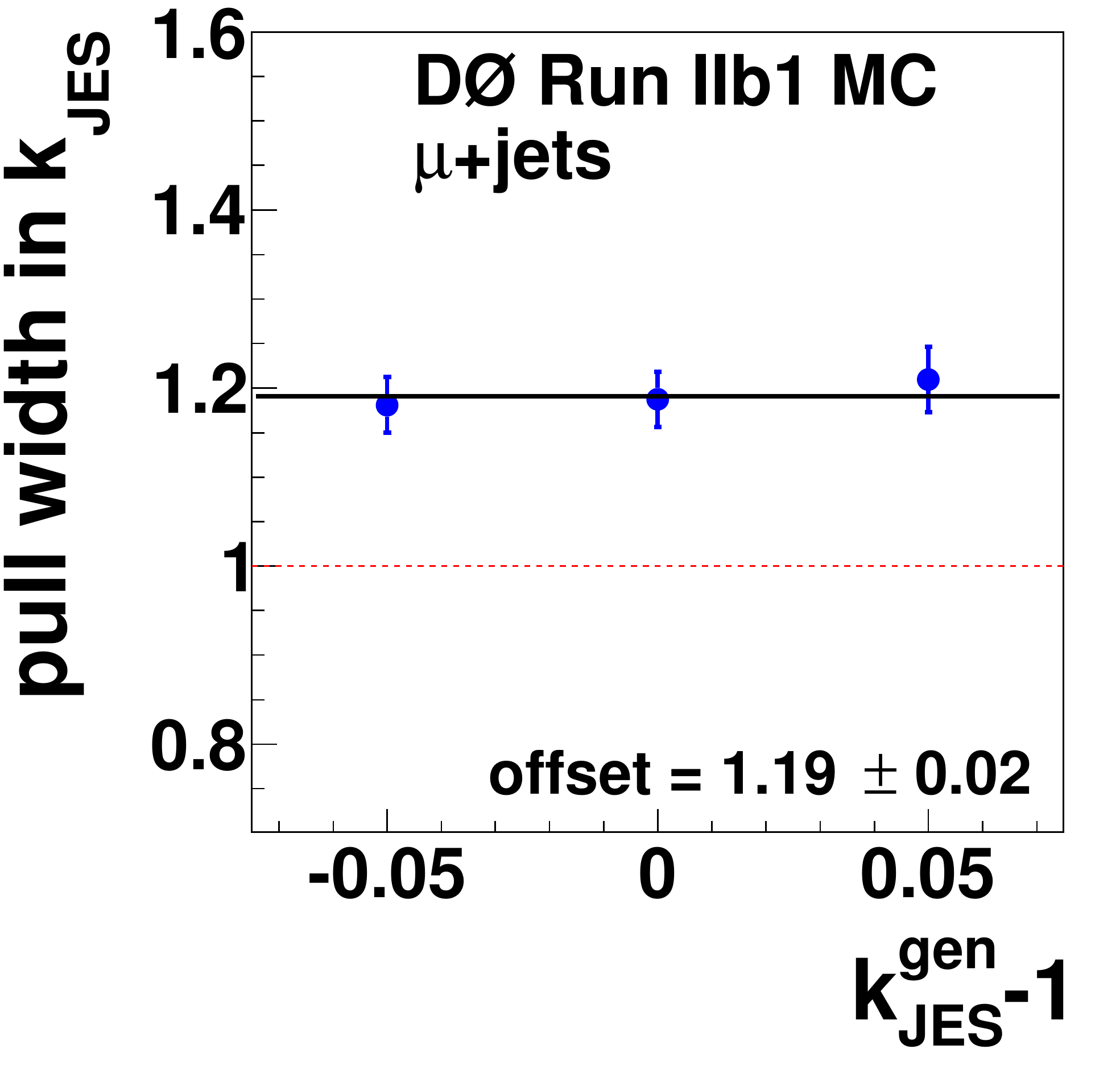}
\put(26,86){\footnotesize\textsf{\textbf{(b)}}}
\end{overpic}
\begin{overpic}[width=0.49\columnwidth]{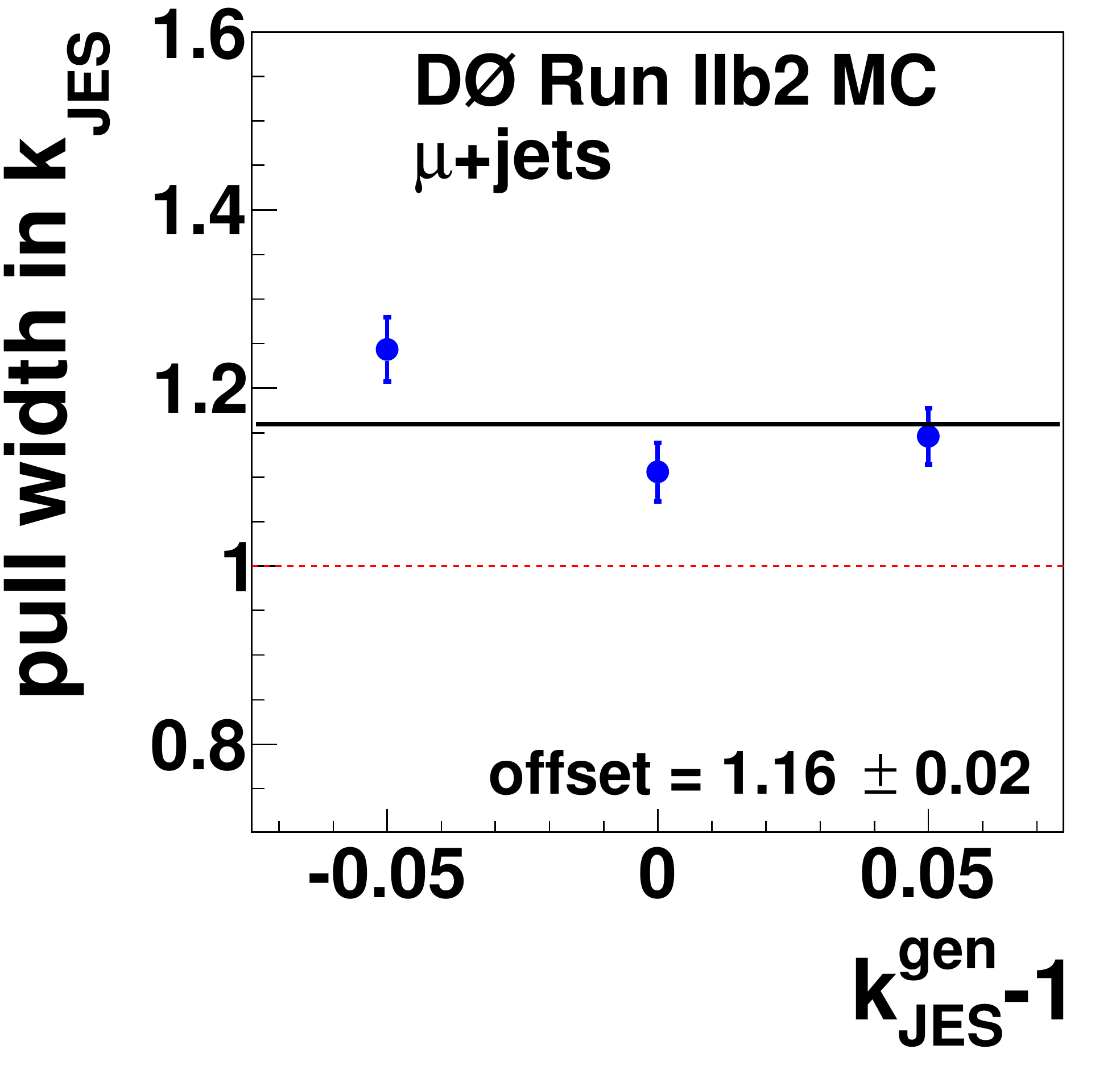}
\put(26,86){\footnotesize\textsf{\textbf{(c)}}}
\end{overpic}
\begin{overpic}[width=0.49\columnwidth]{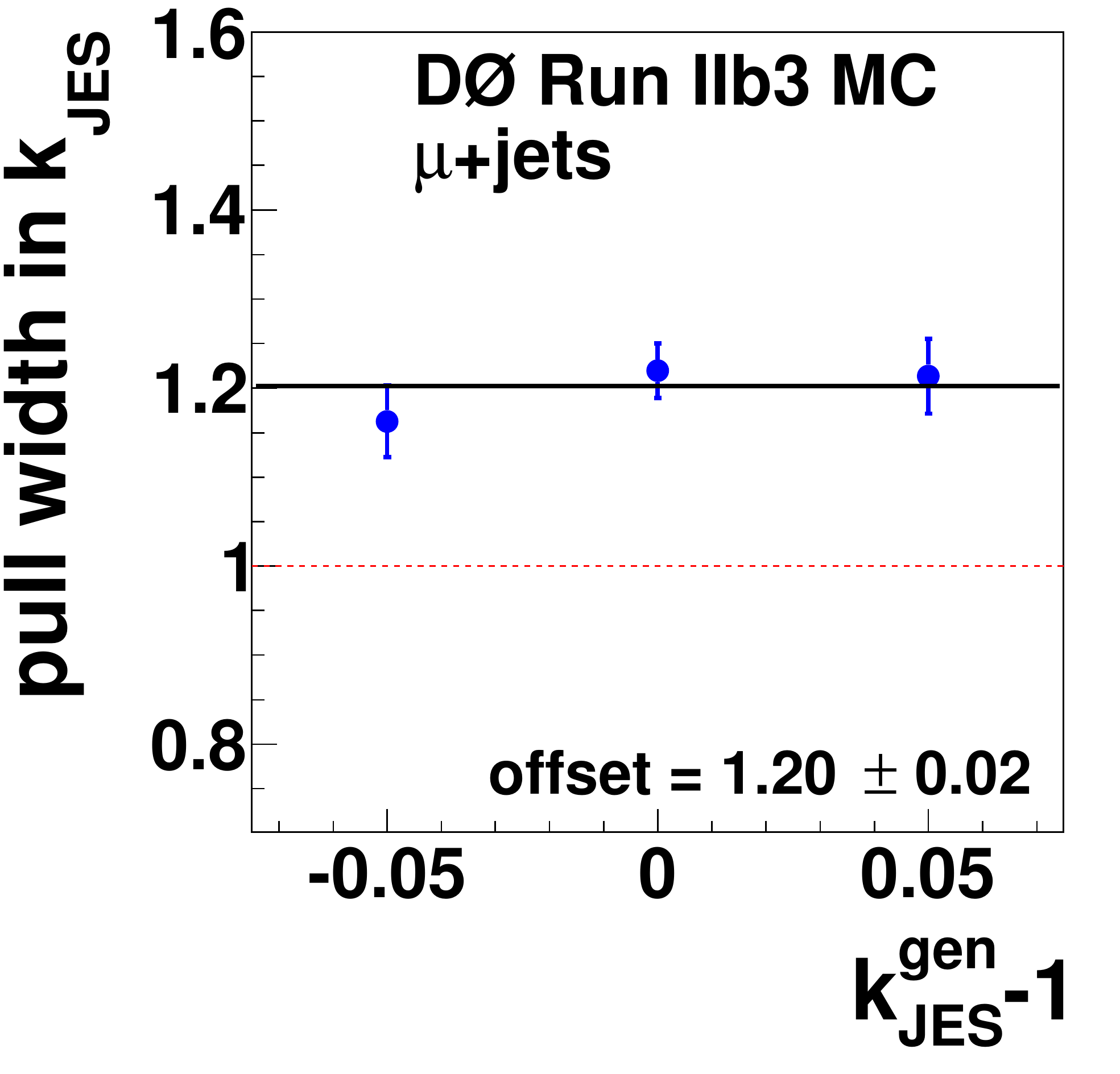}
\put(26,86){\footnotesize\textsf{\textbf{(d)}}}
\end{overpic}
\caption{\label{fig:pullkjesmu}
Same as Fig.~\ref{fig:pullmtem}, but in the \mujets final state, as derived for (a)~Run~IIa, (b)~Run~IIb1, (c)~Run~IIb2, and (d)~Run~IIb3.
}
\end{figure}

To check the validity of the \dmt and \dkjes uncertainties extracted from the ME estimates described in Section~\ref{sec:extract}, we calculate the pulls in \mt ($\pi_{\mt}$) and in \kjes ($\pi_{\kjes}$), according to
\begin{eqnarray}
\pi_{\mt} &=& \frac{\mtfit-\langle\mtfit\rangle}{\dmtfit}\,, \\
\pi_{\kjes} &=& \frac{\kjesfit-\langle\kjesfit\rangle}{\dkjesfit}\,.
\end{eqnarray}
For an ideal response, the pull should follow a Gaussian distribution of unit width centered at zero. In our case, both $\pi_{\mt}$ and $\pi_{\kjes}$ are Gaussian distributed, and centered within uncertainties at zero. For each of the simulated $(\mt,\kjes)$ points, we determine the pull widths from the $\sigma$ parameters of Gaussian fits to $\pi_{\mt}$ and $\pi_{\kjes}$. The average width of $\pi_{\mt}$ is then obtained from a constant fit of the distribution of pull width versus \mtgen, as shown for each epoch in Fig.~\ref{fig:pullmtem} for \ejets and in Fig.~\ref{fig:pullmtmu} for \mujets final states. The average width of $\pi_{\kjes}$ is obtained in the same way and is given for each epoch in Fig.~\ref{fig:pullkjesem} for \ejets and in Fig.~\ref{fig:pullkjesmu} for \mujets final states. Typically, pull widths are  $\approx20\%$ higher than unity, indicating that the uncertainties \dmtfit and \dkjesfit estimated with the ME technique are somewhat underestimated. They are therefore corrected by inflating by the value of the pull width found for the respective epoch and final state.

\subsection{
Results
}
\label{sec:resultmt}

Here, we present the results of our application of the ME technique to the full Run~II data sample corresponding to 9.7~\fb of integrated luminosity. The analysis is performed blinded in \mt by adding a constant unknown offset to the extracted \mt value. The offset is drawn according to a uniform distribution in the range $[-2~\GeV,2~\GeV]$. This interval roughly corresponds to $[-3\sigma_{\rm WA},3\sigma_{\rm WA}]$, where $\sigma_{\rm WA}=0.76~\GeV$ denotes the total uncertainty on the current world average in \mt.

The method to extract \mt and \kjes is described in Section~\ref{sec:extract}. The linear calibrations from Section~\ref{sec:calibmt} are applied to data through the likelihoods $\like(\X_N;\,\mt)$, defined in Eq.~(\ref{eq:likemt}), and $\like(\X_N;\,\kjes)$ by transforming the \mt and \kjes parameters according to
\begin{linenomath}
\begin{eqnarray}
\mt \to \mt'     & = & \frac{(\mt-172.5~\GeV)-o_{\mt}}{s_{\mt}} + 172.5~\GeV,\hspace{10pt}\\
\kjes \to \kjes' & = & \frac{(\kjes-1)-o_{\kjes}}{s_{\kjes}}+1\,,
\end{eqnarray}
\end{linenomath}
where the offset and slope parameters are given in Figs.~\ref{fig:mtem}-\ref{fig:kjesmu}. The direct calibration of the likelihoods $\like(\X_N;\,\mt)$ and $\like(\X_N;\,\kjes)$ takes into account the effect of slopes $\neq1$ when estimating the statistical uncertainties \dmt and \dkjes.
Subsequently, the slope-corrected uncertainties \dmt are calibrated by multiplying them with the average widths of the distributions in $\pi_{\mt}$, summarized  in Fig.~\ref{fig:pullmtem} for \ejets and in Fig.~\ref{fig:pullmtmu} for \mujets final states. Similarly, the uncertainties \dkjes are calibrated with the average widths in $\pi_{\kjes}$, given in Figs.~\ref{fig:pullkjesem} and~\ref{fig:pullkjesmu}.

\begin{figure}
\centering
\begin{overpic}[width=0.99\columnwidth]{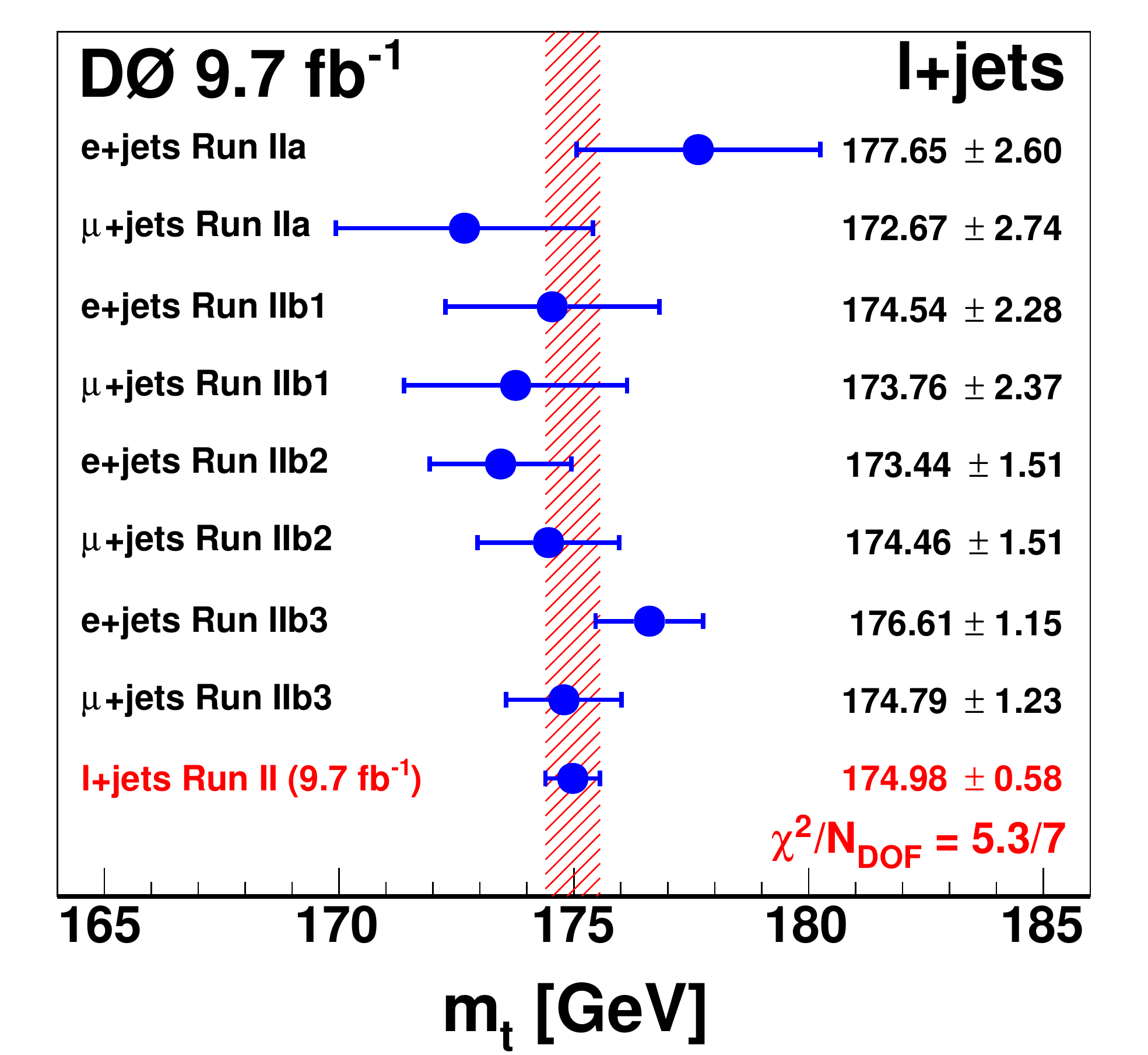}
\end{overpic}
\caption{
\label{fig:mt_overview}
Summary of the measured \mt values with their statistical uncertainties for each data-taking epoch and split according to final state. The red-shaded band indicates the statistical uncertainty on the combined \mt value in Run~II data corresponding to 9.7~\fb of integrated luminosity.
}
\end{figure}

\begin{figure}
\centering
\begin{overpic}[width=0.49\columnwidth]{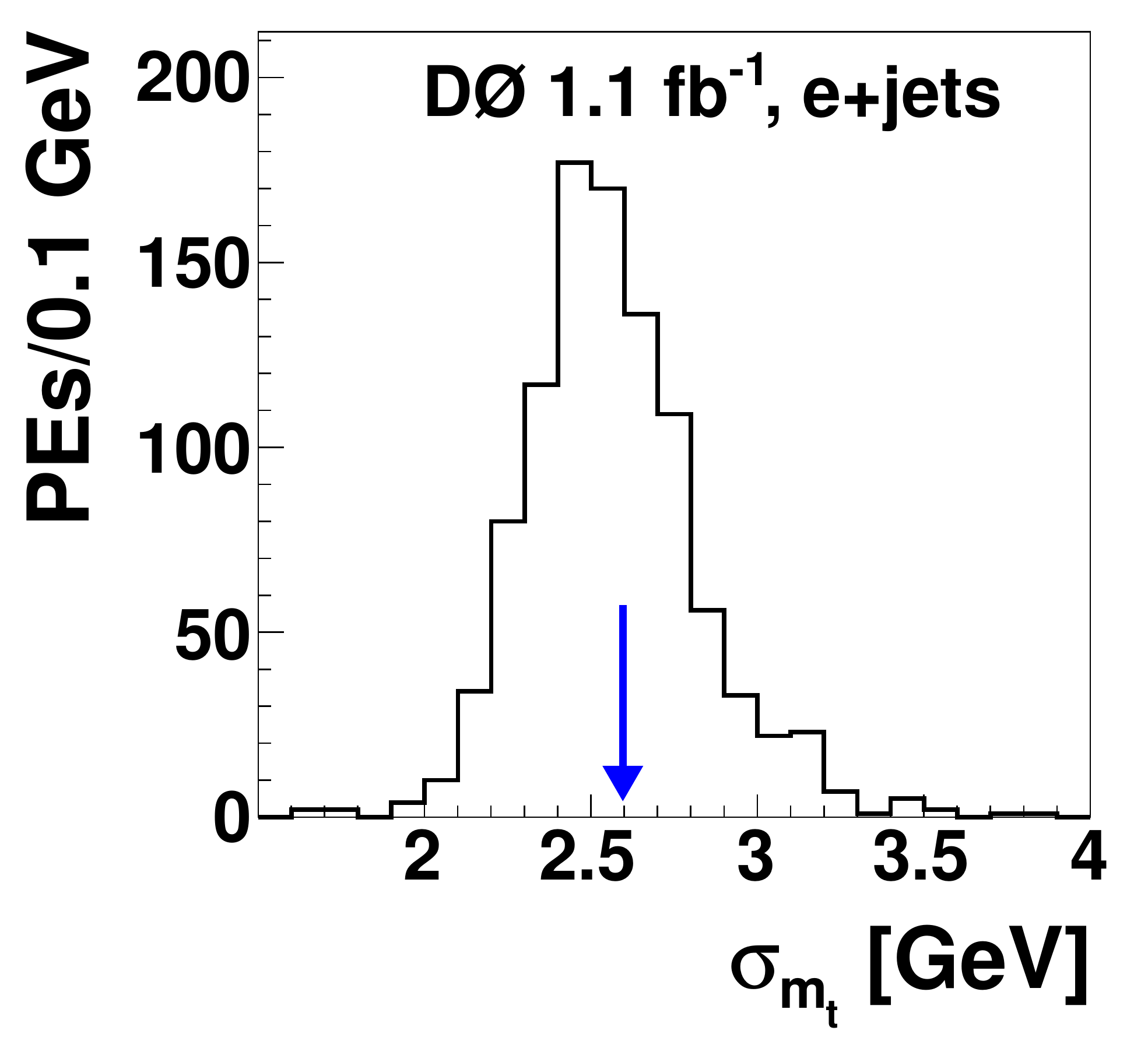}
\put(26,82){\footnotesize\textsf{\textbf{(a)}}}
\end{overpic}
\begin{overpic}[width=0.49\columnwidth]{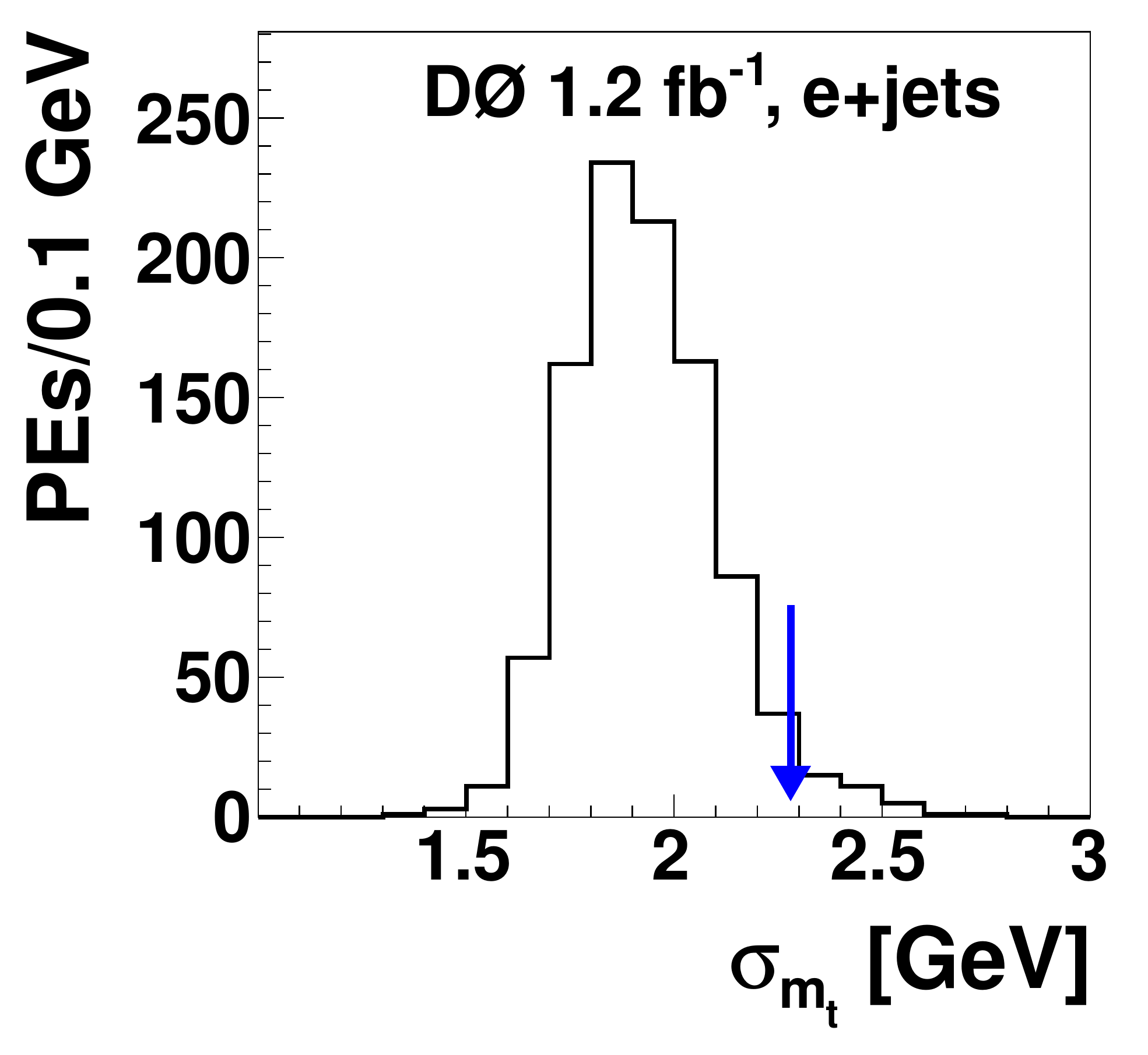}
\put(26,82){\footnotesize\textsf{\textbf{(b)}}}
\end{overpic}
\begin{overpic}[width=0.49\columnwidth]{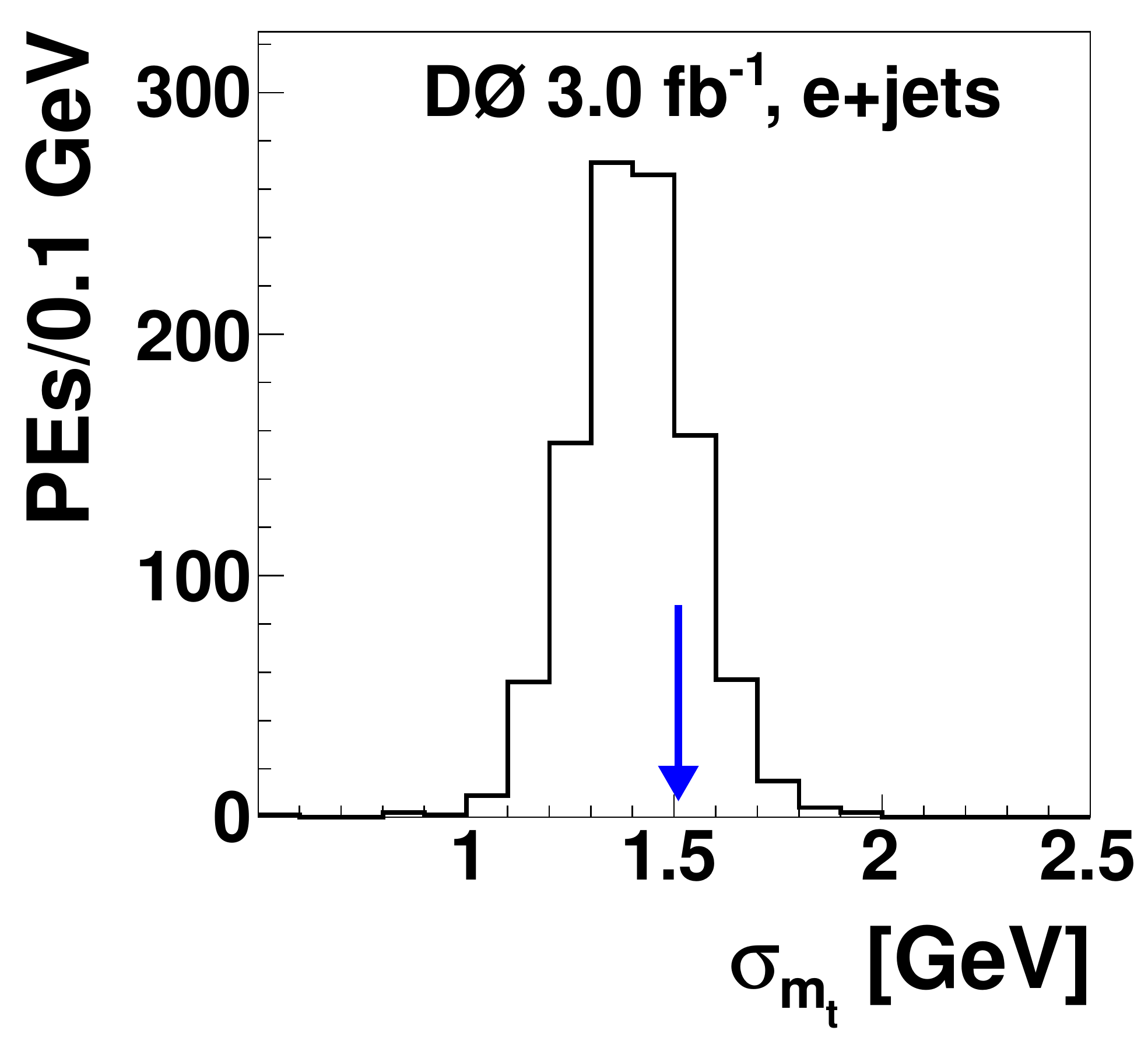}
\put(26,82){\footnotesize\textsf{\textbf{(c)}}}
\end{overpic}
\begin{overpic}[width=0.49\columnwidth]{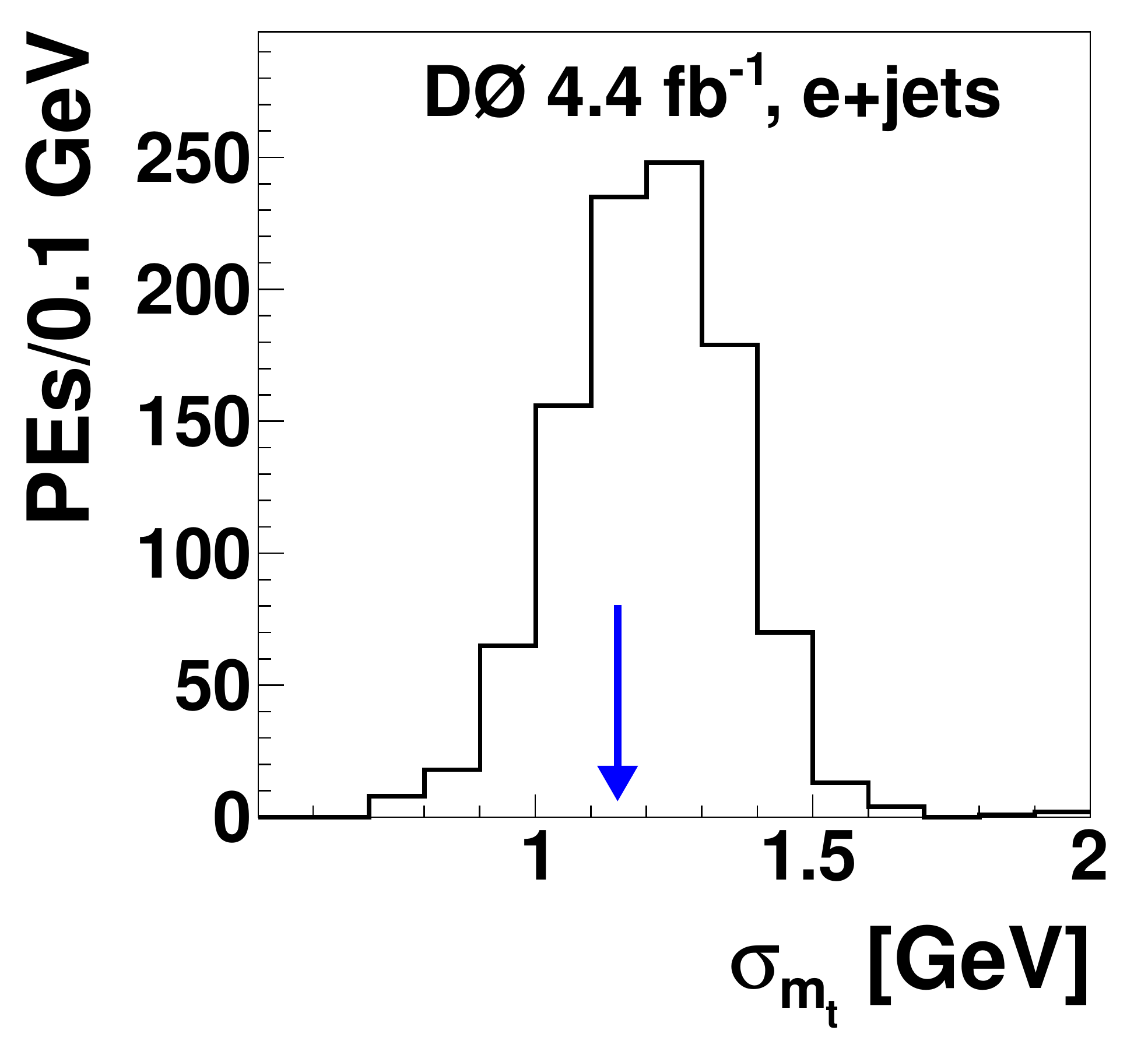}
\put(26,82){\footnotesize\textsf{\textbf{(d)}}}
\end{overpic}
\caption{
\label{fig:uncertem}
Distribution of expected statistical uncertainties \dmt in \ejets final states in (a)~Run~IIa, (b)~Run~IIb1, (c)~Run~IIb2, and (d)~Run~IIb3, obtained in 1000 PEs at $\mtgen=172.5~\GeV$ and $\kjes=1$, after calibrating for  non-unit width of the distribution in $\pi_{\mt}$. The observed statistical uncertainties in data are indicated by arrows.
}
\end{figure}

\begin{figure}
\centering
\begin{overpic}[width=0.49\columnwidth]{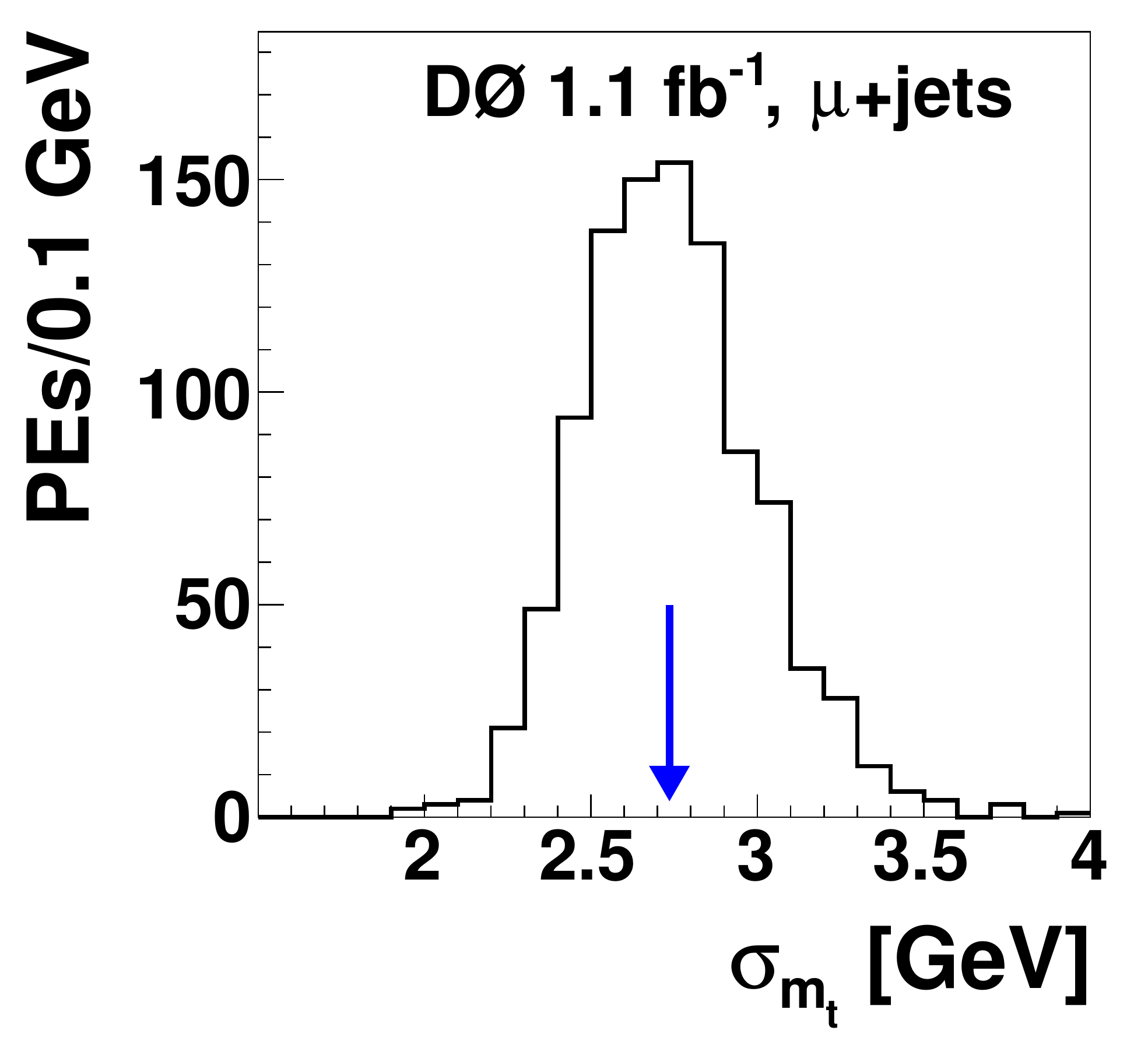}
\put(26,82){\footnotesize\textsf{\textbf{(a)}}}
\end{overpic}
\begin{overpic}[width=0.49\columnwidth]{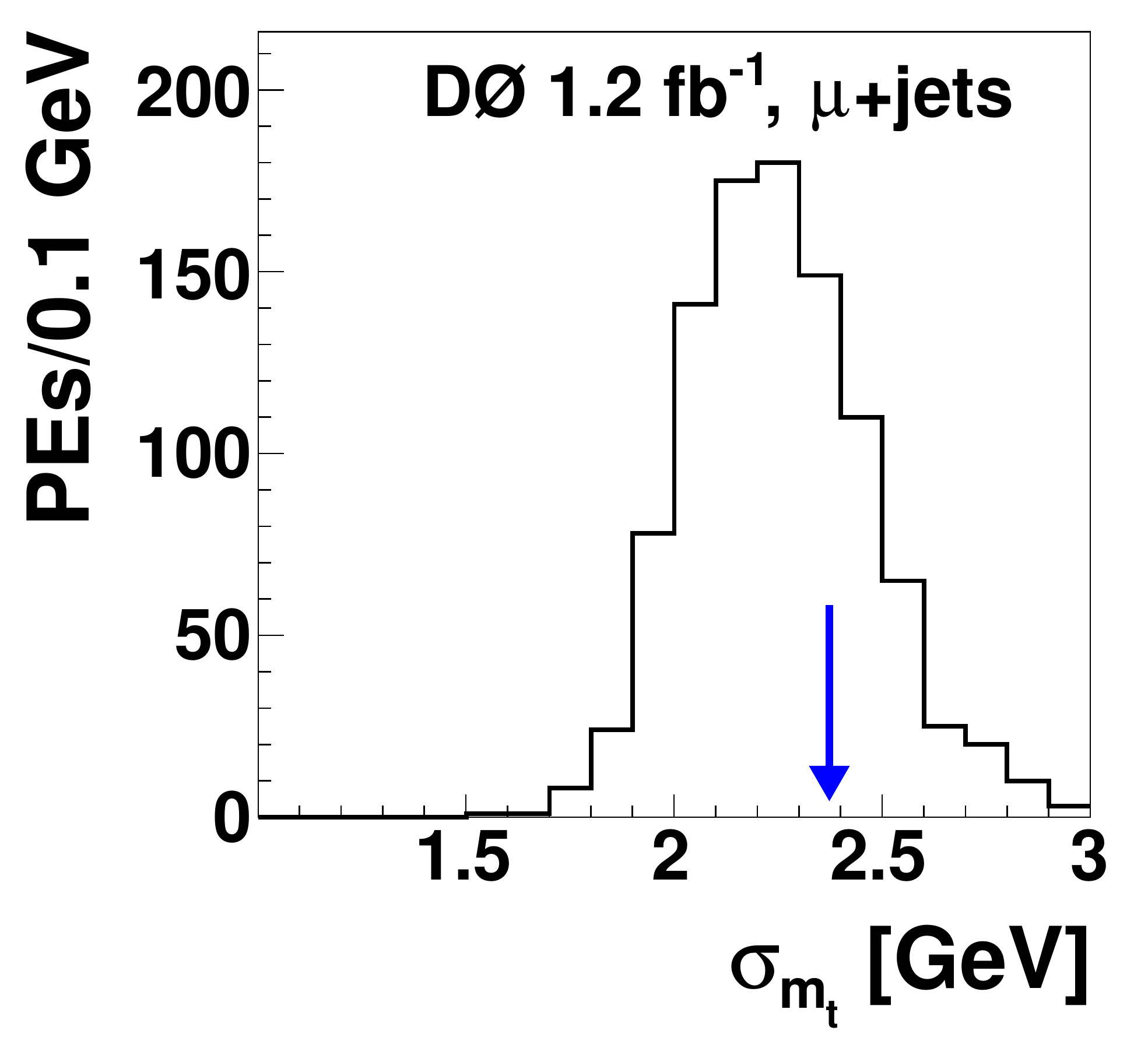}
\put(26,82){\footnotesize\textsf{\textbf{(b)}}}
\end{overpic}
\begin{overpic}[width=0.49\columnwidth]{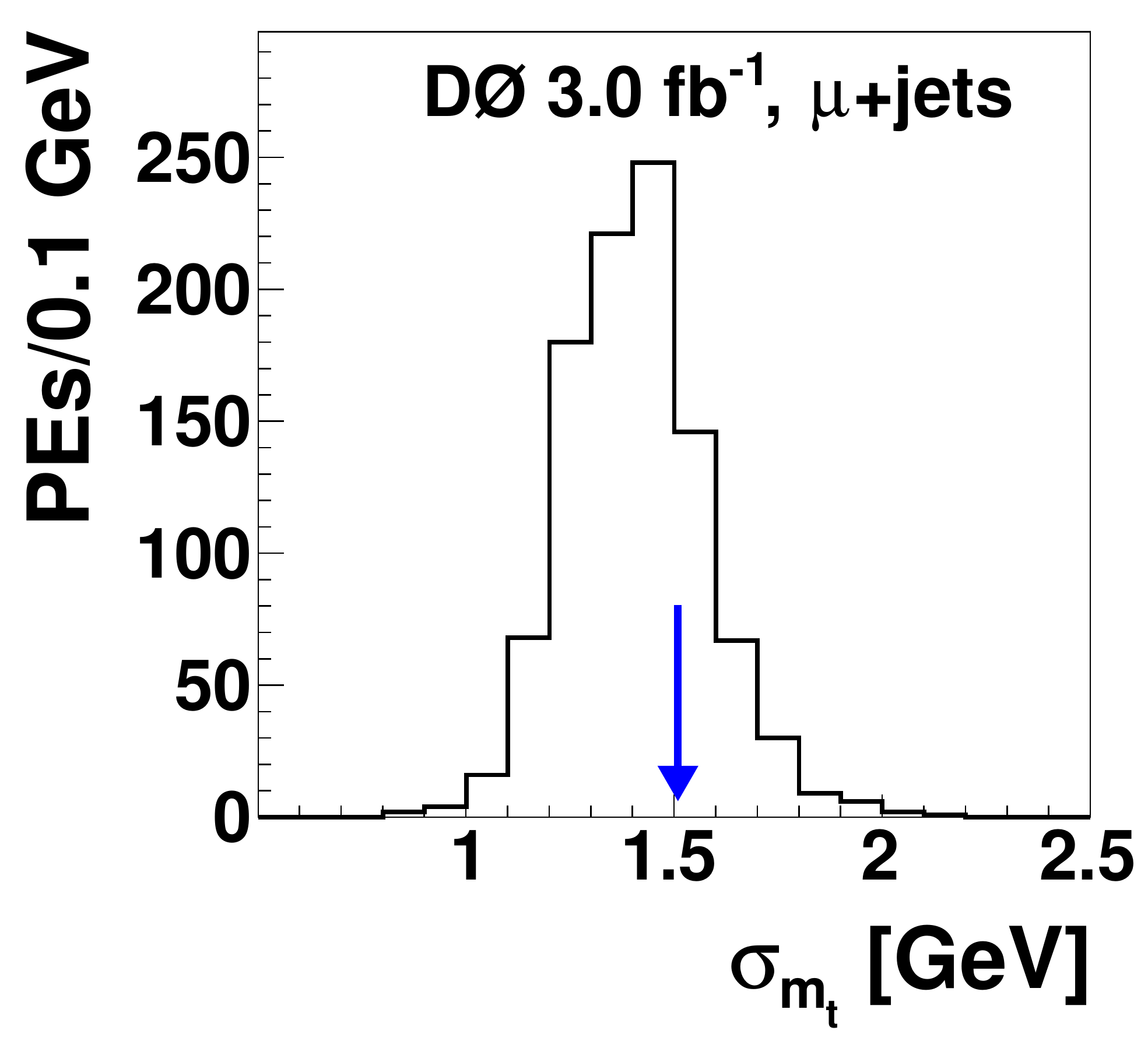}
\put(26,82){\footnotesize\textsf{\textbf{(c)}}}
\end{overpic}
\begin{overpic}[width=0.49\columnwidth]{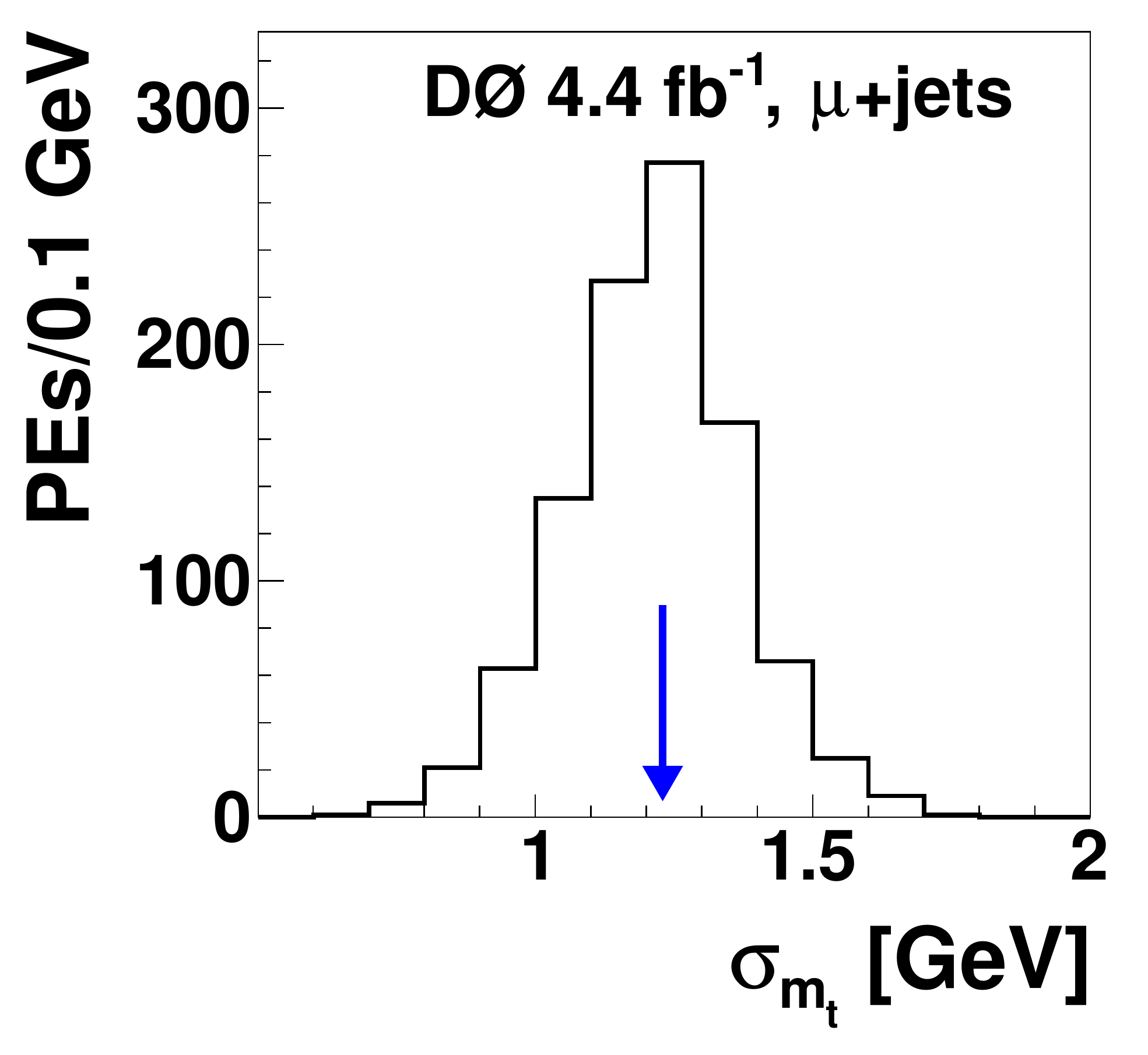}
\put(26,82){\footnotesize\textsf{\textbf{(d)}}}
\end{overpic}
\caption{
\label{fig:uncertmu}
Same as Fig.~\ref{fig:uncertem}, in \mujets final states in (a)~Run~IIa, (b)~Run~IIb1, (c)~Run~IIb2, and (d)~Run~IIb3.
}
\end{figure}

The \mt measurements are summarized in Fig.~\ref{fig:mt_overview}, along with their statistical uncertainties, which include the statistical contribution from \kjes. The observed statistical uncertainty is compared with expectations from pseudo-experiments for $\mtgen=172.5~\GeV$ in Fig.~\ref{fig:uncertem} for \ejets and in Fig.~\ref{fig:uncertmu} for \mujets final states. The observed uncertainties are consistent with expectations from simulation. We obtain the combined \mt for the full Run~II data set of 9.7~\fb by calculating the uncertainty-weighted mean, which is the best linear unbiased estimator in case of Gaussian uncertainties, according to
\begin{equation}
\label{eq:combi}
\langle\mt\rangle    \equiv \frac{ \sum_{i=1}^8\frac{m_{t,i}}{\sigma_{\mt,i}^2} }{ \sum_{i=1}^8\frac{1}{\sigma_{\mt,i}^2} }\,,
\qquad\qquad
\langle\sigma\rangle \equiv \left( \sum_{i=1}^8 \frac{1}{\sigma_{\mt,i}^2} \right)^{-\frac12}  \,,
\end{equation}
where $i$ runs over all final states and epochs. The combined result after unblinding is $\mt=174.98\pm0.58~\mbox{(stat+JES)}~\mbox{GeV}$. To verify that the values measured in each epoch and final state are consistent within their statistical \mbox{(stat+JES)} uncertainties, we calculate the value of the $\chi^2$ distribution for 7 degrees of freedom, which yields a probability of 62\%.

\begin{figure}
\centering
\begin{overpic}[width=0.99\columnwidth]{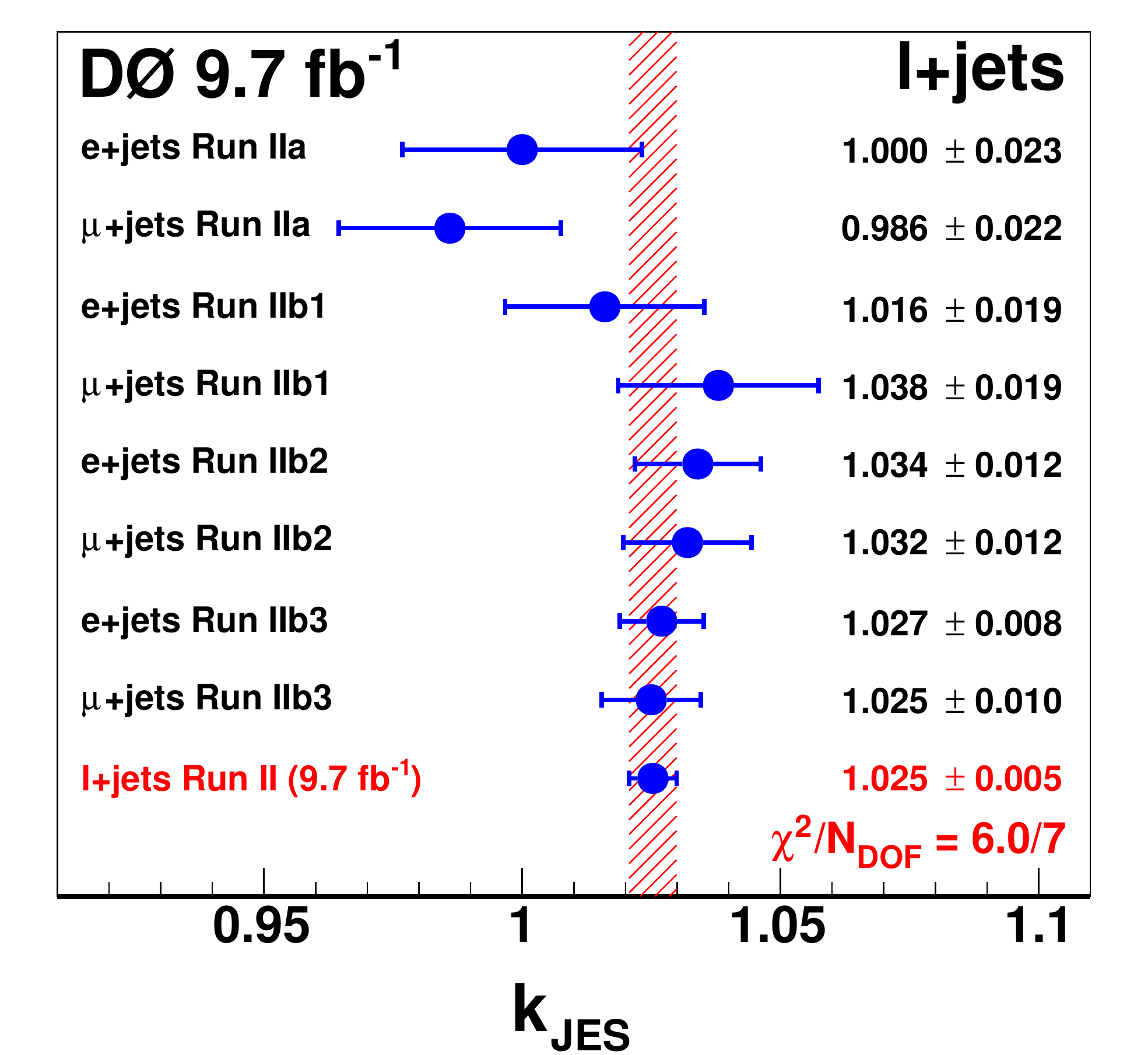}
\end{overpic}
\caption{
\label{fig:kjes_overview}
Same as Fig.~\ref{fig:mt_overview}, but for the \kjes parameter.
}
\end{figure}

The measured \kjes values are shown in Fig.~\ref{fig:kjes_overview} together with their statistical uncertainties. We also compare the observed values of \dkjes with expectations from simulation, and find consistency. We obtain a combined value of $\kjes=1.025\pm0.005~\mbox{(stat+\mt)}$ by applying Eq.~\ref{eq:combi} after replacing $\kjes\leftrightarrow\mt$. The individual measured \kjes are consistent with a $\chi^2$ probability of 54\%. The total JES uncertainty for \ttbar events after all selections is between 2.0\% and 2.1\%, depending on the epoch, consistent with our measured \kjes at the level of $1.3$~standard deviations (SD). Due to the busy environment of \ttbar events containing four jets, a higher fraction of the event's energy is expected in jet cones leading to a higher \kjes according to Eq.~(\ref{eq:kjesdef})), which is not the case in $\gamma+{\rm jet}$ and dijet samples where the JES is derived. 

\begin{figure}
\begin{centering}
\includegraphics[width=0.99\columnwidth]{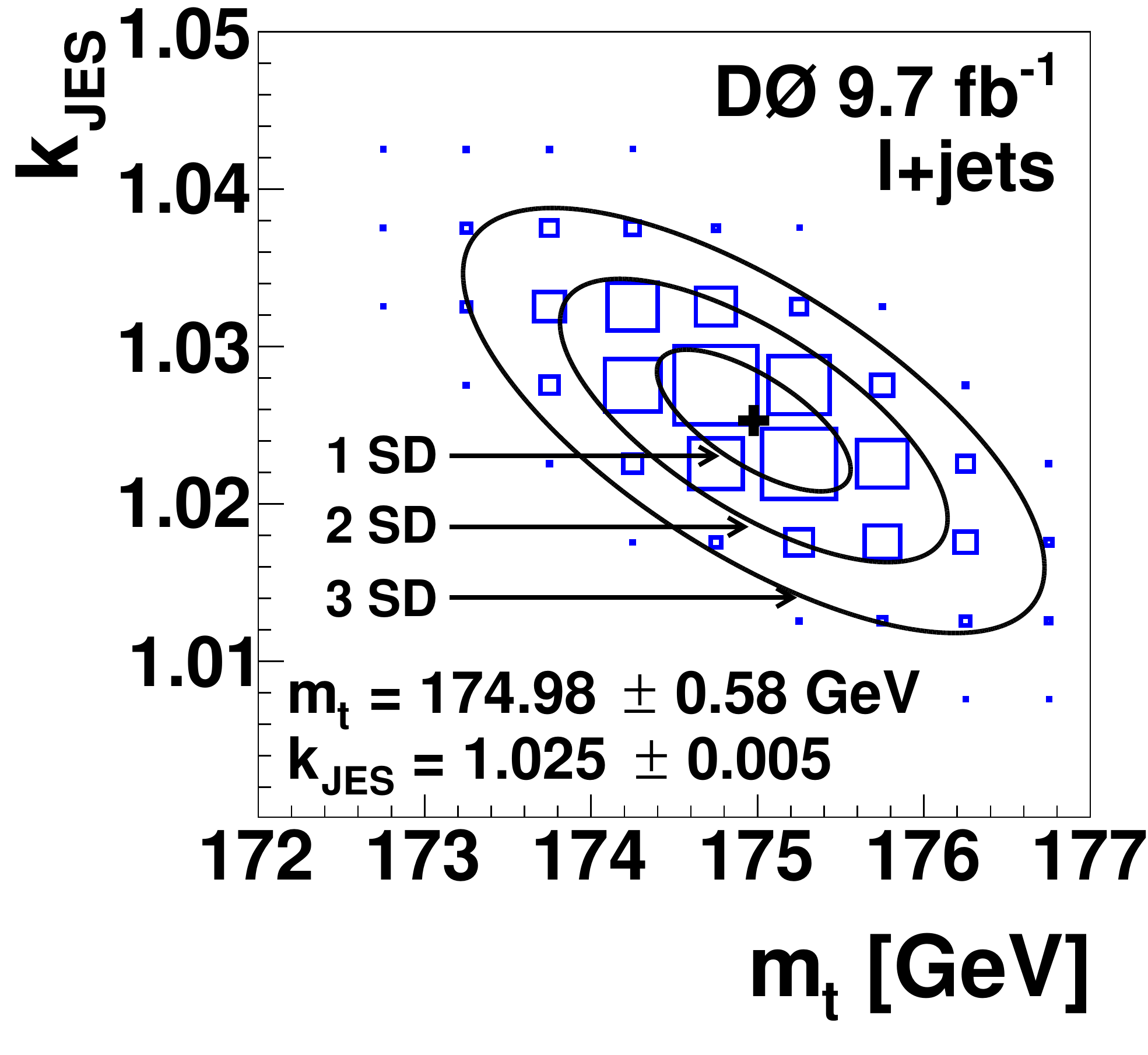}
\par\end{centering}
\vspace{-3mm}
\caption{
\label{fig:like2d}
Two-dimensional likelihood  $\like(\X_N;\,\mt,\kjes)/\like_{\rm max}$ for data. Fitted contours of equal probability are overlaid as solid lines. The maximum is marked with a cross. Note that the bin boundaries do not necessarily correspond to the grid points on which $\like$ is calculated.
\vspace{-3mm}
}
\end{figure}

\begin{figure}
\centering
\begin{overpic}[width=0.49\columnwidth]{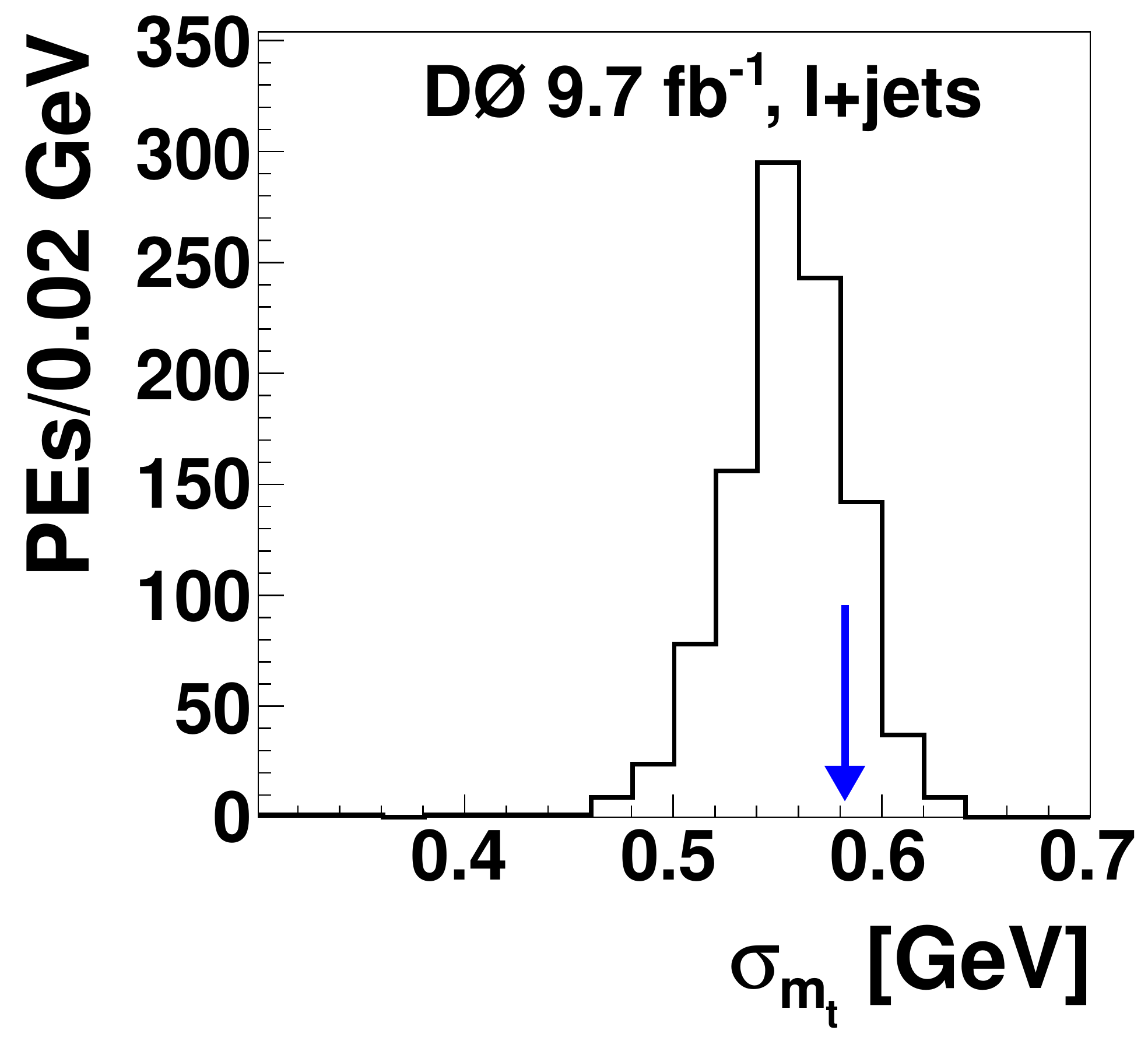}
\put(26,82){\footnotesize\textsf{\textbf{(a)}}}
\end{overpic}
\begin{overpic}[width=0.49\columnwidth]{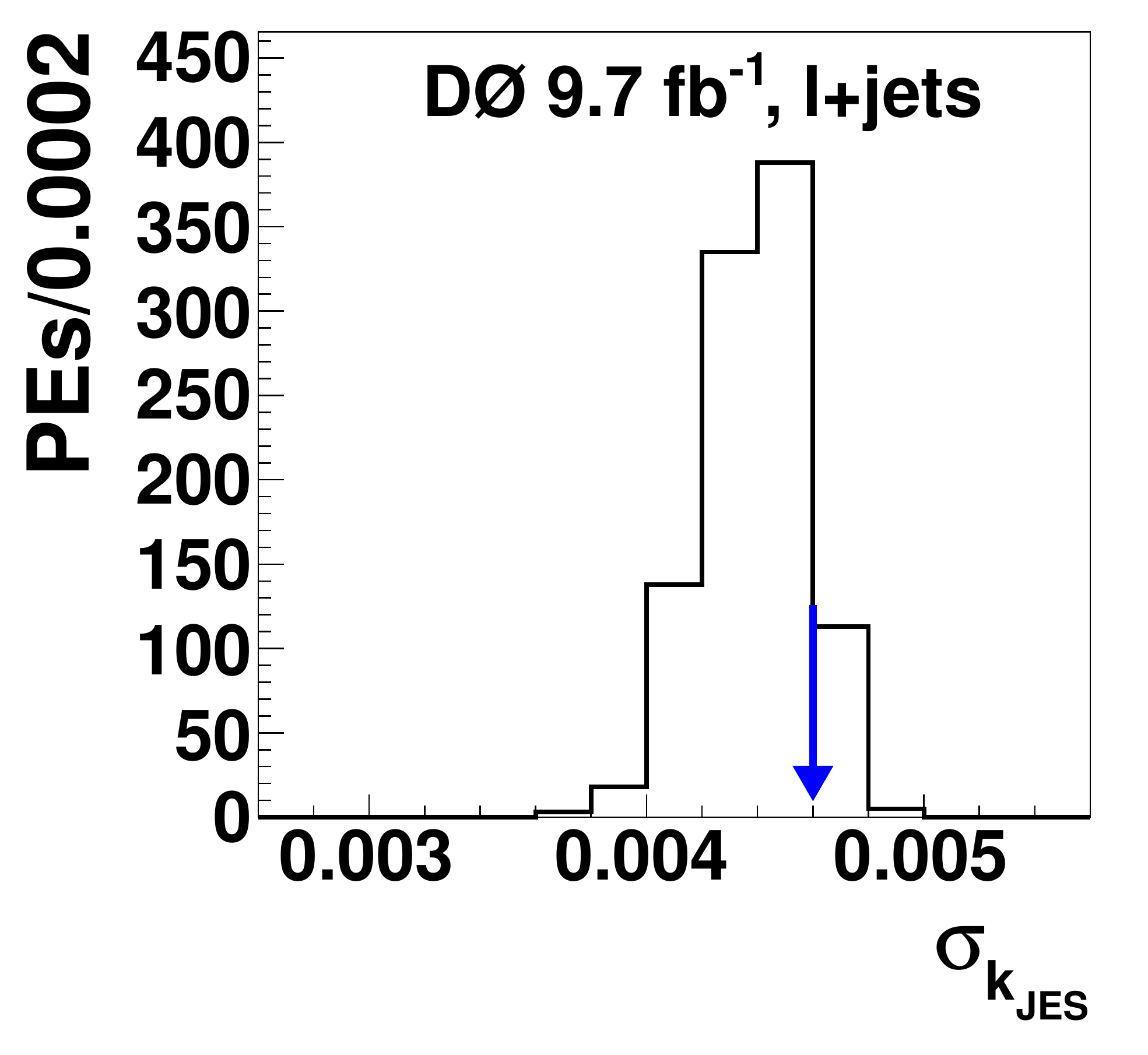}
\put(26,82){\footnotesize\textsf{\textbf{(b)}}}
\end{overpic}
\vspace{-3mm}
\caption{\label{fig:uncert}
(a) Distribution of the expected uncertainty on \mt, as obtained using PEs where the \ttbar contribution is taken at $\mtgen=172.5~\GeV$ and $\kjes=1$, after calibration. The observed statistical uncertainty is indicated by an arrow.
(b)~Same for uncertainty on \kjes.
\vspace{-3mm}
}
\end{figure}

Figure~\ref{fig:like2d} presents the two-dimensional likelihood density in $(\mt,\kjes)$, which is obtained by multiplying the calibrated likelihood densities for each data-taking epoch and each final state. We compare the \mt and \kjes values obtained from the minimum of the double-Gaussian fit to the two-dimensional likelihood density in Fig.~\ref{fig:like2d} and find them to match at the level of $10^{-4}$. In Fig.~\ref{fig:uncert} we compare the observed statistical uncertainties on \mt and \kjes with expectations from PE studies using \ttbar samples for $\mt^{\rm gen}=172.5~\GeV$ and $\kjes^{\rm gen}=1$. The observed uncertainty is consistent with the expectation.

\begin{figure}
\centering
\begin{overpic}[width=0.49\columnwidth]{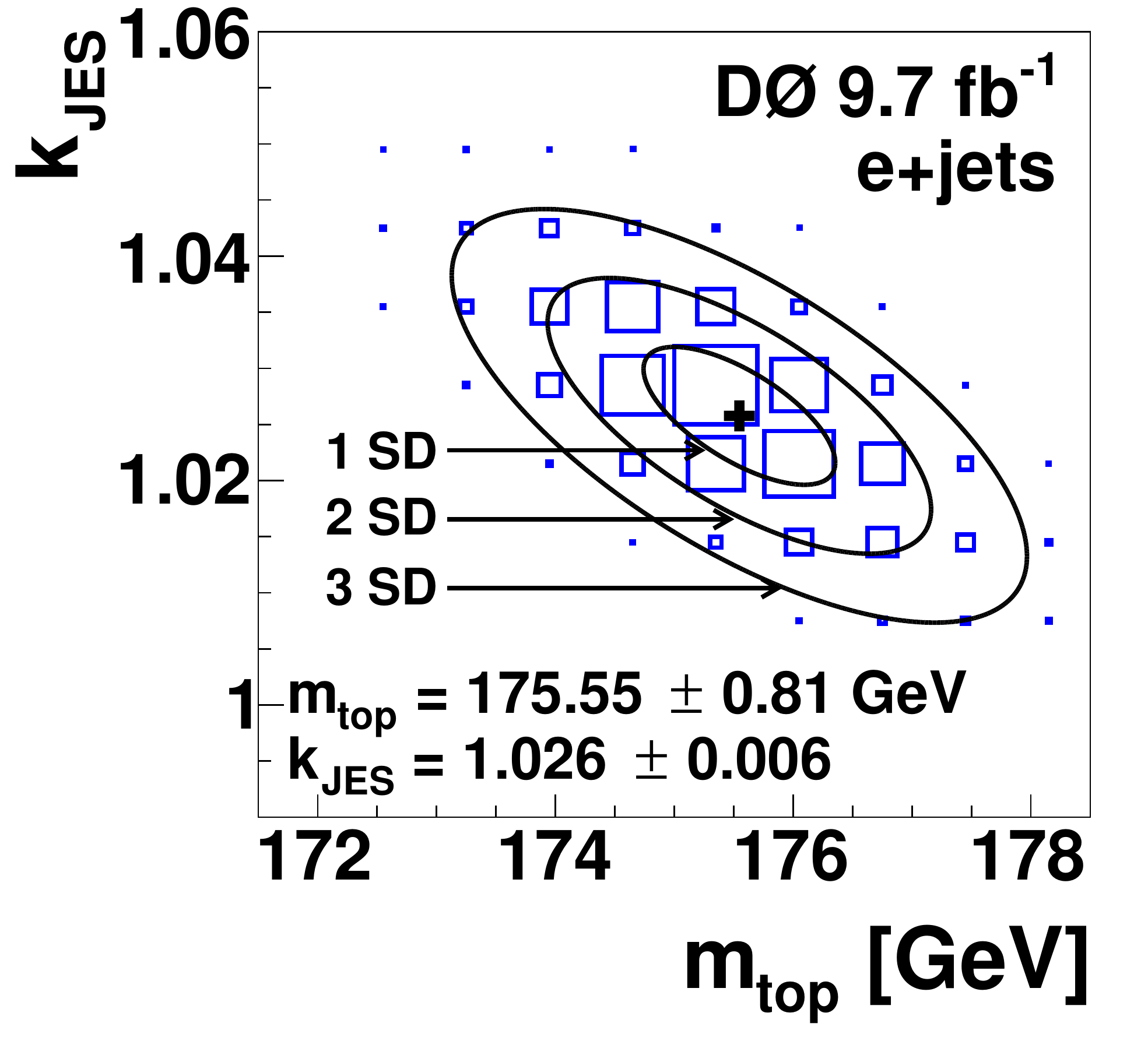}
\put(26,82){\footnotesize\textsf{\textbf{(a)}}}
\end{overpic}
\begin{overpic}[width=0.49\columnwidth]{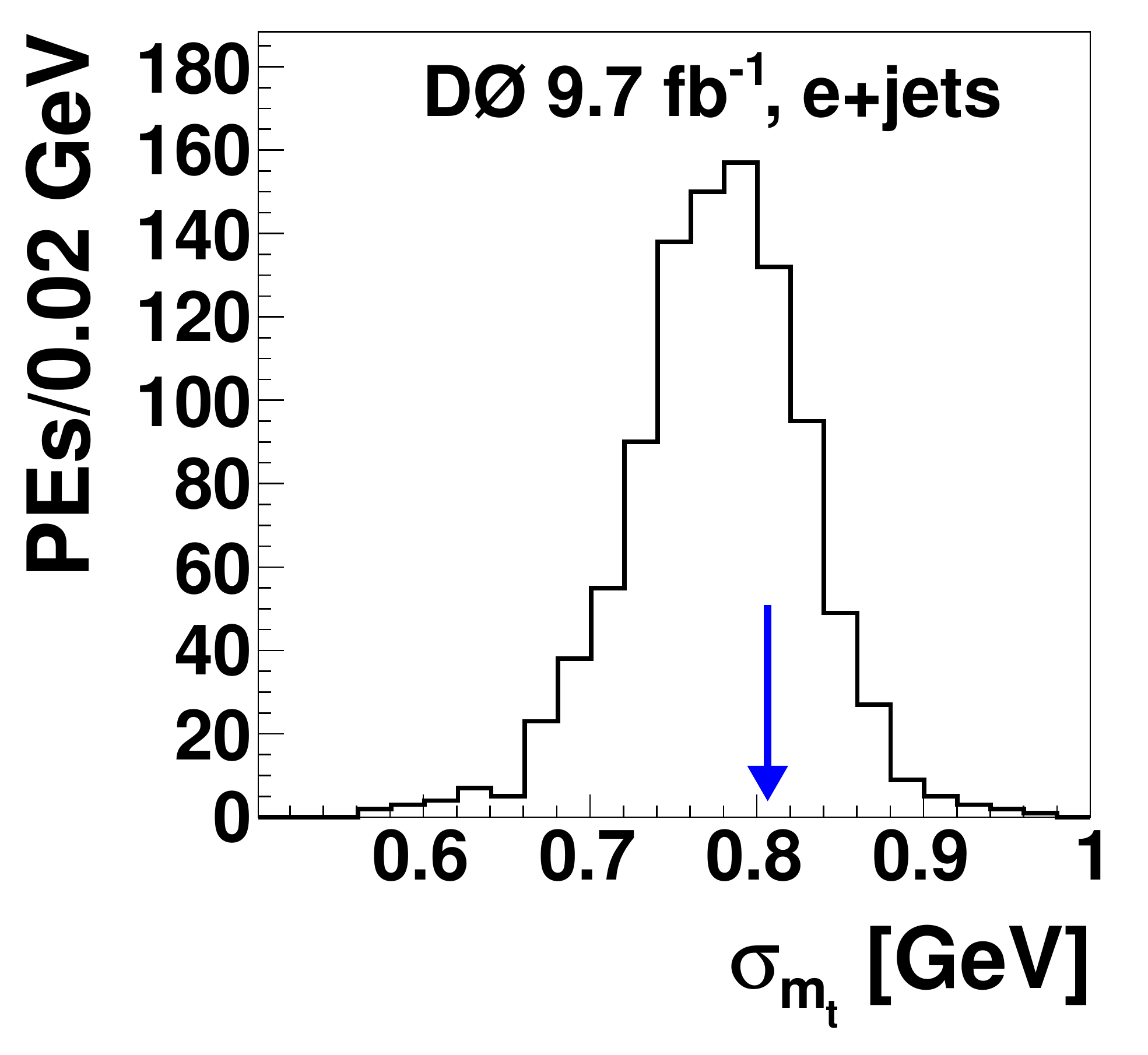}
\put(26,82){\footnotesize\textsf{\textbf{(b)}}}
\end{overpic}
\begin{overpic}[width=0.49\columnwidth]{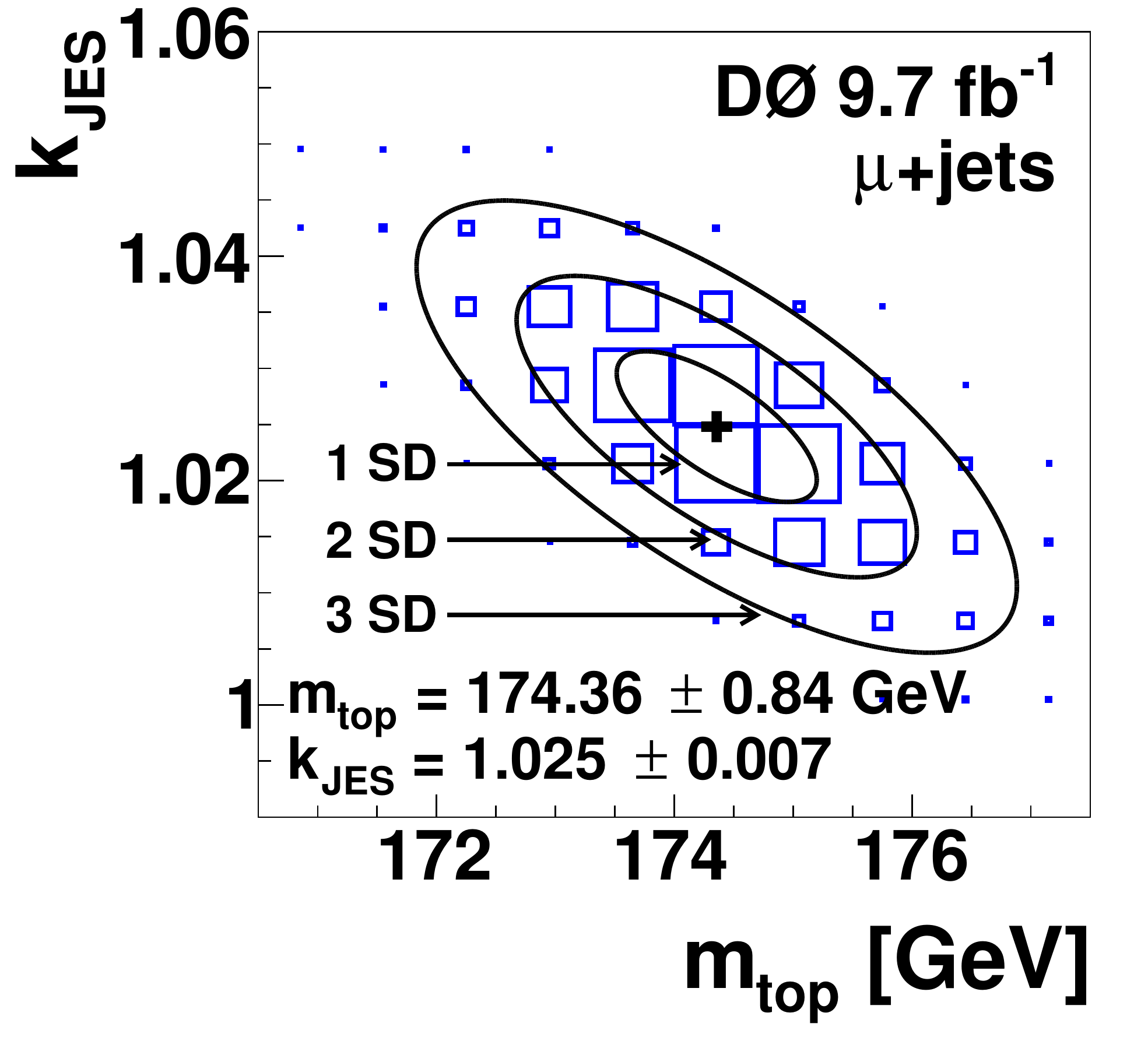}
\put(26,82){\footnotesize\textsf{\textbf{(c)}}}
\end{overpic}
\begin{overpic}[width=0.49\columnwidth]{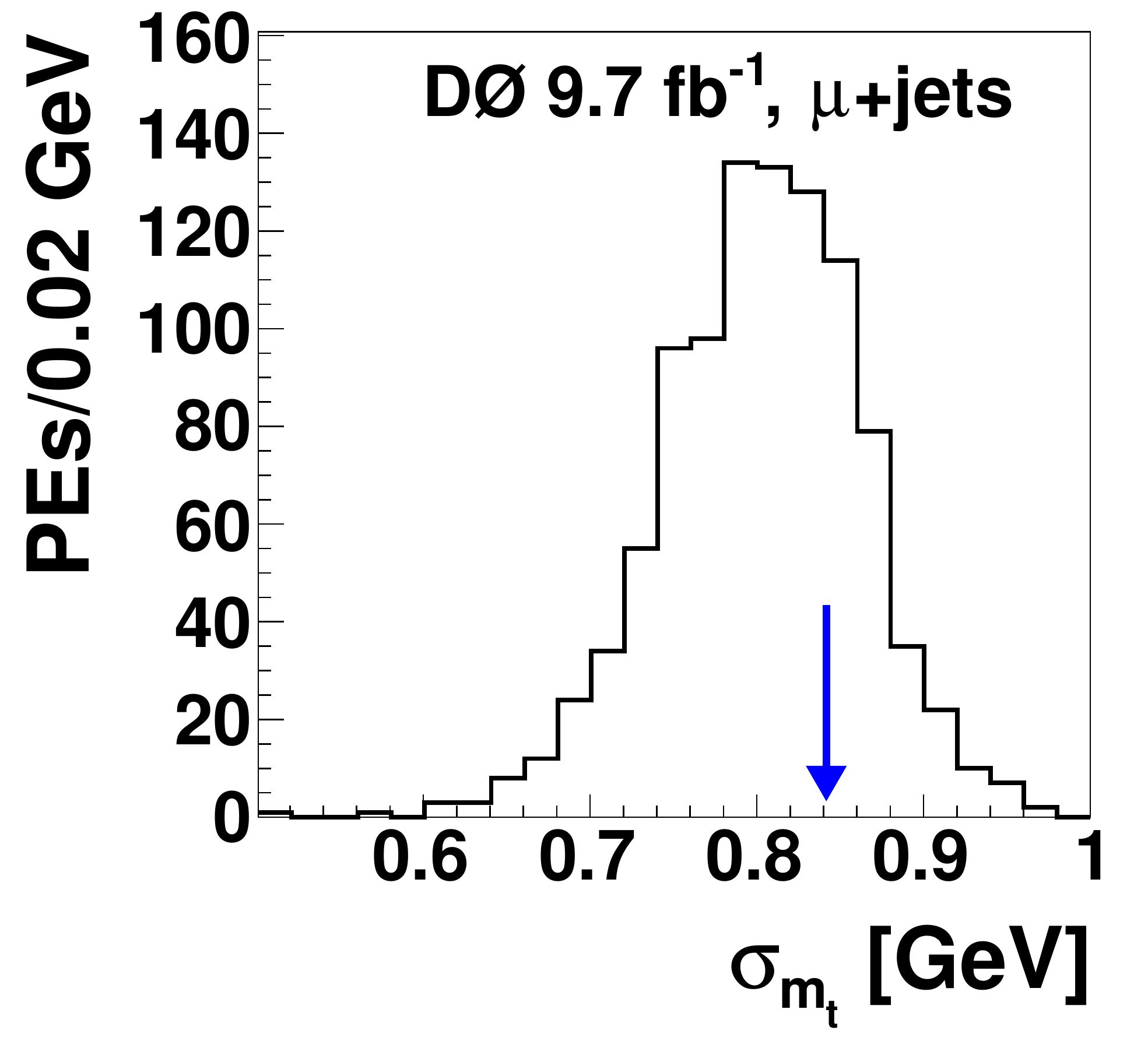}
\put(26,82){\footnotesize\textsf{\textbf{(d)}}}
\end{overpic}
\vspace{-3mm}
\caption{
\label{fig:uncertchnl}
(a) Two-dimensional likelihood  $\like(\X_N;\,\mt,\kjes)/\like_{\rm max}$ for data in the \ejets final state. Fitted contours of equal probability are overlaid as solid lines. The maximum is marked with a cross. 
(b)~Distribution in expected uncertainty on \mt in \ejets final states, obtained using PEs where the \ttbar contribution is taken at $\mtgen=172.5~\GeV$ and $\kjes=1$, after calibration.  The observed statistical uncertainty in data is indicated by the arrow.
(c)~Same as (a), but for the \mujets channel. 
(d)~Same as (b), but for the \mujets channel.
\vspace{-3mm}
}
\end{figure}

The values of \mt and \kjes measured in \ejets and \mujets final states using  9.7~\fb of integrated luminosity are 
\begin{eqnarray*}
\mt^{\ejets}    & = & 175.55\pm0.81~\mbox{(stat+JES)}~\mbox{GeV}\,,\\
\kjes^{\ejets}  & = & 1.026\pm0.006~\mbox{(stat+\mt)}\,,\\
\mt^{\mujets}    & = & 174.36\pm0.84~\mbox{(stat+JES)}~\mbox{GeV}\,,\\
\kjes^{\mujets}  & = & 1.025\pm0.007~\mbox{(stat+\mt)}\,.
\end{eqnarray*}
The corresponding two-dimensional likelihood densities in $(\mt,\kjes)$ and the comparison between the expected statistical uncertainty for each final state and the observation  are shown in Fig.~\ref{fig:uncertchnl}. The results in \ejets and \mujets channels are consistent at the level of 1~SD considering just statistical uncertainties, and the corresponding $p$~value is 0.30.

We repeat the measurement assuming $\kjes=1$, i.e., by evaluating the one-dimensional likelihood $\lumi(\mathbb{X}_N;\mt,\kjes\equiv1)$, and find $\mt^{\kjes\equiv1}=176.88\pm0.41~\GeV$, which is consistent with the final results obtained by maximizing $\lumi(\mathbb{X}_N;\mt,\kjes)$ as a function of $(\mt,\,\kjes)$, considering the measured $\kjes$. Separating the statistical uncertainty \dmt into two components, one from \mt and the other from \kjes, we obtain $\mt=174.98\pm0.41\thinspace({\rm stat})\pm0.41\thinspace({\rm JES})~\GeV$.

In contrast to the previous measurement~\cite{bib:mt36}, we do not use the JES determined in exclusive $\gamma+$jet and dijet events with an uncertainty of $\approx2\%$ to constrain \kjes. 
We follow this strategy because the statistical uncertainty on the measured \kjes value is smaller than the typical uncertainty on the JES and because \kjes relates jet energies at detector level to parton energies, while JES relates jet energies at detector level to jet energies at particle level.

\subsection{
Comparison of data with simulations using the measured $f,~\mt$, and $k_{\rm JES}$ values
}
\label{sec:postfit}

\begin{figure*}[t]
\centering
\begin{overpic}[width=0.99\columnwidth]{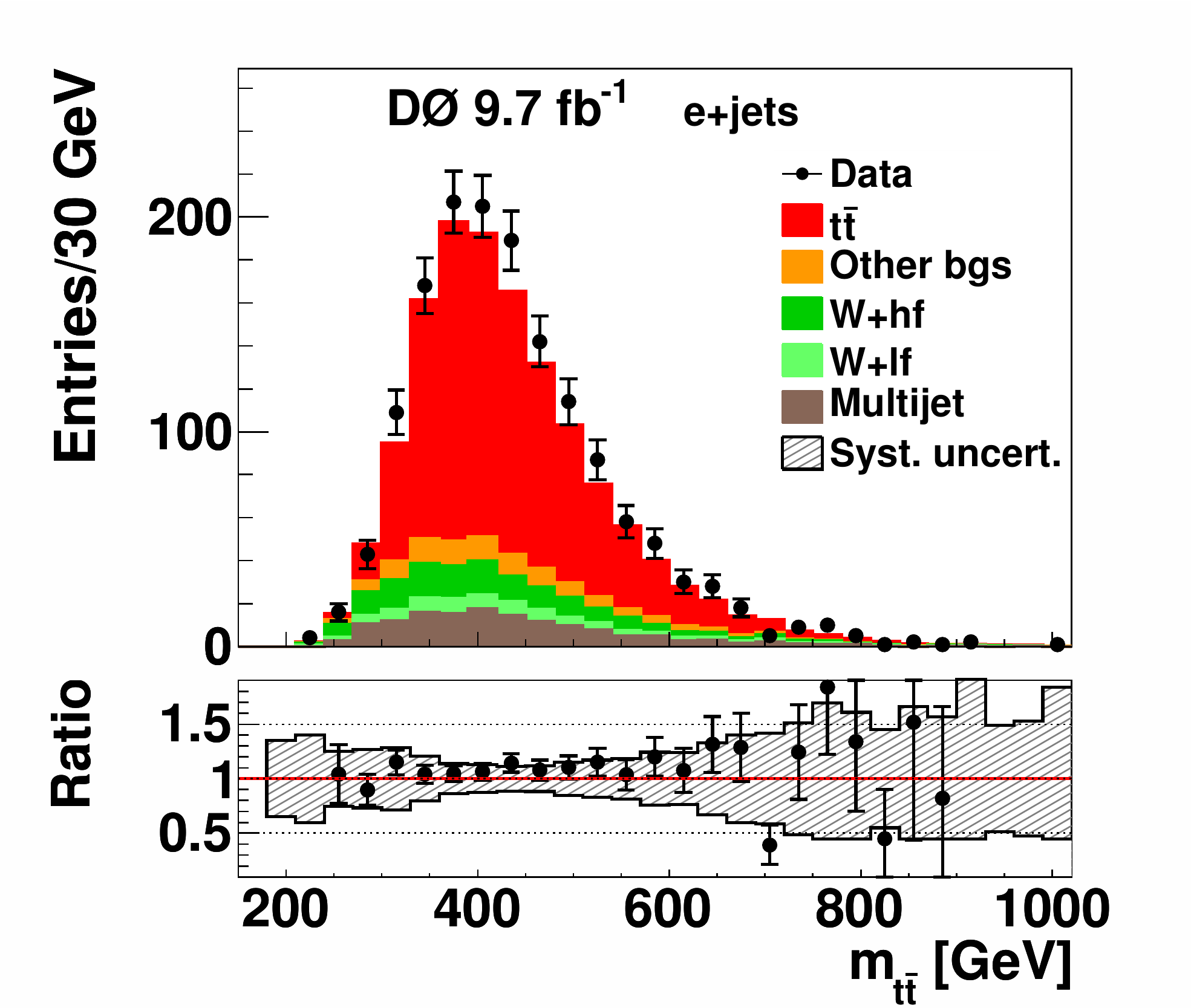}
\put(80.5,73){\large\textsf{\textbf{(a)}}}
\end{overpic}
\begin{overpic}[width=0.99\columnwidth]{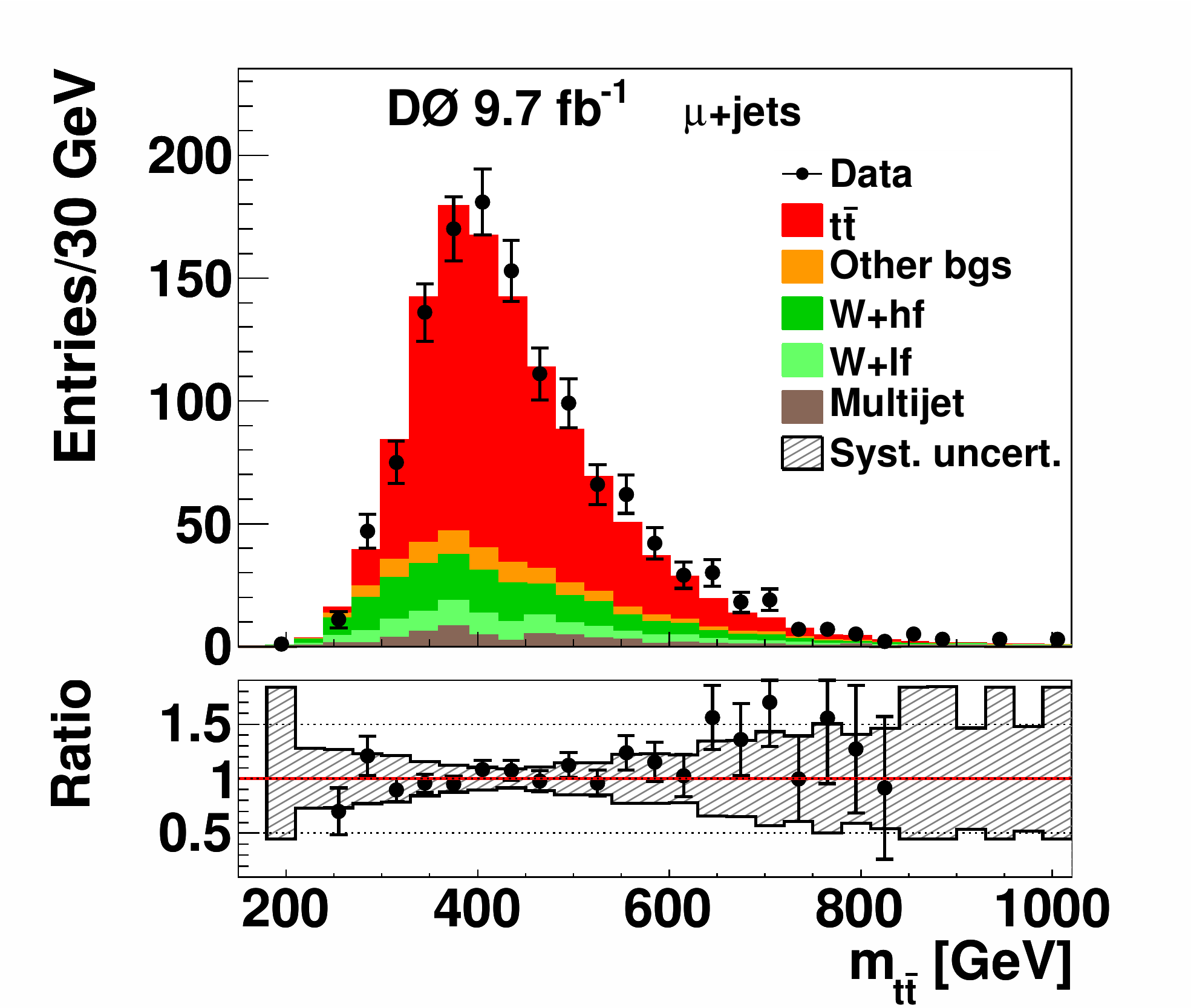}
\put(80.5,73){\large\textsf{\textbf{(b)}}}
\end{overpic}
\begin{overpic}[width=0.99\columnwidth]{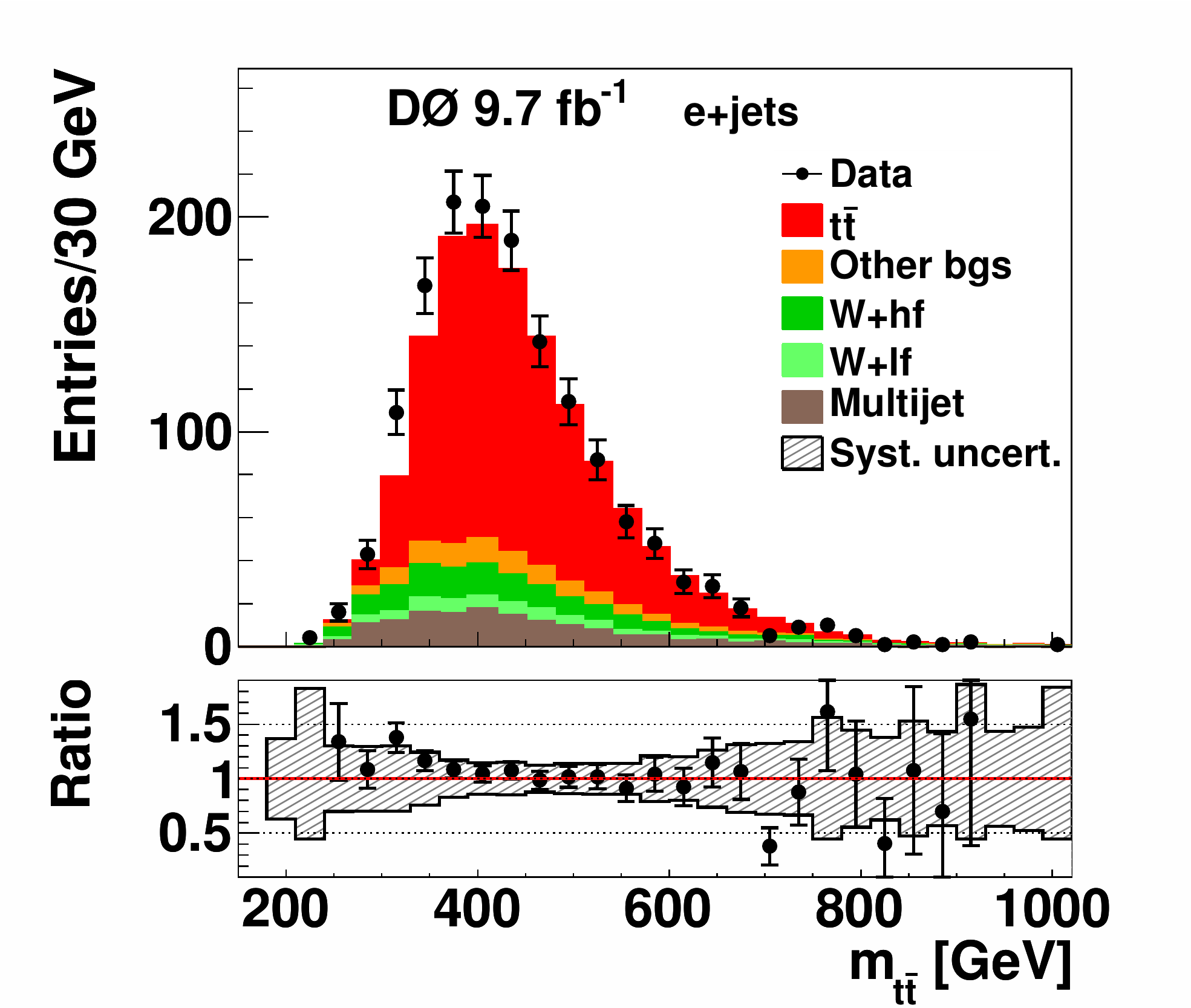}
\put(80.5,73){\large\textsf{\textbf{(c)}}}
\end{overpic}
\begin{overpic}[width=0.99\columnwidth]{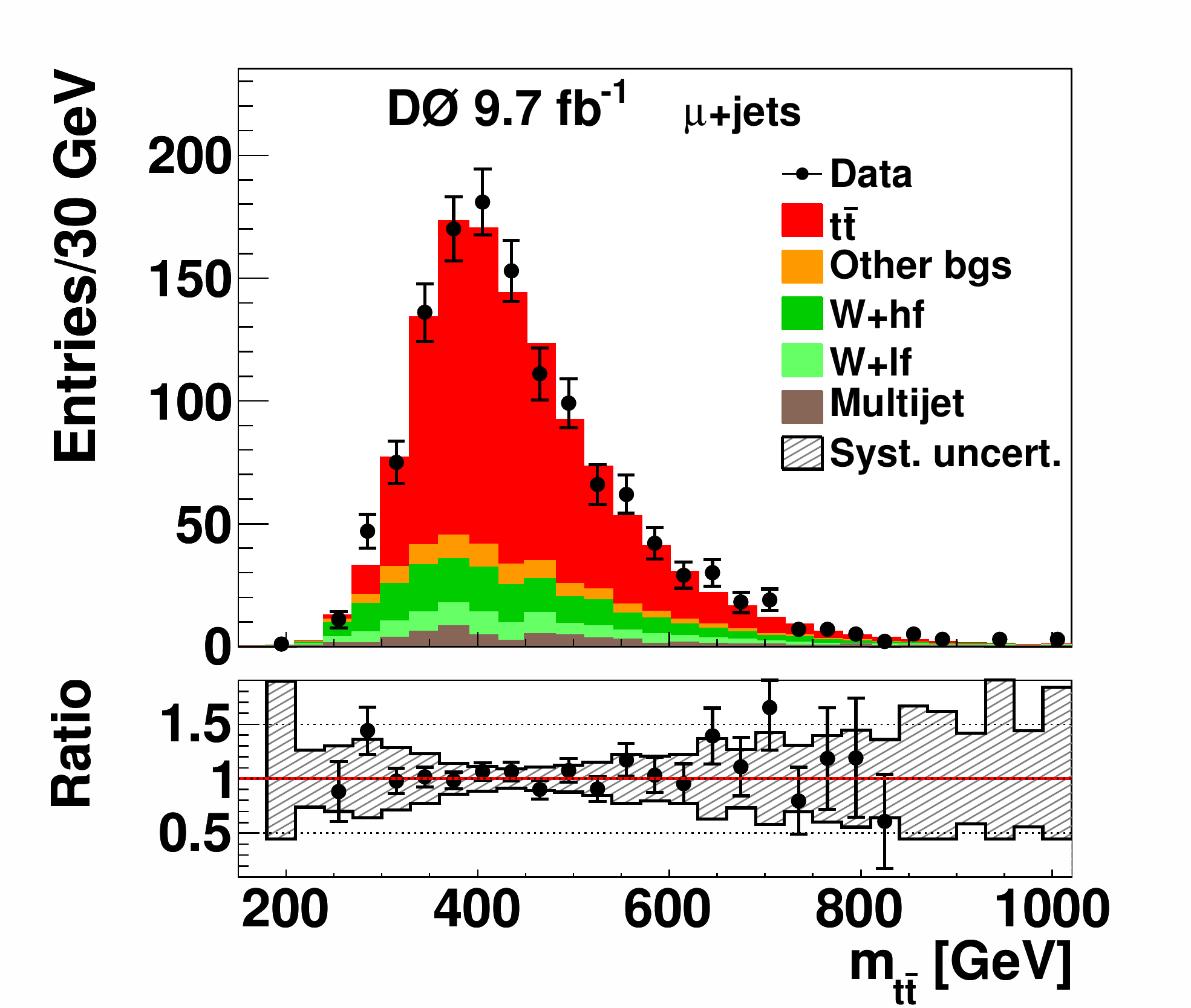}
\put(80.5,73){\large\textsf{\textbf{(d)}}}
\end{overpic}
\vspace{-3mm}
\caption{\label{fig:mtt}
The invariant mass of the \ttbar system in the (a)~\ejets and (b)~\mujets final states, assuming $\mtgen=172.5~\GeV$, $\kjes=1$, and $\sigma_{\ttbar}=7.24~\pb$~\cite{bib:xsec}. The same observable in the (c)~\ejets and (d)~\mujets final states for $\mtgen=175~\GeV$, with energies of jets scaled up by $\kjes=1.025$, corresponding to their measured values. The fractions of signal events, $f$, are set to their measured values of 63\% in the \ejets and 70\% in the \mujets final states in~(c) and~(d) (cf.\ Section~\ref{sec:resultf}).
The category ``Other bgs'' encompasses $\zjets$, $WW$, $WZ$, $ZZ$, and single top quark production, as well as dileptonic \ttbar decays.  The production of a $W$ boson in association with at least one $b$ or $c$ quark is denoted as ``$W+{\rm hf}$'', while ``$W+{\rm lf}$'' stands for all other flavor configurations. The last bin includes overflow events. The ratio shows the number of observed events divided by the number of expected events in a given bin. The band of systematic uncertainty is indicated by the shaded area in the ratio plots, and includes contributions from dominant sources:  JES, jet energy resolution, jet identification, flavor dependence of the detector response, $b$~tagging, lepton identification, lepton momentum scale, trigger, as well as hadronization and higher-order effects for \ttbar events.
\vspace{-3mm}
}
\end{figure*}

In this Section, we demonstrate that our simulations agree with the data for the measured \mt and \kjes parameters. We use a sample of simulated \ttbar events at $\mtgen=175~\GeV$ and scale the energies of jets by $\kjes=1.025$, while adjusting the magnitude of the three-momentum $|\vec p|$ to preserve the invariant mass of the jet. The fraction of signal events, $f$, is set to our measured values of 63\% in the \ejets and 70\% in the \mujets final states (cf.\ Section~\ref{sec:resultf}). We consider the same systematic uncertainties as in Section~\ref{sec:selection}: JES, jet energy resolution, jet identification, flavor dependence of the detector response, $b$~tagging, lepton identification, lepton momentum scale, trigger, as well as hadronization and higher-order effects for \ttbar events.

\begin{figure*}
\vspace{1cm}
\centering
\begin{overpic}[width=0.99\columnwidth]{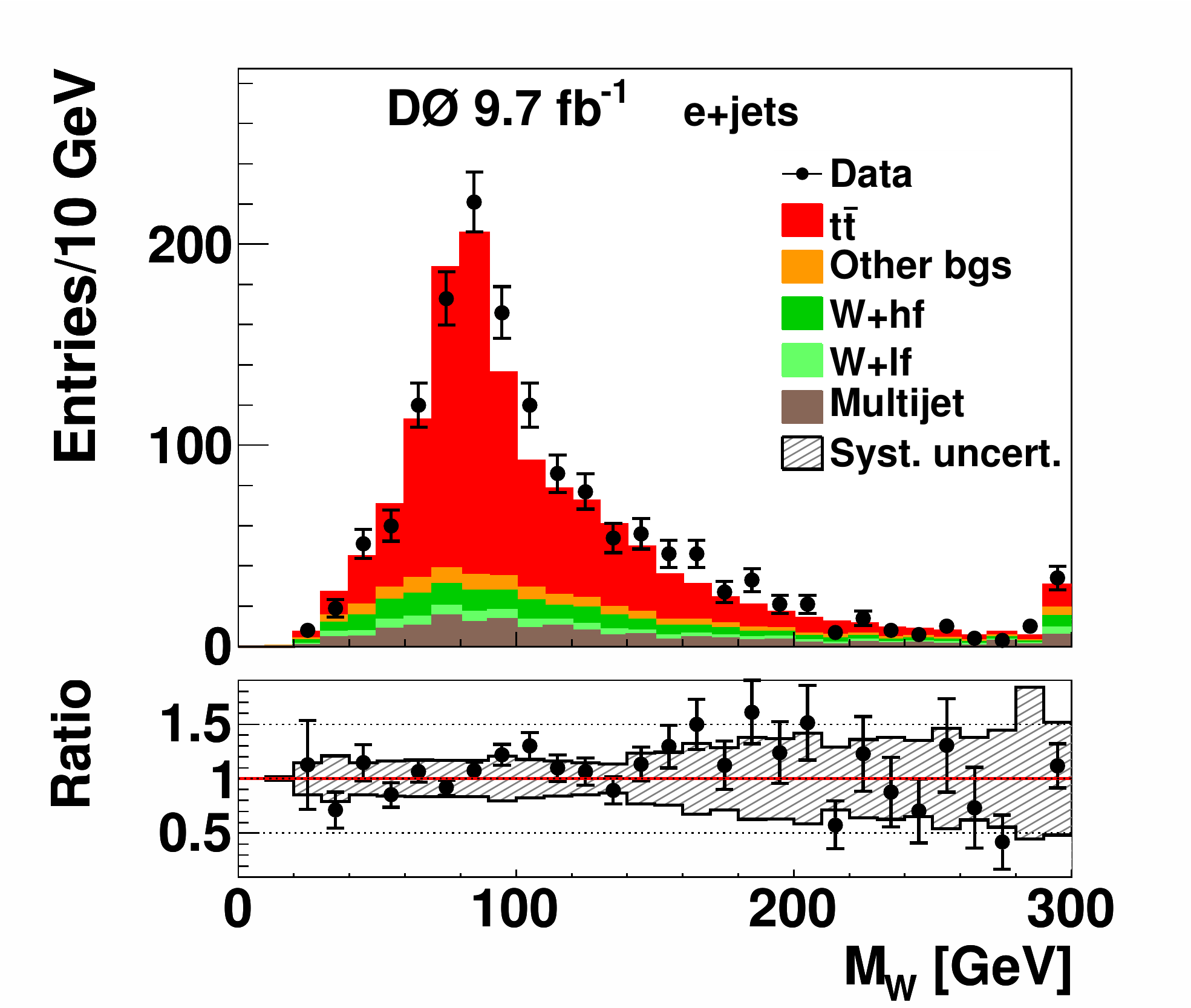}
\put(80.5,73){\large\textsf{\textbf{(a)}}}
\end{overpic}
\begin{overpic}[width=0.99\columnwidth]{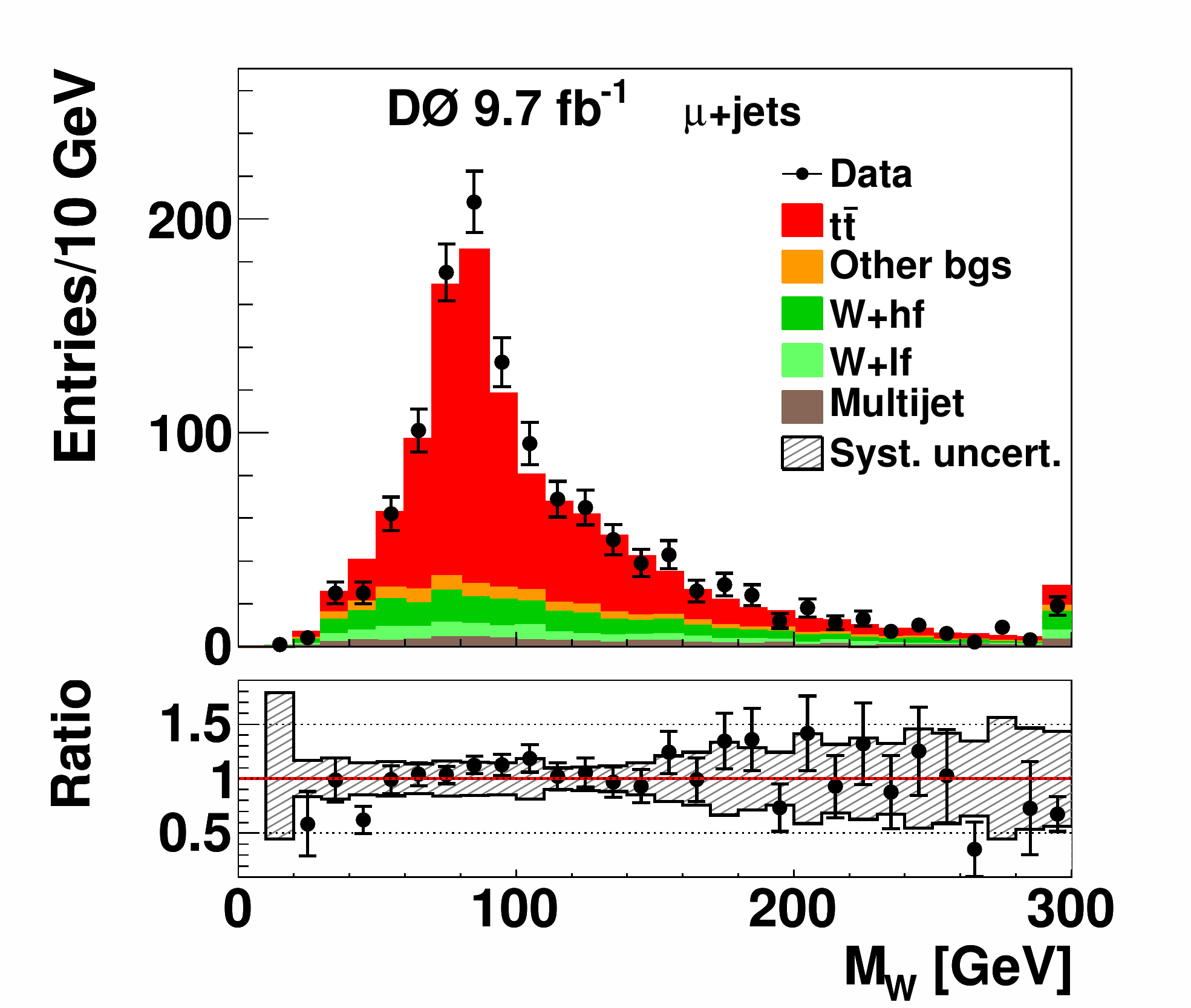}
\put(80.5,73){\large\textsf{\textbf{(b)}}}
\end{overpic}
\begin{overpic}[width=0.99\columnwidth]{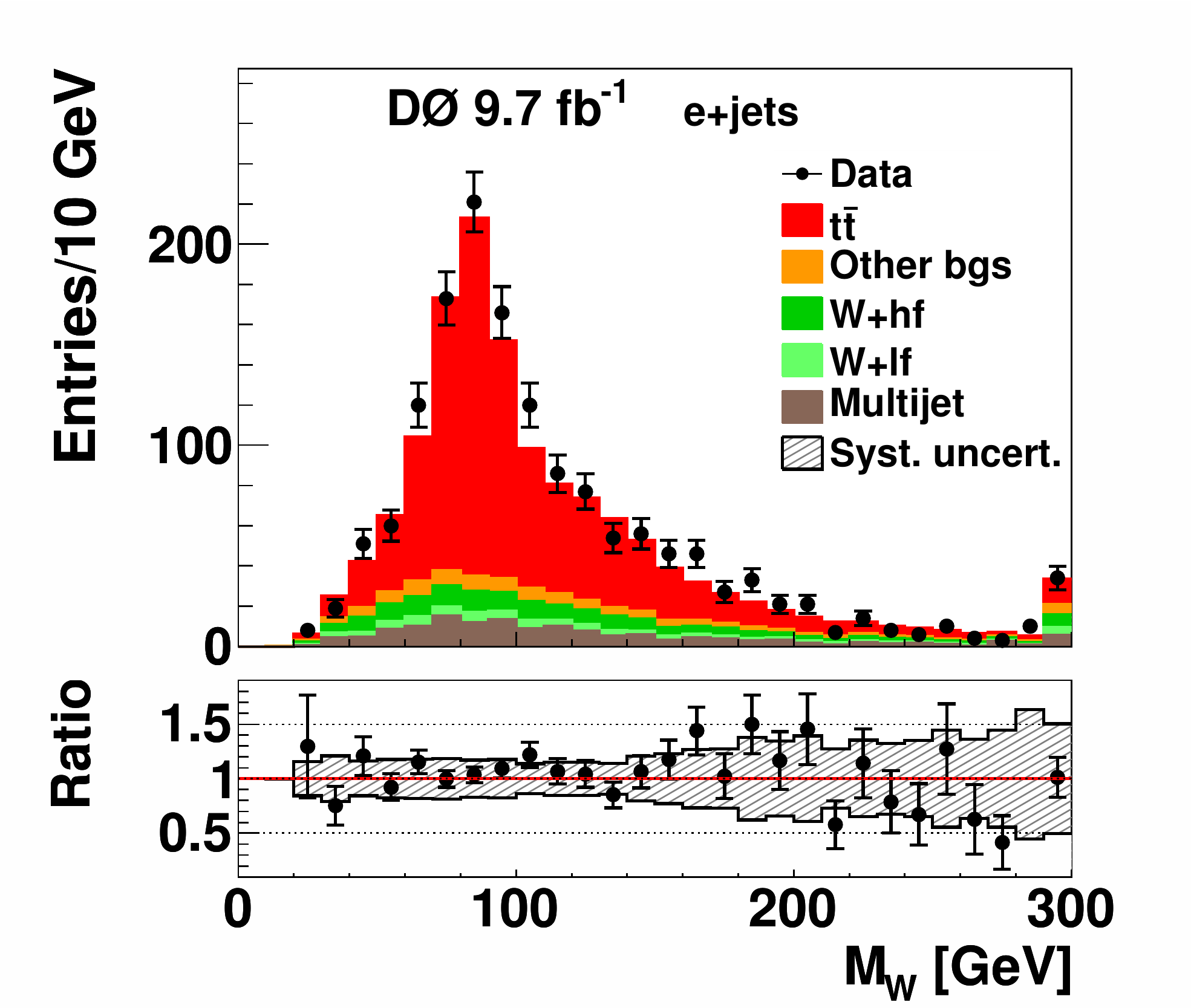}
\put(80.5,73){\large\textsf{\textbf{(c)}}}
\end{overpic}
\begin{overpic}[width=0.99\columnwidth]{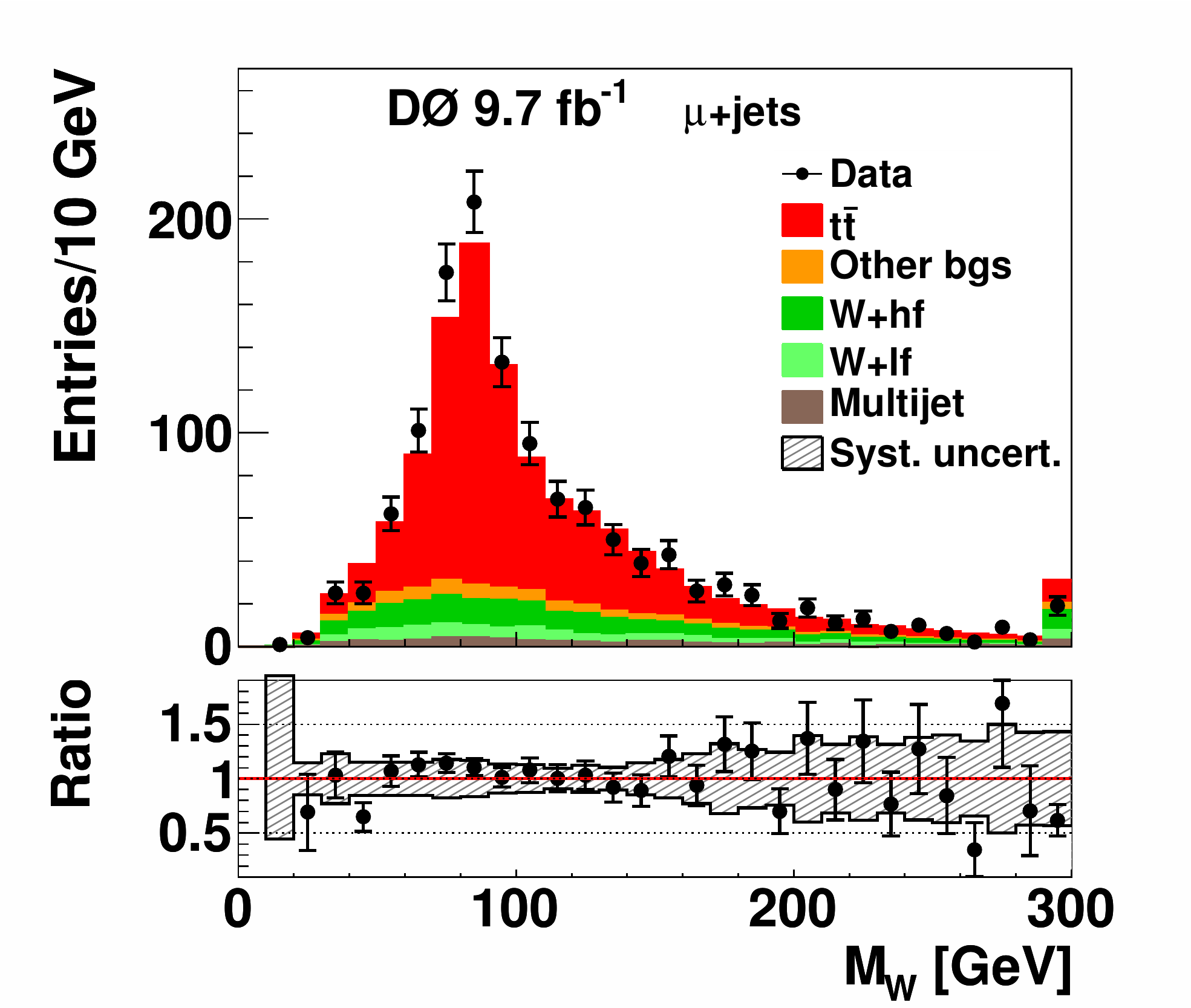}
\put(80.5,73){\large\textsf{\textbf{(d)}}}
\end{overpic}
\caption{\label{fig:mwh}
Same as Fig.~\ref{fig:mtt}, but for the invariant mass of the dijet system matched to the $W$ boson, assuming $\mtgen=172.5~\GeV$, $\kjes=1$, and $\sigma_{\ttbar}=7.24~\pb$~\cite{bib:xsec} in the (a)~\ejets and (b)~\mujets final states, and for $\mtgen=175~\GeV$, $\kjes=1.025$, and $f$ corresponding to their measured values  in the (c)~\ejets and (d)~\mujets final states.
\vspace{2cm}
}
\end{figure*}

Distributions in the invariant mass of the \ttbar system and in the invariant mass of the jet pair matched to one of the $W$ bosons  are shown in Figs.~\ref{fig:mtt}~(c-d) and~\ref{fig:mwh}~(c-d), respectively, together with the predictions from our signal and background models. The observables $m_{\ttbar}$ and $M_W$ are reconstructed with the same algorithm as described in Section~\ref{sec:selection}. As expected, the agreement between data and simulations improves compared to Figs.~\ref{fig:mtt}~(a-b) and~\ref{fig:mwh}~(a-b) when we use the measured values of $f$, \mt, and \kjes.

\begin{figure*}[h]
\begin{centering}
\begin{overpic}[width=0.99\columnwidth]{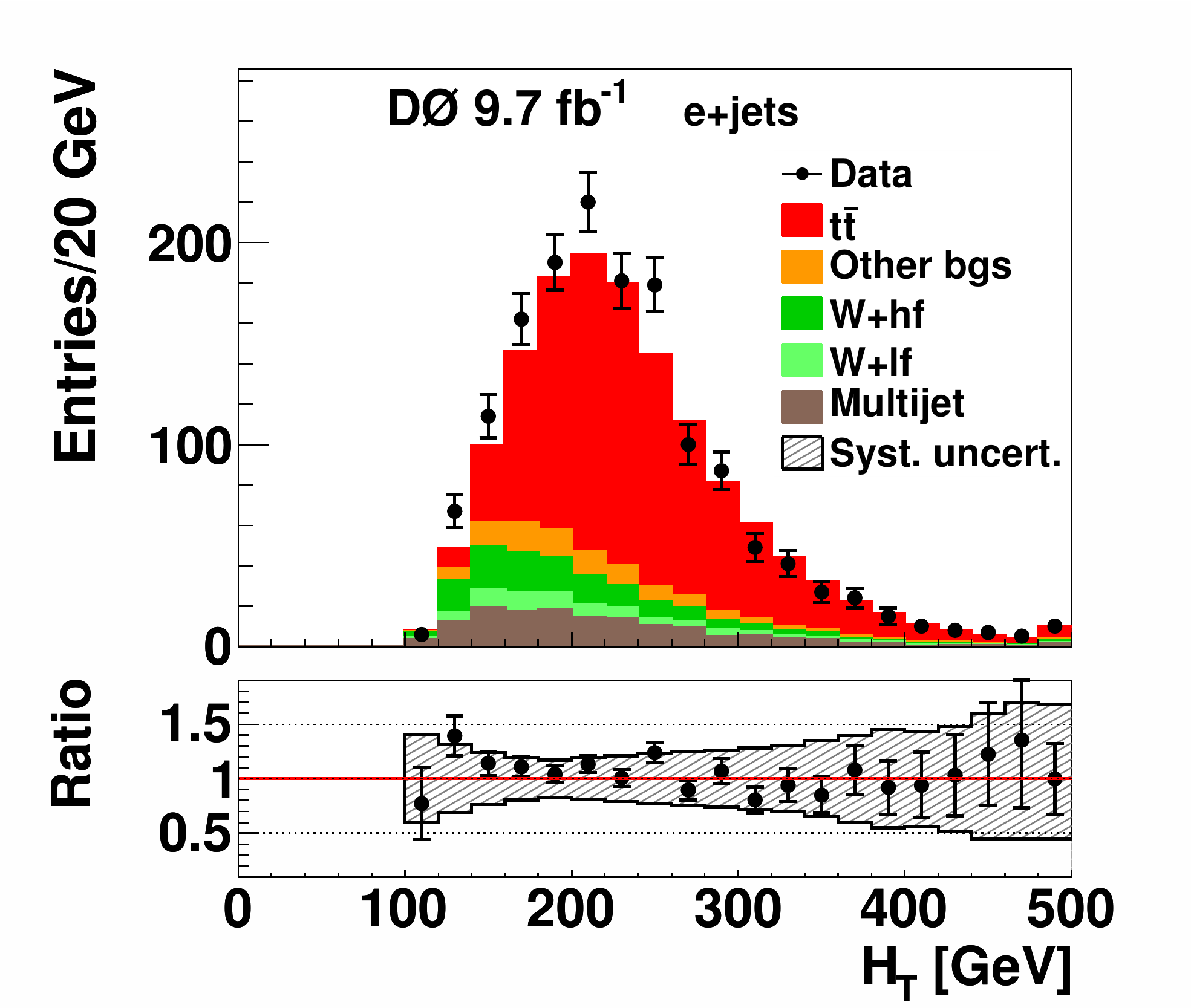}
\put(80.5,73){\large\textsf{\textbf{(a)}}}
\end{overpic}
\begin{overpic}[width=0.99\columnwidth]{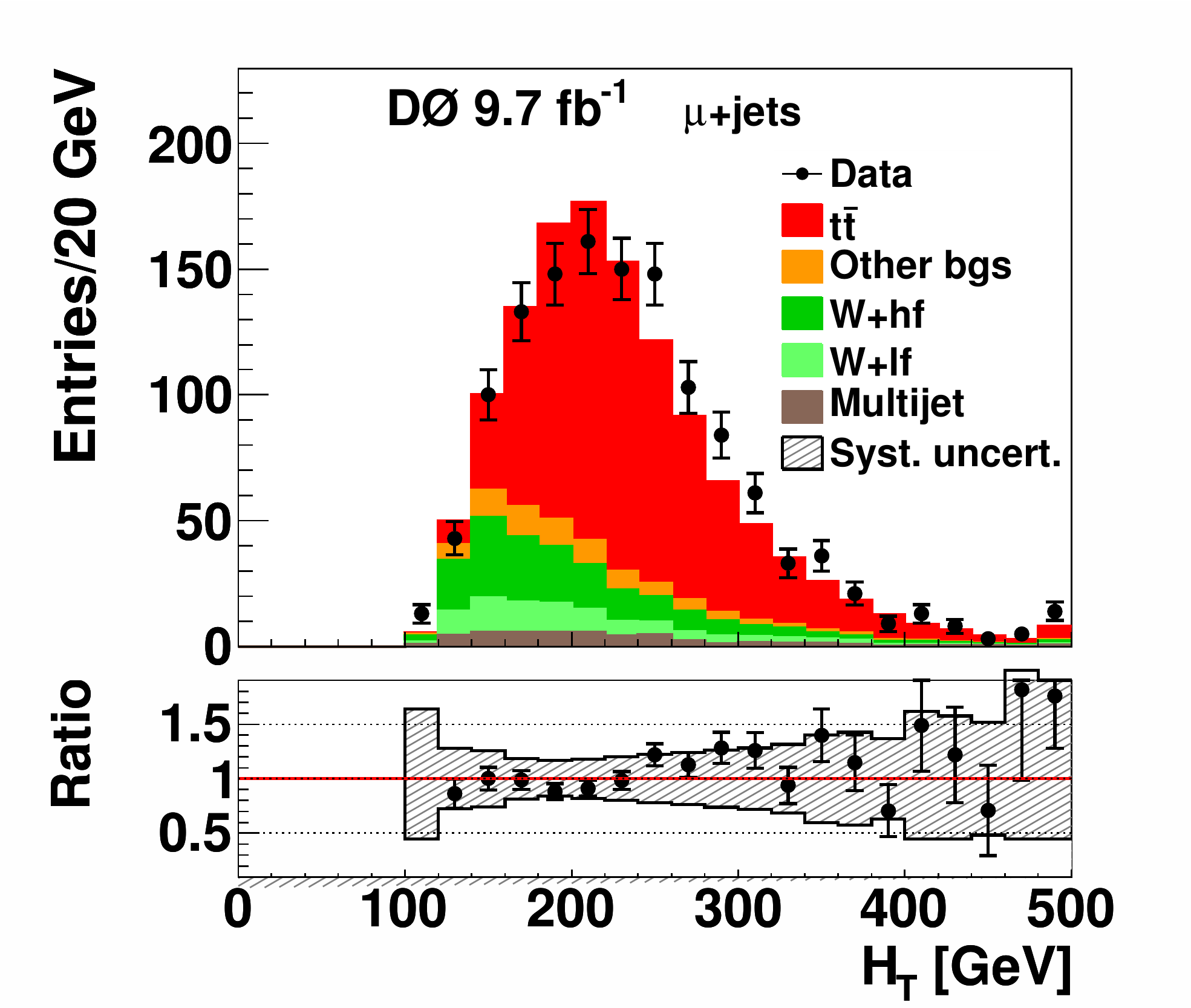}
\put(80.5,73){\large\textsf{\textbf{(b)}}}
\end{overpic}
\begin{overpic}[width=0.99\columnwidth]{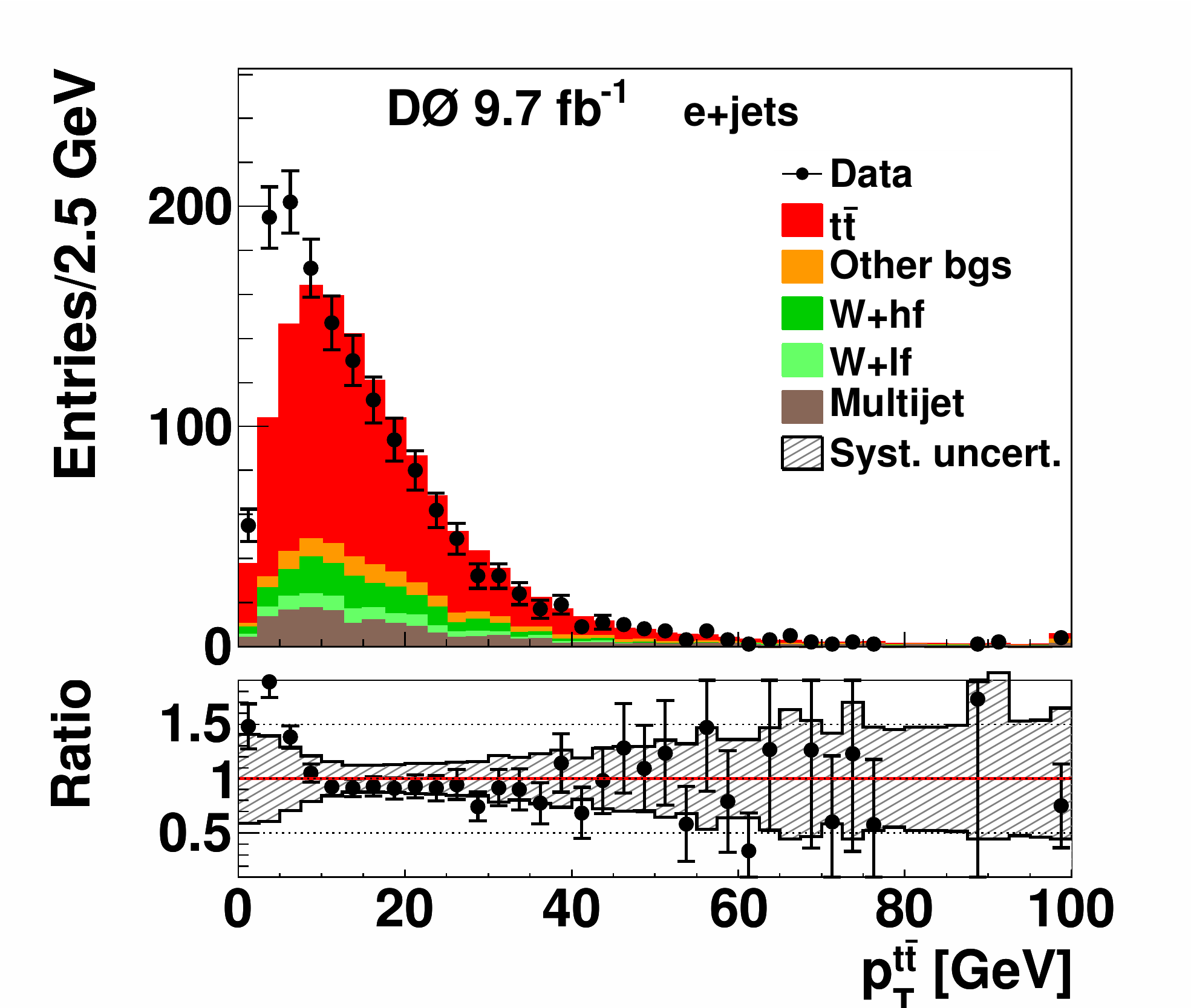}
\put(80.5,73){\large\textsf{\textbf{(c)}}}
\end{overpic}
\begin{overpic}[width=0.99\columnwidth]{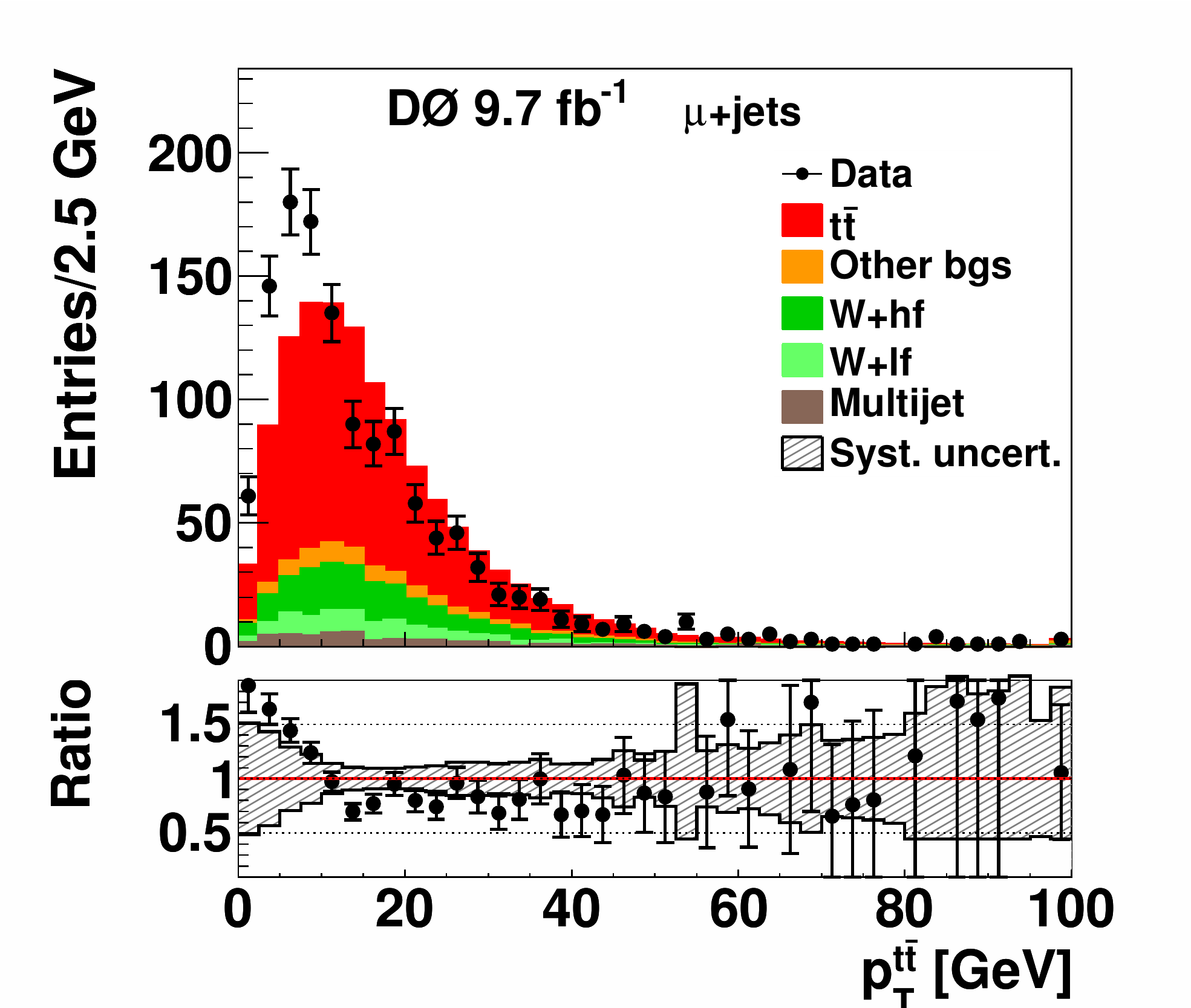}
\put(80.5,73){\large\textsf{\textbf{(d)}}}
\end{overpic}
\begin{overpic}[width=0.99\columnwidth]{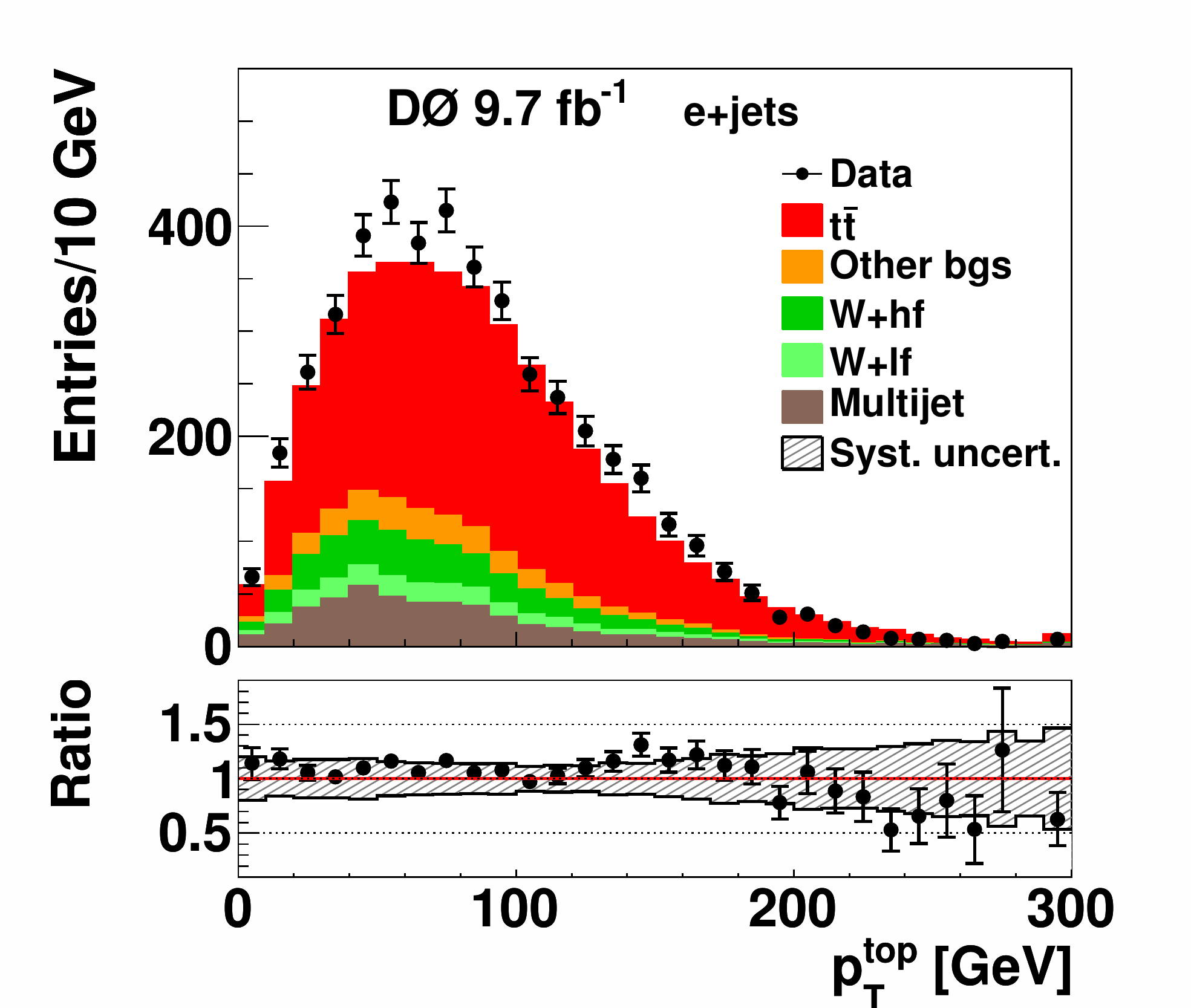}
\put(80.5,73){\large\textsf{\textbf{(e)}}}
\end{overpic}
\begin{overpic}[width=0.99\columnwidth]{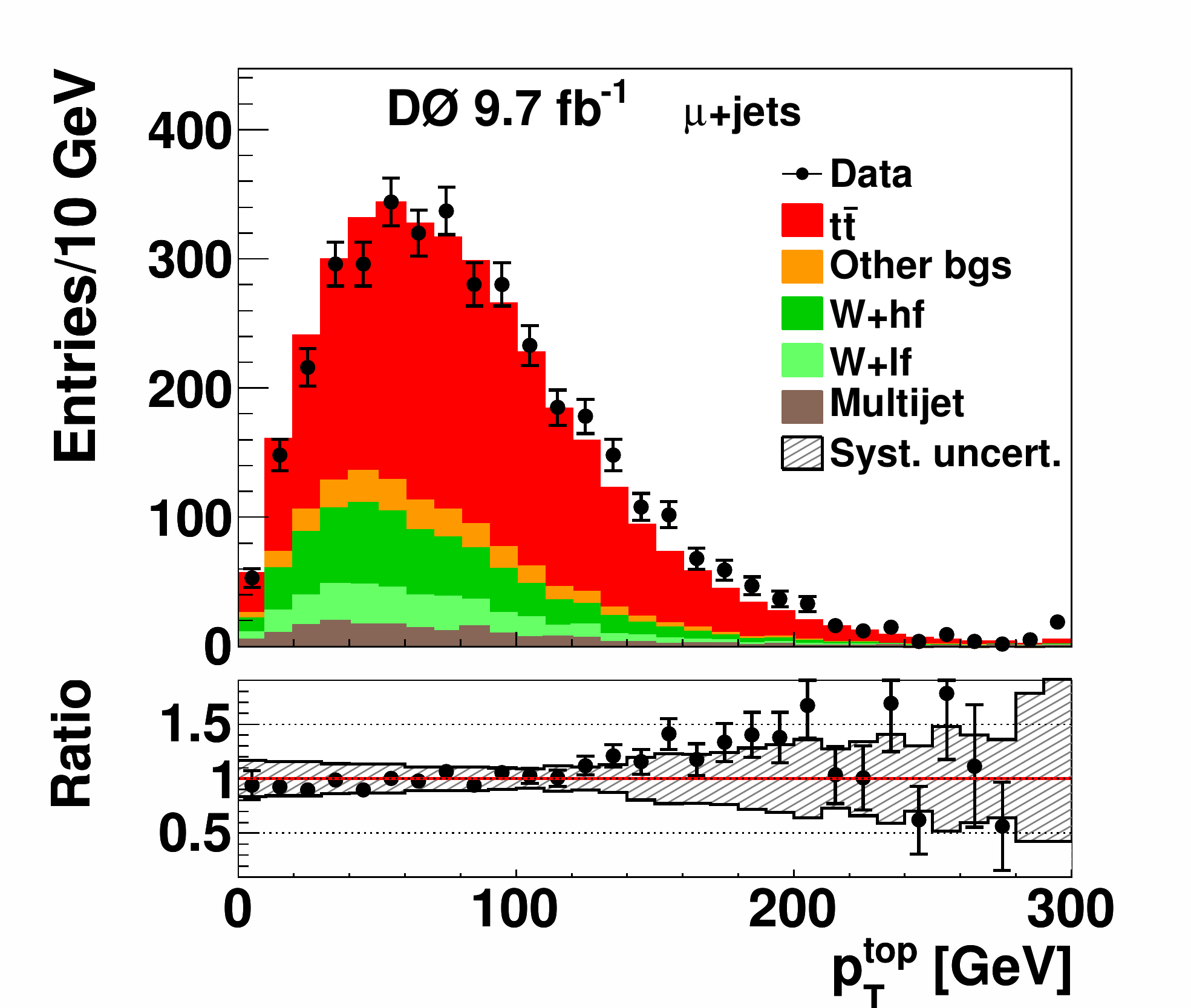}
\put(80.5,73){\large\textsf{\textbf{(f)}}}
\end{overpic}
\par\end{centering}
\caption{
\label{fig:postpttt}
The scalar sum of $E_T$ of the four jets $H_T$ in (a)~\ejets and (b)~\mujets final states, for the transverse momentum of the \ttbar system in (c)~\ejets and (d)~\mujets final states, as well as for the transverse momentum of the top quarks in (a)~\ejets or (b)~\mujets final states. 
All panels are for $\mtgen=175~\GeV$, with energies of jets scaled up by $\kjes=1.025$, corresponding to their measured values. The fractions of signal events, $f$, are set to their measured values of 63\% in the \ejets and 70\% in the \mujets final states in~(c) and~(d) (cf.\ Section~\ref{sec:resultf}). The list of included background contributions and sources of systematic uncertainties considered are identical to those from Fig.~\ref{fig:mtt}.
Panels (e) and (f) have two entries per event, one for each of the $t$ and $\bar t$ quarks.
}
\end{figure*}

In Fig.~\ref{fig:postpttt}, we study the distribution in $H_T$, which is defined as the scalar sum of $E_T$ of the four jets, the transverse momentum of the \ttbar system, and the transverse momentum of the top quarks. The observable $\pt^{\ttbar}$ is given by the transverse coordinates of the four-momentum vector used to calculate $m_{\ttbar}$, while $\pt^{\rm top}$ is obtained from the four-momenta of final state particles used to reconstruct $m_{\ttbar}$. These are combined to give the four-momenta of the $t$ and $\bar t$ quarks that minimize the  $|m_t-m_{\bar t}|$ mass difference.

\begin{figure*}[h]
\begin{centering}
\begin{overpic}[width=0.99\columnwidth]{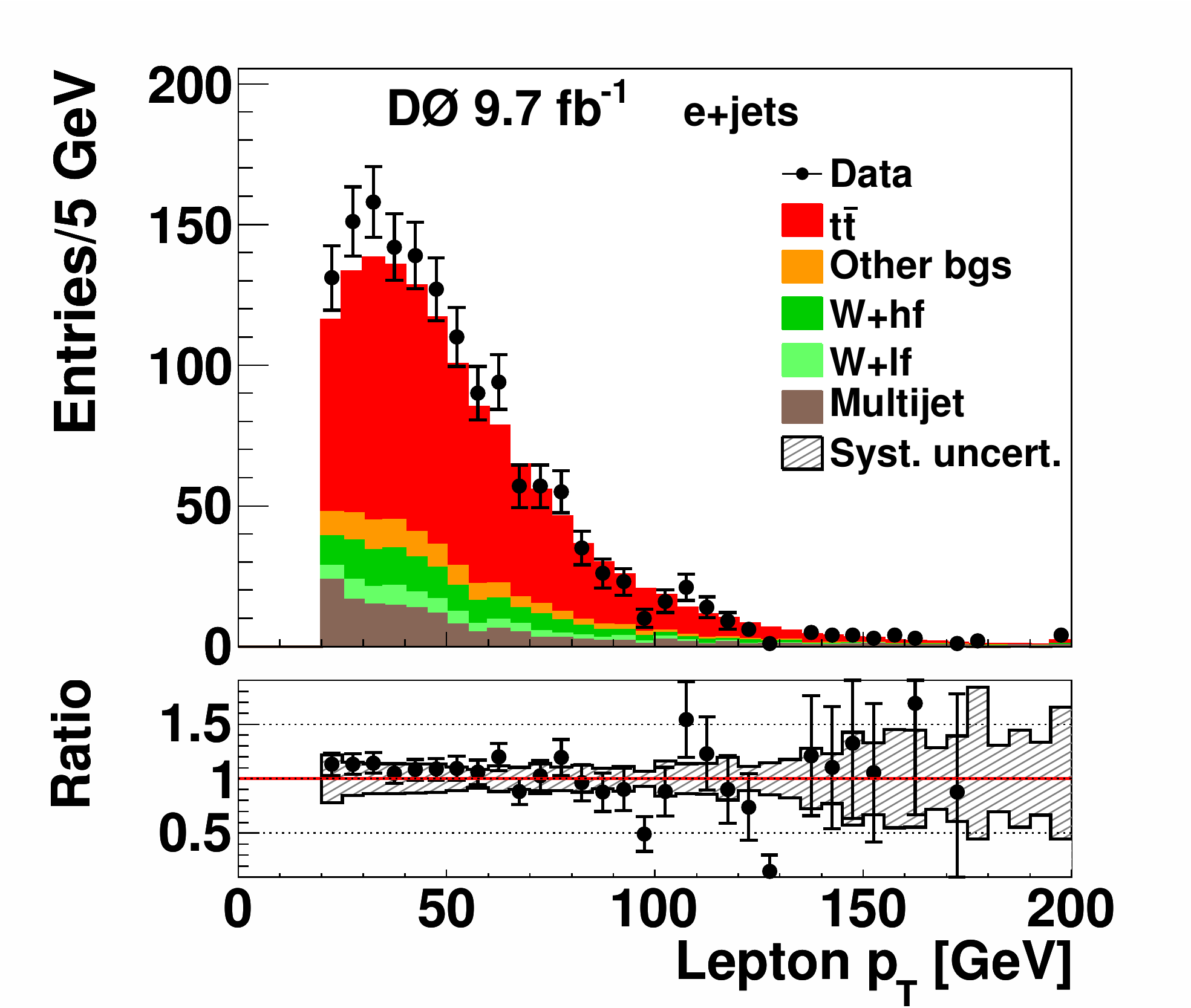}
\put(80.5,73){\large\textsf{\textbf{(a)}}}
\end{overpic}
\begin{overpic}[width=0.99\columnwidth]{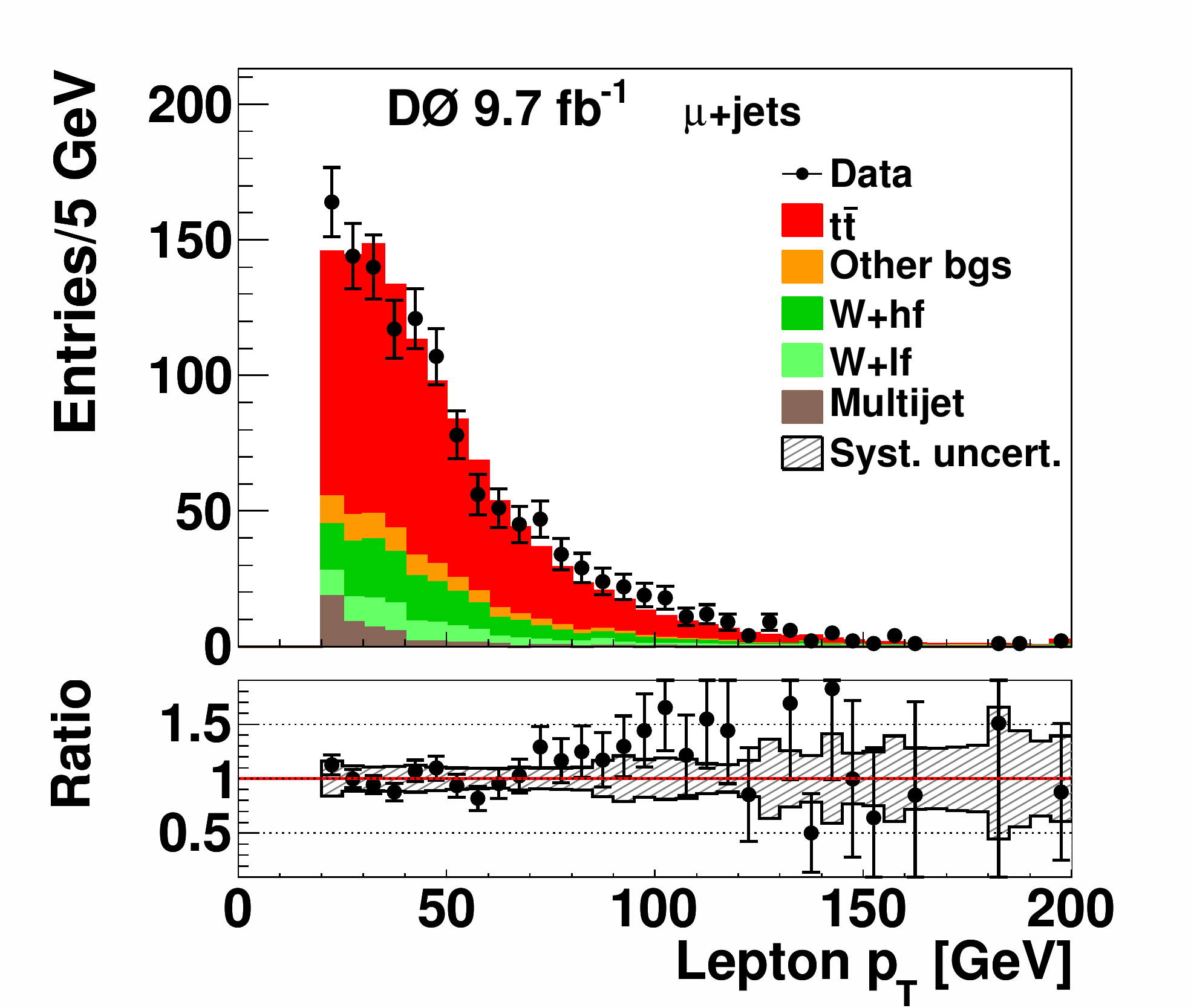}
\put(80.5,73){\large\textsf{\textbf{(b)}}}
\end{overpic}
\begin{overpic}[width=0.99\columnwidth]{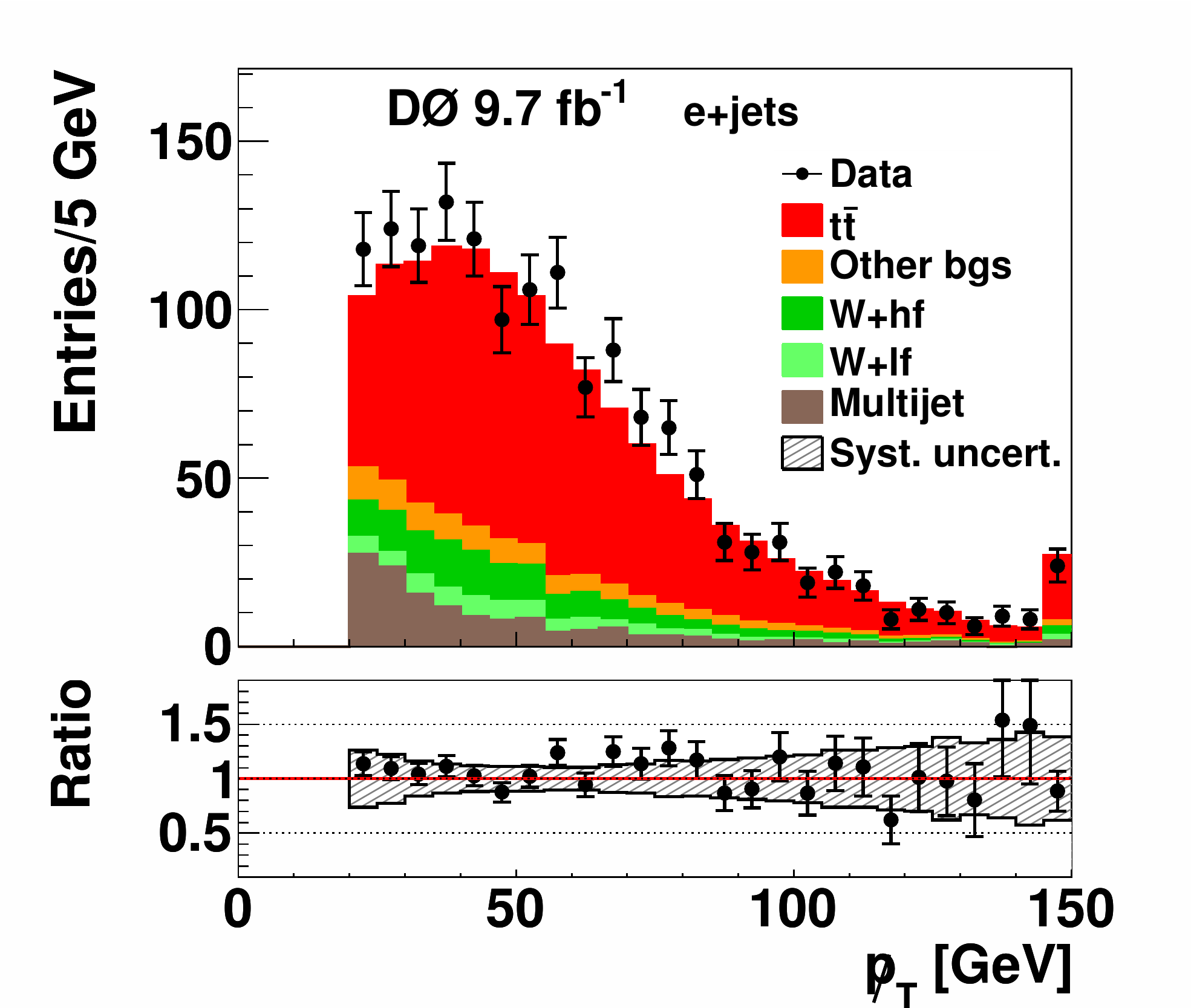}
\put(80.5,73){\large\textsf{\textbf{(c)}}}
\end{overpic}
\begin{overpic}[width=0.99\columnwidth]{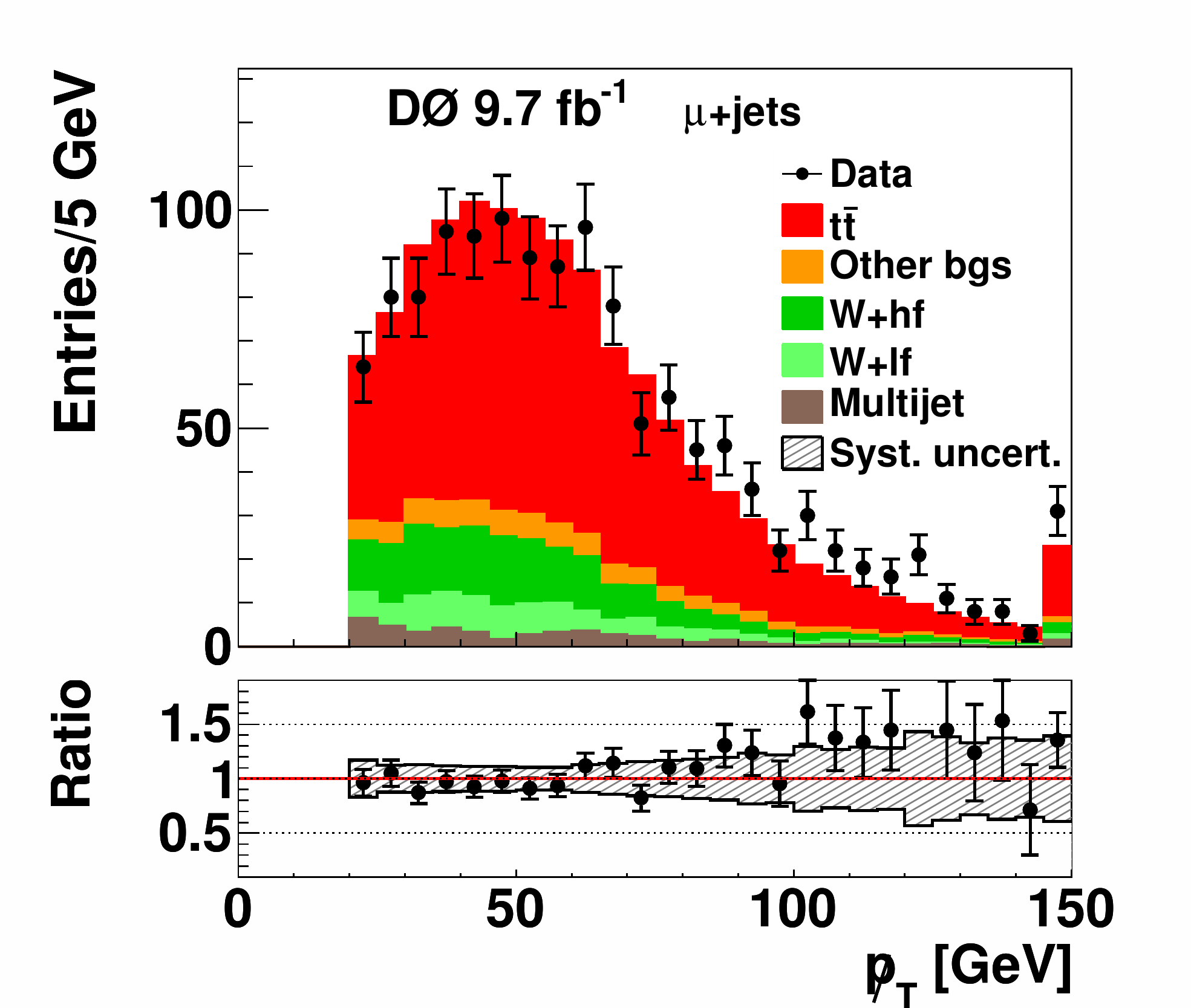}
\put(80.5,73){\large\textsf{\textbf{(d)}}}
\end{overpic}
\begin{overpic}[width=0.99\columnwidth]{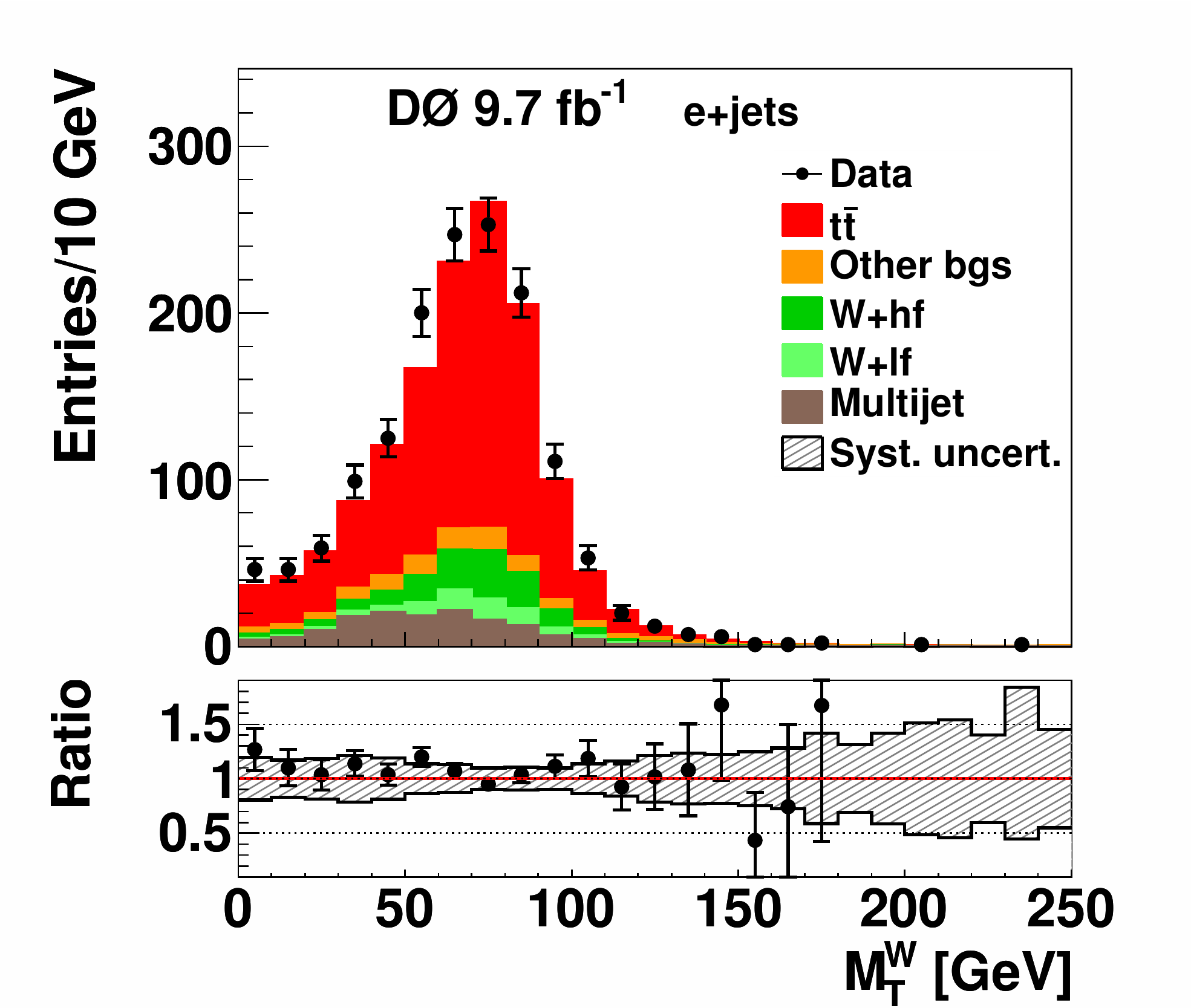}
\put(80.5,73){\large\textsf{\textbf{(e)}}}
\end{overpic}
\begin{overpic}[width=0.99\columnwidth]{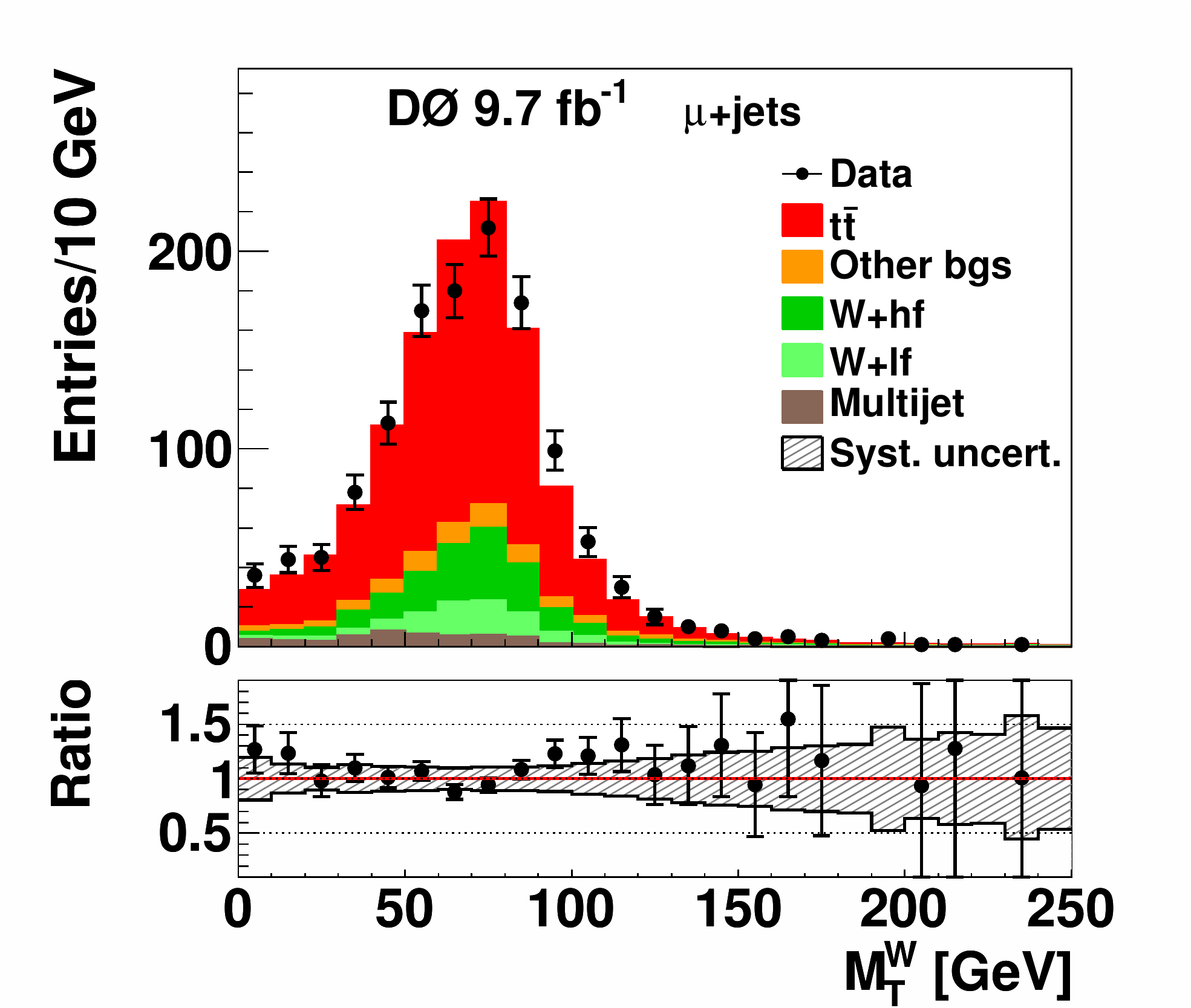}
\put(80.5,73){\large\textsf{\textbf{(f)}}}
\end{overpic}
\par\end{centering}
\caption{
\label{fig:postleppt}
Same as Fig.~\ref{fig:postpttt}, but for the transverse momentum of the (a)~electron or (b)~muon, for the missing transverse momentum in (c)~\ejets and (d)~\mujets final states, as well as for the transverse mass of the $W\to\ell\nu_\ell$ decay in (e)~\ejets and (f)~\mujets final states.
}
\end{figure*}

\begin{figure}
\begin{centering}
\begin{overpic}[width=0.99\columnwidth]{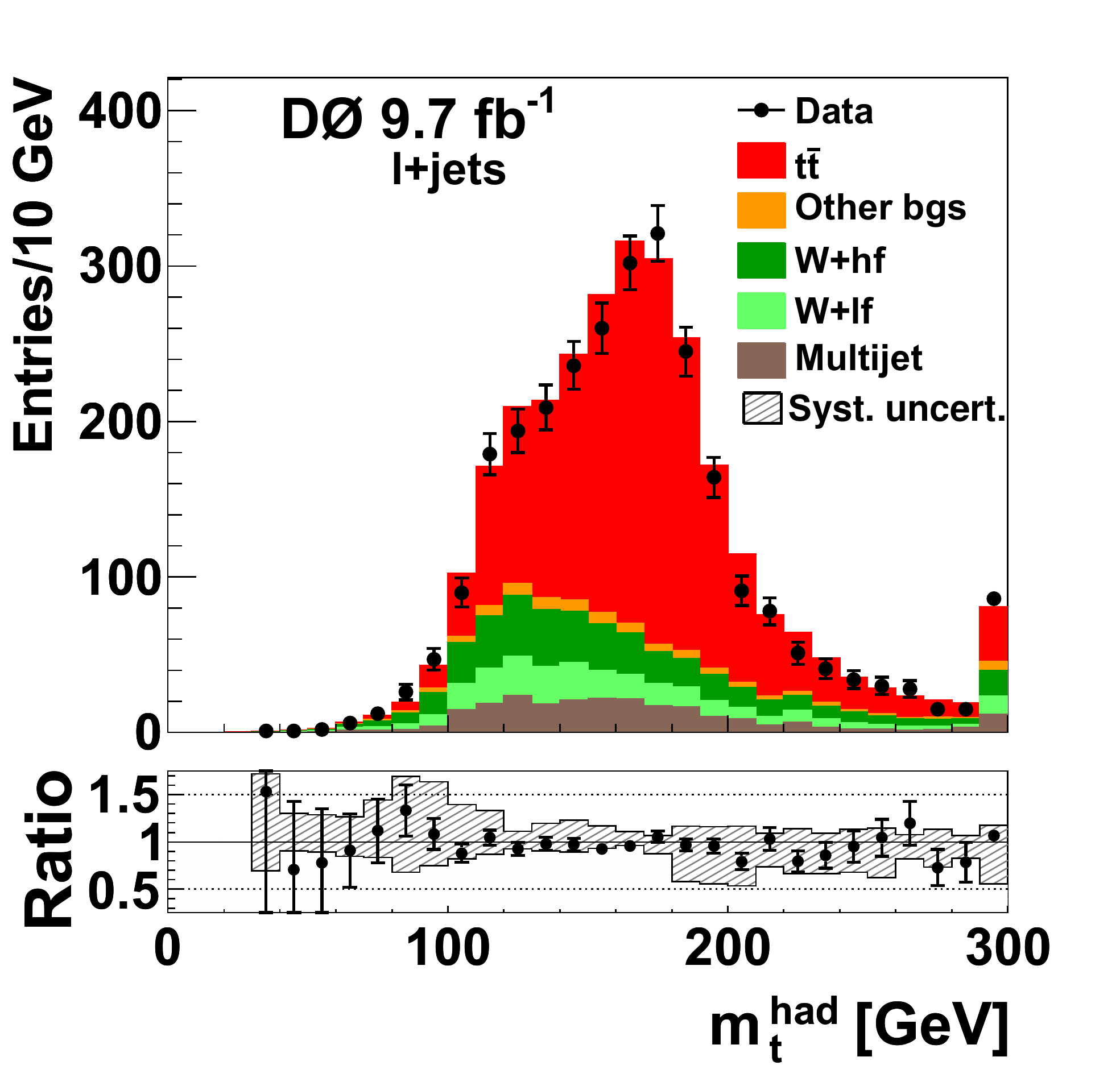}
\end{overpic}
\par\end{centering}
\caption{
\label{fig:postfitmthad}
Same as Fig.~\ref{fig:postpttt}, but for the invariant mass of the trijet system matched to the hadronic decay of the top quark in~\ljets final states.
\vspace{-3mm}
}
\end{figure}

The transverse momenta of the electrons and muons, the distributions of \met, and the transverse mass of the $W\to\ell\nu_\ell$ decay are shown in Fig.~\ref{fig:postleppt}.  The invariant of the trijet system matched to the hadronic decay of the top quark, reconstructed from the same four-momenta used in the reconstruction of $m_{\ttbar}$ and $M_W$, is shown in Fig.~\ref{fig:postfitmthad}. Notable improvement due to reduced backgrounds is observed relative to the Run~I measurement~\cite{bib:mtrunone}.


Good agreement between data and simulations is observed in all variables except $\pt^{\ttbar}$, where \apythia predicts fewer events with $\pt^{\ttbar}<10~\GeV$ than observed in data, for which a systematic uncertainty is evaluated in Section~\ref{sec:isr}. 
The variable $\pt^{\rm top}$, for which agreement within uncertainties is observed, is shown at detector level. However, some discrepancy between measured $\dif\sigma_{\ttbar}/\dif\pt^{\rm top}$ corrected for detector effects, and SM calculations is present in our data~\cite{bib:diffxsec}.

\clearpage

\section{
Systematic uncertainties
}
\label{sec:syst}

Systematic uncertainties are evaluated using sets of PEs constructed from simulated signal and background events, for three categories of sources:
\begin{enumerate}
\item
modeling of signal and background events,
\item
simulation of the detector response, and
\item
analysis procedures and assumptions.
\end{enumerate}
The contributions from these sources listed in Table~\ref{tab:syst} are evaluated individually and combined in quadrature to obtain the total systematic uncertainty. 

All sources of systematic uncertainties are considered fully correlated between \ejets and \mujets final states, except for the uncertainties from the lepton momentum scale and the calibration of the method. In comparison to the previous analysis using 3.6~\fb of integrated luminosity~\cite{bib:mt36}, the number of simulated \ttbar events used for the calibration of the ME technique and for the evaluation of systematic uncertainties has increased significantly since the computation time for \psig has been reduced by two orders of magnitude, as discussed in Section~\ref{sec:psigcalc}. As a result, the statistical components of systematic uncertainties  are $\approx0.05~\GeV$ in each of the first four sources listed in Table~\ref{tab:syst}, namely those from higher-order corrections,  initial and/or final-state radiation, hadronization and the underlying  event (UE), and color reconnection, and are $\approx0.01~\GeV$ for  the heavy-flavor scale factor, residual jet energy scale, flavor-dependent  response to jets, lepton momentum scale, jet energy resolution, jet identification efficiency, and modeling of multijet events. The statistical components are negligible for all the other sources of systematic uncertainty. The uncertainties from the first four sources are evaluated for $\mtgen=172.5~\GeV$ and $\kjesgen=1$ by comparing results for \mt using different models for signal. All other systematic uncertainties are evaluated by applying a recalibration using simulations that reflect an alternative model to data. The statistical components due to the finite number of simulated events are never larger than the net difference between the central and alternative models for any of the sources of systematic uncertainty.

Unless stated otherwise, the systematic uncertainties are evaluated using fully simulated events. Background events are always included, as described in Sections~\ref{sec:calibf} and~\ref{sec:calibmt}. The signal fraction used to compose PEs is kept fixed at the value obtained for that data-taking epoch and final state, as summarized in Table~\ref{tab:postfit}, except in evaluating the systematic uncertainty from $f$. The systematic uncertainties from category (iii), which are associated with analysis procedures and assumptions, are evaluated including the MJ background in the ensemble tests. The estimate of all other systematic uncertainties is not expected to be affected by the inclusion of the MJ background contribution in the ensemble tests, which is therefore neglected. 
We quote signed uncertainties for one-sided sources of uncertainty or sources dominated by a one-sided component in Table~\ref{tab:syst}, indicating the direction of \mt change when using an alternative instead of the central model. Two-sided uncertainties or uncertainties dominated by a two-sided component are marked with a $\pm$ or $\mp$ symbol. For one-sided sources of systematic uncertainty, the full magnitude of the effect is taken in each direction when calculating the total quadrature sum to obtain the total systematic uncertainty, in accordance with Ref.~\cite{bib:combitevprd}.

\begin{table}
\begin{centering}
\caption{
\label{tab:syst}
Summary of uncertainties on the measured \mt. One-sided sources of uncertainty or sources dominated by a one-sided component are signed, indicating the direction of \mt change when using an alternative instead of the central model. Two-sided uncertainties or uncertainties dominated by a two-sided component are marked with a $\pm$ or $\mp$ symbol.
\vspace{2mm}
}
\begin{ruledtabular}
\begin{tabular}{lc}
Source of uncertainty & Effect on \mt (GeV) \\
\hline
\emph{Signal and background modeling:} & \\
~~Higher-order corrections & $+0.15$ \\
~~Initial/final state radiation & $\mp0.06$ \\
~~Transverse momentum of the \ttbar system & $-0.07$ \\
~~Hadronization and underlying event & $+0.26$ \\
~~Color reconnection & $+0.10$ \\
~~Multiple $p\bar p$ interactions & $-0.06$ \\
~~Heavy-flavor scale factor & $\mp0.06$\\
~~Modeling of $b$-quark jet & $+0.09$ \\
~~Parton distribution functions & $\pm0.11$ \\
\emph{Detector modeling:} & \\
~~Residual jet energy scale & $\pm0.21$\\
~~Flavor-dependent response to jets & $\mp0.16$\\
~~Tagging of $b$~jets & $\mp0.10$\\
~~Trigger & $\pm0.01$\\
~~Lepton momentum scale & $\pm0.01$\\
~~Jet energy resolution & $\pm0.07$\\
~~Jet identification efficiency & $-0.01$\\
\emph{Method:} & \\
~~Modeling of multijet events & $+0.04$\\
~~Signal fraction & $\pm0.08$\\
~~MC calibration & $\pm0.07$\\
\hline 
{\em Total systematic uncertainty} & $\pm0.49$\\
{\em Statistical uncertainty} & $\pm0.58$\\
{\em Total uncertainty} & $\pm0.76$\\
\end{tabular}
\end{ruledtabular}
\par\end{centering}
\end{table}

\subsection{Signal and background modeling}
%
\subsubsection{Higher-order corrections}
We calibrate the response of the ME technique using \ttbar events generated with the LO \alpgen program, which accounts for real corrections to the Born level process up to second-order in $\alpha_s$ through the ME, i.e., for 0, 1, or 2 additional light partons. To evaluate the effect of higher-order virtual corrections on \mt, we use signal events generated with the \mcnlo program~\cite{bib:mcnlo}, which calculates the processes $q\bar q\to\ttbar$, $gg\to\ttbar$, and $qg\to\ttbar$ at NLO, as an alternative model. Because \mcnlo is interfaced to \herwig~\cite{bib:herwig} for simulating the contributions from leading-logarithmic (LL) corrections and fragmentation  to the hard process of interest, we use \aherwig events as the central model for this comparison to avoid double-counting the uncertainty from a different showering model. Performing the ensemble tests for $\mt=172.5~\GeV$, we find a shift of $+0.15~\GeV$, which we assign as the systematic uncertainty due to higher-order corrections. 

%
\subsubsection{Initial/final state radiation}
\label{sec:isr}

The modeling of extra jets from initial/final state radiation (ISR/FSR) can affect the measurement of \mt. To evaluate the contribution from this source, we adjust the amount of radiation by changing the {\tt ktfac} parameter~\cite{bib:isrmangano} that defines the renormalization scale in the CKKW scale-setting procedure, from its standard value of ${\tt ktfac}=1$ in our \apythia simulations described in Section~\ref{sec:mc}, as suggested in Ref.~\cite{bib:isrmangano}. We constrain changes in the amount of ISR/FSR by studying the cross section for Drell-Yan events ($q\bar q\to Z/\gamma^*\to\ell^+\ell^-$) measured differentially in the $\phi^*$ variable~\cite{bib:isr}, which is related to the \pt of the $Z/\gamma^*$ boson, and has sensitivity to the amount of QCD radiation. Our studies, documented in Appendix~\ref{app:isr}, indicate that changes in the {\tt ktfac} parameter by a factor of $\pm$1.5 cover the excursion of MC simulations from data, which results in an uncertainty of $\mp0.06~\GeV$ on \mt. Previously, the systematic uncertainty from modeling of ISR/FSR was evaluated using a pure LO generator, \pythia, and changes made in shower-evolution parameters. The new procedure is more accurate, as it uses a larger data sample for deriving the ISR/FSR dependence, and employs the same generator setup as for simulating the default \ttbar sample. 

%
\subsubsection{Transverse momentum of the \ttbar system}
\label{sec:pttt}
The distribution in the transverse momentum of the \ttbar system, $\pt^{\ttbar}$, is not modeled well as shown in Fig.~\ref{fig:postpttt}, and a dedicated systematic uncertainty is assigned. This mismodeling is corrected by reweighting the contribution from \ttbar events differentially in $\pt^{\ttbar}$ to achieve agreement between the simulations and data. Propagating this effect to \mt using ensemble tests, we find a shift of $-0.07~\GeV$.

%
\subsubsection{Hadronization and underlying event}
\label{sec:had}

\begin{figure}
\centering
\begin{overpic}[width=0.49\columnwidth]{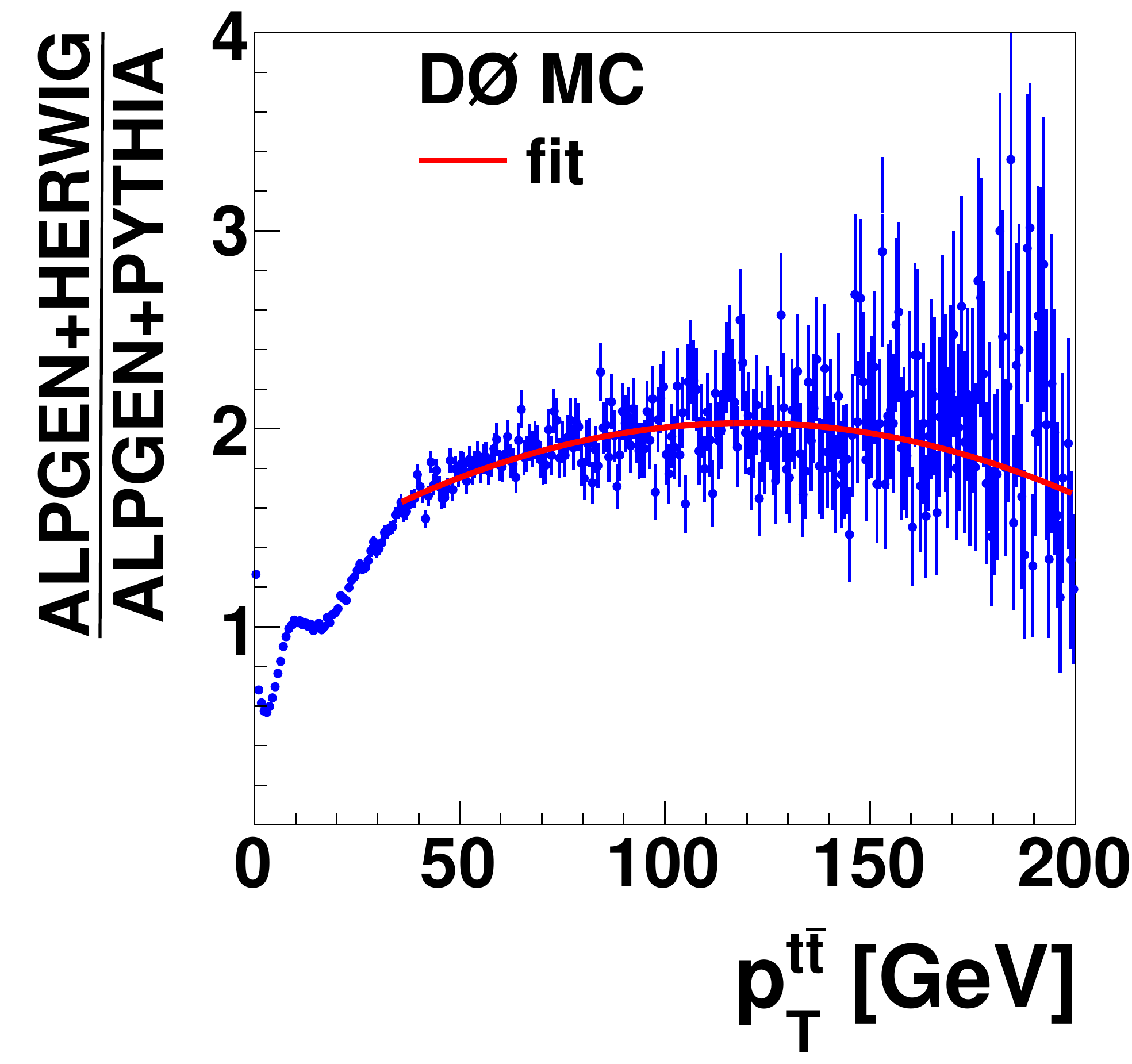}
\put(26,82){\footnotesize\textsf{\textbf{(a)}}}
\end{overpic}
\begin{overpic}[width=0.49\columnwidth]{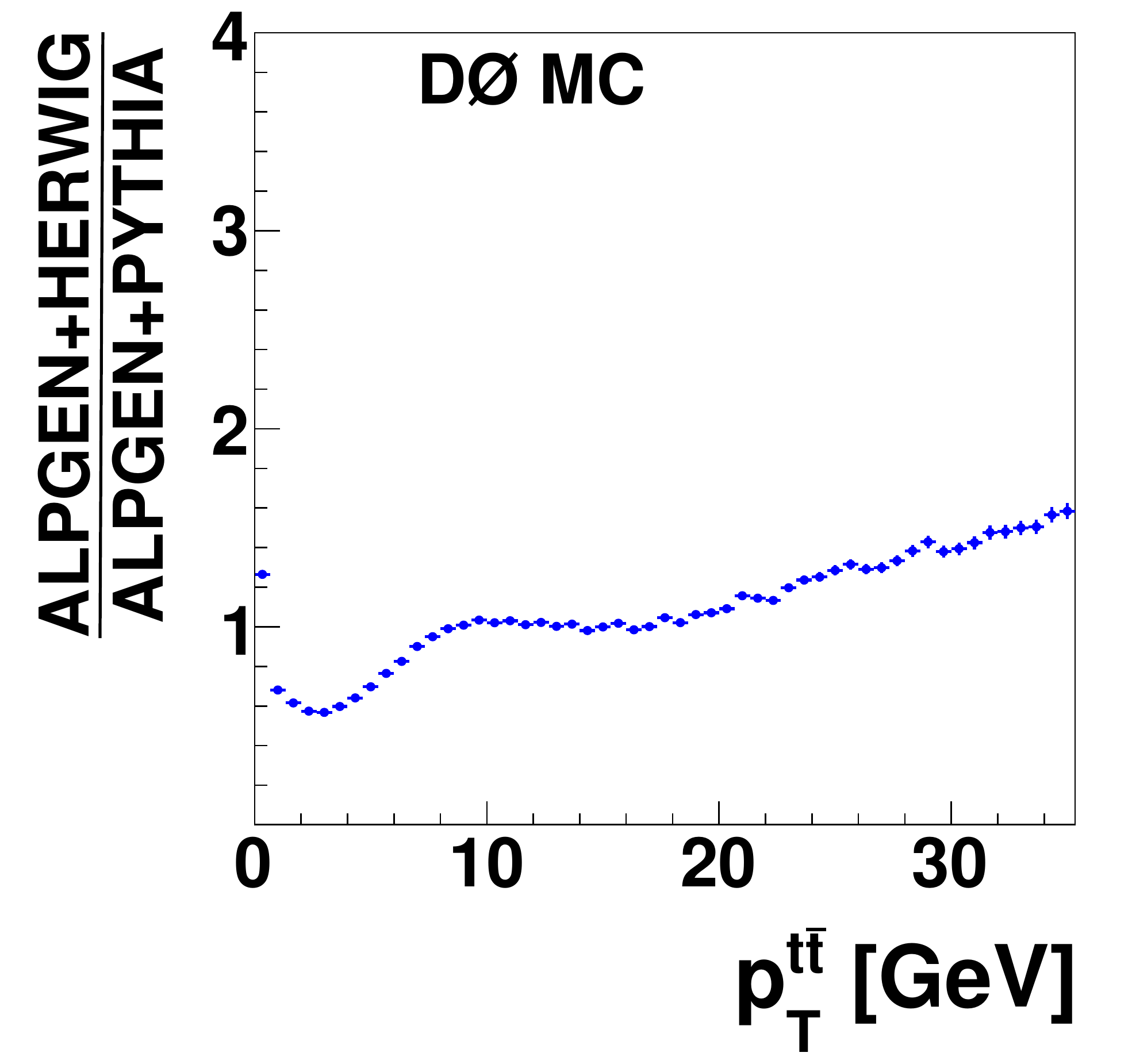}
\put(26,82){\footnotesize\textsf{\textbf{(b)}}}
\end{overpic}
\caption{\label{fig:pttt}
(a) The ratio of predicted differential cross sections from \aherwig to those from \apythia as a function of ${\pt^{\ttbar}}$ at particle level. Also shown is the fit with a second-order polynomial used for reweighting \apythia events in $\pt^{\ttbar}$ for $35~\GeV<\pt^{\ttbar}<200~\GeV$. Below that range the reweighting is performed according to the histogram, and above with a constant factor of $1.67$. Uncertainties are only statistical. 
(b)~Same ratio in the range $\pt^{\ttbar}<35~\GeV$.
}
\end{figure}

The choice of a model for the parton-shower (PS) evolution, the hadronization, and the underlying event (UE) can lead to a bias on the measured \mt value. To evaluate the size of this effect, we compare MC samples where PS evolution, hadronization, and UE are simulated using \pythia with \herwig as an alternative model. In both cases, the \alpgen generator is used to simulate the hard process.
The comparison of \aherwig with \apythia will give a convolution of four terms: 
\begin{enumerate}
\item
a statistical component due to the limited size of the MC samples, 
\item
a difference in kinematic spectra, most notably in $\pt^{\ttbar}$,  between \aherwig and \apythia, 
\item
a difference in the response of the detector to jets simulated with \herwig and \pythia, and
\item\label{en:had}
a difference in the modeling of the PS evolution, hadronization, and UE when using \aherwig and \apythia.
\end{enumerate}
We are interested in the effect from source~(iv),  since (ii) is already accounted for in the systematic uncertainty from modeling of ISR/FSR, and (iii) is taken into account in the systematic uncertainties from modeling of the detector response. We reduce the contributions from sources (i), (ii), and (iii) as follows:
\begin{enumerate}
\item
The contribution from source~(i) is reduced to \mbox{$\approx0.05~\GeV$}, about a factor of five smaller than the previous \mt measurement~\cite{bib:mt36}, by increasing the size of MC samples by a factor of $\approx25$.
\item
Reshuffling of momenta between the hard and soft part of the event~\cite{bib:nason} causes the $\pt^{\ttbar}$ distribution in \aherwig to differ significantly from \apythia, as shown in Fig.~\ref{fig:pttt}. To isolate this effect in the evaluation of the systematic uncertainty, we reweight \apythia in $\pt^{\ttbar}$ to match the \aherwig simulation, which results in a shift of 0.05~\GeV in \mt. The uncertainty from modeling of transverse momentum of the \ttbar system already accounts for the potential impact on \mt of this difference, as discussed in Section~\ref{sec:pttt}.
\item
The JES calibration is derived using \pythia with D0~tune~A~\cite{bib:jes}, and is therefore valid only for this specific configuration. The contribution from source~(iii) would not be present if a dedicated JES derived with \herwig was available. Like the default JES calibration using \pythia, the alternative calibration using \herwig would be affected by similar uncertainties accounted for in the ``detector modeling'' category of Table~\ref{tab:syst}. Including the difference of the JES derived with \pythia and \herwig in the evaluation of the uncertainty from the choice PS evolution, hadronization, and UE models would lead to double-counting.
To eliminate the contribution from the response of the detector, we use particle-level jet energies. The standard selection as described in Section~\ref{sec:selection} is applied at detector level to take into account any potential effects from the trigger or selection requirements, and the detector-level jets are matched to particle-level jets. The matching is done in $(\eta,\phi)$ space, using the requirement of $\Delta R(j_{\rm detector},j_{\rm particle})<R_{\rm cone}/2=0.25$. 
The energy of each detector-level jet is replaced by the energy of the matched particle-level jet, and the magnitude of the three-momentum $|\vec p|$ of the detector-level jet is adjusted to preserve its original  $\eta$, $\phi$, and $m$ parameters. Those modified four-momenta of jets are then used, together with all other detector-level information in the event, to extract \mt with the ME technique just as in data. 
\end{enumerate}

Reducing the contribution from source~(i), and factorizing out contributions from sources~(ii) and~(iii), as described above, we find an effect of $+0.26~\GeV$ from the choice of the model for PS evolution, hadronization, and UE. We verify that $\kjes^{\hbox{\scriptsize{\aherwig}}}$ and $\kjes^{\hbox{\scriptsize{\apythia}}}$ are consistent within uncertainties, as expected at particle level, indicating that the effect of JES has been factorized out.

As a cross-check, we also evaluate the impact of a different PS model and a change in the models for hadronization and UE by comparing \ttbar events simulated with \alpgen and interfaced to either \pythia with \dzero tune~A (the standard MC setup), or \pythia with the Perugia 2011C tune~\cite{bib:perugia}. While the former is based on a $Q^2$ ordered PS model, the latter is $\pt^2$ ordered~\cite{bib:pythia1,bib:pythia2}. Furthermore, the latter is a recent tune from the Perugia 2011 family that includes LHC data, which is recommended for this comparison~\cite{bib:perugia_advice}. Both use the CTEQ6L1 set of PDFs. We verify that the Perugia 2011 family describes the transverse jet shapes sufficiently well and is therefore applicable to Tevatron data dominated by quark jets, despite being tuned including LHC data dominated by wider gluon jets. For simplicity, we carry out the comparison between  \pythia with \dzero tune~A and \pythia with Perugia 2011C at detector level, and find a change of $+0.30~\GeV$ in \mt, which serves as an upper estimate of the effect evaluated with particle-level jet energies, as for the comparison of \aherwig with \apythia.

%
\subsubsection{Color reconnection}

Our standard \pythia tune, \dzero tune A, does not include an explicit model for color reconnection~(CR). Color reconnection is sometimes confused with strong color correlation between the hard scattering process and the underlying event. The latter is an integral part of \pythia \dzero tune A, as not including it would violate confinement. Recently, research has been carried out to include CR in \pythia~\cite{bib:color}. To evaluate the CR effect, we compare identical \alpgen events interfaced to \pythia with two different tunes, Perugia 2011 and Perugia 2011NOCR. While the former tune includes an explicit CR model, the latter does not. The Perugia 2011 tune rather than Perugia 2011C is used as an alternative model because Perugia 2011 uses the same set of CTEQ5L PDFs as the reference model Perugia 2011NOCR, while Perugia 2011C uses CTEQ6L1. This choice of generator setup is advised in Ref.~\cite{bib:perugia_advice}, since Perugia includes a better, more phenomenologically-motivated CR model than the ACRpro tune~\cite{bib:acrpro}. The new tune models the color-string survival probability depending on the distance in rapidity between the beginning and the end of the color string, whereas the previous model uses only an overall approximation. We find an effect of $+0.10~\GeV$ on \mt from this source of systematic uncertainty.

To cross-check the impact of modeling of multiple parton interactions (MPI) on the measurement, we compare events generated with  \alpgen interfaced to \pythia with the Perugia 2011mpiHi and Perugia 2011 tunes, where the former features harder MPI, albeit at a smaller rate, compared to the latter.  
We find an effect of 0.14~GeV, with a statistical component of $\approx0.15~\GeV$ due to limited MC event statistics. Since \herwig and \pythia use different models for MPI, an uncertainty from the choice of a model for MPI is already included in the uncertainty from modeling of hadronization and the UE discussed in Section~\ref{sec:had}. In addition, it is at least partly included in the comparison between the Perugia~2011 and Perugia~2011NOCR tunes discussed above, since the latter, without CR, must simultaneously adjust the amount of MPI to describe data. Therefore, the comparison between Perugia 2011mpiHi and Perugia 2011 is treated as a cross-check.

\subsubsection{Multiple $p\bar p$ interactions}

Multiple $p\bar p$  interactions can potentially influence the measurement of \mt. We reweight the spectrum in instantaneous luminosity  of our simulated MC samples to the instantaneous luminosity profile found in data for the respective data-taking epoch and final state. This default correction is derived for each data-taking stream, while the instantaneously luminosity profile in a subset of this data can differ due to the trigger and selection requirements. 
To evaluate this systematic uncertainty, we obtain an alternative calibration where an additional reweighting in instantaneous luminosity is applied after all selections to bring its spectra in \ttbar simulations and in data into agreement. We find a change in \mt of $-0.06~\GeV$.

\subsubsection{Heavy-flavor scale factor}

As mentioned in Section~\ref{sec:mc}, a heavy-flavor scale factor of $k_{bb,cc}=1.47$ is applied to the $Wb\bar{b}$ and $Wc\bar{c}$ cross sections to correct the heavy-flavor content of the \wjets MC samples to expectations from NLO calculations. The uncertainty on $k_{bb,cc}$ is $\approx15\%$~\cite{bib:diffxsec}. Moreover, there is an additional uncertainty of $\approx12\%$ on the $W+c$ content of the sample~\cite{bib:diffxsec}. These uncertainties are combined linearly and rounded to 30\%. The effect of $k_{bb,cc}$ variations is propagated to \mt using PEs. This introduces variations in the kinematic distributions of the combined contribution from $\wjets$ production. As we are interested only in the effect of the changed background composition, we do not change $f$ when constructing the pseudo-experiments. We find an effect of $\mp0.06~\GeV$ due to this source of systematic uncertainty.

\subsubsection{Modeling of $b$-quark jet}

Uncertainties in the simulation of $b$~quark fragmentation can affect the measurement of $\mt$ through several aspects of the analysis such as $b$~tagging and the TF. Such effects are studied by reweighting the simulated $\ttbar$ events according to other possible fragmentation models for $b$~quarks. The standard \ttbar MC samples consist of events reweighted from the standard \pythia $b$~fragmentation function to a Bowler scheme~\cite{bib:bowler} that has been tuned to LEP (ALEPH, OPAL, and DELPHI) data~\cite{bib:bfrag}. To evaluate the potential bias from this source, \ttbar events are further reweighted to account for differences in SLD~\cite{bib:bfrag} and LEP data, resulting in a shift of $+0.08$~GeV in the measured \mt.  

The detector response to $b$-quark jets can be different in the presence of semileptonic decays of $b$~or $c$ quarks. The incorrect simulation of semileptonic branching ratios in $b$~and $c$ quark decays can thus result in a systematic effect on \mt. We take an uncertainty of 0.05~\GeV determined in Ref.~\cite{bib:mt2006} as the contribution from this source.
Quadratically combining the two sources of uncertainty, we find a shift in \mt of $+0.09~\GeV$.

\subsubsection{Parton distribution functions}

To evaluate the uncertainty from the choice of PDFs, the standard \apythia sample at $\mtgen=172.5~\GeV$ is reweighted to match possible excursions in the PDF parameters provided in CTEQ6M~\cite{bib:cteq}. Ensemble tests are repeated for each of these changes and the total uncertainty is evaluated using the following formula~\cite{bib:cteq}:
\begin{equation}
\delta_{\mt}=\frac{1}{2}\left\{{\displaystyle \sum_{i=1}^{20}[\Delta \mt(S_{i}^{+})-\Delta \mt(S_{i}^{-})]^{2}}\right\}^{\frac12}\,,
\end{equation}
where the sum runs over PDF uncertainties in the positive ($S_{i}^{+})$ and
negative ($S_{i}^{-}$) excursions. We find changes in \mt of $\pm 0.11$\,GeV due to this source of systematic uncertainty. The effect on \mt from replacing the central PDF set from CTEQ6M by that from CTEQ6L1 is much smaller.

\subsection{Detector modeling}

%
\subsubsection{Residual jet energy scale}
\label{sec:rjes}

The overall jet energy scale factor \kjes is fitted simultaneously with \mt, and the corresponding contribution to the statistical uncertainty is included in \dmt. However, \kjes does not account for possible effects from uncertainties in the JES corrections that are differential in $E$ or $\eta$ of the jet.
To estimate the potential bias on \mt from the residual JES uncertainty, we use a variety of approaches:
\begin{enumerate}
\item
As for the 3.6~\fb analysis, we parameterize the dependence of the upper side of the uncertainty of the JES on energy $E$ of the jet, in four bins in $\eta$, as in Ref.~\cite{bib:jes}. 
This parameterization is applied to scale up jet energies and to provide an alternative calibration. A single constant offset is added to the parameterization across all four $\eta$ bins, such that the average correction applied to the energies of all the jets in the sample is zero to avoid double-counting of acceptance effects. We find an effect of $+0.13~\GeV$ on \mt.
\item
We also apply the downward changes, and find an effect of $-0.08~\GeV$ on \mt.
\item
We apply the positive JES changes directly, i.e., the momentum of each jet is increased by its 1 SD uncertainty, without any parametrization. In contrast to (i) and (ii), the jet energies are not rescaled on average, and the method is expected to account for the change in the overall JES through the $\kjes$ parameter in the two-dimensional likelihood fit, at the cost of some double-counting of acceptance effects. We find a change in \mt of $-0.20$~GeV with this procedure, while \kjes changes by $-0.018$, as expected.
\item
The energies of jets are increased using a linear parameterization, which is unity for $E=0~\GeV$ and increases linearly with $E$, such that it tangentially approaches the upper limit of uncertainty of the JES. The linear parameterization is obtained separately for each data-taking epoch, and changes slightly due to changes in the JES uncertainty. No overall rescaling of the jet energies is applied. 
We find $-0.21$~GeV with this approach for the impact on~\mt.
\end{enumerate}
As only one dependence of the true JES on $E$ and $\eta$ of the jet is realized within the uncertainty corridor, we take the maximum envelope of the above uncertainties of $0.21~\GeV$ and assign it as the systematic uncertainty due to the residual JES correction. In previous analyses~\cite{bib:mt36,bib:mt2006}, only approach (i) was used, which corresponds to an uncertainty of 0.13~\GeV using our results.

%
\subsubsection{Flavor-dependent response to jets}
\label{sec:flav}

The standard JES calibration used in this measurement contains a flavor-dependent jet response correction, which is responsible for bringing the simulation of calorimeter response to jets into agreement with that observed in data for a given jet flavor~\cite{bib:jes}. The flavor-dependent correction is given by
\begin{linenomath}
\begin{equation}
\label{eq:fcorr}
F_{\rm corr}=\frac1{\langle F\rangle_{\gamma+{\rm jet}}}\cdot\frac{\sum_i E_i\cdot R_i^{\rm data}}{\sum_i E_i\cdot R_i^{\rm MC}}\,,
\end{equation}
\end{linenomath}
where the index $i$ runs over all constituent particles that belong to a particle jet, and $R_i$ is the detector response to particle $i$, given differentially in \pt and $\eta$ of that particle. The factor ${\langle F\rangle_{\gamma+{\rm jet}}}$ ensures that the flavor-averaged part of the JES calibration, extracted from exclusive $\gamma+\rm jet$ events, is preserved. The correction $F_{\rm corr}$ is largest for $b$-quark jets and gluon jets, and is relatively small for light quark jets, compared to the flavor-averaged JES, as shown in Fig.~34 of Ref.~\cite{bib:jes}. We are in a position to apply such a correction in a straightforward way, since the JES is corrected to particle level both for data and MC. To propagate the effect of the uncertainty on $F_{\rm corr}$ to our measured \mt value, we change $F_{\rm corr}$ by $\pm$1 SD for each data-taking epoch and final state. We apply this factor to the jets in our signal MC samples to get alternative calibrations, resulting in a systematic uncertainty of $\mp0.16~\GeV$ after symmetrization. This uncertainty accounts for the difference in detector response to different jet flavors, in particular $b$-quark jets versus light quark jets.

\begin{figure}
\begin{centering}
\includegraphics[width=0.99\columnwidth]{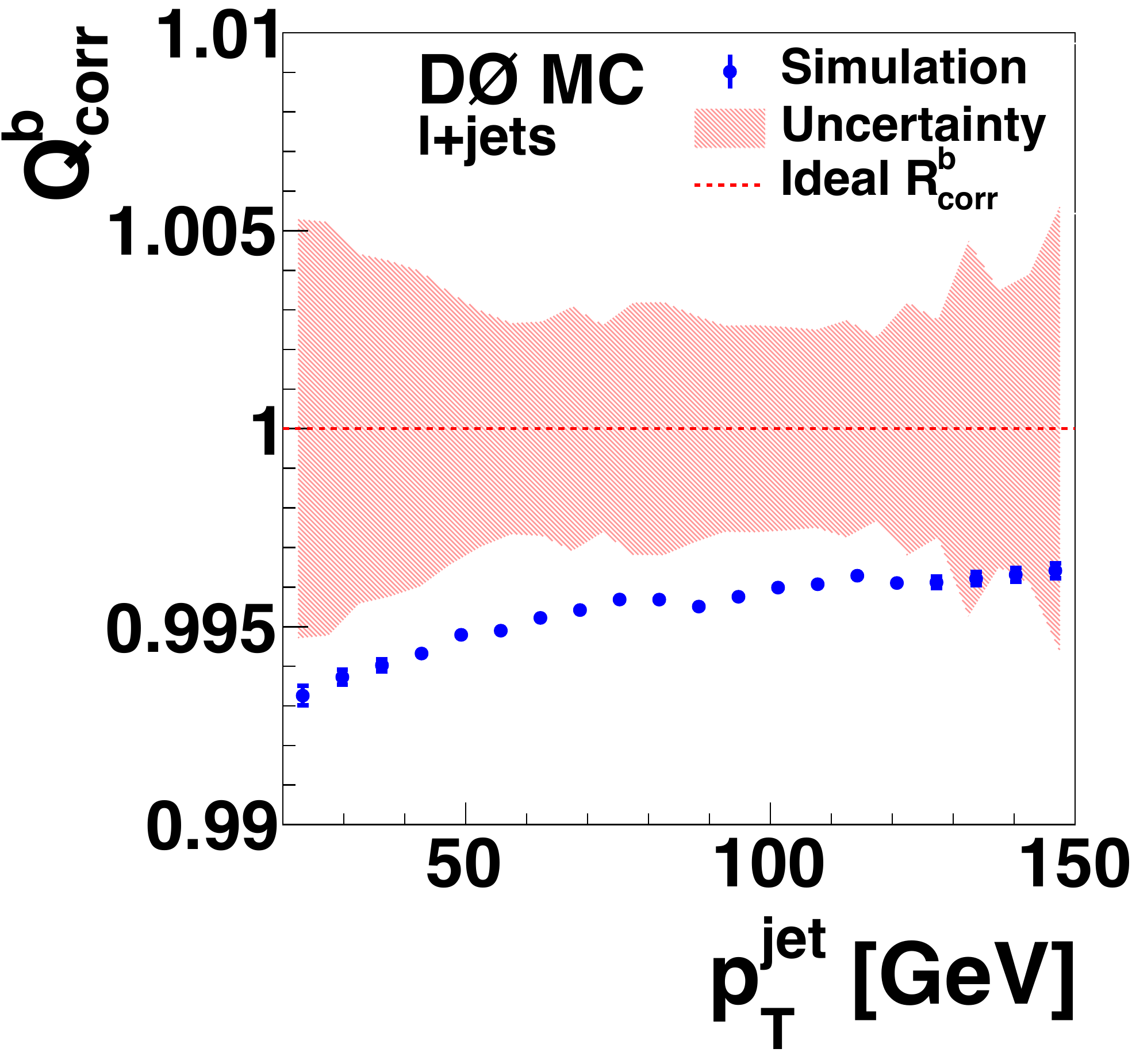}
\par\end{centering}
\caption{
\label{fig:flav}
The ratio $Q_{\rm corr}^b={F_{{\rm corr},\,\hbox{\scriptsize{\aherwig}}}^b}/{F_{{\rm corr},\,\hbox{\scriptsize{\apythia}}}^b}$ for $b$-quark jets in simulated \ttbar events and \ljets final state as a function of the \pt of the jet. The ideal $Q_{\rm corr}^b$ of unity is indicated as the broken line. The uncertainties on the points reflect the finite size of the MC sample. The total uncertainty on the standard flavor-dependent correction for \apythia is indicated as the band around unity.
}
\end{figure}

\begin{figure}
\begin{centering}
\includegraphics[width=0.99\columnwidth]{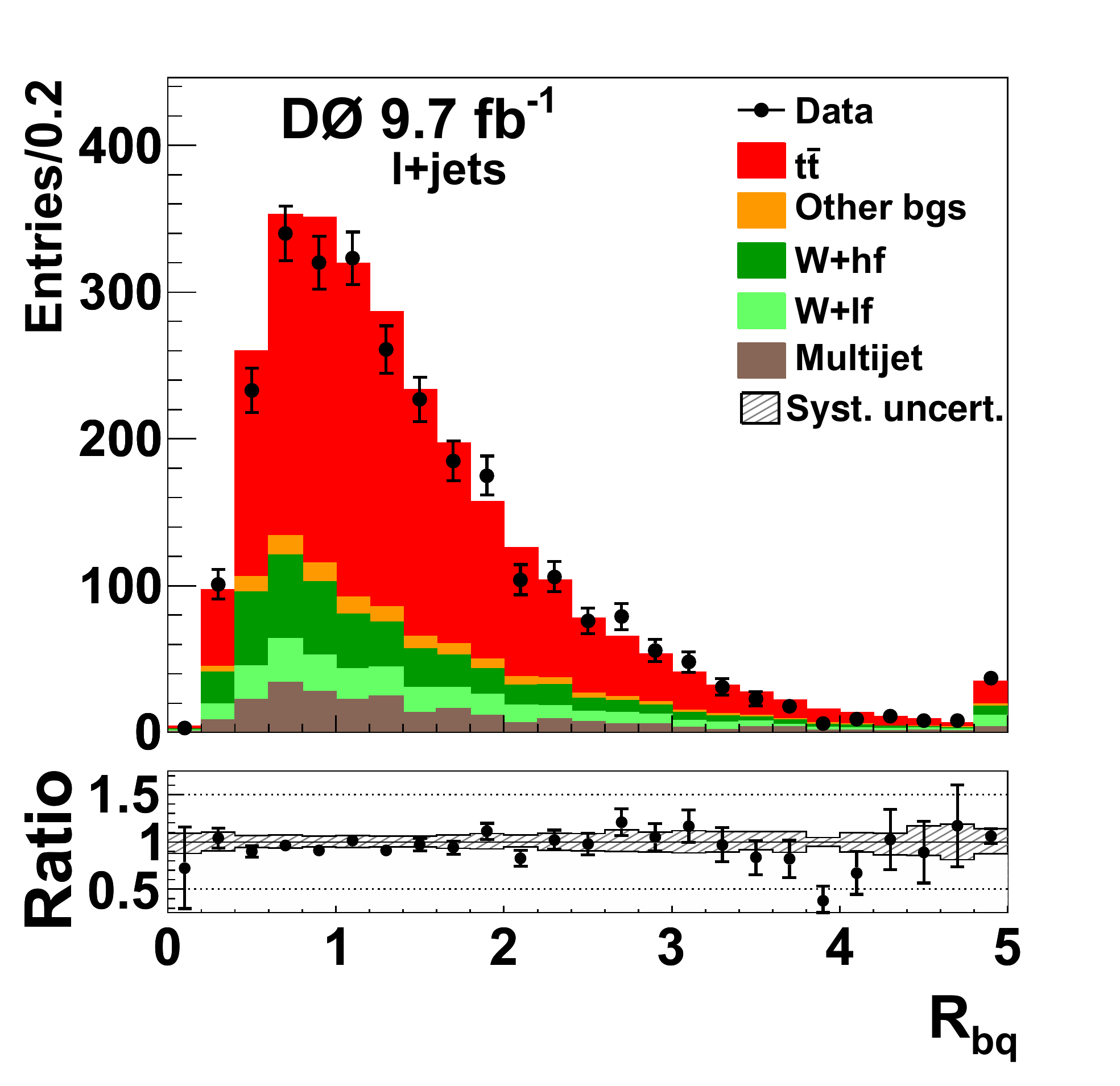}
\par\end{centering}
\caption{
\label{fig:rbl}
The ratio $R_{bq}=({\pt^{b_1}+\pt^{b_2}})/({\pt^{j_1}+\pt^{j_2}})$ for events with $\geq2$ $b$-tagged jets, and $R_{bq}={2p_T^{b}}/(p_T^{j_1}+p_T^{j_2})$ for events with one $b$-tagged jet, after all selections.
The category ``Other bgs'' encompasses $\zjets$, $WW$, $WZ$, $ZZ$, and single top quark production, as well as dileptonic \ttbar decays. The production of a $W$ boson in association with at least one $b$ or $c$ quark is denoted as ``$W+{\rm hf}$'', while ``$W+{\rm lf}$'' stands for all other flavor configurations.  The last bin includes overflow events. The ratio shows the number of observed events divided by the number of expected events in a given bin. The band of systematic uncertainty is indicated as a shaded area in the ratio plots, and includes dominant sources:  JES, jet energy resolution, jet identification, flavor dependence of the detector response, and $b$~tagging.
}
\end{figure}

We perform additional studies beyond those done in Ref.~\cite{bib:jes} to verify that $F_{\rm corr}$ is correctly estimated: 
\begin{enumerate}
\item
The derivation of $F_{\rm corr}$ requires {\em a priori} input for the relative contributions of $b$~quark, gluon, and light quark jets to the samples of $\gamma+{\rm jets}$ and dijet events that are used for its derivation. A variation on the assumption of these relative contributions would result in somewhat different fitted $R_i^{\rm data}$ in Eq.~(\ref{eq:fcorr}). The relative contributions from different jet flavors to the sample are taken from \pythia with D0 tune A. To estimate the effect of this assumption, we exploit the fact that $\gamma+{\rm jets}$ and dijet samples show a rather different sample composition in terms of jet flavor, and obtain $F_{\rm corr}$ in those two samples independently, rather than simultaneously in the standard procedure. We find that the results for $F_{\rm corr}$ in $\gamma+{\rm jets}$ and dijet events are consistent with each other within assigned uncertainties.
\item
Similarly, {\em a priori} input about the particle composition of the jet is needed for the derivation of $F_{\rm corr}$, which is also taken from \pythia with D0 tune A. A variation of the composition of the jet would result in a change to the numerator and the denominator of Eq.~(\ref{eq:fcorr}) because the sum in $i$ runs over all the particles in the jet, and to $R_i^{\rm data}$, which are determined from a fit to data in $\gamma+{\rm jets}$ and dijet events. To estimate an upper limit on the impact of the particle composition on $F_{\rm corr}$ for $b$-quark jets, for which $F_{\rm corr}$ is largest, we use simulated \ttbar events to study the ratio 
\begin{linenomath}
\begin{equation}
\label{eq:qcorr}
Q_{\rm corr}^b=\frac{F_{{\rm corr},\,\hbox{\scriptsize{\aherwig}}}^b}{F_{{\rm corr},\,\hbox{\scriptsize{\apythia}}}^b}\,,
\end{equation}
\end{linenomath}
where $F_{{\rm corr},\,\hbox{\scriptsize{\apythia}}}^b$ is the default correction from Ref.~\cite{bib:jes}, and $F_{{\rm corr},\,\hbox{\scriptsize{\aherwig}}}^b$ takes the particle composition of the jet predicted by \aherwig as input, but uses the same $R_i^{\rm data}$ and $R_i^{\rm MC}$ as in the default case. The ratio $Q_{\rm corr}^b$ is shown for \ljets final states in Fig.~\ref{fig:flav}. The excursion of the ratio from unity is within 0.5\% for the $\pt$ spectrum of $b$-quark jets in \ttbar events.  

The ratio $Q_{\rm corr}^b$ can only provide an upper limit on the impact of the assumption of the particle composition of the jet on the flavor-dependent calorimeter response, as it does not involve a coherent JES calibration using $\gamma+{\rm jet}$ and dijet events. In particular, $R_i^{\rm data}$, which are determined with \pythia, are unchanged, while a calibration of the single particle responses using \herwig would bring $F_{{\rm corr},\,\hbox{\scriptsize{\aherwig}}}^b$ into better agreement with $\gamma+{\rm jet}$ and dijet data, which $F_{{\rm corr},\,\hbox{\scriptsize{\apythia}}}^b$ is also tuned to, resulting in a $Q_{\rm corr}^b$ that is closer to unity. Given these findings, we conclude that the uncertainty on \mt from flavor dependence of the calorimeter response to jets of $\mp0.16~\GeV$ covers the effect from different parton shower and hadronisation models in \pythia and \herwig.  
Taking our upper limit estimate using Eq.~(\ref{eq:qcorr}) at face value would increase the systematic uncertainty on \mt from flavor dependence of the calorimeter response to jets to $\mp0.24~\GeV$, and the total uncertainty of the measurement to 0.78~\GeV. 
\item
We perform another cross check of the energy scale for $b$-quark jets using a similar approach to Ref.~\cite{bib:atlas3D}, by studying the ratio of transverse momenta of $b$-tagged jets to jets without a $b$~tag. For events with $\geq2$ $b$-tagged jets, this ratio is defined as
\begin{equation}
\label{eqn:bjesScale}
R_{bq}=\frac{\pt^{b_1}+\pt^{b_2}}{\pt^{q_1}+\pt^{q_2}}\,,
\end{equation}
where $b_i$ refers to the $b$-tagged jets with the highest $\pt$, and similarly $q_i$ for the other two jets. In events with one $b$-tagged jet, the ratio
\begin{equation}
\label{eqn:bjesScale2}
R_{bq}=\frac{2p_T^{b}}{p_T^{q_1}+p_T^{q_2}}
\end{equation}
is computed. The distribution of $R_{bq}$ after all selections is shown in Fig.~\ref{fig:rbl} for $\mt=175~\GeV$ and $\sigma_{\ttbar}=7.7~$pb, as provided by the ME technique. Good agreement between data and simulations is observed. We extract the overall b quark JES $k_{\rm bJES}$ by performing a template fit to the distribution in $R_{bq}$, and obtain
\begin{equation}
k_{\rm bJES}=1.008 \pm 0.0195\thinspace({\rm stat})\,^{+0.037} _{-0.031}\thinspace({\rm syst})\,.
\end{equation}
The systematic uncertainties include dominant sources like JES, jet energy resolution, jet identification, flavor dependence of the detector response, and $b$~tagging, which provides the main contribution. The measured $k_{\rm bJES}$ value is consistent with unity, and underlines the correctness of $F_{\rm corr}$ from Eq.~\ref{eq:fcorr}.

\end{enumerate}

%
\subsubsection{Tagging of $b$~jets}

Discrepancies in the $b$~tagging efficiency between data and MC simulations can lead to a systematic shift in the extracted \mt. To evaluate the effect of possible discrepancies, the standard corrections to $b$~tagging rates of $u,d,s$ quark jets and gluons, $c$ quark jets, and $b$-quark jets, which depend on $\eta$ and $\pt$, are changed within their uncertainties. We evaluate the uncertainty from those corrections by simultaneously decreasing the $b$~tagging efficiencies for $b$ and $c$ quark jets within their uncertainties, while increasing the efficiencies for jets from all other quarks, and find $\mp0.10~\GeV$ after symmetrizing their impact on \mt.

\subsubsection{Trigger}

To evaluate the impact of the trigger on our analysis, we apply trigger weights that simulate the impact of the lepton plus jet trigger differentially in $\pt,~\eta$, and $\phi$ of the lepton. An alternative calibration with those trigger weights applied is obtained, resulting in  a change in the extracted \mt of $\pm 0.01~\GeV$.

\subsubsection{Lepton momentum scale}

The momentum scale of electrons and muons is known with some finite precision. To evaluate the impact of this effect, we change the momentum scale of a given lepton flavor, while keeping the momentum scale of the other flavor constant, and combine in quadrature the effects found for the two flavors. To evaluate the impact for a given lepton flavor, we assume the worst case, where the momentum scale is shifted in the same direction for all data-taking epochs. We perform this for the up and down changes, and find an effect that is consistent with zero within uncertainties for electrons, and corresponding to 0.01~GeV for muons. Combining the two values in quadrature yields 0.01~GeV, which we quote for this source of systematic uncertainty.

\subsubsection{Jet energy resolution}

An additional smearing of jet energies is applied to all MC samples as part of the JES calibration in order to achieve better agreement of MC simulations and data~\cite{bib:jes}. To evaluate the possible effect of disagreement between data and MC simulations in jet energy resolutions on $\mt$, we produce simulated MC samples where the jet energy resolution correction is changed up and down by its uncertainty. Event probabilities are then re-calculated and ensemble tests are repeated. We find an effect of $\pm0.07~\GeV$ on \mt from jet energy resolution, after symmetrizing the excursions.

\subsubsection{Jet identification efficiency}

The efficiency for the identification of jets is slightly higher in MC simulation than in data, and is therefore corrected as part of the JES calibration~\cite{bib:jes} to achieve better agreement between MC simulations and data in jet identification efficiencies. To evaluate the potential impact from this source on $\mt$, the jet identification efficiencies in the signal MC sample are decreased according to their uncertainties. Subsequently, event probabilities are recalculated and ensemble tests are repeated. The extracted value of $\mt$ using this alternative calibration differs from the central value by $-0.01~\GeV$, which is quoted as systematic uncertainty.

\subsection{Method}

%
\subsubsection{Modeling of multijet events}

Since we require four jets, and at least one of them $b$-tagged, the contribution of the MJ background to our sample of \ttbar candidate events is small (cf.\ Table~\ref{tab:prefit}). Nonetheless, we explicitly account for the next-to-dominant contribution from this background when constructing the PEs for the calibration of the ME technique, as described in Sections~\ref{sec:calibf} and~\ref{sec:calibmt}. This was not done in the previous measurement using 3.6~\fb of integrated luminosity~\cite{bib:mt36}. To evaluate a systematic uncertainty from the modeling of MJ background, we  do not include it in the ensemble tests for an alternative calibration in \mt, which yields a change in \mt of $+0.04~\GeV$.

%
\subsubsection{Signal fraction}

We apply the $f$ values measured in each data-taking epoch and final state, as summarized in Table~\ref{tab:postfit}, to construct the PEs used for calibrating the response of the ME technique in \mt and \kjes. To evaluate the systematic uncertainty from $f$, we use alternative calibrations in \mt and \kjes, where $f$ is changed by $\pm$5\%, simultaneously for all data-taking epochs and final states. The 5\% value is motivated by the magnitude of the systematic uncertainty on the measured \ttbar production cross section in the D0 data~\cite{bib:xsec54}, ignoring the uncertainty from integrated luminosity. We obtain $\pm0.08~\GeV$ as an effect from this source of systematic uncertainty.

%
\subsubsection{MC calibration}

The statistical uncertainties due to the limited size of MC samples that are used to construct PEs to study the response of the ME technique, as discussed in Sections~\ref{sec:calibf} and~\ref{sec:calibmt}, contribute an  uncertainty on the final calibrated value of $\mt$ that is statistical in nature. To determine this contribution, we note the uncertainties on the offset and slope parameters of the linear calibrations in Figs.~\ref{fig:mtem} and~\ref{fig:mtmu}, and propagate them, assuming Gaussian uncorrelated uncertainties, to the extracted \mt values in each of the data-taking epochs and final states. Combining the individual \mt results for the up and down changes and symmetrizing, we find an uncertainty of $\pm0.07~\GeV$ due to the limited size of MC samples.

\subsection{Summary of uncertainties}

Quadratically combining contributions from all sources of systematic uncertainties from Table~\ref{tab:syst}, we obtain a total systematic uncertainty of 0.49~\GeV. The total uncertainty of the measurement of 0.76~\GeV is obtained by combining the total systematic uncertainty and the statistical uncertainty of 0.58~\GeV. The statistical uncertainty consists of two parts: 0.41~\GeV from \mt alone and 0.41~\GeV from \kjes. Considering the latter as a source of systematic uncertainty, as is done for the Tevatron combination~\cite{bib:combitev}, the total systematic uncertainty including the statistical contribution from the {\em in situ} constraint on the JES is 0.64~\GeV.

\section{Comparison with previous measurements}
\label{sec:comparison}

\subsection{Comparison with the Tevatron average}

\begin{figure}
\begin{centering}
\includegraphics[width=0.99\columnwidth]{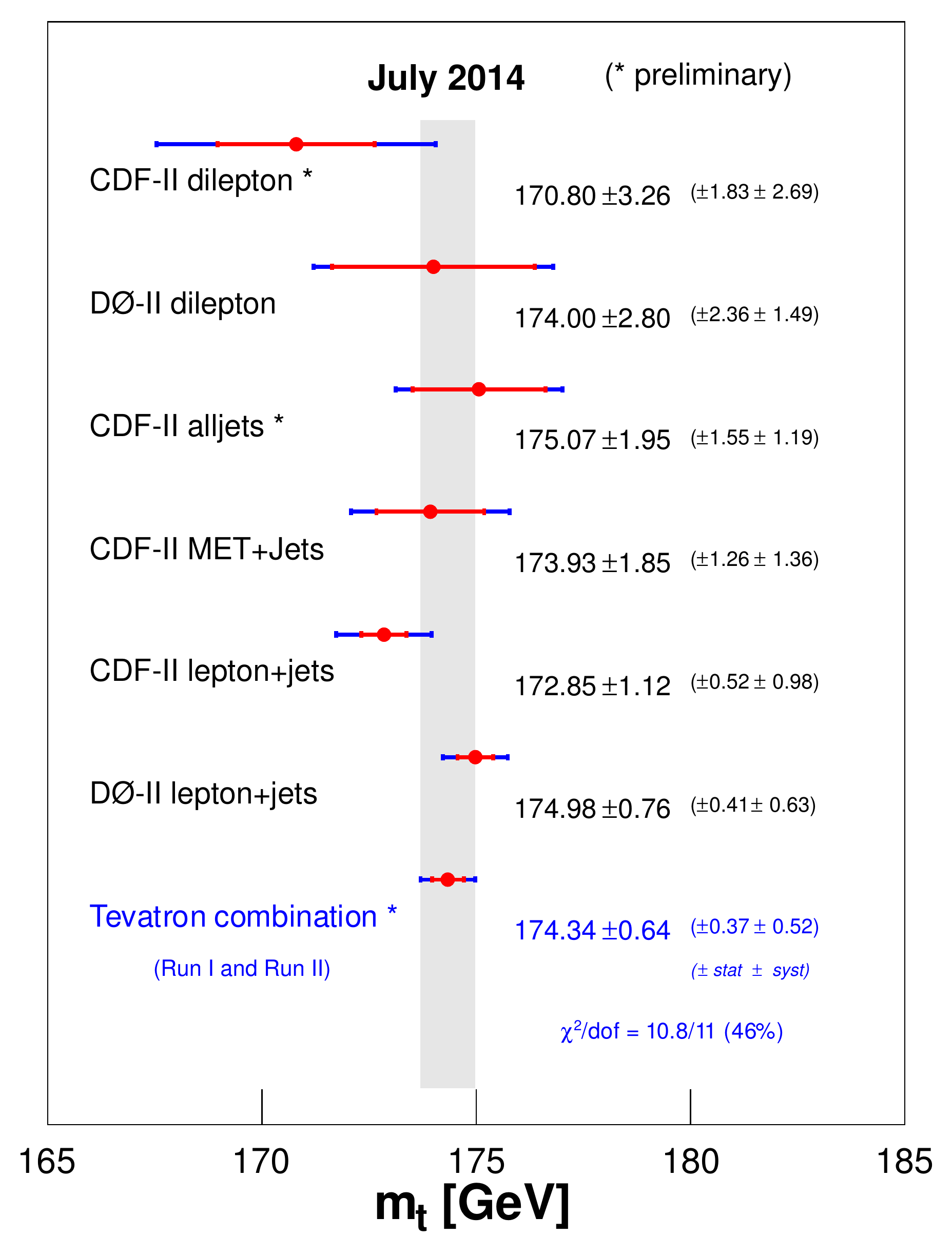}
\par\end{centering}
\caption{
\label{fig:combi}
Summary of the measurements performed in Run~II of the Tevatron which are used as inputs to the Tevatron combination~\cite{bib:combitev}. The inner error bars indicate the statistical uncertainties, while the outer bars represent the total uncertainties. The Tevatron average value of \mt obtained using input measurements from Run~I and Run~II is given at the bottom and its uncertainty is shown by the band. The Figure is adapted from Ref.~\cite{bib:combitev}.
}
\end{figure}

Our result is included in the Tevatron combination from July 2014~\cite{bib:combitev}, which takes into account 10 published and 2 preliminary results from the CDF and D0 collaborations using $p\bar p$ collision data from Run~I and Run~II of the Fermilab Tevatron Collider. Considering correlations between sources of systematic uncertainty as described in detail in Ref.~\cite{bib:combitevprd}, the final result is $\mt=174.34\pm0.64~\GeV$, with a relative weight of 67\% from the D0 measurement. An overview of the input measurements performed using Run~II data is presented in Fig.~\ref{fig:combi}. 
All measurements from Run~I and Run~II are consistent with $\chi^2=10.8$ for 11 degrees of freedom, which corresponds to a $\chi^2$ probability of 46\%.

\subsection{Comparison with the world average}

Our result can be compared with the current world average of $\mt=173.34\pm0.76~\GeV$~\cite{bib:combiworld}, which encompasses 11 measurements from the ATLAS, CDF, CMS, and D0 Collaborations, excluding this measurement.
We do this comparison considering the full uncertainty in the world combination and only the statistical uncertainty in our measurement, which provides a reasonable approximation, given the large correlation among the experiments for most sources of systematic uncertainty. Furthermore, we use the simplifying assumption that there is no correlation between the statistical uncertainty of our new measurement and of the world average. We find consistency at the 1.7~SD level. 
Due to the complicated correlation of systematic uncertainties, a more detailed comparison should be performed in a separate document, with participation from all the collaborations supplying measured \mt values as inputs, and include all updates since the publication of Ref.~\cite{bib:combiworld}.

\section{Summary}
\label{sec:summary}

In summary, we have performed a measurement of the mass of the top quark using the matrix element technique in $\ttbar$ candidate events in lepton$+$jets final states using 9.7~\fb of Run~II integrated luminosity collected by the D0 detector at the Fermilab Tevatron $p\bar p$ Collider. The result, 
\begin{eqnarray*}
\mt &=& 174.98 \pm 0.58\thinspace({\rm stat+JES}) \pm 0.49\thinspace({\rm syst})~\GeV\,,~{\rm or}\\ 
\mt &=& 174.98 \pm 0.76~\GeV\,, 
\end{eqnarray*}
is consistent with the values given by the current Tevatron and world combinations of the top-quark mass~\cite{bib:combitev,bib:combiworld} and achieves by itself a similar precision. With an uncertainty of $0.43\%$, it constitutes the most precise single measurement of the top-quark mass, and is $\approx70\%$ more precise than the next-to-most precise single measurement~\cite{bib:mtcms}. The total systematic uncertainty of our result is smaller than that of any other single measurement.

\section*{Acknowledgements}
%

We thank the staffs at Fermilab and collaborating institutions,
and acknowledge support from the
Department of Energy and National Science Foundation (United States of America);
Alternative Energies and Atomic Energy Commission and
National Center for Scientific Research/National Institute of Nuclear and Particle Physics  (France);
Ministry of Education and Science of the Russian Federation, 
National Research Center ``Kurchatov Institute" of the Russian Federation, and 
Russian Foundation for Basic Research  (Russia);
National Council for the Development of Science and Technology and
Carlos Chagas Filho Foundation for the Support of Research in the State of Rio de Janeiro (Brazil);
Department of Atomic Energy and Department of Science and Technology (India);
Administrative Department of Science, Technology and Innovation (Colombia);
National Council of Science and Technology (Mexico);
National Research Foundation of Korea (Korea);
Foundation for Fundamental Research on Matter (The Netherlands);
Science and Technology Facilities Council and The Royal Society (United Kingdom);
Ministry of Education, Youth and Sports (Czech Republic);
Bundesministerium f\"{u}r Bildung und Forschung (Federal Ministry of Education and Research) and 
Deutsche Forschungsgemeinschaft (German Research Foundation) (Germany);
Science Foundation Ireland (Ireland);
Swedish Research Council (Sweden);
China Academy of Sciences and National Natural Science Foundation of China (China);
and
Ministry of Education and Science of Ukraine (Ukraine).
%

\bibliography{mt_10fb_dzero_ljets}

\begin{thebibliography}{75}
\bibitem[{bib({\natexlab{b}})}]{bib:coor}
\bibinfo{note}{The D0 coordinate system is right-handed, with the $\rm z$~axis
  pointing in the direction of the Tevatron proton beam and the $\rm y$~axis
  pointing upwards. The angles $\phi$ and $\theta$ are the azimuthal and polar
  angles relative to the $\rm x$ and $\rm z$~axes, respectively. We also use as
  an angular variable the pseudorapidity defined by
  $\eta=-\ln[\tan(\theta/2)]$.}
\bibitem[{bib({\natexlab{c}})}]{bib:coor2}
\bibinfo{note}{Note the difference between $x$ and $y$ which denote detector
  and parton level quantities, while $\rm x$, $\rm y$, and $\rm z$ represent
  the geometrical axes defined in Ref.~\cite{bib:coor}.}
\bibitem[{bib({\natexlab{d}})}]{bib:cpuhours}
\bibinfo{note}{Throughout this document, we cite CPU hours for a single core of
  the XEON E5-2620 CPU with a clock rate of 2~GHz and a 64 bit computation.}
\end{thebibliography}

\clearpage

\begin{appendix}

\section{Constraining the amount of radiation using Drell-Yan events}
\label{app:isr}
Because of the limited statistics in \ttbar candidate events recorded by D0 in Run II, it is difficult to use \ttbar\ events directly to study ISR and FSR. We therefore take an alternative approach using the $Z/\gamma^*\to\ell^+\ell^-$ process to experimentally constrain ISR and FSR. In the following, we motivate this approach, briefly review the measurement of $Z/\gamma^*\to\ell^+\ell^-$ production~\cite{bib:isr} that we use to obtain the dependence of ISR and FSR, and present our results.

\subsection{Motivation}
At the Tevatron, \ttbar pairs are dominantly produced via $q\bar q$ annihilation. The ISR will stem therefore dominantly from the process where an initial-state quark radiates a gluon, which can be described by the $P_{q\to qg}$ splitting function. In the LO picture of the \ttbar\ process, the final-state partons are also quarks, and FSR is therefore described by the same $P_{q\to qg}$ splitting function. At the same time, the production of $Z$ bosons at the Tevatron is also dominated by $q\bar q$ annihilation, and ISR in $Z/\gamma^*\to\ell^+\ell^-$ can be also described by the $P_{q\to qg}$ splitting function. A measurement of ISR in $Z/\gamma^*\to\ell^+\ell^-$ events can therefore be used to constrain ISR and FSR in \ttbar\ events. 

The experimental approach outlined above was first proposed in Ref.~\cite{bib:isrcdf}. Beyond Ref.~\cite{bib:isrcdf}, we use the $\phi^*$ variable~\cite{bib:isr}, which is more sensitive to soft ISR. Since this measurement is corrected for detector effects, it can be used to compare directly generated $Z/\gamma^*\to\ell^+\ell^-$ events with data.

\subsection{The measurement of the $Z/\gamma^*\to\ell^+\ell^-$ cross section}

The $\phi^*$ observable used to establish sensitivity to ISR in Ref.~\cite{bib:isr} is defined as
\begin{equation}
\phi^* = \tan\left(\phi_{\rm acop}/2\right)\sin(\theta^*_{\eta})\,,
\end{equation}
where $\phi_{\rm acop}$ is the acoplanarity angle, given by $\phi_{\rm acop} = \pi - \Delta\phi^{\ell^+\ell^-}$, and $\Delta\phi^{\ell^+\ell^-}$ is the difference in azimuthal angle between the two lepton candidates. The variable $\theta^*_{\eta}$ is a measure of the scattering angle of the leptons relative to the proton beam direction in the rest frame of the dilepton system, and is defined as
\begin{equation}
\cos(\theta^{*}_{\eta})=\tanh\left(\frac{\eta^- - \eta^+}2\right)\,,
\end{equation}
where $\eta^-$ and $\eta^+$ are the laboratory pseudorapidities of the negatively
and positively charged lepton, respectively.

The measurement of $\dif\sigma_{Z/\gamma^*\to\ell^+\ell^-}/\dif\phi^*$ is corrected for detector effects, and performed in the fiducial region defined by the presence of the two oppositely charged leptons, as summarized in Table~\ref{tab:fiducial}. The electron four-momentum is defined as the sum of the four-momenta of all electrons and photons in the $\Delta R < 0.2$ cone around the detector-level electron. No cone summation is applied to muons. The measurement is performed in three bins of rapidity of the boson of $0<|y|<1$,  $1<|y|<2$, and  $|y|>2$ in $e^+e^-$, and in two bins of rapidity, $0<|y|<1$ and $1<|y|<2$ in $\mu^+\mu^-$ final states. This ensures that  $P_{q\to qg}$ is probed in different kinematic regimes of the square of the four-momentum transfer $Q^2$.

\begin{table}
\caption{\label{tab:fiducial}
Summary of kinematic requirements used to define the measurement of $\dif\sigma_{Z/\gamma^*\to\ell^+\ell^-}/\dif\phi^*$ in the fiducial region of Ref.~\cite{bib:isr}. 
}
\begin{ruledtabular}
\begin{tabular}{lll}
\multirow{2}{*}{Charged leptons:}     & $\pt > 20~\GeV$ & $|\eta| < 1.1$ or $1.5 < |\eta| < 3.0$ ($e$)\\
           & $\pt > 15~\GeV$ & $|\eta| < 2.0$ ($\mu$)\\
${\ell^+\ell^-}$ system:   & \multicolumn{2}{l}{$60~\GeV < m_{\ell^+\ell^-} < 120~\GeV$}
\end{tabular}
\end{ruledtabular}
\end{table}

\subsection{Results}

\begin{figure}
\centering
\begin{overpic}[width=0.79\columnwidth]{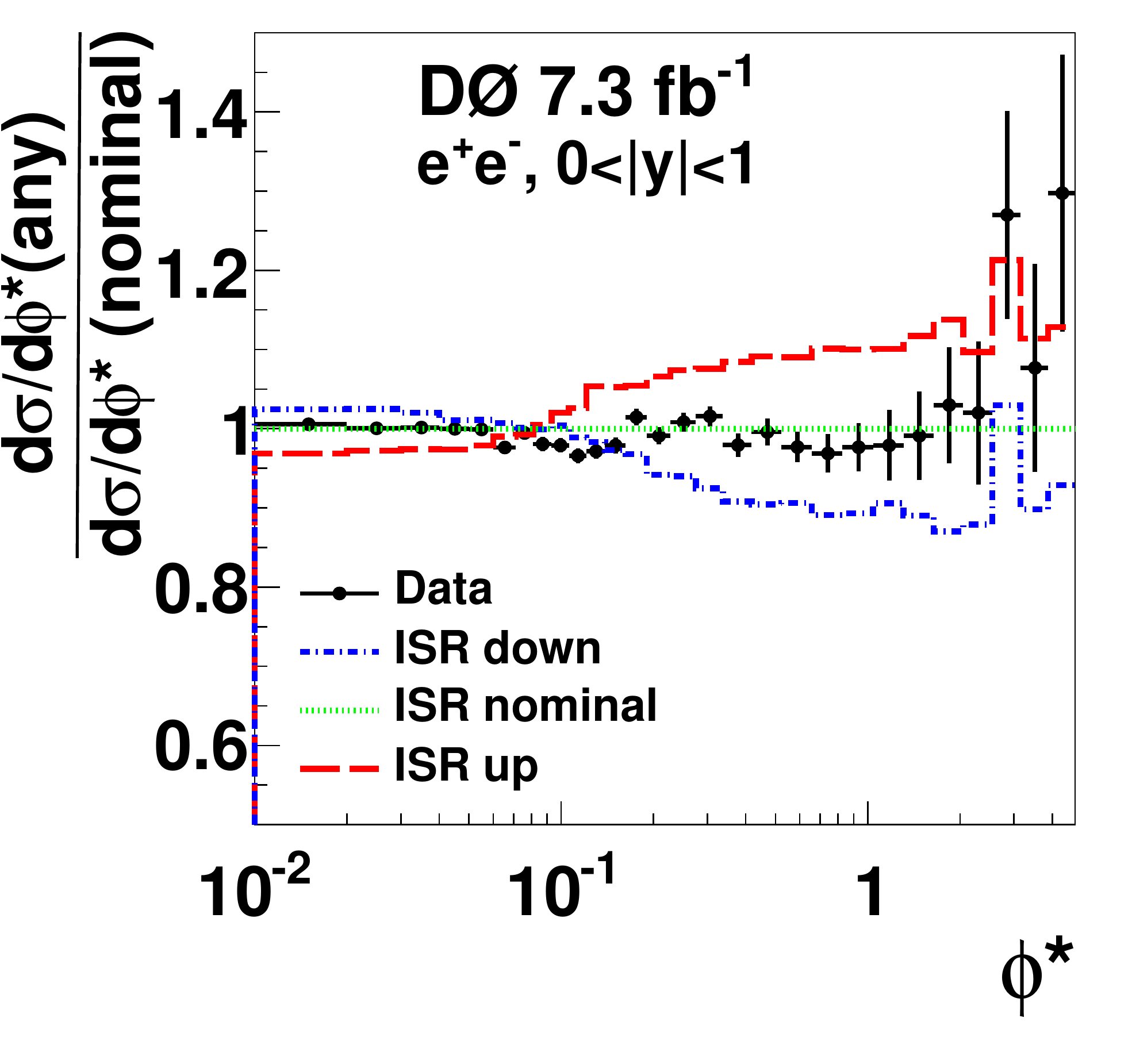}
\put(26,82){\footnotesize\textsf{\textbf{(a)}}}
\end{overpic}
\begin{overpic}[width=0.79\columnwidth]{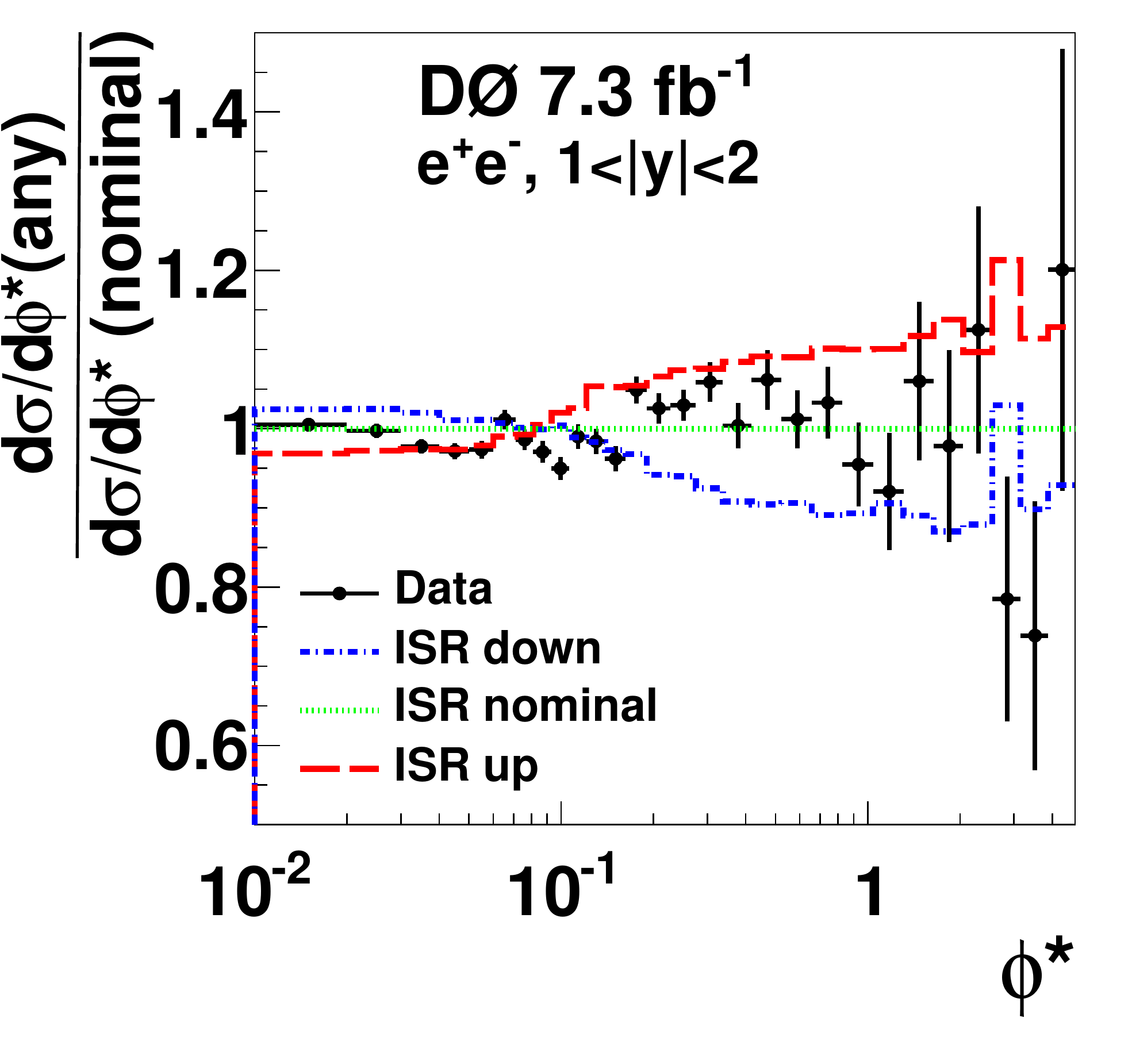}
\put(26,82){\footnotesize\textsf{\textbf{(b)}}}
\end{overpic}
\begin{overpic}[width=0.79\columnwidth]{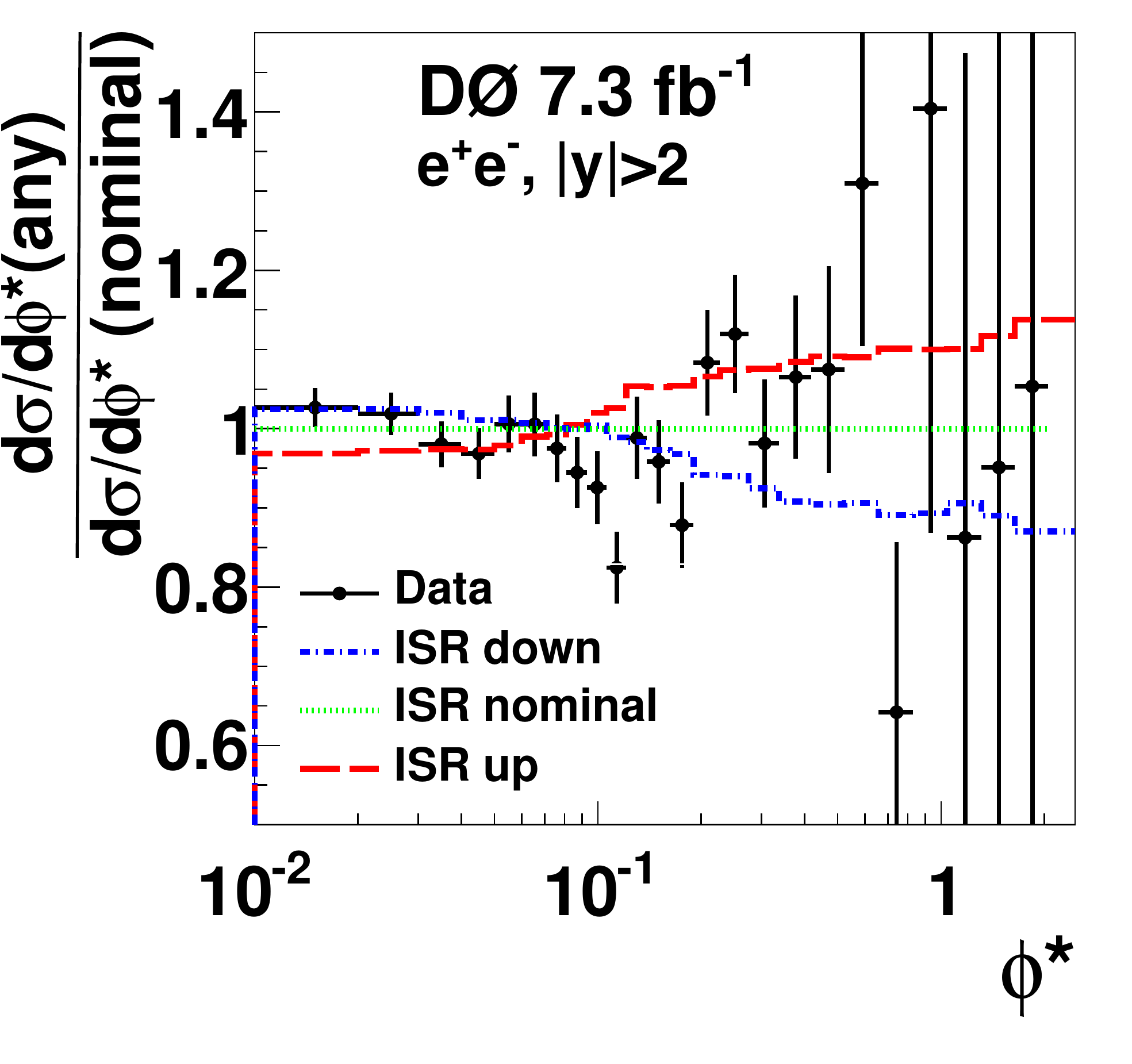}
\put(26,82){\footnotesize\textsf{\textbf{(c)}}}
\end{overpic}
\caption{\label{fig:isree}
(a) Ratio of the unfolded differential cross section $\dif\sigma_{Z/\gamma^*\to e^+e^-}/\dif\phi^*$ and the nominal predictions from \apythia for $0<|y|<1$, where the error bars represent the total uncertainties in the data. The ratios of predictions determined with \apythia where the amount of radiation is changed up and down relative the nominal predictions from \apythia are shown, respectively, as red broken  and blue dash-dotted lines. The statistical uncertainties from the finite number of simulated MC events are negligible compared to the uncertainties in the data.
(b)~Same as (a), but for $1<|y|<2$.
(c)~Same as (a), but for $|y|>2$.
}
\end{figure}

\begin{figure}[h]
\centering
\begin{overpic}[width=0.79\columnwidth]{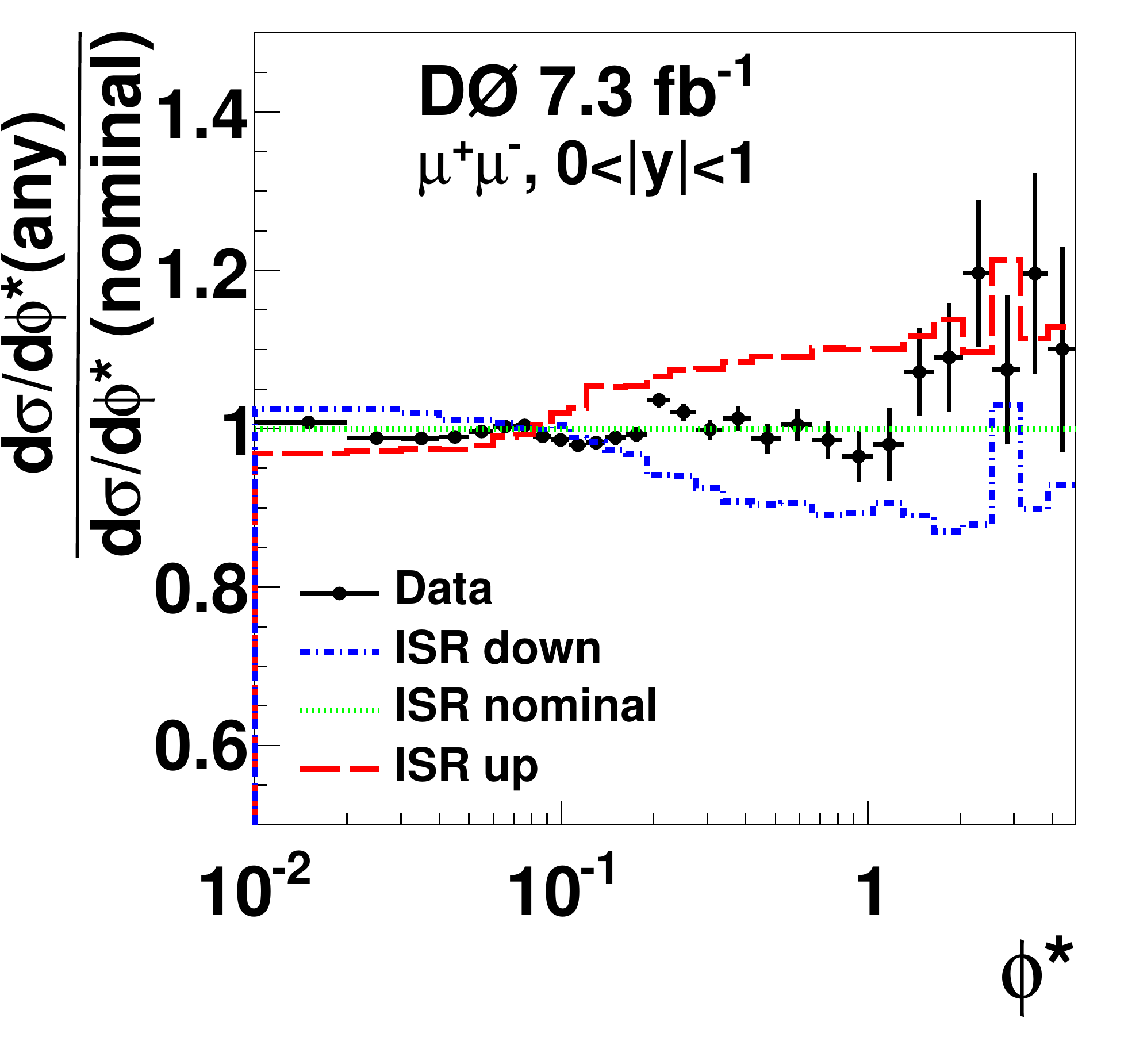}
\put(26,82){\footnotesize\textsf{\textbf{(a)}}}
\end{overpic}
\begin{overpic}[width=0.79\columnwidth]{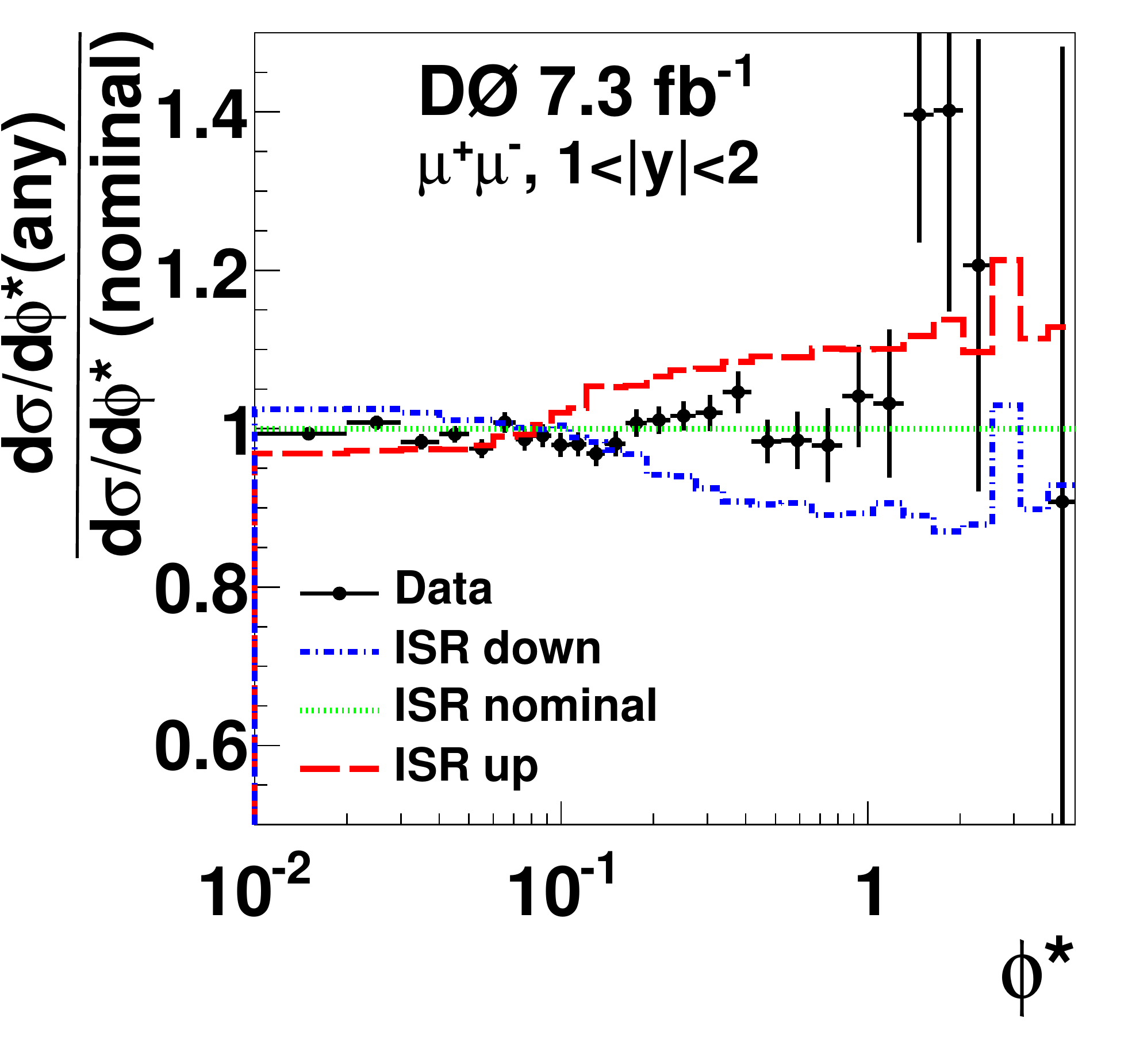}
\put(26,82){\footnotesize\textsf{\textbf{(b)}}}
\end{overpic}
\caption{\label{fig:isrmumu}
(a) Ratio of the unfolded differential cross section $\dif\sigma_{Z/\gamma^*\to \mu^+\mu^-}/\dif\phi^*$ and the nominal predictions from \apythia for $0<|y|<1$, where the error bars represent the total uncertainties in the data. The ratios of predictions obtained with \apythia where the amount of radiation is changed up and down relative the nominal predictions from \apythia are shown, respectively, as red broken  and blue dash-dotted lines. The statistical uncertainties from the finite number of simulated MC events are negligible compared to the uncertainties in the data.
(b)~Same as (a), but for $1<|y|<2$.
}
\end{figure}

As described in Sec.~\ref{sec:mc}, \apythia is used to generate \ttbar MC events, where Feynman diagrams accounting for hard radiation from the initial or final state partons up to second-order in $\alpha_s$ are introduced through the hard ME in \alpgen, i.e., in bins of 0, 1, and 2 additional light partons, which are then combined according to their cross sections. For consistency, we use an identical setup for generating $Z\to\ell^+\ell^-$ events. The ratio of unfolded  $\dif\sigma_{Z/\gamma^*\to\ell^+\ell^-}/\dif\phi^*$ data to the predictions from \apythia at particle level, is shown in Fig.~\ref{fig:isree} for the $e^+e^-$ and in Fig.~\ref{fig:isrmumu} for the $\mu^+\mu^-$ final states. The \apythia model is not able to describe the data over the entire spectrum in $\phi^*$, but the overall agreement is reasonable. In the analysis, we identify up and down changes in the radiation rate that cover the excursions of nominal MC predictions from the data.

Modifying parameters such as $\Lambda_{\rm QCD}$ or $\alpha_s$ adjusting the rate of ISR/FSR in the PS, as is done in stand-alone \pythia, leads to unphysical results for \apythia, since the MLM matching of the PS to the hard ME calculation partly compensates for such effects in the PS~\cite{bib:isrmangano}. Thus, the preferred approach is to change the amount of ISR/FSR through  rescaling of the renormalization scale in the CKKW scale-setting procedure {\tt ktfac} from its default of unity, as suggested in Ref.~\cite{bib:isrmangano}. Our studies indicate that changing the {\tt ktfac} parameter by a factor of $\pm$1.5 provide coverage for the excursions of the nominal MC predictions from the data. This is demonstrated in Fig.~\ref{fig:isree} for the $e^+e^-$ and in Fig.~\ref{fig:isrmumu} for the $\mu^+\mu^-$ final states. Based on these studies, we evaluate the systematic uncertainty from ISR/FSR in \ttbar events by setting the {\tt ktfac} parameter to $2/3$ for the up and $3/2$ for the down point.

\clearpage

%

\onecolumngrid
\section{Transfer function parameters}
\label{app:tf}
In this Appendix we summarize the jet transfer function parameters $a_i$ and $b_i$ defined in Eq.~(\ref{eq:aibi}) as a function of data-taking epoch, jet flavor, and pseudorapidity bin. Jet flavor is denoted as ``light jets'' for $u,d,c,s$ quark jets, ``$b_\mu$ jets'' for $b$-quark jets with a soft muon tag, and ``$b$ jets'' for all other $b$-quark jets. The parameter $a_3$ is fixed to 0 in order to improve the convergence of the double-Gaussian fit function. The parameter values are given in units of GeV~($a_1$), 1 ($b_1$), GeV~($a_2$), 1 ($b_2$), 1~($a_3$), GeV$^{-1}$~($b_3$), GeV~($a_4$), 1 ($b_4$), GeV~($a_5$), and 1 ($b_5$).


\begin{table}[h]
\caption{
\label{tab:aibiIIa}
Jet TF parameters $a_i$ and $b_i$ for Run IIa and $|\eta|\leq0.4$. 
}
\centering
\begin{ruledtabular}
\begin{tabular}{lcccccc}
 & \multicolumn{2}{c}{Light jets} & \multicolumn{2}{c}{$b$ jets} & \multicolumn{2}{c}{$b_\mu$ jets} \\
$i$~~ & $a_{i}$ & $b_{i}$ & $a_{i}$ & $b_{i}$ & $a_{i}$ & $b_{i}$ \\
\hline
1 & $-1.29\times10^{+0}$ & $\phantom{+}2.21\times10^{-2}$ & $\phantom{+}2.92\times10^{+0}$ & $-1.91\times10^{-1}$ & $\phantom{+}6.06\times10^{+0}$ & $-1.40\times10^{-1}$ \\
2 & $\phantom{+}4.04\times10^{+0}$ & $\phantom{+}1.09\times10^{-1}$ & $\phantom{+}2.77\times10^{+0}$ & $\phantom{+}2.07\times10^{-1}$ & $\phantom{+}1.67\times10^{+0}$ & $\phantom{+}1.65\times10^{-1}$ \\
3 & $0$ & $\phantom{+}2.35\times10^{-4}$ & $0$ & $\phantom{+}4.09\times10^{-2}$ & $0$ & $\phantom{+}8.68\times10^{-5}$ \\
4 & $\phantom{+}2.37\times10^{+1}$ & $-1.70\times10^{-1}$ & $-9.96\times10^{+0}$ & $\phantom{+}4.68\times10^{-2}$ & $\phantom{+}4.60\times10^{+1}$ & $-4.15\times10^{-1}$ \\
5 & $\phantom{+}1.89\times10^{+1}$ & $\phantom{+}9.64\times10^{-2}$ & $\phantom{+}3.82\times10^{+0}$ & $\phantom{+}9.23\times10^{-2}$ & $\phantom{+}1.83\times10^{+1}$ & $\phantom{+}1.44\times10^{-1}$
\end{tabular}
\end{ruledtabular}
\end{table}

\begin{table}[h]
\caption{
Jet TF parameters $a_i$ and $b_i$ for Run IIa and $0.4<|\eta|\leq0.8$. 
}
\centering
\begin{ruledtabular}
\begin{tabular}{lcccccc}
 & \multicolumn{2}{c}{Light jets} & \multicolumn{2}{c}{$b$ jets} & \multicolumn{2}{c}{$b_\mu$ jets} \\
$i$~~ & $a_{i}$ & $b_{i}$ & $a_{i}$ & $b_{i}$ & $a_{i}$ & $b_{i}$ \\
\hline
1 & $-6.87\times10^{-1}$ & $\phantom{+}8.60\times10^{-4}$ & $\phantom{+}4.34\times10^{+0}$ & $-2.11\times10^{-1}$ & $\phantom{+}6.88\times10^{+0}$ & $-1.56\times10^{-1}$ \\
2 & $\phantom{+}3.64\times10^{+0}$ & $\phantom{+}1.19\times10^{-1}$ & $\phantom{+}3.13\times10^{+0}$ & $\phantom{+}2.03\times10^{-1}$ & $\phantom{+}1.71\times10^{+0}$ & $\phantom{+}1.65\times10^{-1}$ \\
3 & $0$ & $\phantom{+}2.73\times10^{-4}$ & $0$ & $\phantom{+}3.40\times10^{-2}$ & $0$ & $\phantom{+}9.12\times10^{-5}$ \\
4 & $\phantom{+}2.48\times10^{+1}$ & $-1.73\times10^{-1}$ & $-8.77\times10^{+0}$ & $\phantom{+}2.43\times10^{-2}$ & $\phantom{+}4.71\times10^{+1}$ & $-3.41\times10^{-1}$ \\
5 & $\phantom{+}1.89\times10^{+1}$ & $\phantom{+}9.84\times10^{-2}$ & $\phantom{+}3.45\times10^{+0}$ & $\phantom{+}9.94\times10^{-2}$ & $\phantom{+}2.09\times10^{+1}$ & $\phantom{+}1.07\times10^{-1}$
\end{tabular}
\end{ruledtabular}
\end{table}

\begin{table}[h]
\caption{
Jet TF parameters $a_i$ and $b_i$ for Run IIa and $0.8<|\eta|\leq1.6$. 
}
\centering
\begin{ruledtabular}
\begin{tabular}{lcccccc}
 & \multicolumn{2}{c}{Light jets} & \multicolumn{2}{c}{$b$ jets} & \multicolumn{2}{c}{$b_\mu$ jets} \\
$i$~~ & $a_{i}$ & $b_{i}$ & $a_{i}$ & $b_{i}$ & $a_{i}$ & $b_{i}$ \\
\hline
1 & $\phantom{+}4.53\times10^{+0}$ & $-7.58\times10^{-2}$ & $\phantom{+}1.29\times10^{+1}$ & $-1.58\times10^{-1}$ & $\phantom{+}6.00\times10^{+0}$ & $-1.08\times10^{-1}$ \\
2 & $\phantom{+}3.32\times10^{+0}$ & $\phantom{+}1.55\times10^{-1}$ & $\phantom{+}5.24\times10^{+0}$ & $\phantom{+}2.40\times10^{-1}$ & $\phantom{+}6.04\times10^{+0}$ & $\phantom{+}1.30\times10^{-1}$ \\
3 & $0$ & $\phantom{+}4.28\times10^{-3}$ & $0$ & $\phantom{+}1.76\times10^{-1}$ & $0$ & $\phantom{+}1.00\times10^{-3}$ \\
4 & $\phantom{+}9.02\times10^{+0}$ & $\phantom{+}7.90\times10^{-2}$ & $-5.98\times10^{+0}$ & $-6.28\times10^{-2}$ & $\phantom{+}6.00\times10^{+1}$ & $-7.50\times10^{-1}$ \\
5 & $\phantom{+}1.30\times10^{+1}$ & $\phantom{+}8.39\times10^{-2}$ & $\phantom{+}1.63\times10^{+0}$ & $\phantom{+}1.82\times10^{-1}$ & $\phantom{+}1.00\times10^{+1}$ & $\phantom{+}2.00\times10^{-1}$
\end{tabular}
\end{ruledtabular}
\end{table}

\begin{table}[h]
\caption{
Jet TF parameters $a_i$ and $b_i$ for Run IIa and $1.6<|\eta|\leq2.5$. 
}
\centering
\begin{ruledtabular}
\begin{tabular}{lcccccc}
 & \multicolumn{2}{c}{Light jets} & \multicolumn{2}{c}{$b$ jets} & \multicolumn{2}{c}{$b_\mu$ jets} \\
$i$~~ & $a_{i}$ & $b_{i}$ & $a_{i}$ & $b_{i}$ & $a_{i}$ & $b_{i}$ \\
\hline
1 & $\phantom{+}1.62\times10^{+1}$ & $-2.28\times10^{-1}$ & $\phantom{+}1.07\times10^{+1}$ & $-3.07\times10^{-1}$ & $\phantom{+}3.28\times10^{+1}$ & $-4.28\times10^{-1}$ \\
2 & $\phantom{+}2.78\times10^{+0}$ & $\phantom{+}1.46\times10^{-1}$ & $\phantom{+}3.61\times10^{+0}$ & $\phantom{+}1.23\times10^{-1}$ & $\phantom{+}8.34\times10^{+0}$ & $\phantom{+}7.76\times10^{-2}$ \\
3 & $0$ & $\phantom{+}6.15\times10^{-3}$ & $0$ & $\phantom{+}6.98\times10^{-3}$ & $0$ & $\phantom{+}7.06\times10^{-3}$ \\
4 & $\phantom{+}1.86\times10^{+1}$ & $-1.30\times10^{-3}$ & $\phantom{+}5.35\times10^{+0}$ & $-4.33\times10^{-2}$ & $\phantom{+}2.07\times10^{+1}$ & $-1.31\times10^{-1}$ \\
5 & $\phantom{+}1.47\times10^{+1}$ & $\phantom{+}8.60\times10^{-2}$ & $\phantom{+}1.38\times10^{+1}$ & $\phantom{+}9.66\times10^{-2}$ & $\phantom{+}1.18\times10^{+1}$ & $\phantom{+}1.08\times10^{-1}$
\end{tabular}
\end{ruledtabular}
\end{table}



\begin{table}[h]
\caption{
\label{tab:aibiIIb1}
Jet TF parameters $a_i$ and $b_i$ for Run IIb1 and $|\eta|\leq0.4$. 
}
\centering
\begin{ruledtabular}
\begin{tabular}{lcccccc}
 & \multicolumn{2}{c}{Light jets} & \multicolumn{2}{c}{$b$ jets} & \multicolumn{2}{c}{$b_\mu$ jets} \\
$i$~~ & $a_{i}$ & $b_{i}$ & $a_{i}$ & $b_{i}$ & $a_{i}$ & $b_{i}$ \\
\hline
1 & $-2.17\times10^{+0}$ & $\phantom{+}2.39\times10^{-2}$ & $\phantom{+}8.74\times10^{-1}$ & $-1.86\times10^{-1}$ & $\phantom{+}5.26\times10^{+0}$ & $-1.29\times10^{-1}$ \\
2 & $\phantom{+}4.08\times10^{+0}$ & $\phantom{+}6.66\times10^{-2}$ & $\phantom{+}2.67\times10^{+0}$ & $\phantom{+}1.82\times10^{-1}$ & $\phantom{+}1.45\times10^{+0}$ & $\phantom{+}1.51\times10^{-1}$ \\
3 & $0$ & $\phantom{+}2.77\times10^{-4}$ & $0$ & $\phantom{+}4.41\times10^{-2}$ & $0$ & $\phantom{+}2.13\times10^{-4}$ \\
4 & $\phantom{+}1.27\times10^{+1}$ & $-1.76\times10^{-1}$ & $-8.38\times10^{+0}$ & $\phantom{+}4.36\times10^{-2}$ & $\phantom{+}3.78\times10^{+1}$ & $-5.31\times10^{-1}$ \\
5 & $\phantom{+}1.46\times10^{+1}$ & $\phantom{+}1.73\times10^{-1}$ & $\phantom{+}4.01\times10^{+0}$ & $\phantom{+}5.24\times10^{-2}$ & $\phantom{+}1.85\times10^{+1}$ & $\phantom{+}1.06\times10^{-1}$
\end{tabular}
\end{ruledtabular}
\end{table}

\begin{table}[h]
\caption{
Jet TF parameters $a_i$ and $b_i$ for Run IIb1 and $0.4<|\eta|\leq0.8$. 
}
\centering
\begin{ruledtabular}
\begin{tabular}{lcccccc}
 & \multicolumn{2}{c}{Light jets} & \multicolumn{2}{c}{$b$ jets} & \multicolumn{2}{c}{$b_\mu$ jets} \\
$i$~~ & $a_{i}$ & $b_{i}$ & $a_{i}$ & $b_{i}$ & $a_{i}$ & $b_{i}$ \\
\hline
1 & $-1.65\times10^{+0}$ & $\phantom{+}9.52\times10^{-3}$ & $\phantom{+}2.67\times10^{+0}$ & $-2.07\times10^{-1}$ & $\phantom{+}5.00\times10^{+0}$ & $-1.15\times10^{-1}$ \\
2 & $\phantom{+}3.79\times10^{+0}$ & $\phantom{+}7.95\times10^{-2}$ & $\phantom{+}2.79\times10^{+0}$ & $\phantom{+}1.82\times10^{-1}$ & $\phantom{+}1.33\times10^{+0}$ & $\phantom{+}1.46\times10^{-1}$ \\
3 & $0$ & $\phantom{+}2.71\times10^{-4}$ & $0$ & $\phantom{+}3.65\times10^{-2}$ & $0$ & $\phantom{+}4.99\times10^{-4}$ \\
4 & $\phantom{+}1.61\times10^{+1}$ & $-2.06\times10^{-1}$ & $-7.79\times10^{+0}$ & $\phantom{+}3.40\times10^{-2}$ & $\phantom{+}4.00\times10^{+1}$ & $-8.00\times10^{-1}$ \\
5 & $\phantom{+}1.47\times10^{+1}$ & $\phantom{+}1.72\times10^{-1}$ & $\phantom{+}3.42\times10^{+0}$ & $\phantom{+}6.41\times10^{-2}$ & $0$ & $\phantom{+}4.00\times10^{-1}$
\end{tabular}
\end{ruledtabular}
\end{table}

\begin{table}[h]
\caption{
Jet TF parameters $a_i$ and $b_i$ for Run IIb1 and $0.8<|\eta|\leq1.6$. 
}
\centering
\begin{ruledtabular}
\begin{tabular}{lcccccc}
 & \multicolumn{2}{c}{Light jets} & \multicolumn{2}{c}{$b$ jets} & \multicolumn{2}{c}{$b_\mu$ jets} \\
$i$~~ & $a_{i}$ & $b_{i}$ & $a_{i}$ & $b_{i}$ & $a_{i}$ & $b_{i}$ \\
\hline
1 & $\phantom{+}6.21\times10^{+0}$ & $-1.65\times10^{-1}$ & $\phantom{+}7.16\times10^{+0}$ & $-2.28\times10^{-1}$ & $\phantom{+}8.00\times10^{+0}$ & $-1.33\times10^{-1}$ \\
2 & $\phantom{+}8.73\times10^{-1}$ & $\phantom{+}1.57\times10^{-1}$ & $\phantom{+}4.31\times10^{+0}$ & $\phantom{+}1.78\times10^{-1}$ & $\phantom{+}6.04\times10^{+0}$ & $\phantom{+}1.08\times10^{-1}$ \\
3 & $0$ & $\phantom{+}1.77\times10^{-2}$ & $0$ & $\phantom{+}2.11\times10^{-2}$ & $0$ & $\phantom{+}1.00\times10^{-3}$ \\
4 & $\phantom{+}2.57\times10^{+0}$ & $\phantom{+}2.67\times10^{-2}$ & $-9.69\times10^{+0}$ & $\phantom{+}9.76\times10^{-3}$ & $\phantom{+}6.00\times10^{+1}$ & $-7.50\times10^{-1}$ \\
5 & $\phantom{+}1.01\times10^{+1}$ & $\phantom{+}6.36\times10^{-2}$ & $\phantom{+}3.41\times10^{+0}$ & $\phantom{+}9.95\times10^{-2}$ & $\phantom{+}1.00\times10^{+1}$ & $\phantom{+}2.00\times10^{-1}$
\end{tabular}
\end{ruledtabular}
\end{table}

\begin{table}[h]
\caption{
Jet TF parameters $a_i$ and $b_i$ for Run IIb1 and $1.6<|\eta|\leq2.5$. 
}
\centering
\begin{ruledtabular}
\begin{tabular}{lcccccc}
 & \multicolumn{2}{c}{Light jets} & \multicolumn{2}{c}{$b$ jets} & \multicolumn{2}{c}{$b_\mu$ jets} \\
$i$~~ & $a_{i}$ & $b_{i}$ & $a_{i}$ & $b_{i}$ & $a_{i}$ & $b_{i}$ \\
\hline
1 & $\phantom{+}5.79\times10^{+0}$ & $-1.07\times10^{-1}$ & $\phantom{+}1.69\times10^{+0}$ & $-2.27\times10^{-1}$ & $\phantom{+}2.80\times10^{+1}$ & $-3.99\times10^{-1}$ \\
2 & $\phantom{+}2.98\times10^{+0}$ & $\phantom{+}1.40\times10^{-1}$ & $\phantom{+}3.09\times10^{+0}$ & $\phantom{+}1.27\times10^{-1}$ & $\phantom{+}7.99\times10^{+0}$ & $\phantom{+}7.81\times10^{-2}$ \\
3 & $0$ & $\phantom{+}2.61\times10^{-3}$ & $0$ & $\phantom{+}5.69\times10^{-3}$ & $0$ & $\phantom{+}7.00\times10^{-3}$ \\
4 & $\phantom{+}9.11\times10^{+0}$ & $\phantom{+}4.16\times10^{-2}$ & $-1.08\times10^{+1}$ & $\phantom{+}2.90\times10^{-2}$ & $\phantom{+}9.46\times10^{+0}$ & $-9.70\times10^{-2}$ \\
5 & $\phantom{+}1.72\times10^{+1}$ & $\phantom{+}5.85\times10^{-2}$ & $\phantom{+}1.33\times10^{+1}$ & $\phantom{+}8.71\times10^{-2}$ & $\phantom{+}8.78\times10^{+0}$ & $\phantom{+}1.13\times10^{-1}$
\end{tabular}
\end{ruledtabular}
\end{table}

\afterpage{\clearpage}


\begin{table}[h]
\caption{
\label{tab:aibiIIb2}
Jet TF parameters $a_i$ and $b_i$ for Run IIb2 and $|\eta|\leq0.4$. 
}
\centering
\begin{ruledtabular}
\begin{tabular}{lcccccc}
 & \multicolumn{2}{c}{Light jets} & \multicolumn{2}{c}{$b$ jets} & \multicolumn{2}{c}{$b_\mu$ jets} \\
$i$~~ & $a_{i}$ & $b_{i}$ & $a_{i}$ & $b_{i}$ & $a_{i}$ & $b_{i}$ \\
\hline
1 & $-2.19\times10^{+0}$ & $\phantom{+}2.49\times10^{-2}$ & $\phantom{+}1.22\times10^{+0}$ & $-1.92\times10^{-1}$ & $\phantom{+}1.60\times10^{+0}$ & $-4.93\times10^{-2}$ \\
2 & $\phantom{+}4.23\times10^{+0}$ & $\phantom{+}6.91\times10^{-2}$ & $\phantom{+}2.55\times10^{+0}$ & $\phantom{+}1.87\times10^{-1}$ & $\phantom{+}3.66\times10^{+0}$ & $\phantom{+}8.15\times10^{-2}$ \\
3 & $0$ & $\phantom{+}2.63\times10^{-4}$ & $0$ & $\phantom{+}4.37\times10^{-2}$ & $0$ & $\phantom{+}7.81\times10^{-3}$ \\
4 & $\phantom{+}1.41\times10^{+1}$ & $-1.81\times10^{-1}$ & $-8.57\times10^{+0}$ & $\phantom{+}3.83\times10^{-2}$ & $\phantom{+}7.48\times10^{+0}$ & $-1.91\times10^{-1}$ \\
5 & $\phantom{+}1.48\times10^{+1}$ & $\phantom{+}1.80\times10^{-1}$ & $\phantom{+}3.85\times10^{+0}$ & $\phantom{+}5.67\times10^{-2}$ & $\phantom{+}4.56\times10^{+0}$ & $\phantom{+}1.59\times10^{-1}$
\end{tabular}
\end{ruledtabular}
\end{table}

\begin{table}[h]
\caption{
Jet TF parameters $a_i$ and $b_i$ for Run IIb2 and $0.4<|\eta|\leq0.8$. 
}
\centering
\begin{ruledtabular}
\begin{tabular}{lcccccc}
 & \multicolumn{2}{c}{Light jets} & \multicolumn{2}{c}{$b$ jets} & \multicolumn{2}{c}{$b_\mu$ jets} \\
$i$~~ & $a_{i}$ & $b_{i}$ & $a_{i}$ & $b_{i}$ & $a_{i}$ & $b_{i}$ \\
\hline
1 & $-1.68\times10^{+0}$ & $\phantom{+}1.37\times10^{-2}$ & $\phantom{+}2.19\times10^{+0}$ & $-2.02\times10^{-1}$ & $\phantom{+}6.00\times10^{+0}$ & $-1.46\times10^{-1}$ \\
2 & $\phantom{+}4.12\times10^{+0}$ & $\phantom{+}7.83\times10^{-2}$ & $\phantom{+}2.53\times10^{+0}$ & $\phantom{+}1.86\times10^{-1}$ & $\phantom{+}1.38\times10^{+0}$ & $\phantom{+}1.60\times10^{-1}$ \\
3 & $0$ & $\phantom{+}2.80\times10^{-4}$ & $0$ & $\phantom{+}3.43\times10^{-2}$ & $0$ & $\phantom{+}1.50\times10^{-4}$ \\
4 & $\phantom{+}1.68\times10^{+1}$ & $-2.16\times10^{-1}$ & $-8.54\times10^{+0}$ & $\phantom{+}3.46\times10^{-2}$ & $\phantom{+}3.75\times10^{+1}$ & $-3.59\times10^{-1}$ \\
5 & $\phantom{+}1.56\times10^{+1}$ & $\phantom{+}1.66\times10^{-1}$ & $\phantom{+}3.59\times10^{+0}$ & $\phantom{+}6.47\times10^{-2}$ & $\phantom{+}1.59\times10^{+1}$ & $\phantom{+}1.87\times10^{-1}$
\end{tabular}
\end{ruledtabular}
\end{table}

\begin{table}[h]
\caption{
Jet TF parameters $a_i$ and $b_i$ for Run IIb2 and $0.8<|\eta|\leq1.6$. 
}
\centering
\begin{ruledtabular}
\begin{tabular}{lcccccc}
 & \multicolumn{2}{c}{Light jets} & \multicolumn{2}{c}{$b$ jets} & \multicolumn{2}{c}{$b_\mu$ jets} \\
$i$~~ & $a_{i}$ & $b_{i}$ & $a_{i}$ & $b_{i}$ & $a_{i}$ & $b_{i}$ \\
\hline
1 & $\phantom{+}2.20\times10^{+0}$ & $-2.52\times10^{-2}$ & $\phantom{+}6.75\times10^{+0}$ & $-2.27\times10^{-1}$ & $\phantom{+}8.00\times10^{+0}$ & $-1.39\times10^{-1}$ \\
2 & $\phantom{+}5.77\times10^{+0}$ & $\phantom{+}1.07\times10^{-1}$ & $\phantom{+}4.14\times10^{+0}$ & $\phantom{+}1.78\times10^{-1}$ & $\phantom{+}5.54\times10^{+0}$ & $\phantom{+}1.28\times10^{-1}$ \\
3 & $0$ & $\phantom{+}3.57\times10^{-4}$ & $0$ & $\phantom{+}1.93\times10^{-2}$ & $0$ & $\phantom{+}1.43\times10^{-4}$ \\
4 & $\phantom{+}2.16\times10^{+1}$ & $-1.23\times10^{-1}$ & $-1.09\times10^{+1}$ & $\phantom{+}1.23\times10^{-2}$ & $-4.00\times10^{+1}$ & $-2.22\times10^{-1}$ \\
5 & $\phantom{+}1.91\times10^{+1}$ & $\phantom{+}1.19\times10^{-1}$ & $\phantom{+}3.25\times10^{+0}$ & $\phantom{+}1.03\times10^{-1}$ & $\phantom{+}4.10\times10^{+1}$ & $\phantom{+}2.87\times10^{-1}$
\end{tabular}
\end{ruledtabular}
\end{table}

\begin{table}[h]
\caption{
Jet TF parameters $a_i$ and $b_i$ for Run IIb2 and $1.6<|\eta|\leq2.5$. 
}
\centering
\begin{ruledtabular}
\begin{tabular}{lcccccc}
 & \multicolumn{2}{c}{Light jets} & \multicolumn{2}{c}{$b$ jets} & \multicolumn{2}{c}{$b_\mu$ jets} \\
$i$~~ & $a_{i}$ & $b_{i}$ & $a_{i}$ & $b_{i}$ & $a_{i}$ & $b_{i}$ \\
\hline
1 & $\phantom{+}7.33\times10^{+0}$ & $-1.41\times10^{-1}$ & $\phantom{+}2.55\times10^{+0}$ & $-2.48\times10^{-1}$ & $\phantom{+}9.25\times10^{+0}$ & $-1.81\times10^{-1}$ \\
2 & $\phantom{+}3.16\times10^{+0}$ & $\phantom{+}1.35\times10^{-1}$ & $\phantom{+}4.56\times10^{+0}$ & $\phantom{+}1.13\times10^{-1}$ & $\phantom{+}3.35\times10^{+0}$ & $\phantom{+}1.53\times10^{-1}$ \\
3 & $0$ & $\phantom{+}3.51\times10^{-3}$ & $0$ & $\phantom{+}5.90\times10^{-3}$ & $0$ & $\phantom{+}1.68\times10^{-4}$ \\
4 & $\phantom{+}5.54\times10^{+0}$ & $\phantom{+}4.50\times10^{-2}$ & $-1.21\times10^{+1}$ & $\phantom{+}2.30\times10^{-2}$ & $\phantom{+}2.40\times10^{+1}$ & $\phantom{+}2.36\times10^{-1}$ \\
5 & $\phantom{+}1.60\times10^{+1}$ & $\phantom{+}6.68\times10^{-2}$ & $\phantom{+}1.32\times10^{+1}$ & $\phantom{+}8.86\times10^{-2}$ & $\phantom{+}3.55\times10^{+1}$ & $-1.18\times10^{-1}$
\end{tabular}
\end{ruledtabular}
\end{table}



\begin{table}[h]
\caption{
\label{tab:aibiIIb3}
Jet TF parameters $a_i$ and $b_i$ for Run IIb3 and $|\eta|\leq0.4$. 
}
\centering
\begin{ruledtabular}
\begin{tabular}{lcccccc}
 & \multicolumn{2}{c}{Light jets} & \multicolumn{2}{c}{$b$ jets} & \multicolumn{2}{c}{$b_\mu$ jets} \\
$i$~~ & $a_{i}$ & $b_{i}$ & $a_{i}$ & $b_{i}$ & $a_{i}$ & $b_{i}$ \\
\hline
1 & $-1.87\times10^{+0}$ & $\phantom{+}1.94\times10^{-2}$ & $\phantom{+}4.64\times10^{+0}$ & $-2.02\times10^{-1}$ & $\phantom{+}6.72\times10^{+0}$ & $-1.39\times10^{-1}$ \\
2 & $\phantom{+}4.11\times10^{+0}$ & $\phantom{+}7.20\times10^{-2}$ & $\phantom{+}2.78\times10^{+0}$ & $\phantom{+}2.32\times10^{-1}$ & $\phantom{+}1.56\times10^{+0}$ & $\phantom{+}1.53\times10^{-1}$ \\
3 & $0$ & $\phantom{+}3.39\times10^{-4}$ & $0$ & $\phantom{+}7.93\times10^{-2}$ & $0$ & $\phantom{+}2.60\times10^{-4}$ \\
4 & $\phantom{+}1.70\times10^{+1}$ & $-1.47\times10^{-1}$ & $-6.31\times10^{+0}$ & $\phantom{+}1.41\times10^{-2}$ & $\phantom{+}3.27\times10^{+1}$ & $-3.09\times10^{-1}$ \\
5 & $\phantom{+}1.63\times10^{+1}$ & $\phantom{+}1.60\times10^{-1}$ & $\phantom{+}3.56\times10^{+0}$ & $\phantom{+}6.45\times10^{-2}$ & $\phantom{+}1.32\times10^{+1}$ & $\phantom{+}1.98\times10^{-1}$
\end{tabular}
\end{ruledtabular}
\end{table}

\begin{table}[h]
\caption{
Jet TF parameters $a_i$ and $b_i$ for Run IIb3 and $0.4<|\eta|\leq0.8$. 
}
\centering
\begin{ruledtabular}
\begin{tabular}{lcccccc}
 & \multicolumn{2}{c}{Light jets} & \multicolumn{2}{c}{$b$ jets} & \multicolumn{2}{c}{$b_\mu$ jets} \\
$i$~~ & $a_{i}$ & $b_{i}$ & $a_{i}$ & $b_{i}$ & $a_{i}$ & $b_{i}$ \\
\hline
1 & $-1.70\times10^{+0}$ & $1.67\times10^{-2}$ & $4.22\times10^{+0}$ & $-2.03\times10^{-1}$ & $6.70\times10^{+0}$ & $-1.44\times10^{-1}$ \\
2 & $\phantom{+}4.17\times10^{+0}$ & $\phantom{+}7.83\times10^{-2}$ & $\phantom{+}2.91\times10^{+0}$ & $\phantom{+}2.08\times10^{-1}$ & $\phantom{+}1.64\times10^{+0}$ & $\phantom{+}1.56\times10^{-1}$ \\
3 & $0$ & $\phantom{+}3.12\times10^{-4}$ & $0$ & $\phantom{+}4.86\times10^{-2}$ & $0$ & $\phantom{+}1.98\times10^{-4}$ \\
4 & $\phantom{+}1.89\times10^{+1}$ & $-1.77\times10^{-1}$ & $-7.71\times10^{+0}$ & $\phantom{+}2.99\times10^{-2}$ & $\phantom{+}3.68\times10^{+1}$ & $-2.84\times10^{-1}$ \\
5 & $\phantom{+}1.77\times10^{+1}$ & $\phantom{+}1.55\times10^{-1}$ & $\phantom{+}3.43\times10^{+0}$ & $\phantom{+}6.77\times10^{-2}$ & $\phantom{+}1.66\times10^{+1}$ & $\phantom{+}1.76\times10^{-1}$
\end{tabular}
\end{ruledtabular}
\end{table}

\clearpage

\begin{table}[h]
\caption{
Jet TF parameters $a_i$ and $b_i$ for Run IIb3 and $0.8<|\eta|\leq1.6$. 
}
\centering
\begin{ruledtabular}
\begin{tabular}{lcccccc}
 & \multicolumn{2}{c}{Light jets} & \multicolumn{2}{c}{$b$ jets} & \multicolumn{2}{c}{$b_\mu$ jets} \\
$i$~~ & $a_{i}$ & $b_{i}$ & $a_{i}$ & $b_{i}$ & $a_{i}$ & $b_{i}$ \\
\hline
1 & $\phantom{+}1.56\times10^{+0}$ & $-1.84\times10^{-2}$ & $\phantom{+}1.02\times10^{+1}$ & $-2.61\times10^{-1}$ & $\phantom{+}8.00\times10^{+0}$ & $-1.33\times10^{-1}$ \\
2 & $\phantom{+}5.96\times10^{+0}$ & $\phantom{+}1.08\times10^{-1}$ & $\phantom{+}4.92\times10^{+0}$ & $\phantom{+}1.91\times10^{-1}$ & $\phantom{+}6.04\times10^{+0}$ & $\phantom{+}1.08\times10^{-1}$ \\
3 & $0$ & $\phantom{+}3.87\times10^{-4}$ & $0$ & $\phantom{+}3.34\times10^{-2}$ & $0$ & $\phantom{+}1.00\times10^{-3}$ \\
4 & $\phantom{+}2.12\times10^{+1}$ & $-9.65\times10^{-2}$ & $-9.51\times10^{+0}$ & $-5.50\times10^{-3}$ & $\phantom{+}6.00\times10^{+1}$ & $-7.50\times10^{-1}$ \\
5 & $\phantom{+}2.06\times10^{+1}$ & $\phantom{+}1.08\times10^{-1}$ & $\phantom{+}3.00\times10^{+0}$ & $\phantom{+}1.16\times10^{-1}$ & $\phantom{+}1.00\times10^{+1}$ & $\phantom{+}2.00\times10^{-1}$
\end{tabular}
\end{ruledtabular}
\end{table}

\hbox{}

\begin{table}[h]
\caption{
Jet TF parameters $a_i$ and $b_i$ for Run IIb3 and $1.6<|\eta|\leq2.5$. 
}
\centering
\begin{ruledtabular}
\begin{tabular}{lcccccc}
 & \multicolumn{2}{c}{Light jets} & \multicolumn{2}{c}{$b$ jets} & \multicolumn{2}{c}{$b_\mu$ jets} \\
$i$~~ & $a_{i}$ & $b_{i}$ & $a_{i}$ & $b_{i}$ & $a_{i}$ & $b_{i}$ \\
\hline
1 & $\phantom{+}6.59\times10^{+0}$ & $-1.06\times10^{-1}$ & $-3.29\times10^{-1}$ & $-1.80\times10^{-1}$ & $\phantom{+}2.35\times10^{+1}$ & $-3.25\times10^{-1}$ \\
2 & $\phantom{+}2.77\times10^{+0}$ & $\phantom{+}1.37\times10^{-1}$ & $\phantom{+}9.64\times10^{-1}$ & $\phantom{+}1.47\times10^{-1}$ & $\phantom{+}6.73\times10^{+0}$ & $\phantom{+}1.00\times10^{-1}$ \\
3 & $0$ & $\phantom{+}2.01\times10^{-3}$ & $0$ & $\phantom{+}3.62\times10^{-3}$ & $0$ & $\phantom{+}3.65\times10^{-3}$ \\
4 & $\phantom{+}1.26\times10^{+1}$ & $\phantom{+}2.99\times10^{-2}$ & $-6.32\times10^{+0}$ & $\phantom{+}1.59\times10^{-2}$ & $\phantom{+}9.75\times10^{+0}$ & $-6.83\times10^{-2}$ \\
5 & $\phantom{+}1.78\times10^{+1}$ & $\phantom{+}5.45\times10^{-2}$ & $\phantom{+}1.47\times10^{+1}$ & $\phantom{+}7.99\times10^{-2}$ & $\phantom{+}1.29\times10^{+1}$ & $\phantom{+}9.45\times10^{-2}$
\end{tabular}
\end{ruledtabular}
\end{table}

\clearpage

\end{appendix}

\end{document}